\documentclass[11pt,a4paper]{article}
\usepackage{jheppub}

\usepackage{amssymb,amsmath,amsthm,graphics,graphicx}
\usepackage{hyperref} 
\usepackage{latexsym}
\usepackage{bm}
\usepackage{braket}
\usepackage{color}
    \definecolor{darkgreen}{rgb}{0,0.5,0}
    \definecolor{darkred}{rgb}{0.5,0,0}
    \definecolor{darkblue}{rgb}{0,0,0.6}
    \definecolor{purple}{rgb}{0.4,.2,0.7}
    
\allowdisplaybreaks

\def\be{\begin{equation}}
\def\ee{\end{equation}}
\def\beq{\begin{eqnarray}}
\def\eeq{\end{eqnarray}}

\newcommand{\eg}{{\it e.g.,}\ }
\newcommand{\ie}{{\it i.e.\,}\ }

\def\lsim{\mathrel{\rlap{\lower4pt\hbox{\hskip1pt$\sim$}}
    \raise1pt\hbox{$<$}}}
\def\gsim{\mathrel{\rlap{\lower4pt\hbox{\hskip1pt$\sim$}}
    \raise1pt\hbox{$>$}}}
\def\be{\begin{equation}}
\def\ee{\end{equation}}
\def\bea{\begin{eqnarray}}
\def\eea{\end{eqnarray}}
\newcommand{\dd}{\mathrm{d}}
\newcommand{\LL}{\mathcal{L}}
\newcommand{\DD}{\mathcal{D}}

\begin{document}





\title{Eigenvalue repulsions and quasinormal mode spectra of Kerr-Newman: an extended study}
\author[a]{\'Oscar J.~C.~Dias,}
\author[b]{Mahdi~Godazgar,}
\author[c]{Jorge~E.~Santos}

\affiliation[a]{STAG research centre and Mathematical Sciences, Highfield Campus, University of Southampton, Southampton SO17 1BJ, UK}
\affiliation[b]{School of Mathematical Sciences, Queen Mary University of London, Mile End Road, London E1 4NS, UK.}
\affiliation[c]{DAMTP, University of Cambridge, Wilberforce Road, Cambridge CB3 0WA, UK}

\emailAdd{ojcd1r13@soton.ac.uk}
\emailAdd{m.godazgar@qmul.ac.uk}
\emailAdd{jss55@cam.ac.uk}

\abstract{
The frequency spectra of the gravito-electromagnetic perturbations of the Kerr-Newman (KN) black hole with the slowest decay rate have been computed recently. It has been found that KN has two families $-$  the photon sphere and the near-horizon families $-$ of quasinormal modes (QNMs), which display the interesting phenomenon of eigenvalue repulsion.
 The perturbation equations, in spite of being a coupled system of two PDEs, are amenable to an analytic solution using the method of separation of variables in a near-horizon expansion around the extremal KN black hole. This leads to an analytical formula for the QNM frequencies that provides an excellent approximation to the numerical data near-extremality. In the present manuscript we provide an extended study of these properties that were not detailed in the original studies. This  includes: 1) a full derivation of a gauge invariant system of two coupled PDEs that describes the perturbation equations \cite{Dias:2015wqa}, 2) a derivation of the eikonal frequency approximation \cite{Zimmerman:2015trm,Dias:2021yju} and its comparison with the numerical QNM data, 3) a derivation of the near-horizon frequency approximation  \cite{Dias:2021yju}  and its comparison with the numerical QNMs,  and 4) more details on the phenomenon of eigenvalue repulsion (also known as  \emph{level repulsion},  \emph{avoided crossing} or \emph{Wigner-Teller effect}) and a first principles understanding of it that was missing in the previous studies. Moreover, we provide  the frequency spectra of other KN QNM families of interest to demonstrate that they are more damped than the ones we discuss in full detail.  
}

\maketitle


\section{Introduction}

When a black hole is (moderately) perturbed, it typically relaxes back to equilibrium by emitting gravitational waves with damped characteristic frequencies $-$ the quasinormal mode (QNM) frequencies $-$ that depend on the conserved charges of the black hole. It follows that these QNM frequencies may be used to determine the mass and angular momentum of a black hole. In fact, this is one way of measuring the mass and angular momentum of the final black hole \cite{LIGOScientific:2021sio} that emerges from the black hole binary coalescences observed in gravitational wave detector experiments  \cite{PhysRevLett.116.061102,LIGO,Virgo,LIGOScientific:2020tif,LIGOScientific:2020ibl,LIGOScientific:2021djp}.

Astrophysical black holes are expected to be described by Einstein gravity; more specifically, by its Kerr solution parametrized by the mass $M$ and angular momentum $J\equiv M a$ (where $a$ is the rotation parameter) \cite{Kerr:1963ud}. Therefore, all LIGO-Virgo~\cite{LIGO,Virgo} observations of events compatible with black hole binaries~\cite{LIGOScientific:2020tif,LIGOScientific:2020ibl,LIGOScientific:2021djp} have been described so far 
mainly under the working assumption that the coalescing objects can be modelled by the Kerr solution or parametrically small deviations thereof \cite{LIGOScientific:2021sio}.   
However, to discuss the physical interpretation of the observed data, we might also want to consider black hole solutions of the Einstein-Maxwell theory that have an electric charge $Q$, in addition to $M$ and $J$.\footnote{For recent theoretical studies discussing  black hole binary coalescence of charged rotating black holes see \cite{Pina:2022dye,Zi:2022hcc}.} In this case, the uniqueness theorems \cite{Robinson:2004zz,Chrusciel:2012jk} guarantee that the Kerr-Newman black hole (KN BH) \cite{Newman:1965my,Adamo:2014baa} parametrized by $M$, $J$ and $Q$ is the unique, most general, analytic, stationary asymptotically flat electro-vacuum black hole of Einstein-Maxwell theory. The Kerr~\cite{PhysRevLett.116.061102}, Reissner-Nordstr\"om (RN)~\cite{1916AnP...355..106R,1918KNAB...20.1238N} and Schwarzschild~\cite{Schwarzschild:1916uq} black holes are then viewed as limiting cases of KN with $Q=0$, $a=0$ and $Q=a=0$, respectively.

Although astrophysical black holes are expected to quickly lose any electric charge that they may have~\cite{Gibbons, Znajek}, one should nevertheless study the properties of KN black holes and compute their quasinormal mode frequencies.
With this theoretical information at hand, we will be better equipped to analyse and interpret observational data to unequivocally establish that the observed system has no charge (or even to compute its charge in the lucky but unlikely event of observing a system during the short timescale where the discharge has not yet occurred).  
Furthermore, the QNM spectra of KN might be of interest for other interpretations of observational data and for applications in both ground and space-based gravitational wave  detectors~\cite{LIGO,Virgo,Kagra,ET,CE,LISA}. For example, it can be used to model  gravitational wave emission~\cite{AstroConstraints}, and it might even be useful  for  constraining some dark matter models~\cite{2016JCAP...05..054C} and modified gravity models~\cite{Bozzola:2020mjx}.
For these reasons, in this manuscript we conclude a series of papers, started in \cite{Dias:2015wqa,Dias:2021yju}, that compute the main families of QNMs of the KN BH and identify their key properties. 

The QNM spectra of Schwarzschild, RN and Kerr black holes were determined many decades ago \cite{Regge:1957td,Zerilli:1974ai,Moncrief:1974am,Chandrasekhar:1975zza,Moncrief:1974gw,Moncrief:1974ng,Newman:1961qr,Geroch:1973am,Teukolsky:1972my,Detweiler:1980gk,Chandra:1983,Leaver:1985ax,Whiting:1988vc,Onozawa:1996ux,Berti:2005eb,Berti:2003jh,Yang:2012pj} (see review \cite{Berti:2009kk}).
This was possible at a relatively small computational cost because for these black holes the QNM spectrum turns out (remarkably) to be encoded in a single separable equation that effectively yields a pair of angular and radial ODEs that one can solve as an eigenvalue problem. For Schwarzschild and RN black holes this is known as the (odd mode) Regge-Wheeler and (even mode) Zerilli equations  \cite{Regge:1957td,Zerilli:1974ai,Moncrief:1974am}, while for the Kerr black hole this is known as the Teukolsky equation \cite{Teukolsky:1972my}. The existence of such a simplification allows one to find the QNM spectra,  and in doing so, to establish evidence in favour of the linear mode stability of these solutions and to ultimately motivate a formal proof of the linear mode stability of the Kerr solution \cite{Whiting1989}.\footnote{\label{footNLkerr}Even though the nonlinear stability of Kerr remains an open problem (see \cite{Dafermos:2010hb,Dafermos:2014cua,Dafermos:2014jwa, Dafermos:2017yrz, Dafermos:2016uzj, Ma:2017bxq, Ma:2020lqj, Klainerman:2017nrb, Dafermos:2021cbw} for recent progress), it is also believed to be stable beyond the linear level.}

The state of affairs is very different in the Kerr-Newman case. Generic gravito-electromagnetic perturbations of KN are no longer described by a single separable equation. Thus, initial hints about the QNM spectra of KN were obtained only within perturbation theory about the RN or Kerr black holes: perturbative results in the small rotation parameter $a$ about RN were discussed in \cite{Pani:2013ija,Pani:2013wsa}, and perturbative results in the small charge parameter $Q$ around Kerr were computed in \cite{Mark:2014aja}.

To make further progress and compute the KN QNM spectrum for generic $Q$ and $J$, one must solve the perturbed Einstein-Maxwell equation which is a coupled partial differential equation (PDE) system. Na\"ively, one expects to find a system of nine coupled PDEs. However, working in the so-called phantom gauge, Chandrasekhar reduced the problem to the study of `just' two coupled PDEs \cite{Chandra:1983} (see also \cite{Mark:2014aja}). Despite this significant progress, finding the QNM spectrum and addressing the problem of the linear mode stability of the KN BH has remained an open problem for several decades. Further progress was made in \cite{Dias:2015wqa} where it was shown that generic gravito-electromagnetic perturbations of KN (except for those that change the mass and angular momentum of the solution) are described by a coupled system of two PDEs for two gauge invariant Newman-Penrose (NP) fields. Upon gauge fixing, these reduce to the coupled PDE system originally found by Chandrasekhar~\cite{Chandra:1983,Mark:2014aja}. 
Moreover, in \cite{Dias:2015wqa} a numerical search of KN modes was finally performed in regions of the KN parameter space that could be more prone to developing an {\it instability}, finding none and thus providing evidence for the linear mode stability of KN (further supported by the non-linear time evolution study of~\cite{Zilhao:2014wqa}). 
More recently, in \cite{Dias:2021yju}, the numerical code of \cite{Dias:2015wqa} was made computationally more efficient and extended to compute the frequency spectra, across the {\it full} KN 2-parameter space, of the most dominant (\ie with slowest decay rate) gravito-electromagnetic QNM family. These are the modes that reduce --- in Chandrasekhar's notation~\cite{Chandra:1983} --- to the $Z_2$ (i.e.\ gravitational), $\ell=m=2$, $n=0$ modes 
in the Schwarzschild limit ($a=Q=0$), where the harmonic number $\ell$ gives the number of zeros of the eigenfunction along the polar direction and $n$ is the radial overtone. 
In the process, \cite{Dias:2021yju} found that KN has not one but two main families of $Z_2$ $\ell=m=2$ QNMs which were coined the {\it photon sphere} ($\mathrm{PS}$), and the {\it near-horizon} ($\mathrm{NH}$) families, although the sharp distinction between the PS and NH modes is unambiguous only for small rotation $a$, i.e., when the KN  black hole is close to the Reissner-Nordstr\"om family. Quite remarkably, \cite{Dias:2021yju} further found that as we evolve along the KN parameter space, the imaginary part of the frequency of these two PS and NH families intersect each other (however, the real part of the frequency is very similar for the PS and NH modes and, typically, does not display crossings). Sometimes this intersection of the imaginary part of the frequencies is a simple crossover where the modes simply trade dominance but, other times this interaction is much more intricate and displays a behaviour that suggests repulsions between the PS and NH modes. These ``eigenvalue repulsions" were unexpected since they are not observed in the QNM spectra of neither Kerr nor Reissner-Nordstr\"om.\footnote{\label{ft:ds} More recently, eigenvalue repulsions were also found in rotating de Sitter black holes where, besides the PS and NH modes, one has a third QNM family associated to the cosmological constant \cite{Davey:2022vyx}. With hindsight, they are also observed in the de Sitter Reissner-Nordstr\"om black hole study of \cite{Dias:2020ncd}.} As a result of these repulsions,  well away from the RN limit of the KN solution,  the PS and NH families lose their individual identities and instead combine to yield what is more appropriately described as a PS$-$NH family of QNMs and its radial overtones. 

In the current manuscript we complement and complete the studies of \cite{Dias:2015wqa,Dias:2021yju} in five main ways:

\begin{enumerate}
\item We use the Newman-Penrose (NP) formalism to derive the aforementioned coupled system of two PDEs for two gauge invariant NP variables, first presented in \cite{Dias:2015wqa}, that describes the most general gravito-electromagnetic perturbations of KN (except for those that change the mass and angular momentum of the solution)  and that reduces, upon gauge fixing, to the Chandrasekhar PDE system~\cite{Chandra:1983,Mark:2014aja}. This derivation was only very briefly sketched in \cite{Dias:2015wqa} but we now give a detailed derivation of it in Section~\ref{sec:Pert}. We also take the opportunity to revisit a simple proof of isospectrality of the Schwarzschild and RN QNM spectra \cite{Dias:2015wqa}.

\item We can envisage solving the perturbation equations for the two gauge invariant NP fields in a WKB analysis at large $|m|=\ell\gg 1$. Similar to the Schwarzschild and Kerr cases, the leading order contribution of this analysis, known as the eikonal or geometric optics limit $|m|=\ell\to \infty$, is expected to be closely connected to the properties of unstable null circular orbits revolving around the KN black hole. In Section~\ref{sec:PSeikonal} we will compare this eikonal result with the numerical data for photon sphere modes to conclude that the eikonal frequency indeed provides a relatively good approximation to the PS frequencies that gets better as $m$ grows.     

\item There is a second class of QNMs that have eigenfunctions that, near-extremality, are very localized around the event horizon and quickly decay to zero away from the horizon. These are the near-horizon modes or the PS$-$NH modes that were already mentioned above. This suggests doing a `{\it poor-man's}' matched asymptotic expansion (MAE) whereby we take the {\it near-horizon limit} of the perturbed equations to find the near-region solution and match with a {\it vanishing} far-region wavefunction in the overlapping region where both solutions are valid. Remarkably, this can be done because the perturbation equations, in spite of being a coupled system of two PDEs, can be solved analytically in the near-horizon region around the extremal (zero temperature) KN black hole using the {\it method of separation of variables}. Ultimately, this is possible because the near-horizon limit of the extremal KN BH is a warped circle fibred over $AdS_2$ (Anti-de Sitter in 1+1 dimensions) and thus its perturbations can be decomposed as a sum of known radial $AdS_2$ harmonics.  The system of 2 coupled PDEs for the gauge invariant NP fields in the near-horizon region of the near-extremal KN geometry separates into a system of 2 decoupled radial ordinary differential equations (ODEs) and a coupled system of 2 angular ODEs. We can solve this near-horizon system, match it with the trivial far-region, and obtain an analytical expression for the NH and PS$-$NH frequencies. The final expression was presented in \cite{Dias:2021yju} but not the long derivation that leads to it.  We will present this detailed derivation in Section~\ref{sec:NHanalytics} and show that it provides an excellent approximation to the numerical frequencies when we are close to extremality.

\item In the Reissner-Nordstr\"om background, there are exactly two distinct sectors of QNMs: the aforementioned PS and NH families (and their radial overtones). However, as we move away from this limit in the KN parameter space we find that this clear distinction between the two families is lost and the two families and their overtones combine in an intricate way to form what is more appropriately described as PS$-$NH modes and their radial overtones. This happens because the phenomenon of {\it eigenvalue repulsion} occurs. These eigenvalue repulsions were already reported in \cite{Dias:2021yju} but in Section~\ref{sec:repulsion} we will give a detailed description of these eigenvalue repulsions in the KN QNM spectra, and we will see how the frequency gaps between different QNM families develop and evolve. No less important, we will provide a \emph{first principles} understanding of this phenomenon that was not discussed in \cite{Dias:2021yju}. For that we will start  by pointing out that eigenvalue repulsion is common in some eigenvalue problems of quantum mechanical systems where it is also known as  \emph{level repulsion},  \emph{avoided crossing} or \emph{Wigner-Teller effect} \cite{Landau1981Quantum,Cohen-Tannoudji:1977}. In Section~\ref{sec:repulsionTextbook} we will start by reviewing (following $\S 79$ of the Landau-Lifshitz textbook \cite{Landau1981Quantum}) the simplest quantum mechanical two-level system with a self-adjoint Hamiltonian that exhibits avoided crossing. We will then extend the discussion of avoided crossing to the case where the perturbed Hamiltonian of the system is not self-adjoint, as is the case with the KN QNM system. Having understood that level repulsions should be present in the QNM spectra of KN, in Section~\ref{sec:repulsionSub} we will give a detailed description of eigenvalue repulsions in the frequency spectra of KN. The analysis of Section~\ref{sec:repulsion} together with the one of Section~\ref{sec:NHanalytics} will allow us to conclude that the \emph{complex} frequencies $\omega$ of KN have level crossing (i.e.\ \emph{both} the real and imaginary parts of the PS and NH modes \emph{cross} each other) exactly at one, \emph{and only one}, point in the 2-dimensional KN parameter space (we collect strong evidence to claim that this is the point at extremality where the PS modes reach $\rm{Im} (\omega)=0$, which will be represented by a $\star$ in Fig.~\ref{Fig:PS-extremality}). In all other KN black holes we either have no crossovers of the imaginary and real parts of the frequency or the imaginary part of the PS and NH frequencies cross, but not the real part of the frequencies. These features are in agreement with the predicted properties of the eigenvalue spectra of a 2-dimensional parameter space system with avoided crossing, as explained in Section~\ref{sec:repulsionTextbook}. This analysis will also explain why avoided crossing is not observed in the 1-parameter family of Kerr solutions. Ultimately, the intricate QNM spectra of KN emerges from the fact that level crossing occurs only at one point but the system reacts to avoid crossings at other points. This leads to the observed elaborate features/repulsions when one is approaching the level crossing point $\star$ of the system.  

\item After revisiting  in Section~\ref{sec:DominantSpectra} the properties of the $Z_2$ $\ell=m=2$ KN QNMs (first presented in  \cite{Dias:2021yju}) that are expected to be the least damped ones, in Section~\ref{sec:FullSpectra} we will present the frequencies of some other relevant gravito-electromagnetic modes of KN. This will give solid, explicit, evidence that the $Z_2$ $\ell=m=2$ QNM is indeed the mode with the slowest decay rate in KN (as with the Schwarzschild, RN and Kerr black holes). 

\end{enumerate}

\section{Derivation of the gauge invariant perturbation equations for KN}\label{sec:Pert}

In subsection~\ref{sec:KN} we briefly review the Kerr-Newman black hole solution. Then, in subsection~\ref{sec:coupled}, we detail how  the Newman-Penrose (NP) formalism can be used to derive a coupled system of two PDEs for two gauge invariant NP variables \cite{Dias:2015wqa} 
 that describes the most general gravito-electromagnetic perturbations of KN (except for those that change the mass and angular momentum of the solution). Finally, in subsection~\ref{sec:BCs} we discuss the boundary conditions that allow one to solve the final eigenvalue problem to find the QNM frequencies of KN.

\subsection{KN black hole: an algebraically special Petrov type D solution}\label{sec:KN}

The KN BH solution with  mass $M$, angular momentum $J\equiv M a$ and charge $Q$ is most commonly expressed in standard Boyer-Lindquist coordinates $\{t,r,\theta,\phi\}$ (time, radial, polar, azimuthal coordinates)~\cite{Newman:1965my,Adamo:2014baa}, in which the metric takes the form
\begin{eqnarray}\label{KNsoln}
ds^2&=&-\frac{\Delta}{\Sigma} \left(\dd t-a \sin^2\theta \dd \phi  \right)^2+\frac{\Sigma }{\Delta }\,\dd r^2 + \Sigma \,\dd \theta^2 
+ \frac{\sin ^2\theta}{\Sigma }\left[\left(r^2+a^2\right)\dd \phi -a \dd t \right]^2, \nonumber\\
A&=& \frac{Q \,r}{\Sigma}\left(\dd t-a \sin^2\theta \dd \phi \right),
\end{eqnarray}
with $\Delta = r^2 -2Mr+a^2+Q^2$ and $\Sigma=r^2+a^2 \cos^2\theta$. 

Roots of the function $\Delta$, namely
\begin{equation}\label{roots}
 r_\pm=M\pm\sqrt{M^2-a^2-Q^2},
\end{equation}
correspond to the inner and outer event horizons, respectively.  Physically, one is most interested in the outer event horizon ($r=r_+$), which is a Killing horizon generated by the Killing vector
\begin{equation}
 K=\partial_t +\Omega_H \partial_\phi\,,
\end{equation}
with angular velocity $\Omega_H$ and temperature $T_H$ given by  
\begin{equation}\label{KNthermo}
 \Omega_H= \frac{a}{r_+^2+a^2} \,, \qquad 
T_H = \frac{1}{4 \pi  r_+}\frac{r_+^2-a^2-Q^2}{r_+^2+a^2 }\,,
\end{equation}
where we have used \eqref{roots} to express $M$ as a function of $r_+,a$ and $Q$.
If $r_-=r_+$, \ie  $a=a_{\hbox{\footnotesize ext}}$, the KN BH has a regular extremal (``ext") configuration with $T_H^{\hbox{\footnotesize ext}} =0$, and maximum angular velocity $\Omega_H^{\hbox{\footnotesize ext}}$
\begin{equation}\label{KNext}
 \Omega_H^{\hbox{\footnotesize ext}} =\frac{a_{\hbox{\footnotesize ext}}}{M^2+a_{\hbox{\footnotesize ext}}^2}\, \qquad a_{\hbox{\footnotesize ext}}=\sqrt{M^2-Q^2}\,.
\end{equation}

Here, we are interested in linear gravito-electromagnetic perturbations about the KN background.  Following Teukolsky \cite{Teukolsky:1972my, Teukolsky:1973ha}, we work within the Newman-Penrose (NP) formalism \cite{Newman:1961qr}.  We will not review the NP formalism here, but instead refer the reader to Chapter 7 of \cite{Stephani:2003tm} for a comprehensive review.  Suffice it to say that the NP formalism starts with a complex null frame or tetrad\footnote{There is a spinor version of the NP formalism.  However, here, we deal only with the Lorentzian version.} and uses this tetrad to transform all quantities of interest (connection coefficients, Ricci, Weyl and Maxwell field strength components) into   complex scalars.  In such a manner, the Weyl tensor, for example is transformed into a set of five complex scalars: $\Psi_a$ ($a=0,1 \cdots, 4$) or the Maxwell field strength into a set of three complex scalars: $\Phi_a$ ($a=0,1,2$) \cite{Chandra:1983,Stephani:2003tm}.  Furthermore, the existence of a NP frame in which a certain combination of the Weyl scalars vanishes determines the Petrov type of the background solution.

Teukolsky \cite{Teukolsky:1972my, Teukolsky:1973ha} showed that on an algebraically special \emph{vacuum} background, which is defined to be one in which there exists a null frame so that $\Psi_0=\Psi_1=0$, the linear perturbations of the background may be expressed in terms of a \emph{decoupled} equation
\begin{equation}
 \mathcal{O} \Psi_0^{(1)} = 0,
\end{equation}
where $\mathcal{O}$ is some linear second-order differential operator and $\Psi_0^{(1)}$ is the gauge-invariant perturbed value of $\Psi_0$.

Fortunately, the Kerr BH, which was of principal interest for Teukolsky, is algebraically special.  In fact it is Petrov type D (i.e.\ a NP frame exists such that the only non-vanishing Weyl scalar is $\Psi_2$).  Furthermore, given global and hidden \cite{Carter:1968ks} symmetries of the Kerr BH, the coordinate dependence of the perturbations \emph{separate} leading to a single ODE.  Thus, the combined simplification of decoupling and separability on the Kerr BH allows one to study its linearized mode perturbations \cite{Teukolsky:1972my, Teukolsky:1973ha,Berti:2009kk} and prove its linear mode stability \cite{Whiting:1988vc} (see also footnote~\ref{footNLkerr}). 

Like its vacuum cousin the Kerr BH, the KN BH is also Petrov type D.  In particular, in a NP null frame $\{{\bf e}_{(a)}\}=\{{\bm \ell}, {\bm n}, {\bm m},  \bar{{\bm m}}\}$ with ($a=1,2,3,4$) adapted to the principal null directions, given by
\begin{gather}
{\bm \ell} = \left(\frac{r^2+a^2}{\Delta} \right) \frac{\partial}{\partial t} 
 + \frac{\partial}{\partial r} + \frac{a}{\Delta } \frac{\partial}{\partial \phi}, \qquad
{\bm n} = \frac{1}{2 \Sigma} \left( (r^2+a^2) \frac{\partial}{\partial t} -\Delta \frac{\partial}{\partial r}+ a \frac{\partial}{\partial \phi} \right)  \notag \\
{\bm m} = \frac{i}{\sqrt{2}\, \bar{r}} \left( a \sin \theta \frac{\partial}{\partial t}-i \frac{\partial}{\partial \theta}+ \frac{1}{\sin \theta} \frac{\partial}{\partial \phi}\right),
 \label{NPframe}
\end{gather}
where $\bar{r} = r+ia\cos \theta$,~\footnote{The standard notation for the complex conjugation in the NP formalism is to use a bar.  We will stick to this notation as far as NP quantities are concerned.  However, this should not be confused with $\bar{r}$ defined here, whose complex conjugation ($\bar{r}^*$) we shall denote with a star.} the only non-zero Weyl scalar is\footnote{Recall that the 5 complex Weyl scalars $\Psi_a$ in the NP formalism encode the information in the 10 independent components $C_{\mu \nu\rho \sigma}$ of the Weyl tensor,
\begin{eqnarray} && \Psi_0 = -C_{1313} =  - C_{\mu\nu\alpha\beta}\, \ell^\mu
m^\nu
\ell^\alpha m^\beta, \qquad  \Psi_1 = -C_{1213} = - C_{\mu\nu\alpha\beta}\, \ell^\mu n^\nu \ell^\alpha
m^\beta, \nonumber \\ &&  \Psi_2 = -C_{1342} = - C_{\mu\nu\alpha\beta}\, \ell^\mu  m^\nu
\bar{m}^\alpha n^\beta, \qquad
 \Psi_3 = -C_{1242} = - C_{\mu\nu\alpha\beta}\, \ell^\mu n^\nu
\bar{m}^\alpha n^\beta, \nonumber \\
&& \Psi_4 = -C_{2424} = - C_{\mu\nu\alpha\beta}\, n^\mu \bar{m}^\nu
n^\alpha \bar{m}^\beta\,,\label{WeylScalars}
\end{eqnarray}
and the 3 complex NP scalars $\Phi_a$ encode the information in the 6 independent components of the anti-symmetric Maxwell field strength, $F=\mathrm{d} A$,
\begin{eqnarray} \label{MaxScalars}
&& \Phi_0 = F_{13} =  F_{\alpha\beta}\,  \ell^\alpha m^\beta, \  \Phi_1 = \frac{1}{2}\left(F_{12}+F_{43} \right) = \frac{1}{2}  F_{\alpha\beta} ( \ell^\alpha n^\beta + \bar{m}^\alpha
m^\beta) , \  \Phi_2 = F_{42}=F_{\alpha\beta}\,   \bar{m}^\alpha n^\beta.
\end{eqnarray}  
}
\begin{equation} \label{Weylsca}
 \Psi_2 =  \frac{Q^2- M \bar{r}}{\bar{r} {\bar{r}^*}{}^3}\,.
\end{equation}
Moreover, the only non-zero Maxwell scalar is
\begin{equation} \label{Maxsca}
 {\Phi_1} = \frac{Q}{2 {\bar{r}^*}{}^2}.
\end{equation}
However, importantly, the decoupling result of Teukolsky does not apply to the KN BH, since it is a non-vacuum solution.  In fact, such a decoupling result does not seem possible for the KN BH (see e.g.\ \cite{Chandra:1983}).  The best that can be done is to derive a gauge-invariant coupled PDE system \cite{Dias:2015wqa}, which we now derive and which reduces to the Chandrasekhar system \cite{Chandra:1983} under a particular gauge choice.  

\subsection{Derivation of the gauge invariant perturbation equations}\label{sec:coupled}

To discuss generic perturbations of the Kerr-Newman black hole one needs to find the perturbed Einstein-Maxwell equation which, {\it a priori} is a system of nine coupled PDEs. Although, a decoupling result cannot be obtained for the perturbations on the KN BH background, one can still reduce this perturbation system to a simpler set of two gauge invariant coupled PDEs \cite{Dias:2015wqa} that, after gauge fixing, reduces to the Chandrasekhar coupled system of two PDEs~\cite{Chandra:1983,Mark:2014aja}. In this section we give the details of the derivation of this system of PDEs.

The linearised perturbations on any background satisfy the linearised Einstein equation on that background.  Therefore, any perturbation equation that we derive must ultimately come from some operator acting on this linearised Einstein equation \cite{Wald:1978vm}.  

There are two common ways of deriving the Teukolsky equations: the original method relies on a particular manipulation of the NP Bianchi equations, which now comprise the non-trivial content of the Einstein equations \cite{Teukolsky:1972my}.  Another method is a more straightforward contraction of the Penrose wave equation
\begin{equation}
\Box R_{\mu \nu \rho \sigma} + R_{\mu \nu}{}^{\tau \lambda} R_{\rho \sigma \tau \lambda}
+ 2 R_{\mu}{}^{\tau}{}_{\rho}{}^{\lambda} R_{\nu \tau \sigma \lambda}
- 2 R_{\mu}{}^{\tau}{}_{\sigma}{}^{\lambda} R_{\nu \tau \rho \lambda} = 0
\end{equation}
into the NP null frame  ($R_{\mu \nu \rho \sigma}$ is the Riemann tensor) \cite{Ryan:1974nt}.  While the second method is more prescriptive and does not require much guesswork as to which equations to look at and how to manipulate them, the former method requires less calculation, once a strategy  has been determined.  Therefore, we shall derive the coupled equations using the Bianchi equations, which, of course, coincide with those derived from the Penrose wave equation.

We derive the equations as follows. First, let us settle the notation. In this section all equations labelled as $(7.\rm{xx})$ refer to equation $(\rm{xx})$  in chapter 7 of \cite{Stephani:2003tm}.  Since these are long equations we do not reproduce them here and simply refer the reader to that reference. Further recall that the fundamental quantities in the NP formalism needed to study perturbations are the directional derivative operators \cite{Chandra:1983,Stephani:2003tm},\footnote{In the NP formalism, $\Delta$ is used to denote $n \cdot \nabla$.  However, since we are already using $\Delta$ in the definition of the KN BH metric \eqref{KNsoln}, in order to avoid confusion, we denote $\hat{\Delta} \equiv n \cdot \nabla$.}
\begin{equation}\label{DirectDeriv}
 D=\ell^\mu\nabla_\mu\,,\qquad \hat{\Delta}=n^\mu\nabla_\mu\,,\qquad
\delta=m^\mu\nabla_\mu\,,\qquad
\bar{\delta}=\bar{m}^\mu\nabla_\mu\,,
\end{equation}
and the 12 complex spin coefficients defined from linear
combinations of the 24 background Ricci rotation connection coefficients 
$\gamma_{cab} = e_{(c)}^{\:\:\:\: \mu} e_{(b)}^{\:\:\:\: \nu} \nabla_\nu e_{(a)\,\mu}$ \cite{Chandra:1983,Stephani:2003tm},
\begin{eqnarray}
\hspace{-0.5cm} && \kappa=\gamma_{311}=0, \quad
\sigma=\gamma_{313}=0, \quad \nu=\gamma_{242}=0, \quad
\lambda=\gamma_{244}=0, \quad
\epsilon=\frac{1}{2} (\gamma_{211}+\gamma_{341})=0, \nonumber \\
\hspace{-0.5cm}  &&
 \mu=\gamma_{243}=-\frac{\Delta}{2 \Sigma \bar{r}^*}, \quad \rho=\gamma_{314}= -\frac{1}{\bar{r}^*}, \quad
\gamma=\frac{1}{2} (\gamma_{212}+\gamma_{342})=- \frac{\Delta - r(r-M) \bar{r}^*}{2 \Sigma^2} \bar{r}, \nonumber \\
&&
\tau=\gamma_{312}=- \frac{i \, a\, \sin \theta }{\sqrt{2} \Sigma}, \quad
 \alpha=\frac{1}{2} (\gamma_{214}+\gamma_{344})= -i \frac{\Sigma - 2a^2- r \bar{r}}{2 \sqrt{2} a \sin \theta \Sigma^2} \bar{r}^2, \quad  \pi=\gamma_{241}=\frac{i \, a\, \sin \theta }{\sqrt{2} \bar{r}^*{}^2}, \nonumber \\
&& \beta=\frac{1}{2} (\gamma_{213}+\gamma_{343})= \frac{\cot \theta \bar{r}^*}{2 \sqrt{2} \Sigma}\,. \label{spincoef}
\end{eqnarray}
Their complex conjugates (denoted by a bar) correspond to the replacement
$3\leftrightarrow 4$ in $\gamma_{cab}$.\footnote{KN is Petrov type D so, from the Goldberg-Sachs, one must have $\kappa=\sigma=\nu=\lambda=0$. Moreover, one has $\epsilon=0$  because we have chosen $\ell$ to be tangent to an affinely parametrized null geodesic $\ell^{\mu} \nabla_{\mu} \ell_{\nu} =0$.}
 
Consider the expression
\begin{equation}
\bar{\delta} (7.32d) -\hat{\Delta} (7.32c)
\end{equation}
as a first order perturbative equation. Let us consider the left hand side of this expression,\footnote{Note that in equations (7.xx), the derivative terms are written on the left hand side, while the rest of the terms are placed on the right.} which involves second order in derivative quantities:~\footnote{\label{foot1}Generally, we will use the notation that NP scalars with superscript $^{(0)}$ refer to scalars in the KN background and the superscript $^{(1)}$ to first order perturbations of the scalar.  However, in the equations below, for brevity, we suppress the superscripts.  From the expressions above for the background NP scalars it should be clear what is a background quantity and what is a first order perturbed quantity.} 
\begin{equation}
(\hat{\Delta} D - \bar{\delta} \delta) \Psi_4 + (\bar{\delta} \hat{\Delta} - \hat{\Delta} \bar{\delta}) \Psi_3 
+2 (\bar{\delta} \hat{\Delta} + \hat{\Delta} \bar{\delta}) (\Phi_2\bar{\Phi}_1) 
\end{equation}
Now, the operator acting on $\Psi_3$ is a commutation operator.  Therefore, we can use the equation $\overline{(7.6c)}$ to rewrite it in terms of first order in derivative quantities, which can themselves be turned into zeroth order in derivative quantities using equations $(7.32c)$ and $(7.32d)$.  However, the third term involving $\Phi_2$ cannot be similarly simplified.  At best we can use the commutation relations to rewrite $\bar{\delta} \hat{\Delta}$ in terms of $\hat{\Delta} \bar{\delta}$.  Therefore, it is clear already at this stage that a decoupled equation is not going to be possible on the KN background  and at best we can only hope to derive a coupled equation involving $\Psi_4$ and $\Phi_2$.  Further simplifying the first order in derivative terms using equations $(7.32f), (7.32g), (7.32j)$ and $(7.32k)$ and using various NP equations $(7.21a)$--$(7.21r)$, as well as the Maxwell equations $(7.22)$--$(7.25)$, gives
\begin{gather}
\left\{ (\hat{\Delta} + 3 \gamma - \bar{\gamma} + 4 \mu + \bar{\mu}) (D-\rho) 
- (\bar{\delta} + \bar{\beta} + 3 \alpha - \bar{\tau} + 4 \pi)(\delta + 4 \beta - \tau) - 3 \Psi_2 + 4 \Phi_1 \bar{\Phi}_1  \right\} \Psi_4 \notag\\
\hspace{35mm} - 4 \bar{\Phi}_1 \left\{ (\hat{\Delta} + 3 \gamma - \bar{\gamma} + 2\mu )(\bar{\delta} +2 \alpha) + (2 \pi + \bar{\tau})(\hat{\Delta} + 2 \gamma) \right\} \Phi_2 \notag \\
\hspace{25mm}+8 \left\{ (\hat{\Delta}  + 3 \gamma - \bar{\gamma}) \lambda +(\bar{\tau} - \pi) \nu \right\} (\Phi_1\bar{\Phi}_1)=0.
\label{eqn:coup1}
\end{gather}
At this stage we find that perturbed quantities $\lambda$ and $\nu$ are obstructions to a coupled equation involving $\Psi_4$ and $\Phi_2$.  However, inspecting the Bianchi equations closely, we find that $\lambda$ appears in equation $(7.32c)$ with coefficient $3\Psi_2 + 2 \Phi_{1} \bar{\Phi}_1$, while $\nu$ appears in $(7.32d)$ with coefficient $3\Psi_2 - 2 \Phi_{1} \bar{\Phi}_1$.  Thus, we can solve for $\lambda$ and $\nu$ in terms of differential operators on perturbed quantities $\Psi_3$, $\Psi_4$ and $\Phi_2$.  Significantly, the operator on $\lambda$ in equation \eqref{eqn:coup1} and the form of $(7.32c)$ means that $\Psi_3$ will end up with a second order derivative $\hat{\Delta} \bar{\delta}$, the same operator that acts on $\Phi_2$ in equation \eqref{eqn:coup1}.  Thus, we have the possibility of defining a perturbed quantity involving a particular combination of $\Phi_2$ and $\Psi_3$ such that this quantity couples with $\Psi_4$.  

Studying the form of the equations, it is not too difficult to conclude that such a quantity can be defined and is of the form
\begin{equation} \label{pertQuan}
\varphi_{-1} = 2 \Phi_1 \Psi_3 - 3 \Psi_2 \Phi_2.
\end{equation}
The resulting equation is of the form
\begin{align}
&\Bigg\{ (\hat{\Delta} + 3 \gamma - \bar{\gamma} + 4 \mu + \bar{\mu}) (D-\rho) 
- (\bar{\delta} + \bar{\beta} + 3 \alpha - \bar{\tau} + 4 \pi)(\delta + 4 \beta - \tau) - 3 \Psi_2  \notag \\
& \hspace{12mm}+ 4 \Phi_1 \bar{\Phi}_1 \Big[ 1- 2(\hat{\Delta} + 3 \gamma - \bar{\gamma} )\left(\frac{D-\rho}{3\Psi_2 + 2 \Phi_{1} \bar{\Phi}_1}\right) -  \frac{2(\bar{\tau}-\pi)}{3\Psi_2 - 2 \Phi_{1} \bar{\Phi}_1} (\delta +4\beta -\tau)\Big]  \Bigg\} \varphi_{-2}  \notag\\
&+ 4 \Phi_1 \Bigg\{ (\hat{\Delta} + 3 \gamma - \bar{\gamma} + 2\mu )\left(\frac{\bar{\delta} +2 \alpha+6\pi}{3\Psi_2 + 2 \Phi_{1} \bar{\Phi}_1}\right) + \frac{(\bar{\tau}-  \pi)}{3\Psi_2 - 2 \Phi_{1} \bar{\Phi}_1}(\hat{\Delta} + 2 \gamma+6\mu) \Bigg\} \varphi_{-1} =0,
\label{eqn:coup2}
\end{align}
where $\varphi_{-2} = \Psi_4$.

The second coupled equation is derived in a similar manner, except that it is now easier, because we know that the perturbed quantity that couples to $\Psi_4$ is $\varphi_{-1}$ as defined in \eqref{pertQuan}.  Thus we begin by considering
\begin{equation}
( \hat{\Delta} D - \bar{\delta} \delta) (2 \Phi_1 \Psi_3 - 3 \Psi_2 \Phi_2),
\end{equation}
using the fact that an equation for $( \hat{\Delta} D - \bar{\delta} \delta)\Psi_3 $ may be obtained from 
\begin{equation*}
D (7.32d) - \delta (7.32c)
\end{equation*}
and an equation for $( \hat{\Delta} D - \bar{\delta} \delta)\Phi_2 $ may be obtained from
\begin{equation*}
\hat{\Delta} (7.23) -\bar{ \delta} (7.25).
\end{equation*}
The strategy used to simplify the resulting equation is very similar to that used to derive equation \eqref{eqn:coup2}.  Therefore, without going through the details, we give the resulting coupled equation:
\begin{align}
&\Bigg\{ (\hat{\Delta} + 3 \gamma+ \bar{\gamma} +5 \mu + \bar{\mu}) (D-4\rho) 
- (\bar{\delta} +  \alpha+ \bar{\beta}  - \bar{\tau} + 5 \pi)(\delta + 2 \beta -4 \tau)  \notag \\
& \hspace{8mm}+ 4 \Phi_1 \bar{\Phi}_1 \Big[(D-4\rho+\bar{\rho})\left(\frac{\hat{\Delta} + 2 \gamma+6\mu}{3\Psi_2 - 2 \Phi_{1} \bar{\Phi}_1}\right) \notag\\
& \hspace{40mm}+ (\delta +3\beta -\bar{\alpha}  - 4\tau -\bar{\pi}) \left(\frac{\bar{\delta}+2\alpha+6\pi}{3\Psi_2 + 2 \Phi_{1} \bar{\Phi}_1} \right) \Big]  \Bigg\}\varphi_{-1}  \notag\\
&- 8 (\Phi_1)^2 \bar{\Phi}_1 \Bigg\{ (D -2\rho+\bar{\rho} )\left(\frac{\delta +4\beta-\tau}{3\Psi_2 - 2 \Phi_{1} \bar{\Phi}_1}\right) \notag\\
& \hspace{45mm} + (\delta +3\beta -\bar{\alpha}  - 2\tau +\bar{\pi})\left(\frac{D-\rho }{3\Psi_2 - 2 \Phi_{1} \bar{\Phi}_1}\right) \Bigg\} \varphi_{-2}  =0.
\label{eqn:coup3}
\end{align}
In summary, we have derived from the NP equations two coupled PDEs, \eqref{eqn:coup2} and \eqref{eqn:coup3}, satisfied by $\varphi_{-2} =\Psi_4$ and $\varphi_{-1}= 2 \Phi_1 \Psi_3 - 3 \Psi_2 \Phi_2$, which are invariant under infinitesimal diffeomorphisms and tetrad rotations \cite{Chandra:1983}.

These  NP scalars  $\varphi_{-2}$ and $\varphi_{-1}$ are the ones relevant for the study of perturbations that are outgoing at future null infinity and regular at the future horizon.   Note that we could have equally derived a set of coupled equations satisfied by $\Psi_0$ and $2 \Phi_1 \Psi_1 - 3 \Psi_2 \Phi_0$; the positive spin counterparts of $\varphi_{-2}$ and $\varphi_{-1}$.  Such equations are simply obtained via Geroch-Held-Penrose (GHP) transformations \cite{Geroch:1973am} of equations \eqref{eqn:coup2} and \eqref{eqn:coup3} and are relevant for perturbations outgoing at past null infinity.  

We can now substitute the background values of the NP quantities into equations \eqref{eqn:coup2} and \eqref{eqn:coup3} (recall footnote~\ref{foot1}).  Since $\partial_t,\partial_\phi$ are Killing vector fields of the KN background, its gravito-electromagnetic perturbations can be Fourier decomposed as $e^{-i \omega t} e^{i m \phi}$, where  $\omega$ and $m$ are the frequency and azimuthal quantum number of the mode.  Moreover, we rescale 
the perturbed quantities, $\{ \varphi_{-2},\varphi_{-1}\} \to \{ \psi_{-2},\psi_{-1}\} $ as \cite{Dias:2015wqa}:
\begin{eqnarray}\label{gauging}
&&\psi_{-2}= \left(\bar{r}^*\right)^4 \Psi_4^{(1)},  \nonumber\\
&& \psi_{-1}=\frac{\left(\bar{r}^*\right)^3}{2\sqrt{2}\Phi_1^{(0)}} \left(2\Phi_1^{(0)}\Psi_3^{(1)} -3 \Psi_2^{(0)}\Phi_2^{(1)}\right).
\end{eqnarray}
Having done this, we obtain the following coupled system of two PDEs  (first presented in \textbf{\cite{Dias:2015wqa}}):\footnote{There is a set of two coupled PDEs --- related to \eqref{ChandraEqs} by a Geroch-Held-Penrose \cite{Geroch:1973am} transformation --- for the quantities $\psi_{2}$ and $\psi_{1}$ that are the positive spin counterparts of \eqref{ChandraEqs}; however these would be relevant if we were interested in perturbations that were outgoing at the past null infinity.}
\begin{eqnarray}\label{ChandraEqs}
&& \left(\mathcal{F}_{-2}+ Q^2 \mathcal{G}_{-2}\right) \psi_{-2}  + Q^2 \mathcal{H}_{-2}  \psi_{-1} =0 \,, \nonumber \\
&& \left(\mathcal{F}_{-1} +Q^2 \mathcal{G}_{-1}\right)\psi_{-1} + Q^2  \mathcal{H}_{-1} \psi_{-2}=0  \,, 
\end{eqnarray}
where the second order differential operators $\{\mathcal{F},\mathcal{G},\mathcal{H}\}$ are given by 
\begin{eqnarray}\label{def:opsFGH}
\mathcal{F}_{-2}&=&\Delta\DD_{-1}^\dagger\DD_0 +\LL_{-1}\LL_2^\dagger -6i \omega\bar{r} \,,
\nonumber \\
\mathcal{G}_{-2}&=&\Delta \DD_{-1}^\dagger \alpha_-\bar{r}^*\DD_0  -3\Delta \DD_{-1}^\dagger \alpha_- 
- \LL_{-1}\alpha_+ \bar{r}^* \LL_2 ^\dagger  +3 \LL_{-1} \alpha_+ i a \sin \theta \,, 
\nonumber \\
\mathcal{H}_{-2}&=&-\Delta \DD_{-1}^\dagger \alpha_- \bar{r}^* \LL_{-1}  -3 \Delta \DD_{-1}^\dagger \alpha_- i a \sin \theta 
-\LL_{-1} \alpha_+ \bar{r}^* \Delta \DD_{-1}^\dagger  -3\LL_{-1} \alpha_+ \Delta  \,,
\nonumber \\
\mathcal{F}_{-1}&=&\Delta\DD_1\DD_{-1}^\dagger +\LL_2^\dagger\LL_{-1}-6i \omega\bar{r} \,,
\\
\mathcal{G}_{-1}&=& - \DD_0 \alpha_+ \bar{r}^* \Delta \DD_{-1}^\dagger  -3 \DD_0 \alpha_+ \Delta 
 +\LL_2^\dagger \alpha_- \bar{r}^* \LL_{-1}  +3 \LL_2^\dagger \alpha_- i a\sin\theta \,,
\nonumber \\
\mathcal{H}_{-1}&=& -\DD_0 \alpha_+ \bar{r}^* \LL_2^\dagger  +3 \DD_0 \alpha_+ i a \sin \theta  
 -\LL_2^\dagger \alpha_- \bar{r}^* \DD_0  +3 \LL_2^\dagger \alpha _- \,,
\nonumber
\end{eqnarray}
with 
$\alpha_\pm \equiv \left[3(\bar{r}^2M-\bar{r} Q^2)\pm Q^2\bar{r}^*\right]^{-1}$,  and the radial and angular Chandrasekhar operators  \cite{Chandra:1983} are defined
\begin{eqnarray}\label{def:DL}
&& \DD_j = \partial_r+\frac{i K_r}{\Delta}+2j\frac{(r-M)}{\Delta}, \qquad K_r=am-(r^2+a^2)\omega; \nonumber \\
 && \LL_j = \partial_\theta+K_{\theta}+j\cot\theta, \qquad K_{\theta}=\frac{m}{\sin\theta}-a\omega\sin\theta. 
\end{eqnarray}
The complex conjugate of these operators, namely $\DD_j^\dagger $ and $\LL_j^\dagger$, can be obtained from $\DD_j$ and $\LL_j$ via the replacement $K_r \to - K_r$ and $K_{\theta} \to - K_{\theta}$, respectively.

Note that fixing a gauge in which $\Phi_{0}^{(1)}=\Phi_{1}^{(1)}=0$, \eqref{ChandraEqs} reduces to the Chandrasekhar coupled PDE system \cite{Chandra:1983} (see also the derivation in \cite{Mark:2014aja}).  Finally, note that in the limit $Q\to 0$ equations \eqref{ChandraEqs}  decouple yielding the familiar Teukolsky equation for Kerr  \cite{Teukolsky:1972my}.

Before finishing this section, we take the opportunity to discuss a property of QNMs of Schwarzschild and RN black holes that has raised a lot of attention in the literature. This is the fact that the spectra of the Regge-Wheeler (aka odd or axial)~ \cite{Regge:1957td}, and Zerilli (aka even or polar)~ \cite{Zerilli:1974ai} QNMs is isospectral (i.e.\ these two QNM families have exactly the same frequency)~\cite{Chandra:1983}. In the Schwarzschild limit, \eqref{ChandraEqs}  decouples and we can look independently at the gravitational $\psi_{-2}$ or  electromagnetic  $\psi_{-1}$ perturbations. 
In particular, the decoupled equation for $\psi_{-2}$ ($\psi_{-1}$) corresponds to the original Teukolsky master equation \cite{Teukolsky:1972my} for gravitational (electromagnetic) perturbations in Kerr with $a=0$. Thus, we can use the Teukolsky master equation to study the QNMs of the Schwarzschild black hole, instead of the Regge-Wheeler$-$Zerilli (RWZ) formalism. The two must give the same spectrum. 
The Teukolsky formulation has a single gauge invariant variable $\psi_{-2}$ that must translate into two gauge invariant variables in the RWZ formulation, namely Regge-Wheeler's  $\Phi_{\rm v}$ and Zerilli's $\Phi_{\rm s}$ eigenfunctions. Refs. \cite{Chandra:TekRWZ,Sasaki:1981,Dias:2013sdc} give the unique differential map that allows one to derive $\Phi_{\rm v}$ and $\Phi_{\rm s}$ from $\psi_{-2}$: see \eg equation (4.16) of \cite{Dias:2013sdc} (which holds for any cosmological constant). Isospectrality is the statement that $\Phi_{\rm v}$ and $\Phi_{\rm s}$ have the same QNM spectrum. Since $\Phi_{\rm v}$ and $\Phi_{\rm s}$ are constructed from the \emph{same} Teukolsky NP gauge invariant variable $\psi_{-2}$, it follows that the eigenfrequencies of the Regge-Wheeler and Zerilli QNMs must necessarily be the same. This proves the isospectral property of QNMs in Schwarzschild and RN black holes and shows that this property is only non-trivial when viewed from the perspective of the Regge-Wheeler$-$Zerilli formalism.\footnote{The proof given in \cite{Dias:2015wqa} and revisited here is for $\psi_{-2}$ in the Schwarzschild black hole but it extends trivially to $\psi_{-1}$ modes and the RN background.}

\subsection{Boundary conditions of the problem}\label{sec:BCs}

To have a well-posed boundary value problem we must supplement the coupled PDE system \eqref{ChandraEqs} with appropriate (physical) boundary conditions. 
At spatial infinity, we require only outgoing waves, and at the future event horizon, we keep only regular modes in ingoing Eddington-Finkelstein coordinates. Moreover, we must require regularity at the north (south) pole $\theta =\pi\,(-\pi)$. In this subsection, we state what the conditions that these boundary conditions impose on the fields $\{ \psi_{-1}, \psi_{-2}\}$ are.\footnote{The reader interested on a more detailed discussion of boundary conditions in perturbation problems about asymptotically flat backgrounds can see e.g.\ \cite{Dias:2010maa,Dias:2011jg,Dias:2014eua}.}

Recall that $\omega$ and $m$ are the frequency and azimuthal quantum number $m$ of the linear mode perturbations, respectively. The $t -\phi$ symmetry of the KN BH allows one to consider only modes with  Re$(\omega)\geq 0$, as long as we study both signs of $m$. 
Then, to solve the coupled PDEs \eqref{gauging}, we need to impose physical boundary conditions. 
At spatial infinity, a Frobenius analysis of \eqref{ChandraEqs} yields two independent solutions that at leading order behave as $C_{\pm}e^{\pm i\omega r}$. Imposing the boundary condition $C_-=0$, \ie allowing only outgoing waves yields the decay:
\begin{equation}\label{BC:inf}
\psi_{s}{\bigl |}_\infty \!\simeq\!e^{i \omega r} r^{-(2s+1)+i \omega \,\frac{r_+^2+a^2+Q^2}{r_+}} \!\! \left( \!\! \alpha_{s}(\theta)+\frac{\beta_{s}(\theta)}{r}+\cdots\!\!\right), \nonumber
\end{equation}
where $s=-2,-1$, and $\beta_{s}(\theta)$ is a function of $\alpha_{s}(\theta)$ and its derivative fixed by expanding \eqref{ChandraEqs} at spatial infinity.

At the horizon, the boundary condition must be such that only ingoing modes are allowed. A Frobenius analysis at this boundary gives two independent solutions,
\begin{equation}\label{bcH0}
\psi_{s}{\bigl |}_H \sim  A_{\rm{in}}\left(r-r_+\right)^{-s-i\frac{ \omega -m \Omega _H}{4\pi  T_H}}\left[1+\mathcal{O}\left(r-r_+\right)\right]+A_{\rm{out}}\left(r-r_+\right)^{s+i\frac{ \omega -m \Omega _H}{4\pi  T_H}}\left[1+\mathcal{O}\left(r-r_+\right)\right],
\end{equation}
where $A_{\rm{in}}, A_{\rm{out}}$ are arbitrary amplitudes and $\Omega_H,T_H$ are the angular velocity and temperature defined in \eqref{KNthermo}.
To impose the correct boundary condition, we introduce the ingoing Eddington-Finkelstein coordinates $\{v,r,x,\widetilde{\phi} \}$, which extend the solution through the horizon. These are defined via
\begin{equation}\label{EF}
t=v-\int \frac{r^2+a^2}{\Delta}\,\mathrm{d}r\,,\qquad \phi=\widetilde{\phi}-\int \frac{a}{\Delta}\,\mathrm{d}r\,.
\end{equation}
The boundary condition is determined by the requirement that the metric and Maxwell field perturbations are regular in these ingoing Eddington-Finkelstein coordinates. This happens if and only if  $\psi_{s}(r)$ behaves as $\psi_{s}|_H \sim \psi_{s}^{\hbox{\tiny EF}}|_H\left(r-r_+\right)^{-s-i \frac{\omega -m \Omega _H}{4\pi  T_H}}$  where $\psi_{s}^{\hbox{\tiny EF}}(r)$ is a smooth function.
Thus, we must set $A_{\rm{out}}=0$ in \eqref{bcH0}. To conclude, at the horizon, a Frobenius analysis whereby we require only regular modes in ingoing Eddington-Finkelstein coordinates, yields the expansion 
\begin{equation}\label{BC:H}
\psi_{s}{\bigl |}_H \!\simeq\! \left(r-r_+\right)^{-s-\frac{i (\omega -m\Omega_H)}{4 \pi T_H}}[ a_{s}(\theta)+ b_{s}(\theta)(r-r_+) +\cdots ], \nonumber
\end{equation}
where $b_{s}(\theta)$ is a function of $a_s(\theta)$ and its derivative.

At the north (south) pole $x\equiv \cos\theta =1\,(-1)$, regularity dictates that the fields must behave as 
($\varepsilon = 1$ for $|m|\geq 2$, while $\varepsilon = -1$ for $|m|=0,1$ modes)
\begin{equation}\label{BC:N}
\psi_{s}{\bigl |}_{\hbox{\tiny N,(S)}} \hspace{-1pt} \simeq  (1\mp x)^{\varepsilon^{\frac{1\pm 1}{2}} \frac{s+|m|}{2}} \hspace{-1pt} \left[ A^{\pm}_{s}(r)+B^{\pm}_{s}(r)(1\mp x)+\cdots \right],  \nonumber
\end{equation}
where $B^{+}_{s}(r)$($B^{-}_{s}(r)$) is a function of $A^{+}_{s}(r)$($A^{-}_{s}(r)$) and its derivatives along $r$, whose exact form is fixed by expanding \eqref{ChandraEqs} around the North (South) pole.

 The PDE system \eqref{ChandraEqs} subject to the above boundary conditions that describe the gravito-electromagnetic QNMs of the KN black hole with parameters $\{M,a,Q\}$ has a useful scaling symmetry. When  we scale the metric and Maxwell field strength as $g_{\mu \nu}\to \Lambda^2 g_{\mu \nu}$ and $F_{\mu \nu}\to \Lambda F_{\mu \nu}$, for an arbitrary constant $\Lambda$, the equations of motion are left invariant.
This means  we can scale out one of the 3 parameters of the solution. Therefore, we can work with the adimensional parameters $\{\tilde{a},\tilde{Q}\}\equiv \{a/M,Q/M\}$ (or $\{a/r_+,Q/r_+\}$) and $\tilde{\omega}\equiv \omega M$. 
To find the frequency spectra of KN BHs we thus `just' need to scan a 2-dimensional space. 

To solve the PDE problem numerically, we use a pseudospectral method that searches directly for specific QNMs using a Newton-Raphson root-finding algorithm. We refer the reader to the review~\cite{Dias:2015nua} and ~\cite{Dias:2009iu,Dias:2010eu,Dias:2010maa,Dias:2010gk,Dias:2011jg,Dias:2010ma,Dias:2011tj,Cardoso:2013pza,Dias:2014eua,Dias:2018etb} for details. The exponential convergence of the method, and the use of quadruple precision, guarantee that the results are accurate up to, at least, the eighth decimal place.

\section{Two families of QNMs: photon sphere and near-horizon modes} \label{sec:AnalyticalPSNH}

The frequency spectra of gravito-electromagnetic perturbations of KN has two main families of QNMs: 1) the {\it photon sphere} (PS), and 2) the {\it near-horizon} (NH) families. Each of these families can dominate the frequency spectra (\ie have lower $|\mathrm{Im}\,\tilde{\omega}|$) depending on the region of the parameter space we look at.
These two families are the natural extension to the rotating case ($a\neq 0$) of the PS and NH families of QNMs that are present in the Reissner-Nordstr\"om case, although this sharp distinction between the two families is unambiguous only for small rotation $a$ parameter. 
There are particular regimes of parameter space where the frequency of each of these two families can be captured by perturbative expansions (WKB expansion and/or a near-horizon matched asymptotic expansion). This allows us to identify these two families of QNMs (thus providing the basis for their nomenclature), while providing also analytical formulae that give  good approximations to the actual frequencies. Therefore, in this section we discuss in detail two useful perturbative analyses. In subsection~\ref{sec:PSeikonal} we consider a large $m$ WKB expansion that identifies the PS QNMs, while, in subsection~\ref{sec:NHanalytics} we describe a simple but efficient matched asymptotic expansion that captures the NH modes.

Before considering the perturbative analyses, it is enlightening to identify the two families of QNMs in the simplest black hole where they co-exist. This is the  Reissner-Nordstr\"om (RN) black hole.
This  identification will be our guide once we delve into the parameter phase space of KN away from the RN limit. It will also allow us to speculate about expectations for the KN QNM spectra that will then be discussed in the next subsections~\ref{sec:PSeikonal}-\ref{sec:NHanalytics} and in section~\ref{sec:repulsion}.

\begin{figure}[th]
\centering
\includegraphics[width=.52\textwidth]{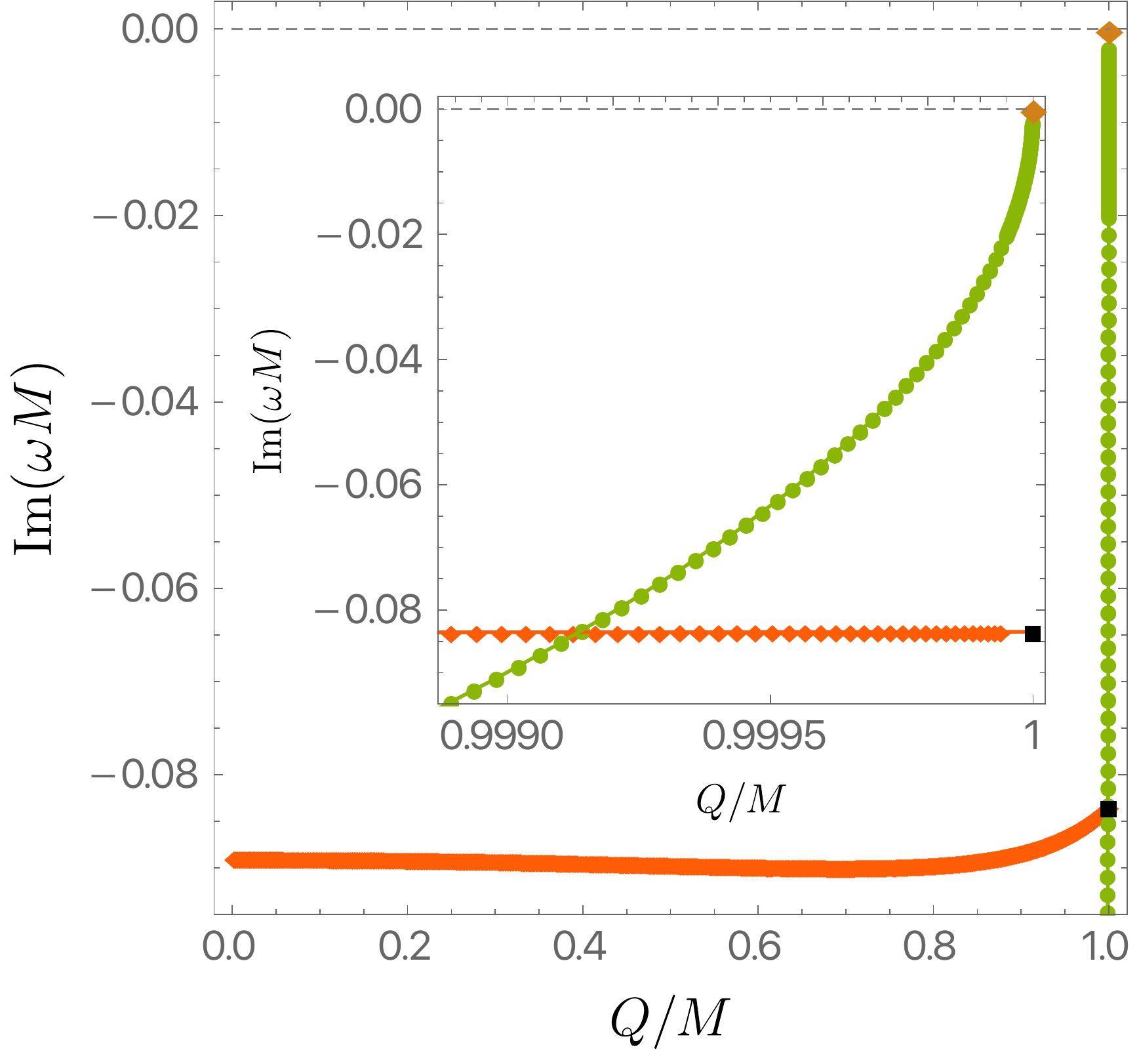}
\hspace{0.2cm}
\includegraphics[width=.44\textwidth]{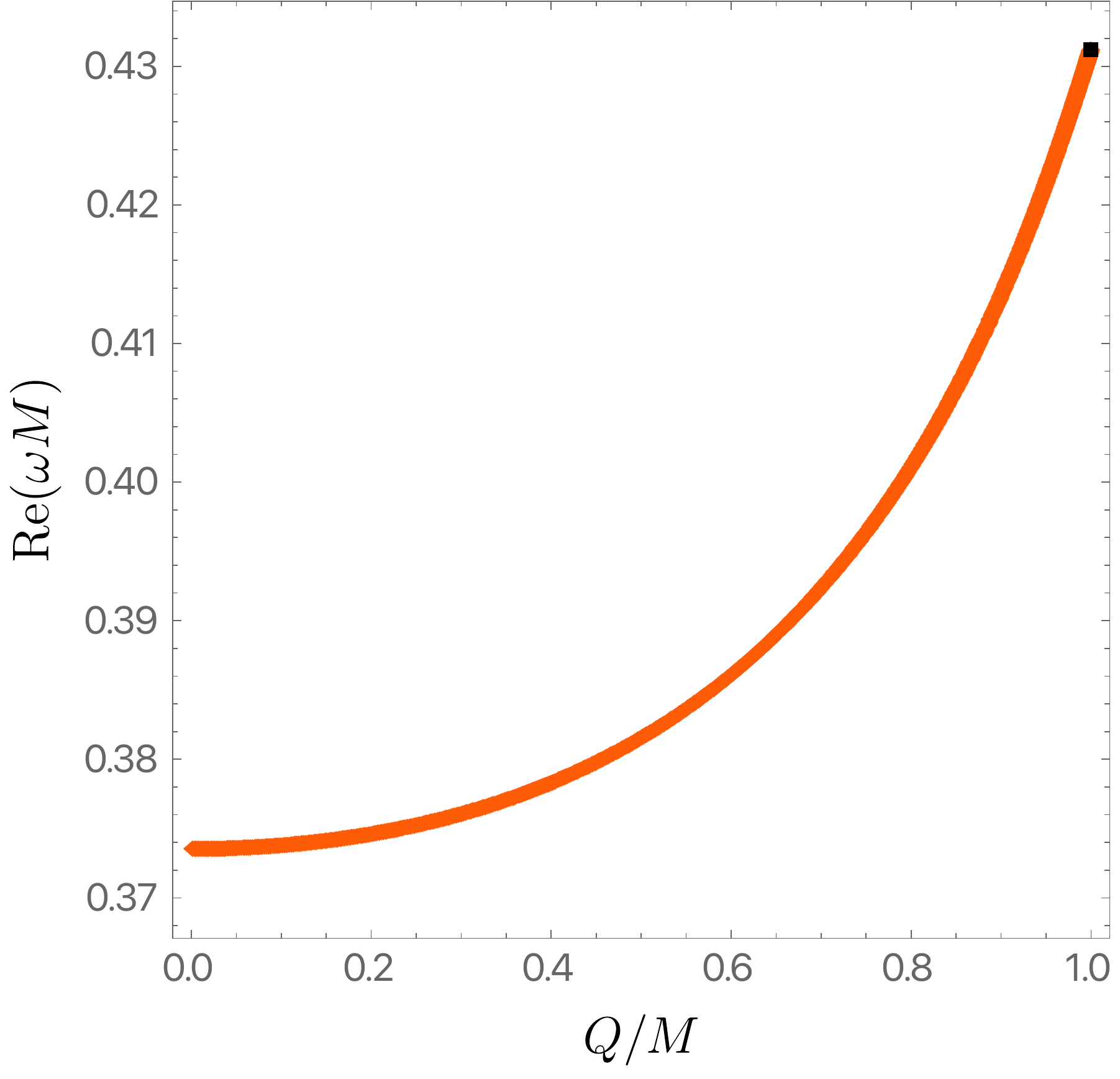}
\caption{Photon sphere (PS; orange diamonds) and near-horizon (NH; green disks) gravitational QNMs (Z$_2$) for the Reissner-Nordstr\"om BH ($a=0$) with $\ell=m=2, n=0$. In the RN case the PS and NH modes are unambiguously identified.
This data was obtained by solving the Regge-Wheeler$-$Zerilli ODE for RN \cite{Regge:1957td,Zerilli:1974ai} and it matches the data for KN with $a=0$, obtained by solving the coupled set of two PDEs for $\{\psi_{-2},\psi_{-1}\}$, which validates our KN numerics.
The black square with $\tilde{\omega}\simeq 0.431341 -0.0834603\,i$ is obtained by solving the Regge-Wheeler$-$Zerilli ODE {\it directly} at extremality (where we have to impose regular boundary conditions on a degenerate horizon); the non-extremal frequencies approach this value as $\tilde{Q}\to 1$ which is yet a further check of our numerics. 
{\bf Left panel:} Imaginary part of the (dimensionless) frequency as a function of the (dimensionless) charge. The inset plot shows the region where PS and NH modes intersect: above $\tilde{Q}=\tilde{Q}_c^{\hbox{\tiny RN}}\sim 0.9991342$, the NH modes have lower $|\mathrm{Im}\,\tilde{\omega}|$ but this quantity grows very large very quickly for $\tilde{Q}<\tilde{Q}_c^{\hbox{\tiny RN}}$ where the PS mode is comfortably the dominant one.  {\bf Right panel:} Real part of the frequency as a function of charge. The NH mode has $\mathrm{Re}\,\tilde{\omega}=0$ in the RN limit (and only in this limit) and is not shown. Thus, the real part of the PS and NH frequencies have no crossing (unlike the imaginary part).}
\label{Fig:RN}
\end{figure}

In Fig.~\ref{Fig:RN} we plot the  frequency spectra for $\ell=m=2,n=0$ gravitational ($Z_2$) QNMs  in the RN BH,\footnote{This figure partially reproduces the top-left panel of  later Fig.~\ref{Fig:spectraFix-a} where the charge is however measured in units of $r_+$.} which are the QNMs with slowest decay rate in RN and  KN as we will demonstrate in section~\ref{sec:FullSpectra}. We see that there are two clearly distinct families of QNMs: 1) the PS family (orange diamonds) which dominates for a wide range of charge, namely for $\tilde{Q}<\tilde{Q}_c^{\hbox{\tiny RN}}\sim 0.9991342$ (and reduces to the Schwarzschild QNM when $Q=0$ \cite{Chandra:1983,Leaver:1985ax}), and 2) the NH family (green circles) which becomes the slowest decaying mode for $\tilde{Q}_c^{\hbox{\tiny RN}}< \tilde{Q}\leq 1$ and approaches  $\mathrm{Im}\,\tilde{\omega}=0$ at extremality  as best seen in the inset plot (in RN these NH modes have  $\mathrm{Re}\,\tilde{\omega}=0$ for any $\tilde{Q}$). 

By continuity, once  rotation is turned on but with small $\tilde{a}=a/M$, we expect the KN spectra to be similar to Fig.~\ref{Fig:RN} (NH modes should still approach extremality with $\mathrm{Im}\,\tilde{\omega}=0$ but this time, as we confirm later, with $\mathrm{Re}\,\tilde{\omega}=m \tilde{\Omega}_H^{\hbox{\footnotesize{ext}}}\neq 0$). Moreover, it also seems reasonable to expect the existence of a line $-$ let us denote it as $\tilde{Q}=\tilde{Q}_c(\tilde{a})$ $-$ that describes the intersection of the PS and NH surfaces and that eventually extends from $\tilde{Q}_c(\tilde{a}=0)=\tilde{Q}_c^{\hbox{\tiny RN}}$ (identified in Fig.~\ref{Fig:RN}) all the way up towards extremality. However, and interestingly, our full numerical results will prove that our expectations are {\it only partially} correct.
Indeed, the KN frequency spectra for fixed  but small $\tilde{a}$ is similar to Fig.~\ref{Fig:RN}. In particular, for $0 \leq \tilde{Q}<\tilde{Q}_c(\tilde{a})$, the PS family has the lowest  $|\mathrm{Im}\,\tilde{\omega}|$ and for $\tilde{Q}_c(\tilde{a}) <\tilde{Q}\leq \tilde{Q}_{\hbox{\footnotesize ext}}$ it is the NH QNM that has slowest decay rate. Moreover, keeping $a/a_{\hbox{\footnotesize ext}}$ small, these two families trade dominance along their intersection line $\tilde{Q}=\tilde{Q}_c(\tilde{a})$ with a {\it simple crossover} in the imaginary part of the frequency like the one observed in the inset plot of the left panel of Fig.~\ref{Fig:RN} (the real part of the frequencies display no crossing as is clear in the right panel of Fig.~\ref{Fig:RN} for $a=0$).
 However, as we keep increasing the rotation $a/a_{\hbox{\footnotesize ext}}$, we find that we enter a region of the parameter space (a window of $\tilde{Q}$) where an {\it unexpected} change occurs: instead of having simple crossovers in $\mathrm{Im}(\tilde{\omega})$ where the PS and NH families  should intersect, one starts observing intricate {\it eigenvalue repulsions} in $\mathrm{Im}(\tilde{\omega})$ that will be discussed in section~\ref{sec:repulsion} and associated Figs.~\ref{Fig:spectraFix-a}-\ref{Fig:spectraFix-a2}, and the sharp distinction between PS and NH modes is lost (in this region PS and NH modes have similar $\mathrm{Re}(\tilde{\omega})$ with no crossings). So much so that the QNM families will now be a combination (to be made precise later) of the two old modes in what will be more properly denoted as PS-NH families and their radial overtones.

After this brief summary of the findings to come, let us discuss the eikonal and near-horizon analytical descriptions of the KN modes.

\subsection{Photon sphere modes in the eikonal limit: analytical formula for the frequencies} \label{sec:PSeikonal}

In the eikonal or geometric optics limit $\ell\sim |m| \gg 1$, where a WKB approximation holds, there are QNM  frequencies $-$ known as ``photon sphere"(PS) QNMs $-$  that are closely related to the properties of the unstable circular photon orbits in the equatorial plane of the KN black hole. Namely, the real part of the PS frequency is proportional to the Keplerian frequency $\Omega_c$ of the circular null orbit and the imaginary part of the PS frequency scales with the Lyapunov exponent $\lambda_L$ of the orbit \cite{Goebel:1972,Ferrari:1984zz,Ferrari:1984ozr,Mashhoon:1985cya,Bombelli:1991eg,Cornish:2003ig,Cardoso:2008bp,Dolan:2010wr,Yang:2012he,Stuchlik1991}. The latter describes how quickly a null geodesic congruence on the unstable circular orbit increases its cross section under infinitesimal radial deformations. 

The PS modes with  an eikonal limit that we will consider are those with $\ell=m$ or $\ell=-m$. This includes the $\ell=m=2, n=0$ modes that have the slowest decay rate and that we typically display as orange diamond curves/surfaces (e.g.\  in Fig.~\ref{Fig:RN} and Figs.~\ref{Fig:spectraFix-a}-\ref{Fig:spectraFix-a2}, among others). And these PS modes of the KN BH are those that reduce to the well-known QNM frequencies of Schwarzschild BH in the limit $Q\to 0$ and $a\to 0$ (typically identified as a dark-red point in our figures) first studied by Chandrasekhar (see Table V, page 262 of \cite{Chandra:1983}). Therefore, in this subsection we  use geometric optics to compute an analytical approximation (to be denoted as $\omega^{\hbox{\tiny eikn}}_{\hbox{\tiny PS}}$) for the frequency of these PS modes in the KN background. A similar analysis was originally done in \cite{Zimmerman:2015trm,Li:2021zct}. Although, the final analytical formula for the PS QNM frequencies is strictly valid in the WKB limit $\ell\sim |m| \to \infty$, in practice we find that it matches  reasonably well the PS frequencies even for values as small as $\ell=|m|=2$. Therefore, the eikonal limit allows us to identify the nature of this QNM family and, furthermore, it provides a check on our numerics. 

The geodesic equation, describing the motion of pointlike particles around a KN BH, leads to a set of quadratures. {\it A priori} this is perhaps an unexpected result since KN only possesses two Killing fields,  $K=\partial/\partial_t$ and $\xi=\partial/\partial_\phi$. We seem to be one Killing field short of an integrable system. However, there is another conserved quantity $-$ the Carter constant $-$ associated to a Killing tensor $K_{ab}$, which saves the day \cite{Chandra:1983}.

The most direct way to identify this integrable structure is to consider the Hamilton-Jacobi equation \cite{Chandra:1983}:
\be\label{HJeq}
\frac{\partial S}{\partial x^\mu}\frac{\partial S}{\partial x^\nu}g^{\mu\nu}=0\,,
\ee
where $S$ is known as the principal function.  One can obtain the motion of null particles by noting that, according to Hamilton-Jacobi theory, the principal function and the particle momenta are related by
\be \label{HJ:defP}
\frac{\partial S}{\partial x^\mu}\equiv p_\mu\quad \text{and}\quad p^\mu =\frac{\mathrm{d}x^\mu}{\mathrm{d}\tau}\,,
\ee
with $\tau$ denoting an affine parameter of the null geodesic.

We can then take a separation \emph{ansatz} of the form ($x=\cos\theta$, where $\theta$ is the polar angle)
\be \label{HJ:ansatz}
S=-e\,t+j\,\phi+R(r)+X(x)\,,
\ee
where the constants $e$ and $j$ are the conserved charges associated with the Killing fields $K$ and $\xi$ via\footnote{For massive particles, these coincide with the energy and angular momentum of the particle, but for massless particles $e$ and $j$ have no physical meaning since they can be rescaled. The ratio $j/e$, however, is invariant under such rescalings.}
\be
e\equiv - K_\mu \dot{x}^\mu\qquad \text{and}\qquad j\equiv \xi_\mu \dot{x}^\mu\,,
\label{eq:conserved}
\ee
where the dot ( $\dot{}$ ) denotes a derivative with respect to the affine parameter $\tau$.

Substituting the ansatz \eqref{HJ:ansatz} into the Hamilton-Jacobi equation \eqref{HJeq} for null geodesics yields a coupled system of ordinary differential equations for $R(r)$ and $X(x)$ (the prime $^{\prime}$ denotes a derivative w.r.t.\ the argument, $r$ or $x$, respectively)
\begin{eqnarray} 
&& \Delta ^2 R'^2 -\left[e \left(r^2+a^2\right)-a j\right]^2+\Delta  \left[\mathcal{Q}+(j-a e)^2\right]=0\,,
\nonumber \\ \label{HJ:RadAng}
&& X'^2-\frac{(j-a e)^2+\mathcal{Q}}{1-x^2 }+\frac{\left[ a e \left(1-x^2\right)-j\right]^2}{ \left(1-x^2\right)^2}=0\,,
\end{eqnarray}
where $\mathcal{Q}$ is a separation constant known as the Carter constant. 

From \eqref{HJ:defP}, \ie $\dot{x}^{\mu}=g^{\mu \nu }\frac{\partial S}{\partial x^{\mu }}$, one further has
\begin{eqnarray} 
&& \hspace{-0.4cm}\dot{t}=\frac{\left(r^2+a^2\right) \left[e \left(r^2+a^2\right)-a j \right]+a \Delta  \left[ j-a e \left(1-x^2\right)\right]}{\Delta  \left(r^2+a^2 x^2\right)},
\nonumber \\ \label{HJ:tphi}
&& \hspace{-0.4cm} \dot{\phi }=\frac{\left(1-x^2\right) a\left[e \left(r^2+a^2\right)-a j\right]+\Delta  \left[j-a e \left(1-x^2\right)\right]}{\Delta  \left(1-x^2\right) \left(r^2+a^2 x^2\right)}. 
\end{eqnarray}

We are interested in matching the behaviour of null geodesics with that of QNMs with large values of $\ell=|m|$, so we can restrict attention to the equatorial plane where $x=0$. From \eqref{HJ:RadAng}, such geodesics exist only if at $\tau=0$ one has $X(0)=\dot{X}(0)=0$ and $\mathcal{Q}=0$. Defining the geodesic impact parameter  
\be
b\equiv\frac{j}{e}\,,
\ee
the equation \eqref{HJ:RadAng} governing the radial motion now gives
\be
\label{HJ:geodesic}
\dot{r}^2=V(r;b)\,,
\ee
where the potential is
\be \label{HJ:pot}
V(r;b)=\frac{j^2}{b^2}\left(1+\frac{a^2-b^2}{r^2}+\frac{2 M (b-a)^2}{r^3}-\frac{Q^2 (b-a)^2}{r^4}\right).
\ee

\begin{figure}[b]
\centering
\includegraphics[width=.52\textwidth]{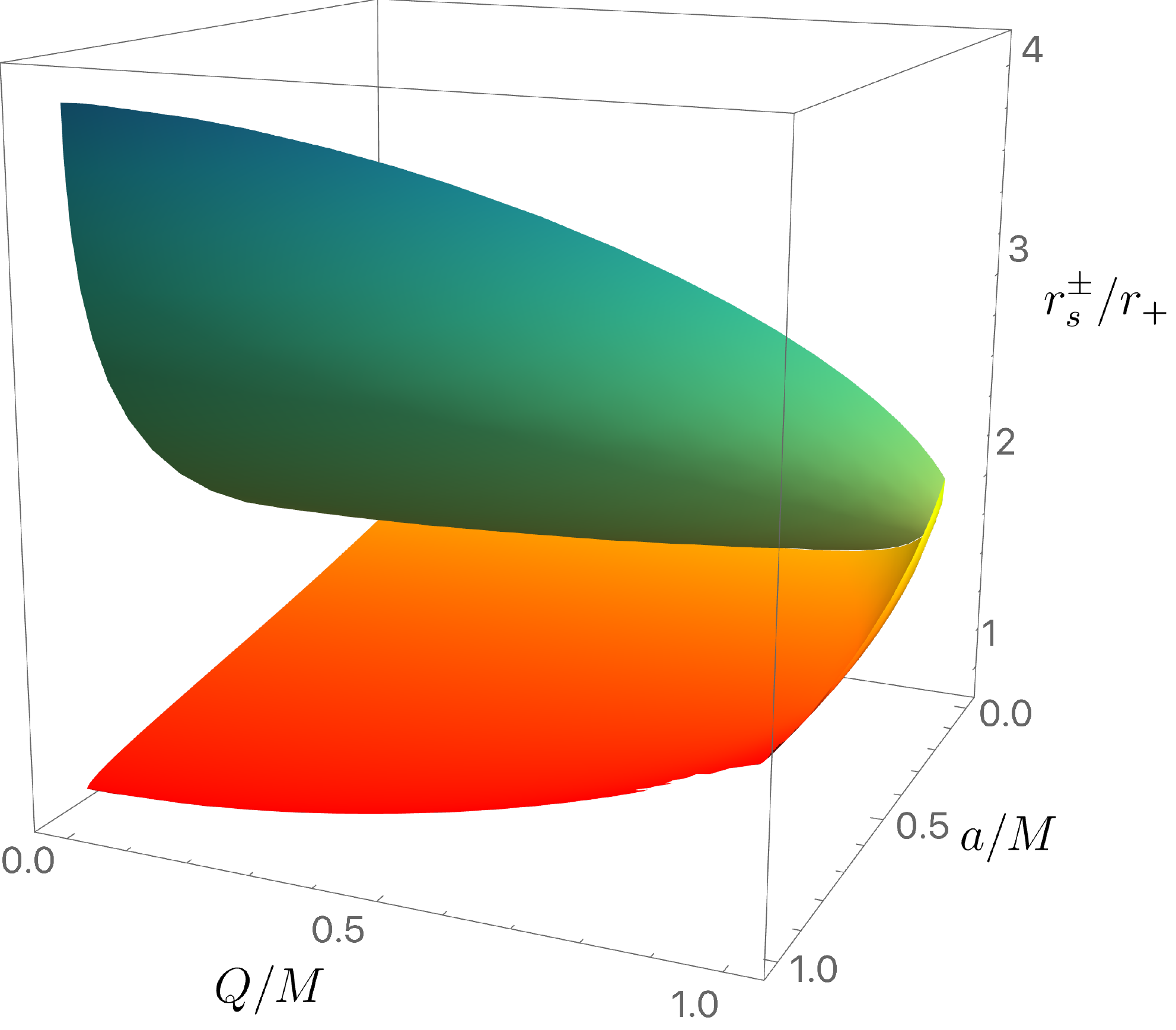}
\caption{The radii $r_s^{\pm}$ (with $r_s^{+}\geq r_s^{-}\geq r_+$) of the two unstable circular orbits in the equatorial plane of the KN black hole that ultimately yield the co-rotating $m=\ell$ (in the $r_s^{-}$ case) and the counter-rotating $m=-\ell$  (in the $r_s^{+}$ case) PS QNM frequencies in the eikonal limit. For $a=0$, one has $r_s^{+}=r_s^{-}$, and at $(\tilde{Q},\tilde{a})=(0,1)$ one has  $r_s^{-}=r_+$.}
\label{Fig:PSradii}
\end{figure}  

We are now interested in finding the photon sphere (region where null particles are trapped on circular unstable orbits), \emph{i.e.}\ the values of $r=r_s$ and $b=b_s$, such that
\be\label{HJ:CircOrb}
V(r_s,b_s)=0\quad\text{and}\quad \left.\partial_r V(r,b)\right|_{r=r_s,b=b_s}=0.
\ee
From the first equation we get
\be\label{PS:bs}
b_s(r_s)=\frac{r_s^2 \sqrt{\Delta(r_s)}+a \left(Q^2-2 M r_s\right)}{r_s^2-2 M r_s+Q^2}\,,
\ee
which we insert in the second equation of  \eqref{HJ:CircOrb} to get a fourth order polynomial equation for $r_s$:
\be \label{PS:4thPolyn}
4\left[r_s^2+2 a \left(\sqrt{\Delta  r_s}+a\right)\right]^2-\left(3 M r_s+\sqrt{9 M^2 r_s^2-8 Q^2\left[r_s^2+2 a \left(\sqrt{\Delta  r_s}+a\right)\right]}\right)^2=0\,,
\ee
where $\Delta(r)$ is defined  below \eqref{KNsoln} and we are interested in solutions with $r_s>r_+$.
Alternatively, we can solve \eqref{HJ:CircOrb} to get the black hole parameters $M$ and $Q$ that have circular orbits with radius $r_s$ and impact parameter $b_s$, namely
\be\label{PS:MQrsbs}
M=\frac{r_s \left(b_s^2-a^2-2 r_s^2\right)}{\left(b_s-a\right)^2}\,, \qquad Q=\frac{r_s\sqrt{b_s^2-a^2-3 r_s^2}}{\sqrt{\left(b_s-a\right){}^2}}.
\ee
There are two real roots $r_s$ higher than $r_+$ which are in correspondence with two PS modes: the {\it co-rotating} one (with $m=\ell$) that maps to the eikonal orbit with radius $r_s=r_s^-$ and $b_s>0$ (and that has the lowest $|\mathrm{Im}\,\tilde{\omega}|$) and the {\it counter-rotating} mode with $m=-\ell$ which is in correspondence with the orbit with radius $r_s=r_s^+$ and $b_s<0$, with $r_s^+\geq r_s^- \geq r_+$.  The two real roots $r_s^{\pm}$ higher than $r_+$ are displayed in 
Fig.~\ref{Fig:PSradii}. 

\begin{figure}[b]
\centering
\includegraphics[width=.48\textwidth]{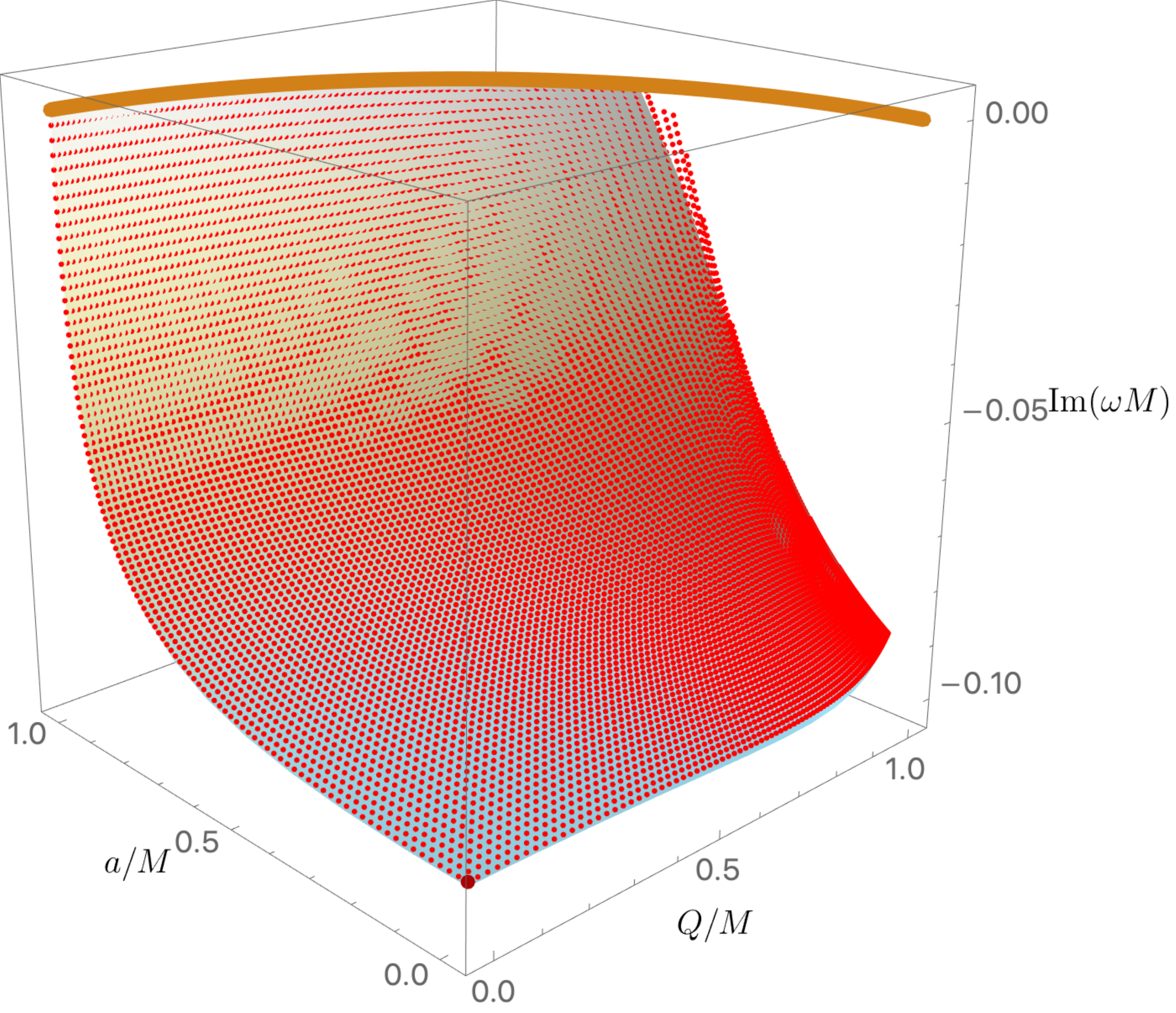}
\hspace{0.0cm}
\includegraphics[width=.5\textwidth]{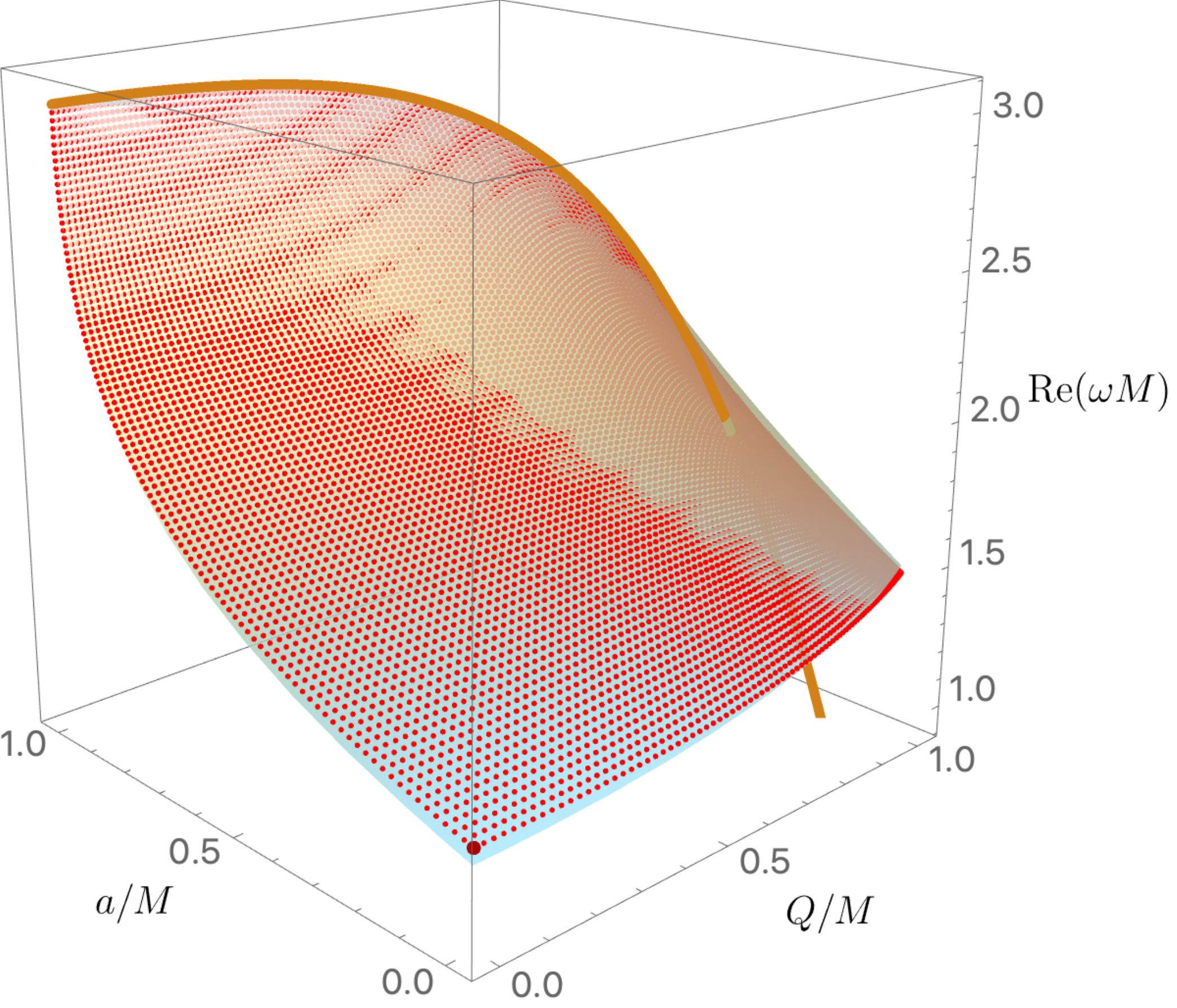}
\caption{Comparing the eikonal  prediction $\omega^{\hbox{\tiny eikn}}_{\hbox{\tiny PS}}$ (light blue surface) with the actual numerical frequencies (orange points) for co-rotating PS modes with $m=\ell=6, n=0$. The former is given by \eqref{PS:bs}-\eqref{PS:eikonal} with $r_s=r_s^{-}$ of Fig.~\ref{Fig:PSradii} and $b_s>0$. The brown curve at extremality has $\mathrm{Im}\,\tilde{\omega}=0$ and $\mathrm{Re}\,\tilde{\omega}=m\tilde\Omega_H^{\hbox{\footnotesize{ext}}}$. So, it turns out that $m=6$ seems to be already within the WKB validity $|m| \gg 1$. The dark-red point at $(\tilde{Q},\tilde{a})=(0,0)$ coincides with the Schwarzschild QNM, $\tilde{\omega}\simeq 1.21200982 - 0.09526585\, i $, first computed in \cite{Chandra:1983}.}
\label{Fig:PSwkb-m6}
\end{figure}  

We can finally compute the orbital angular velocity (also known as Kepler frequency) of the null circular photon orbit, that is simply given by
\be \label{PS:Omega_c}
\Omega_c \equiv \frac{\dot{\phi}}{\dot{t}}=\frac{1}{b_s}\,,
\ee
where we used \eqref{HJ:tphi} evaluated at $r=r_s$ and $b=b_s$.
We can also compute the largest Lyapunov exponent $\lambda_L$, measured in units of $t$, associated with infinitesimal fluctuations around photon orbits with $r(\tau)=r_s$. This can be  done by perturbing the geodesic equation (\ref{HJ:geodesic}) with the potential \eqref{HJ:pot} evaluated on an orbit with impact parameter $b=b_s$ and setting $r(\tau)=r_s+\delta r(\tau)$. One finds that small deviations decay exponentially in time as $\delta r \sim e^{-\lambda_L t}$ with Lyapunov exponent given  by
\begin{eqnarray}
\lambda_L &=& \sqrt{\frac{1}{2}\frac{V''(r,b)}{\dot{t}(\tau )^2}}\bigg|_{r=r_s,b=b_s} \nonumber\\
&=& \frac{1}{b_s r_s^2}\frac{\left | r_s^2+a^2-a b_s\right |}{\left | b_s-a\right |} \sqrt{6 r_s^2+a^2-b_s^2}\,.
\end{eqnarray}

One finally obtains the approximate spectrum of the photon sphere family of QNMs in the WKB limit $\ell=|m|\gg 1$ using \cite{Goebel:1972,Ferrari:1984zz,Ferrari:1984ozr,Mashhoon:1985cya,Bombelli:1991eg,Cornish:2003ig,Cardoso:2008bp,Dolan:2010wr,Yang:2012he} 
\begin{eqnarray} \label{PS:eikonal}
\omega^{\hbox{\tiny eikn}}_{\hbox{\tiny PS}} &\simeq& m\,\Omega_c-i\left(n+\frac{1}{2}\right)\lambda_L \nonumber \\
&\simeq& \frac{m}{b_s}  -i\,\frac{n+1/2}{b_s r_s^2}\frac{\left | r_s^2+a^2-a b_s\right |}{\left | b_s-a\right |} \sqrt{6 r_s^2+a^2-b_s^2}\,,  
\end{eqnarray}
where $n=0,1,2,\ldots$ is the radial overtone. This is the eikonal approximation for the PS modes we were looking for. Note that this expression is blind to the spin of the perturbation, \ie it is the same for scalar and gravito-electromagnetic perturbations. The eikonal analysis, although only based on a geodesic analysis, gives the same result as a leading order $|m|=\ell\to \infty$ WKB analysis of the wave perturbation equations. Although the eikonal frequency is independent on the spin of the perturbation, the higher order frequency corrections in the $1/m$ WKB expansion should certainly depend on the spin of the perturbation. 

\begin{figure}[th]
\centering
\includegraphics[width=.49\textwidth]{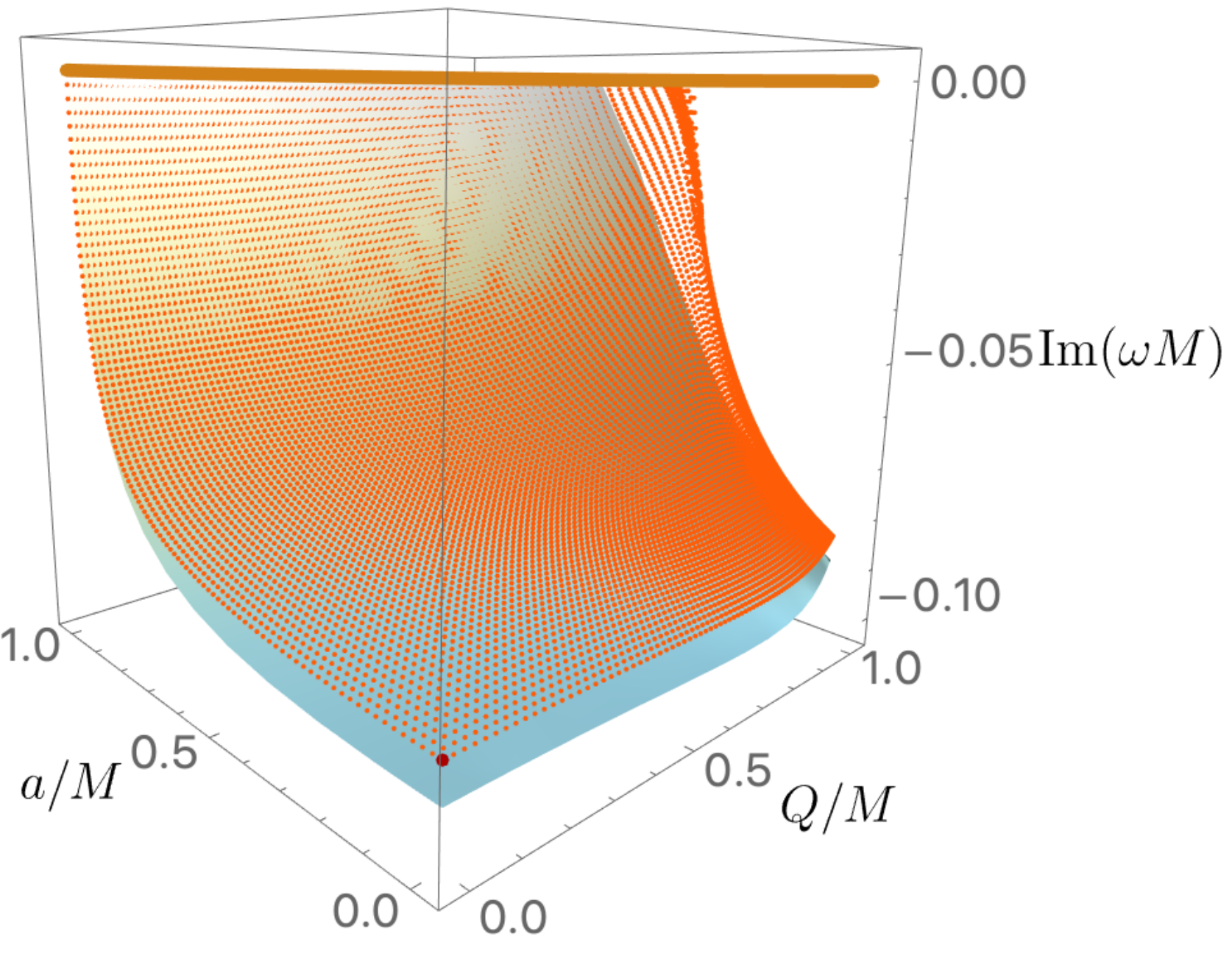}
\hspace{0.0cm}
\includegraphics[width=.49\textwidth]{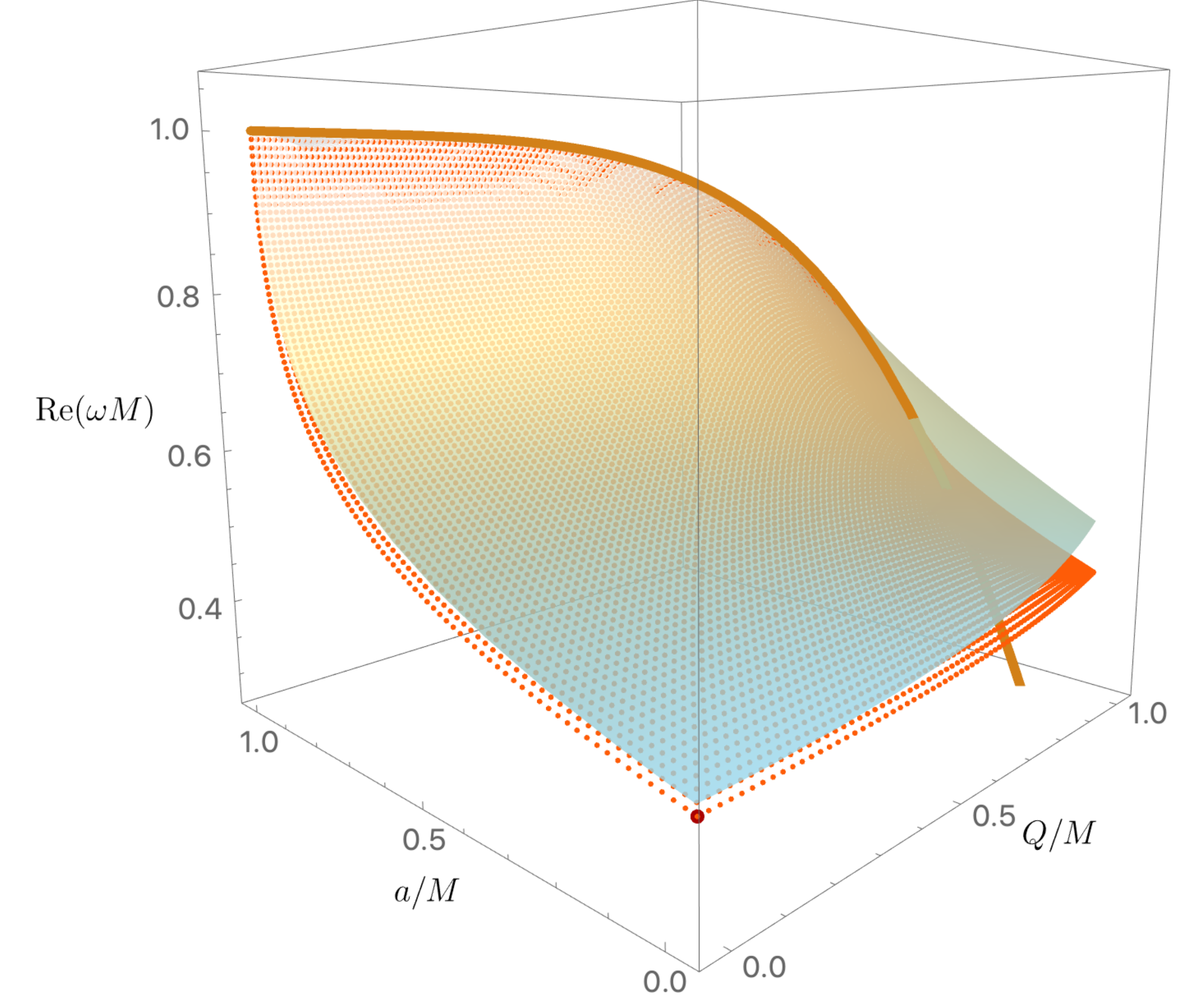}
\caption{Comparing the eikonal  prediction $\omega^{\hbox{\tiny eikn}}_{\hbox{\tiny PS}}$ (light blue surface) with the actual numerical frequencies (orange points) for co-rotating PS modes with $m=\ell=2, n=0$. The former is given by \eqref{PS:bs}-\eqref{PS:eikonal} with $r_s=r_s^{-}$ of Fig.~\ref{Fig:PSradii} and $b_s>0$. The brown curve at extremality has $\mathrm{Im}\,\tilde{\omega}=0$ and $\mathrm{Re}\,\tilde{\omega}=m\tilde\Omega_H^{\hbox{\footnotesize{ext}}}$. Although $m=2$ is certainly outside the regime of validity of the geometrics optics approximation, $|m| \gg 1$, it turns out that the approximation \eqref{PS:eikonal} proves to be reasonably good. The dark-red point at $(\tilde{Q},\tilde{a})=(0,0)$ coincides with the Schwarzschild QNM, $\tilde{\omega}\simeq 0.37367168 - 0.08896232\, i $, first computed in \cite{Chandra:1983,Leaver:1985ax}.}
\label{Fig:PSwkb-m2}
\end{figure}  
\begin{figure}[h]
\centering
\includegraphics[width=.46\textwidth]{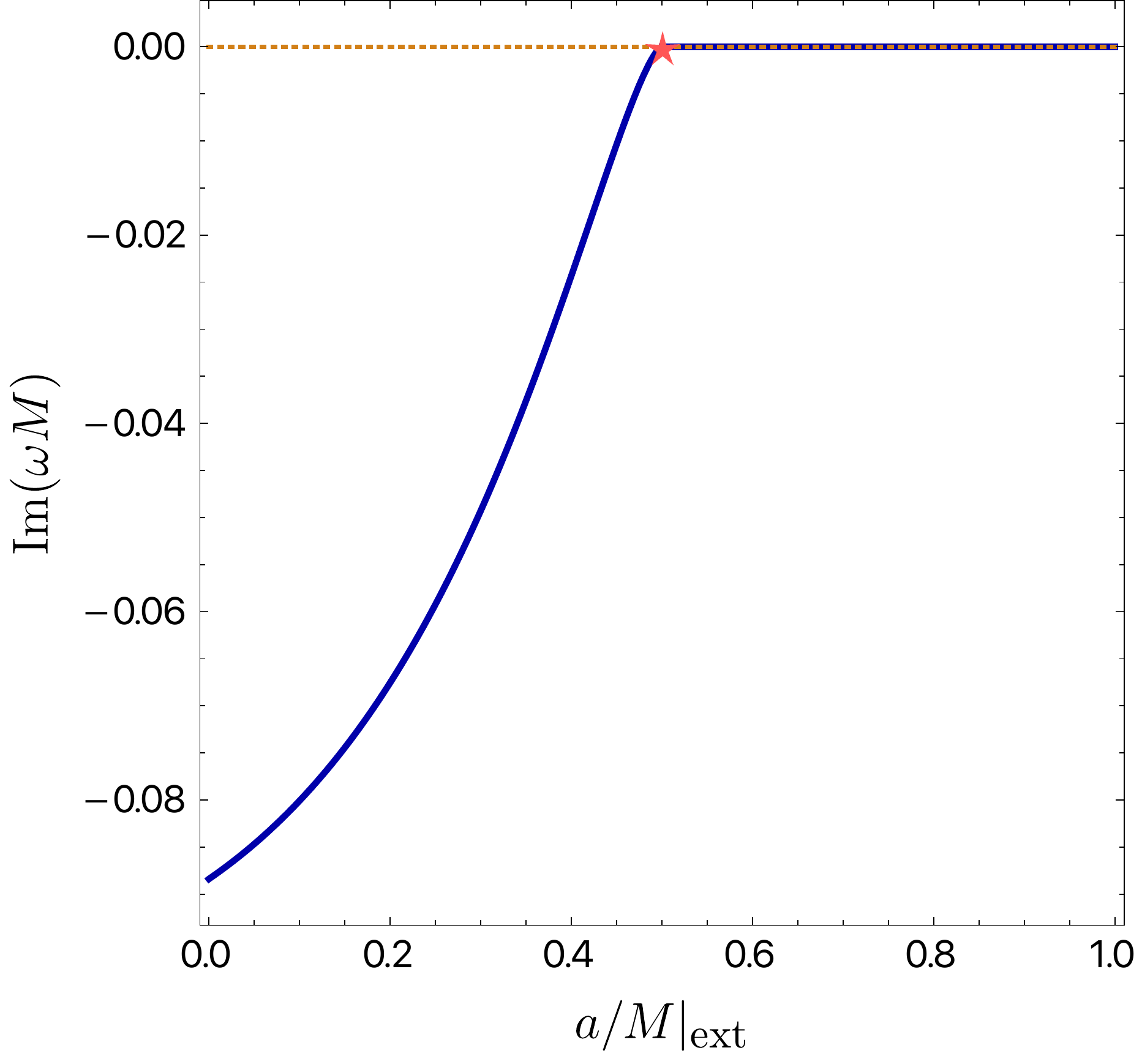}
\hspace{0.5cm}
\includegraphics[width=.44\textwidth]{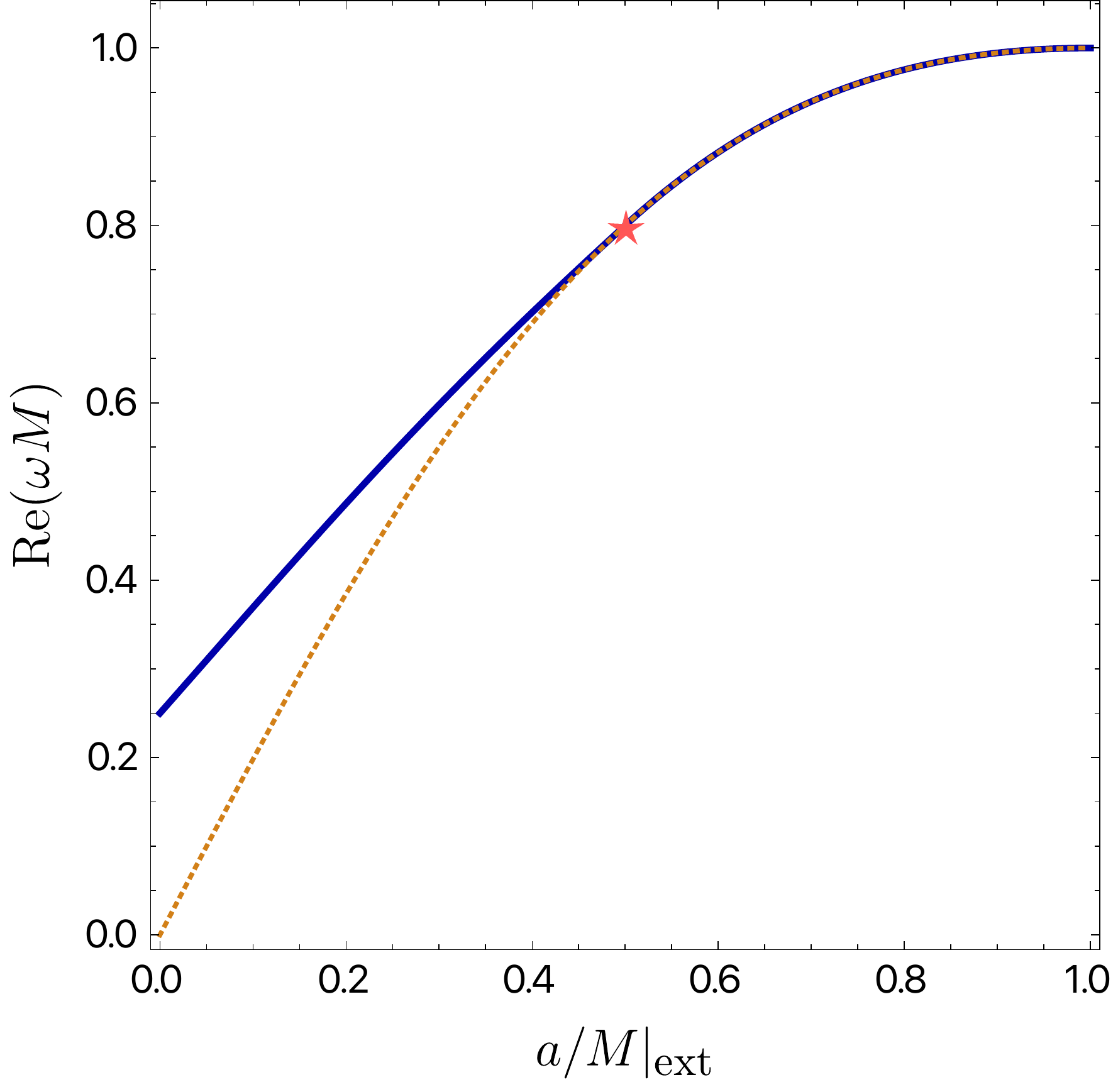}
\caption{The eikonal  prediction for \eqref{PS:eikonal} for $\omega^{\hbox{\tiny eikn}}_{\hbox{\tiny PS}}$  evaluated at extremality (dark-blue line). The dotted brown line has $\mathrm{Im}\,\tilde{\omega}=0$ and $\mathrm{Re}\,\tilde{\omega}=m\tilde{\Omega}_H$ (they correspond to the solid brown lines in Figs.~\ref{Fig:PSwkb-m6}-\ref{Fig:PSwkb-m2}). The red $\star$ point is at $\tilde{a}_{\hbox{\footnotesize ext}}=\tilde{a}_\star^{\hbox{\tiny eikn}}=\frac{1}{2}$.  
}
\label{Fig:PSwkb-m2-extremality}
\end{figure}  

\begin{figure}[th]
\centering
\includegraphics[width=.51\textwidth]{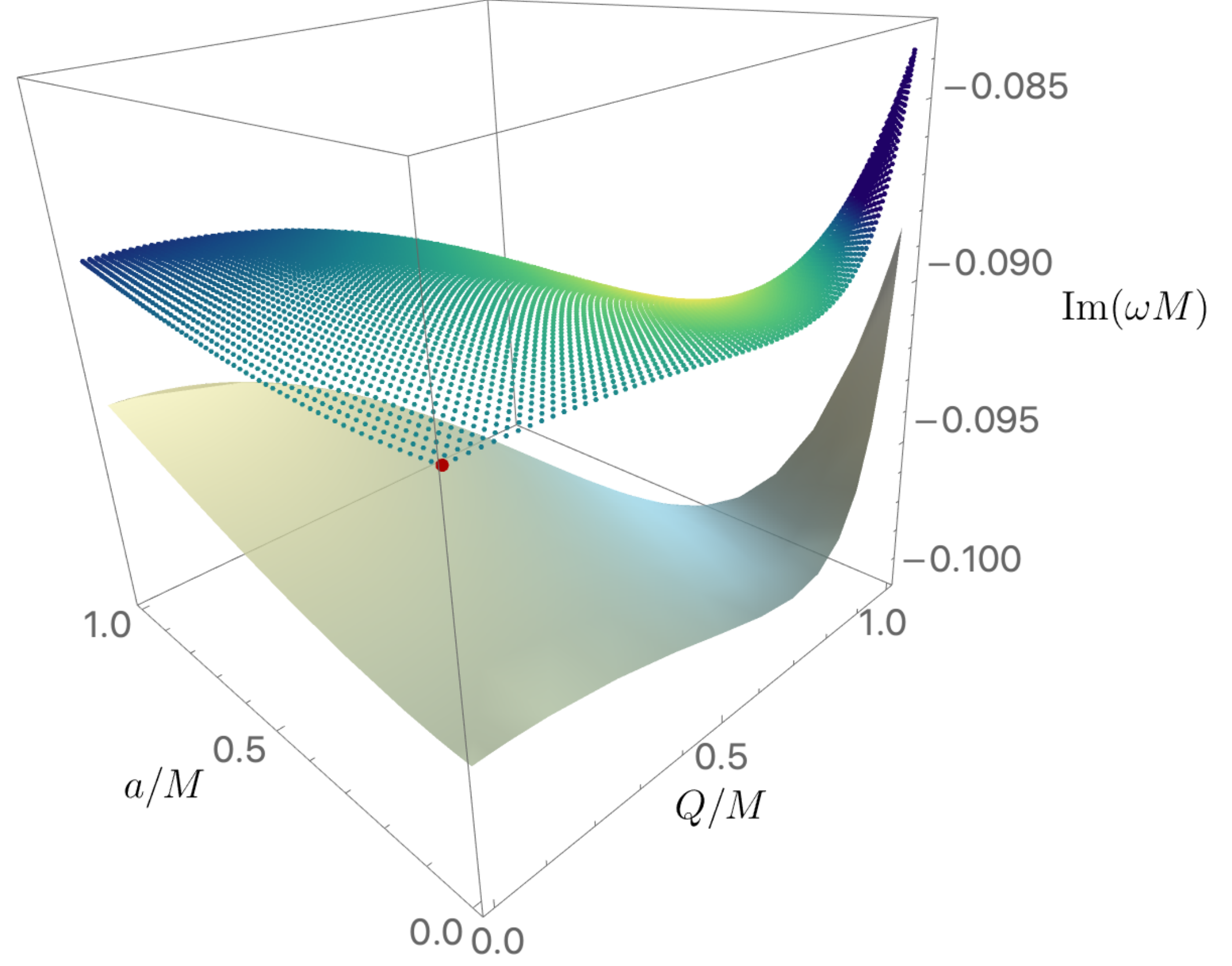}
\hspace{0.0cm}
\includegraphics[width=.47\textwidth]{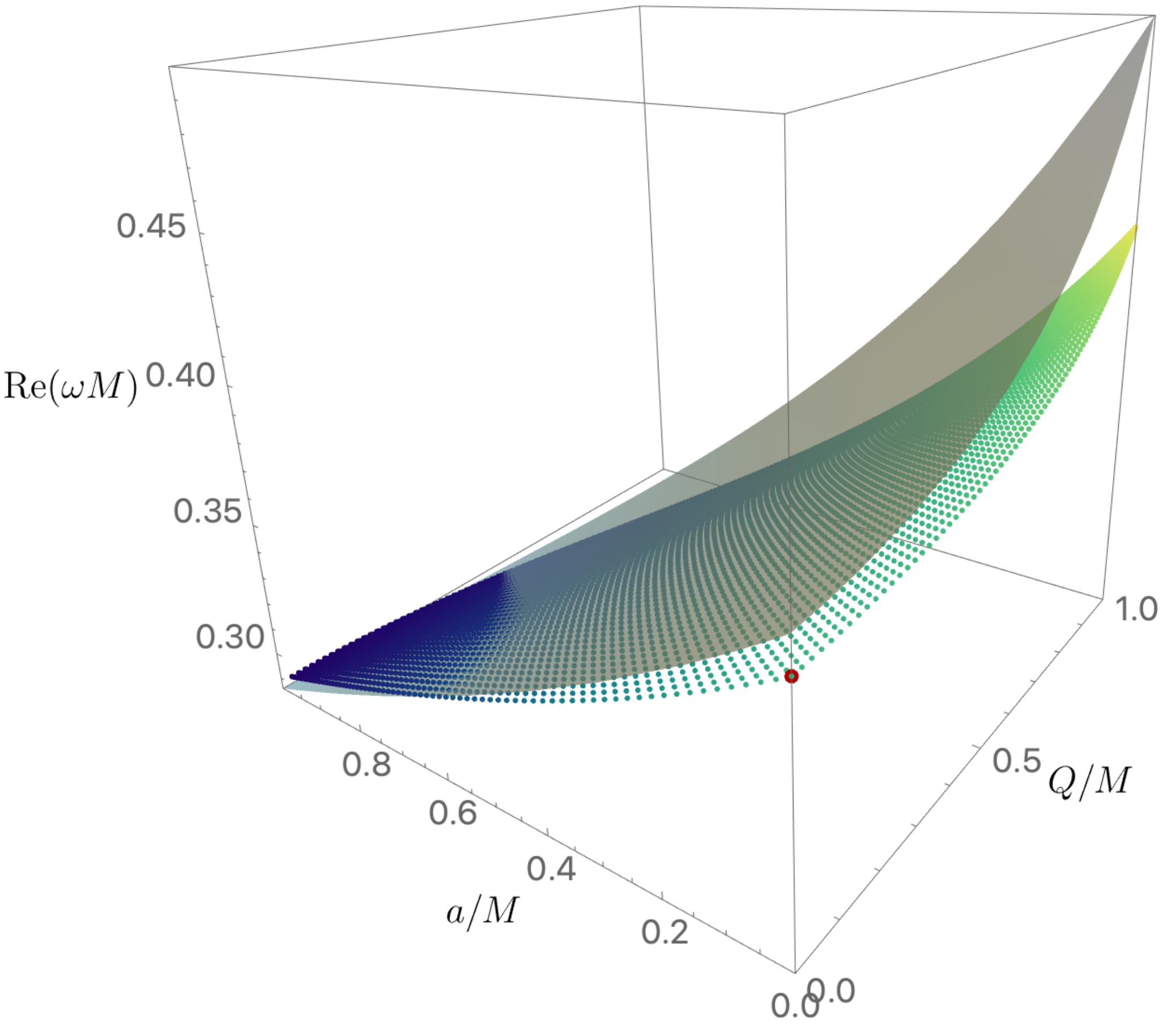}
\caption{Comparing the eikonal  prediction $\omega^{\hbox{\tiny eikn}}_{\hbox{\tiny PS}}$ (light blue surface) with the actual numerical frequencies (blue/green) for counter-rotating PS modes with $m=-\ell=-2, n=0$. The former is given by \eqref{PS:bs}-\eqref{PS:eikonal} with $r_s=r_s^{+}$ of Fig.~\ref{Fig:PSradii} and $b_s<0$. Although $m=-2$ is certainly outside the regime of validity of the geometrics optics approximation, $|m|\gg 1$, it turns out that it already gives a good qualitative approximation for the shape of the PS QNM surface. The dark-red point at $(\tilde{Q},\tilde{a})=(0,0)$ coincides with the Schwarzschild QNM, $\tilde{\omega}\simeq 0.37367168 - 0.08896232\, i $, first computed in \cite{Chandra:1983,Leaver:1985ax}.}
\label{Fig:PSwkb-m2neg}
\end{figure}

\begin{figure}[th]
\centering
\includegraphics[width=.51\textwidth]{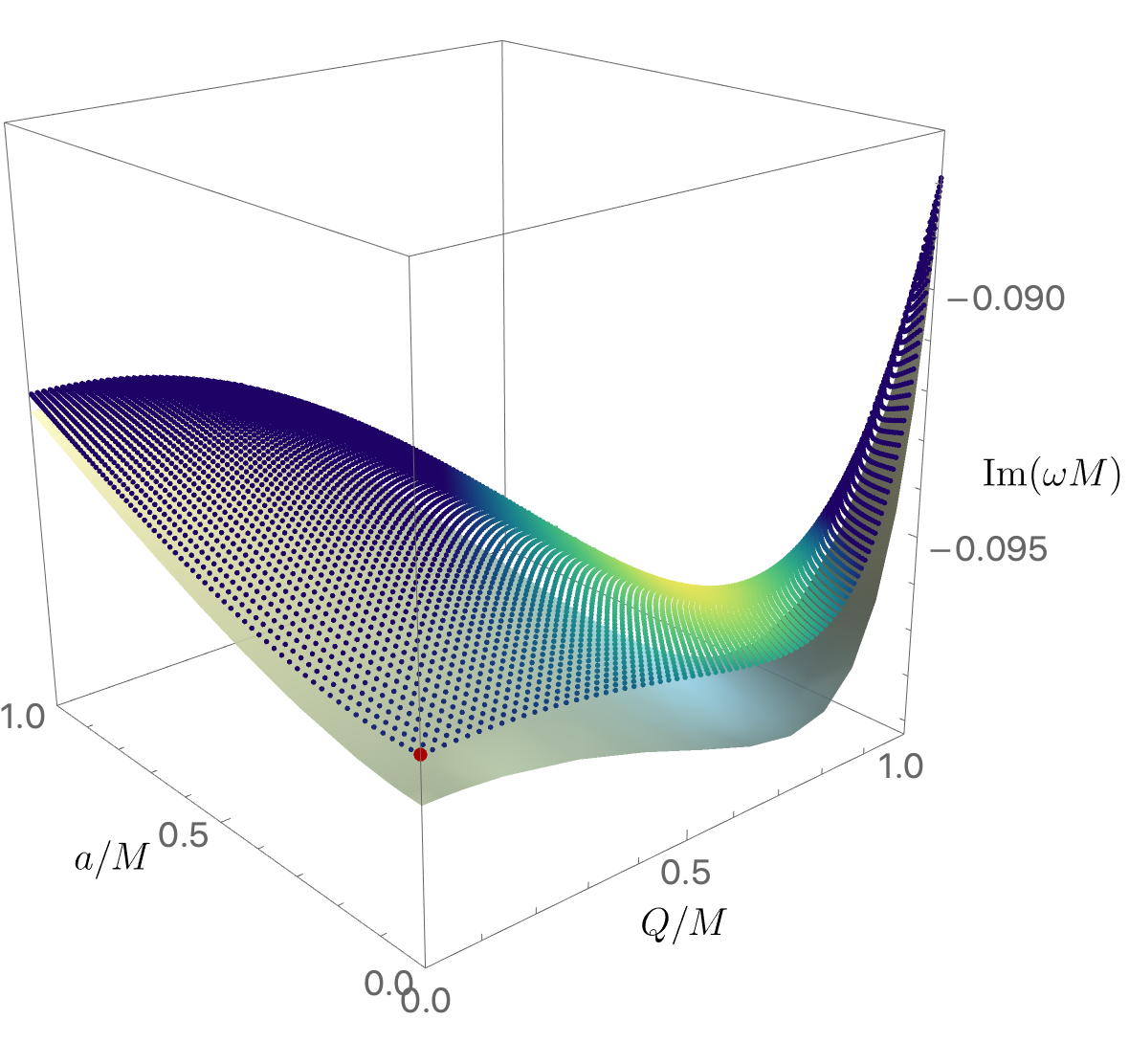}
\hspace{0.0cm}
\includegraphics[width=.47\textwidth]{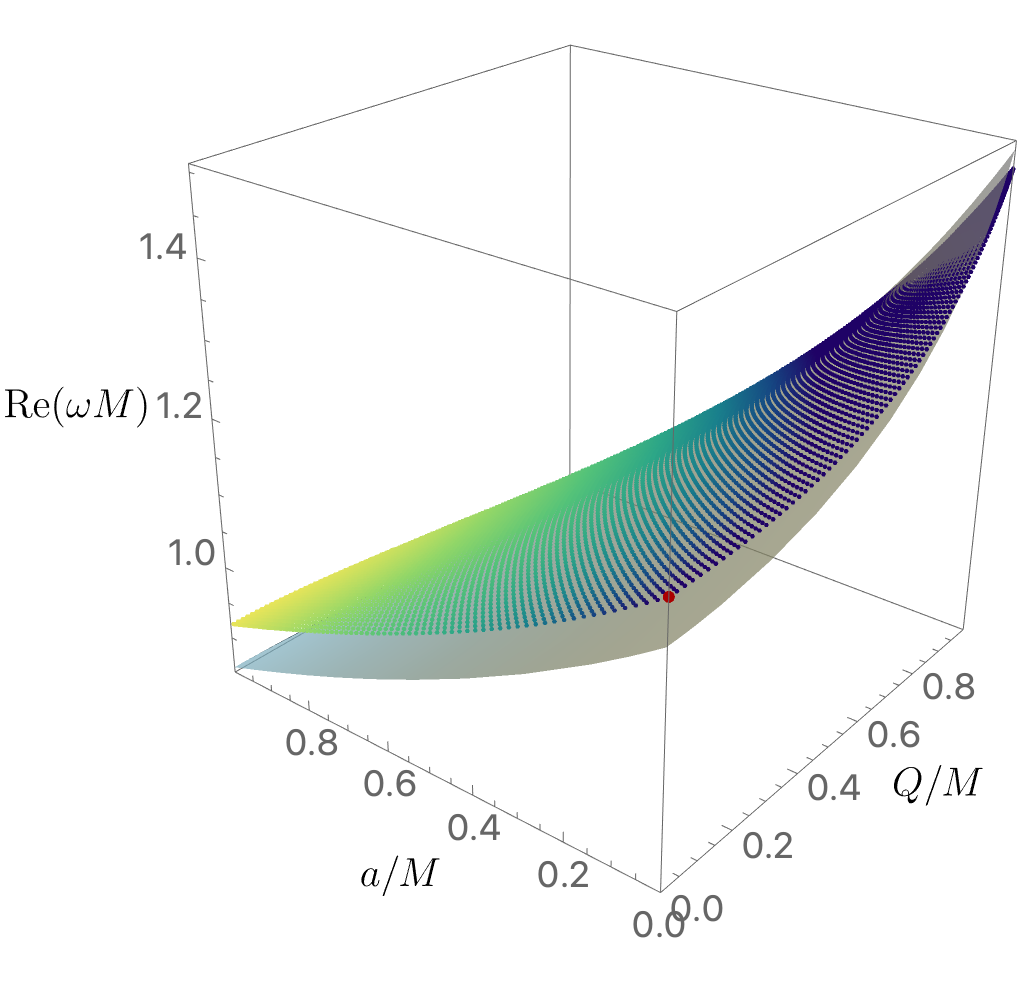}
\caption{Comparing the eikonal  prediction $\omega^{\hbox{\tiny eikn}}_{\hbox{\tiny PS}}$ (light blue surface) with the actual numerical frequencies (blue/green) for counter-rotating PS modes with $m=-\ell=-6, n=0$. The former is given by \eqref{PS:bs}-\eqref{PS:eikonal} with $r_s=r_s^{+}$ of Fig.~\ref{Fig:PSradii} and $b_s<0$. Although $m=-6$ is still outside the regime of validity of the geometrics optics approximation, $|m|\gg 1$, comparing the $m=-2$ case of Fig.~\ref{Fig:PSwkb-m2neg} with the $m=-6$ mode we see that as $|m|$ increases the eikonal approximation quickly starts proving to be a better quantitative approximation. The dark-red point at $(\tilde{Q},\tilde{a})=(0,0)$ coincides with the Schwarzschild QNM, $\tilde{\omega}\simeq 1.21200982 - 0.09526585\, i $ first computed in \cite{Chandra:1983,Leaver:1985ax}.}
\label{Fig:PSwkb-m6neg}
\end{figure}  

Recall that \eqref{PS:eikonal} is strictly valid in the geometric optics limit, $|m|\gg 1$, with corrections to $\mathrm{Im}\,\tilde{\omega}$ and $\mathrm{Re}\,\tilde{\omega}$ of order $\mathcal{O}\left(1/|m|\right)$ and  $\mathcal{O}\left(1\right)$, respectively. However, Fig.~\ref{Fig:PSwkb-m6} compares  
\eqref{PS:eikonal} (light blue surface) with the actual numerical frequency (red dots) of the co-rotating PS modes with $m=\ell=6, n=0$ and already finds an excellent agreement (of course this agreement will improve for $m>6$). Moreover, as Fig.~\ref{Fig:PSwkb-m2} demonstrates, the eikonal approximation \eqref{PS:eikonal} (light blue surface) still provides a reasonably good qualitative approximation to the numerical co-rotating PS modes (orange dots) even in the $m=\ell=2, n=0$ case. Taken together, this identifies the PS QNM family and validates our numerics.  

As shown in Fig.~\ref{Fig:PSwkb-m2-extremality}, an important feature of the eikonal frequency \eqref{PS:eikonal} (the solid dark-blue line) is that it is in good agreement with $\mathrm{Im}\,\tilde{\omega}=0$ and $\mathrm{Re}\,\tilde{\omega}=m\tilde{\Omega}_H$ (dotted brown line) for $\tilde{a}_{\hbox{\footnotesize ext}}>\tilde{a}_\star^{\hbox{\tiny eikn}}$, but not so for   $\tilde{a}_{\hbox{\footnotesize ext}}<\tilde{a}_\star^{\hbox{\tiny eikn}}$  with $\tilde{a}_\star^{\hbox{\tiny eikn}}=\frac{1}{2}$ (recall that $\tilde{a}_{\hbox{\footnotesize ext}}=0$ and $\tilde{a}_{\hbox{\footnotesize ext}}=1$ in the RN and Kerr limits, respectively). This transition point $\star$ is indeed  observed in the numerical data of Fig.~\ref{Fig:PSwkb-m6} (for $m=6$) and Fig.~\ref{Fig:PSwkb-m2} (for $m=2$), where the values $\mathrm{Im}\,\tilde{\omega}=0$ and $\mathrm{Re}\,\tilde{\omega}=m\tilde{\Omega}_H$ are represented by the  solid brown lines.  As expected, the eikonal quantitative value of $\tilde{a}_\star^{\hbox{\tiny eikn}}=\frac{1}{2}$ is not yet a good approximation for $m=2$, where numerically we find $\tilde{a}_\star\simeq 0.360$, but it becomes a better approximation as $m$ increases. For example, for $m=6$ one has $\tilde{a}_\star\simeq 0.463$ and for $m=10$ one has $\tilde{a}_\star\simeq 0.480$. 
  We come back to this issue in the discussion of Fig.~\ref{Fig:WKBlambda2-m}.

For completeness, in Fig.~\ref{Fig:PSwkb-m2neg} we turn our attention to the counter-rotating PS modes, and compare the eikonal prediction \eqref{PS:eikonal} (grey surface) with the numerical data (blue/green points) for the counter-rotating PS modes $Z_2$ $m=-\ell=-2, n=0$. Although $m=-2$ is certainly outside the regime of validity of the geometric optics approximation, $|m|\gg 1$, it turns out that it gives a relatively good approximation for the qualitative shape of the PS QNM surface (although less than for the $m>0$ case). As expected, the quantitative eikonal prediction improves considerably as $m$ grows more negative in the same way as for the $m>0$ case. This is illustrated for the PS modes $Z_2$ $m=-\ell=-6, n=0$ modes in 
Fig.~\ref{Fig:PSwkb-m6neg}.

\subsection{Near-horizon family of QNMs: analytical formula for the frequencies} \label{sec:NHanalytics}

Near-extremality, there is a family of KN wavefunctions that are very localized near the horizon and quickly decay to zero away from it. This suggests doing a `{\it poor-man's}' matched asymptotic expansion (MAE) whereby we take the {\it near-horizon limit} of the perturbed equations \eqref{ChandraEqs}, which can be  solved analytically, to find the near-region solution and then match it with a {\it vanishing} far-region wavefunction in the overlapping region where both solutions are valid.\footnote{Ideally, we would also solve the far-region equations to obtain the next-to-leading order nonvanishing far-region solution but in the KN background we cannot do it analytically.} In fact, motivated by the result that the near-horizon limit of the extremal KN BH corresponds to a warped circle fibred over $AdS_2$ (Anti-de Sitter) \cite{Bardeen:1999px}, the perturbations of which can be decomposed as a sum of known radial  $AdS_2$ harmonics, we can attempt to use {\it separation of variables}.  It turns out that this can indeed be done, and the system of 2 coupled PDEs for $\{\psi_{-2},\psi_{-1}\}$ {\it separates} into a system of two {\it decoupled} radial ODEs and a {\it coupled} system of two angular ODEs. This is a non-trivial, remarkable property of perturbations on KN.

At extremality, the modes with slowest decay rate (independently of belonging to the NH or PS families or, as we will introduce and discuss later, to the PS$-$NH family) always approach $\mathrm{Im}\,\tilde{\omega}=0$ and $\mathrm{Re}\,\tilde{\omega}=m \tilde{\Omega}_H^{\hbox{\footnotesize{ext}}}$
and the near-horizon matched asymptotic expansion analysis that we perform below produces a prediction for the frequencies of these modes that will prove to be an excellent approximation  near-extremality.

After this preliminary outline, we are ready to formulate and perform in detail our matched asymptotic expansion to find the NH family of QNMs.
The near-region is defined as the region $\frac{r}{r_+}-1 \ll 1$ and its wavefunction must be regular, in particular, at the horizon. The far-region is the zone $\frac{r}{r_+}-1\gg \sigma$, with $\sigma=1 - \frac{r_{-}}{r_{+}}$ being an off-extremality parameter,  and its wavefunction must obey the outgoing boundary condition at $r\to +\infty$. The two wavefunctions must be simultaneously valid $-$ and thus the free parameters of the two regions must be matched  $-$ in the matching region $\sigma \ll \frac{r}{r_+}-1\ll 1$. We can guarantee that the latter overlap region exists if we take $\sigma$ $-$ which is our expansion parameter $-$ to be small, i.e.\ if $\sigma\ll 1$ and we are thus near-extremality $r_-\lesssim r_+$.

 Under these conditions, in the near-region $\frac{r}{r_+}-1 \ll 1$ we want to simultaneously zoom in around the horizon and approach extremality. For that we first introduce the dimensionless quantities 
\be\label{def:sigma}
y = 1- \frac{r}{r_{+}}, \qquad \sigma = 1 - \frac{r_{-}}{r_{+}}\,,
\ee
where recall that $r=r_-$ and $r=r_+$ are the Cauchy and event horizon locations, respectively, that satisfy $\Delta=0$ with $\Delta = r^2 -2Mr+a^2+Q^2$  as defined in \eqref{KNsoln}. Equivalently, we can also write $\Delta=(r-r_-)(r-r_+)$. Equating these two expressions and their derivatives we can express $M$ and $Q$ as a function of $(r_-,r_+,a)$:
\be
M=\frac{1}{2} \left(r_-+r_+\right)\,,\qquad Q=\sqrt{r_- r_+-a^2}\,.
\ee
From \eqref{def:sigma}, one sees that for $y\ll 1$ one is close to the event horizon and for $\sigma\ll 1$ the Cauchy  and event horizons are very close, {\it i.e.}\ one is close to extremality. 
Next, we take the limit $\sigma \to 0$. From previous works on QNMs of RN, Kerr, KN \cite{Teukolsky:1974yv,Hod:2008zz,Hod:2014uqa,Yang:2012he,Yang:2012pj,Yang:2013uba,Zimmerman:2015trm,Hod:2015xlh,Dias:2015wqa,Dias:2021yju} and even de Sitter black holes \cite{Cardoso:2017soq,Dias:2018etb,Dias:2018ufh}, when we Fourier decompose the modes as $e^{-i\omega t}e^{i m\phi}$, the near-horizon modes are expected to saturate the superradiant bound $\omega=m\Omega_H$ at extremality (this will be further confirmed by our numerical results). Therefore, we expand the frequency about this bound via the redefinition
\be\label{NH:freqRedefA}
\omega =  m \Omega_H^{\hbox{\footnotesize ext}}  + \sigma \,\delta\omega +\mathcal{O}(\sigma^2)\,.
\ee 
Our task is to find $\delta\omega$. In \eqref{NH:freqRedefA} and hereafter, $a$ and $\Omega_H$ in expressions always refer to their extremal values, $a_{\hbox{\footnotesize ext}} $  and $\Omega_H^{\hbox{\footnotesize ext}}$, although we drop the super/subscripts `ext' for brevity.

In these near-extremality conditions, we are ready to find the near-horizon solution of the KN gravito-electromagnetic perturbation equations  \eqref{ChandraEqs}. We substitute
\be \label{fieldRedef}
 \psi_{-2}=\Sigma_{-2}\,,\qquad  \psi_{-1}=\frac{1}{\sigma}\,\Sigma_{-1}\,,
\ee
together with  \eqref{def:sigma}-\eqref{NH:freqRedefA} into the set of two coupled PDEs \eqref{ChandraEqs}, and keep only the {\it leading order} terms in the $\sigma$ expansion.
After this near-horizon/near-extremal procedure, we still have a set of two coupled PDEs, but this time for $\{\Sigma_{-2},\Sigma_{-1}\}$ and they are expected to only capture the properties of the solution in the near-horizon region of the full near-extremal solution. 

Next we attempt to separate variables. In the KN system $-$ described by a coupled pair of PDEs $-$ this might seem bound to fail. So it is enlightening to make a small diversion from our exposition to explain the motivation for even considering this possibility. 
An extremal KN BH has $Q=\sqrt{r_+^2-a^2}$. Similar to the Kerr case \cite{Bardeen:1999px}, the near-horizon limit of the extremal KN black hole (NHEKN) can be obtained by performing the coordinate transformations $(t,r,x,\phi)\to(T,Z,x,\Phi)$ with
\be
t=\frac{r_+^2+a^2}{r_+}\frac{T}{\varepsilon}\,,\qquad r=r_+\left(1+\frac{\varepsilon}{Z}\right), \qquad \phi=\frac{a}{r_+}\frac{T}{\varepsilon}+\Phi
\ee
in the KN solution \eqref{KNsoln} and taking the limit $\varepsilon\to 0$ (recall that $x=\cos\theta$). This yields the near-horizon geometry of the extremal KN solution (which also solves the original Einstein-Maxwell equation):
\begin{subequations}\label{NHEKN}
\begin{align}
& ds^2 \big |_{\hbox{\tiny NHEKN}}=\left(r_+^2+a^2 x^2 \right)\left[ 
\frac{-\dd T^2+\dd Z^2}{Z^2}+\frac{\dd x^2}{1-x^2} \right. \notag \\
& \hspace{50mm}\left. +\frac{1-x^2}{\left(r_+^2+a^2 x^2\right)^2}
\left(\left(r_+^2+a^2\right)\dd \Phi +\frac{2 a r_+ \dd T}{Z}\right)^2
\right],\\
& A |_{\hbox{\tiny NHEKN}}= \sqrt{r_+^2-a^2} \,\dd T\,.
\end{align}
\end{subequations}
Surfaces of constant $x$ are warped $AdS_3$ geometries; that is they correspond to a
circle fibred over $AdS_2$ (parametrized by $T$ and $Z$) with warping parameter
$\frac{1-x^2}{\left(r_+^2+a^2 x^2\right)^2}$. The isometry group is $SL(2,R) \times U(1)$. 
Consequently, perturbations in NHEKN can be expanded in terms of the $AdS_2$ harmonics and thus they separate into a radial and angular part.
This observation is relevant for our purposes because, returning to the full KN geometry, it suggests that near-extremality and near the horizon the two coupled PDEs for $\{\Sigma_{-2},\Sigma_{-1}\}$ might be amenable to a solution by separation of variables. 

With this strong motivation at hand, we return to the coupled system of two PDEs for $\{\Sigma_{-2},\Sigma_{-1}\}$ described above and we attempt the separation {\it ans\"atze}
\be\label{NH:separationAnsatz}
\Sigma_{-2}(y,x)=Y_1(y)X_1(x)\,, \qquad \Sigma_{-1}(y,x)=Y_2(y)X_2(x)\,.
\ee
Introducing the adimensional quantities $\hat{a}=a/r_+$ and $\delta\hat{\omega}=\delta\omega\, r_+$ this yields the two equations:
\begin{subequations}\label{NH:towardsSep}
\begin{align}
& \frac{\mathrm{ODE}_{Y_1}(Y_1;m,\delta\hat{\omega},\lambda_1)}{Y_1} \notag\\
& \qquad +\bigg\{\frac{\rho_{12}(X_2;m)}{X_1} \frac{1}{Y_1} \bigg[ y (y+1)Y_2'- \left(
1+2 y \left(1-\frac{i \hat{a}  m}{1+\hat{a}^2}\right) -i \left(1+\hat{a}^2\right) \delta\hat{\omega}\right)Y_2 \bigg]  \nonumber \\[2mm]
&\hspace{3.4cm} -\frac{\left(1-x^2\right) \left[1-\hat{a} ^2 \left(4-3 x^2\right)-2 i \hat{a}  x \left(2+\hat{a}^2\right) \right]}{1+ \hat{a}^2  \left(2-3 x^2\right)+2 i \hat{a} x \left(1+2\hat{a}^2\right)}
\frac{\mathcal{A}_1(X_1;m,\lambda_1)}{X_1} \bigg\}=0\,,  \label{NH:towardsSep1} \\[2mm]
& \frac{\mathrm{ODE}_{Y_2} (Y_2;m,\delta\hat{\omega},\lambda_2)}{Y_2} +\bigg\{
\frac{\rho_{21}(X_1;m)}{X_2} \frac{1}{Y_2} \bigg[  
Y_1'-\frac{i \left(2 \hat{a}  m y+\left(1+\hat{a}^2\right)^2 \delta\hat{\omega}\right)}{\left(1+\hat{a}^2\right) y (y+1)}Y_1 \bigg]    \nonumber \\
& \hspace{3.4cm} -\frac{\left(1-x^2\right) \left[1+\hat{a} ^2 \left(2-3 x^2\right)+2 i \hat{a} x \left(1+2\hat{a}^2\right)\right]}{1-\hat{a}^2(4 - 3 x^2) -2 i \hat{a}  x \left(2+\hat{a}^2\right)}
\frac{\mathcal{A}_2(X_2;m,\lambda_2)}{X_2} \bigg\}=0\,,\label{NH:towardsSep2}
\end{align}
\end{subequations}
where $\lambda_1$ and $\lambda_2$ are the two separation constants of the problem that only depend on $m$ and $\hat{a}=\hat{a}_{\hbox{\footnotesize ext}}$.
Furthermore,
\begin{subequations}\label{NH:definitions1}
\begin{align}
& \rho_{12}(X_2;m)=\frac{1}{\left(1+\hat{a}^2\right) \sqrt{1-x^2} (\hat{a}  x-i) \Big[1+\hat{a}  \Big(2 \hat{a} +x \left(4 i \hat{a} ^2-3 \hat{a}  x+2 i\right)\Big)\Big]^2 } 
\nonumber \\
&\hspace{0.8cm}
\bigg\{ 
2 \sqrt{1-\hat{a} ^2} (\hat{a}  x+i)^2  \bigg[m \Big(2 i \hat{a} ^2-\hat{a}  x \left(4 \hat{a} ^2+3 i \hat{a}  x+2\right)+i\Big) \left(\hat{a} ^2 x^2+1\right)^2 \nonumber \\
&\hspace{0.8cm} +\left(1+\hat{a}^2\right) \bigg(-4 \hat{a} ^5 x^2 \left(3 x^2-4\right) 
-i \hat{a} ^4 x \left(6 x^4-29 x^2+22\right)-\hat{a} ^3 \left(3 x^2-5\right) \left(5 x^2-2\right) \nonumber \\
&\hspace{0.8cm} +19 i \hat{a} ^2 x \left(x^2-1\right)+\hat{a}  \left(7 x^2-5\right)-i x\bigg)\bigg] X_2 \nonumber \\
& -\!2 \left(1+\hat{a}^2\right)\!\! \sqrt{1-\hat{a} ^2} \left(1-x^2\right) \left(\hat{a} ^2 x^2+1\right) (\hat{a}  x+i)^2 \bigg[\hat{a}  \Big(2 i \hat{a} -x \left(4 \hat{a} ^2+3 i \hat{a}  x+2\right)\!\!\Big)\!+i\bigg]X_2' \!
\bigg\}, \\
& \rho_{21}(X_1;m)=\frac{2 \left(1-\hat{a} ^2\right)^{3/2}}{\left(1+\hat{a}^2\right) \sqrt{1-x^2} \left(\hat{a} ^2 x^2+1\right) (\hat{a}  x+i) \Big[1-4 \hat{a} ^2+3 \hat{a} ^2 x^2-2 i \left(\hat{a} ^2+2\right) \hat{a}  x\Big]^2}
\nonumber \\
&\hspace{0.8cm}\bigg\{
\bigg[i m \left(-4 \hat{a} ^2+3 \hat{a} ^2 x^2-2 i \left(\hat{a} ^2+2\right) \hat{a}  x+1\right) \left(\hat{a} ^2 x^2+1\right)^2
 \nonumber \\
&
\hspace{0.8cm} +2 \left(1+\hat{a}^2\right) \bigg(3 \hat{a}  \left(\hat{a} ^2-1\right)
+\hat{a} ^3\left(1-\hat{a}^2\right) x^4-i \hat{a} ^2 \left(\hat{a} ^2+5\right) x^3 \nonumber \\
& \hspace{6.5cm}
-\left(\hat{a} ^5+10 \hat{a} ^3+\hat{a} \right) x^2+i x \left(2 \hat{a} ^4+5 \hat{a} ^2-1\right)\bigg)\bigg] X_1 \nonumber \\
&
\hspace{0.8cm} + \left(1-x^2\right)\left(1+\hat{a}^2\right) \left(\hat{a} ^2 x^2+1\right) \bigg[-4 i \hat{a} ^2+3 i \hat{a} ^2 x^2+2 \left(\hat{a} ^2+2\right) \hat{a}  x+i\bigg] X_1'
\bigg\},
\end{align}
\end{subequations}
\begin{subequations}\label{NH:definitions2}
\begin{align}
&\mathcal{A}_1(X_1;m,\lambda_1)=
X_1''+\frac{2 x \left[2 i \hat{a} ^4 \left(1-3 x^2\right)-3 \hat{a} ^3 x \left(2-x^2\right)-3 i \hat{a} ^2 x^2-3 \hat{a}  x+i\right]}{\left(1-x^2\right) (\hat{a}  x-i)\left[1+\hat{a} ^2 \left(2-3 x^2\right)+2 i \left(1+2 \hat{a}^2\right) \hat{a}  x\right]}\,X_1' \notag \\[2mm]
&\hspace{100mm} +U_1(x;m,\lambda_1) X_1\,,
\\
&\mathcal{A}_2(X_2;m,\lambda_2)=  X_2''+\frac{1}{x}\left(\frac{1-\hat{a} ^2 \left(3 x^2+4\right)}{1-4 \hat{a} ^2+3 \hat{a} ^2 x^2-2 i \left(\hat{a} ^2+2\right) \hat{a}  x}+\frac{5-7 x^2}{1-x^2}-\frac{2 (3+i \hat{a}  x)}{\hat{a} ^2 x^2+1}\right)\, X_2' \notag \\[2mm]
&\hspace{100mm} +U_2(x;m,\lambda_2) X_2\,,
\end{align}
\end{subequations}
with
{\footnotesize \begin{subequations}\label{NH:definitions3}
\begin{align}
&U_1(x;m,\lambda_1)= \notag \\
&\frac{2 x}{\left(1+2 \hat{a}^2\right) \left(1-x^2\right)^2 (i-\hat{a}  x) \left[1-\hat{a} ^2 \left(4-3 x^2\right)-2 i  \hat{a}  x \left(\hat{a} ^2+2\right)\right] \left[1+\hat{a} ^2 \left(2-3 x^2\right)+2 i \hat{a} x \left(1+2 \hat{a}^2 \right) \right]}
\nonumber \\
&\hspace{0.1cm}
\bigg(-12 \hat{a}  \left(1-\hat{a}^2\right)^2 \left(6 \hat{a} ^4+5 \hat{a} ^2+1\right)+9 \left(4 \hat{a} ^9-11 \hat{a} ^7+\hat{a} ^5\right) x^6-3 i \hat{a} ^4 \left(32 \hat{a} ^6-60 \hat{a} ^4-69 \hat{a} ^2+7\right) x^5
\nonumber \\
&\hspace{0.2cm}
+\hat{a} ^3 \left(-64 \hat{a} ^8-76 \hat{a} ^6+561 \hat{a} ^4+154 \hat{a} ^2-35\right) x^4+i \hat{a} ^2 \left(224 \hat{a} ^8-316 \hat{a} ^6-477 \hat{a} ^4-2 \hat{a} ^2+31\right) x^3
\nonumber \\
&\hspace{0.2cm}
+\hat{a}  \left(64 \hat{a} ^{10}+144 \hat{a} ^8-454 \hat{a} ^6-121 \hat{a} ^4+87 \hat{a} ^2+10\right) x^2-i \left(128 \hat{a} ^{10}-96 \hat{a} ^8-182 \hat{a} ^6+43 \hat{a} ^4+51 \hat{a} ^2+2\right) x\bigg) 
\nonumber \\
&
+\frac{m^2 x}{\left(1+\hat{a}^2\right)^2 \left(1+2 \hat{a}^2\right) \left(1-x^2\right)^2 \left[1-\hat{a}^2 \left(4-3 x^2\right)-2 i  \hat{a}  x \left(\hat{a} ^2+2\right)\right]} 
\nonumber \\
& \hspace{0.1cm}
\bigg( 6 i \left(1-\hat{a} ^2\right) \left(1+2 \hat{a}^2\right) \hat{a} -3 \left(1+2 \hat{a}^2\right) \hat{a} ^6 x^5+2 i \left(2 \hat{a} ^4+5 \hat{a} ^2+2\right) \hat{a} ^5 x^4+\left(8 \hat{a} ^6-10 \hat{a} ^4-19 \hat{a} ^2+3\right) \hat{a} ^2 x^3
\nonumber \\
& \hspace{0.2cm}
-2 i \left(-4 \hat{a} ^6-18 \hat{a} ^4-6 \hat{a} ^2+1\right) \hat{a}  x^2-\left(8 \hat{a} ^2+1\right) \left(1-2 \hat{a} ^2 \left(1+\hat{a}^2\right)\right) x 
\bigg)
\nonumber \\
&
+\frac{4 m x}{\left(1+\hat{a}^2\right) \left(1+2 \hat{a}^2\right) \left(1-x^2\right)^2 \left[1-\hat{a}^2 \left(4-3 x^2\right)-2 i  \hat{a}  x \left(\hat{a} ^2+2\right)\right] \left[1+\hat{a} ^2 \left(2-3 x^2\right)+2 i \hat{a} x \left(1+2 \hat{a}^2 \right) \right]} 
\nonumber \\
& \hspace{0.1cm}
\bigg(
1-2 \hat{a} ^2 \left(3-6 \hat{a} ^6+23 \hat{a} ^4+21 \hat{a} ^2\right)-6 i \left(7 \hat{a} ^4+4 \hat{a} ^2+7\right) \hat{a} ^5 x^5-\left(20 \hat{a} ^6+30 \hat{a} ^4+132 \hat{a} ^2+61\right) \hat{a} ^4 x^4
\nonumber \\
&\hspace{0.2cm}
+2 i \left(25 \hat{a} ^6+66 \hat{a} ^4+99 \hat{a} ^2+26\right) \hat{a} ^3 x^3+3 \left(12 \hat{a} ^8+22 \hat{a} ^6+84 \hat{a} ^4+48 \hat{a} ^2+5\right) \hat{a} ^2 x^2+9 \left(2 \hat{a} ^8+\hat{a} ^6\right) x^6  
\nonumber \\
&\hspace{0.2cm}
-2 i x \left(6 \hat{a} ^9+53 \hat{a} ^7+75 \hat{a} ^5+27 \hat{a} ^3+\hat{a} \right) 
\bigg)  +\frac{\hat{a} ^2 \left(2-3 x^2\right)+2 i \left(2 \hat{a} ^3+\hat{a} \right) x+1}{\left(1-x^2\right) \left[1-\hat{a} ^2 \left(4-3 x^2\right)-2 i \hat{a}  x \left(\hat{a} ^2+2\right)\right]}\,\lambda_1\,,
\\[3mm]
&U_2(x;m,\lambda_2) = \notag \\
&\frac{-\hat{a} ^2 \left(-36 x^4-23 x^2+40\right)-15 i \hat{a}  \left(x^2+3\right) x+13 x^2+10}{3 x^2 \left(1-x^2\right) \left(\hat{a} ^2 \left(3 x^2-4\right)-2 i \left(\hat{a} ^2+2\right) \hat{a}  x+1\right)}
-\frac{-5 i \hat{a}  x^3-35 x^2+7 i \hat{a}  x+25}{3 x^2 \left(1-x^2\right) \left(\hat{a} ^2 x^2+1\right)}
\nonumber \\
&\hspace{0.1cm} 
+\frac{73 x^4-105 x^2+24}{8 x^2 \left(1-x^2\right)^2}-\frac{2}{x^2 (\hat{a}  x+i)^2}
+\frac{27 \left(-\left(\hat{a} ^2 \left(2-x^2\right)\right)-2 i \hat{a}  x+1\right)}{8 \left(1-x^2\right) \left(\hat{a} ^2 \left(2-3 x^2\right)+2 i \left(1+2 \hat{a}^2\right) \hat{a}  x+1\right)}
\nonumber \\
&\hspace{0.1cm} 
+\frac{2 m}{\left(1+\hat{a}^2\right) \left(1-x^2\right)^2 \left[1-\hat{a}^2 \left(4-3 x^2\right)-2 i  \hat{a}  x \left(\hat{a} ^2+2\right)\right] \left(\hat{a} ^2 \left(2-3 x^2\right)+2 i \left(1+2 \hat{a}^2\right) \hat{a}  x+1\right)}
\nonumber \\
&\hspace{0.1cm} 
\bigg(9 i \left(3 \hat{a} ^3-2 \hat{a} ^5-\hat{a} \right)+9 \hat{a} ^6 x^7-3 i \hat{a} ^5 \left(5 \hat{a} ^2+7\right) x^6-\hat{a} ^4 \left(4 \hat{a} ^4+70 \hat{a} ^2+7\right) x^5+i \hat{a} ^3 \left(55 \hat{a} ^4+79 \hat{a} ^2+10\right) x^4
\nonumber \\
&\hspace{0.2cm} 
+3 \hat{a} ^2 \left(4 \hat{a} ^6+42 \hat{a} ^4+11\right) x^3+i \hat{a}  \left(-42 \hat{a} ^6-38 \hat{a} ^4-35 \hat{a} ^2+7\right) x^2+\left(-36 \hat{a} ^6+34 \hat{a} ^4-26 \hat{a} ^2+1\right) x\bigg)
\nonumber \\
&
-\frac{m^2}{\left(1+\hat{a}^2\right)^2 \left(1-x^2\right)^2 \left(\hat{a} ^2 \left(2-3 x^2\right)+2 i \left(1+2 \hat{a}^2\right) \hat{a}  x+1\right)} 
\nonumber \\
&\hspace{0.1cm} 
\bigg(4 i \hat{a} ^7 x^5+\hat{a} ^6 x^4 \left(2-3 x^2\right)-2 i \hat{a} ^5 x \left(-x^4-8 x^2+4\right)-\hat{a} ^4 \left(17 x^4-32 x^2+16\right)+4 i \hat{a} ^3 x \left(5 x^2-3\right)
\nonumber \\
&\hspace{0.2cm} 
+\hat{a} ^2 \left(6-5 x^2\right)+2 i \hat{a}  x+1\bigg)
-\frac{1-\hat{a} ^2 \left(4-3 x^2\right)-2 i \left(\hat{a} ^2+2\right) \hat{a}  x}{4 \left(1+\hat{a}^2\right)^2 \left(1-x^2\right) \left[1+\hat{a} ^2 \left(2-3 x^2\right)+2 i \left(1+2 \hat{a}^2\right) \hat{a}  x\right]}\,\lambda_2\,. 
\end{align}
\end{subequations} }
\medskip

\noindent Finally, in \eqref{NH:towardsSep}, $\mathrm{ODE}_{Y_1}(Y_1;m,\delta\hat{\omega},\lambda_1)$ and $\mathrm{ODE}_{Y_2}(Y_2;m,\delta\hat{\omega},\lambda_2)$ are two second order differential operators acting on $Y_1$ and $Y_2$ with the property that
\begin{subequations}\label{NH:radialODEs}
\begin{align}
& \mathrm{ODE}_{Y_1}(Y_1;m,\delta\hat{\omega},\lambda_1)=0 \quad\ \Leftrightarrow \quad y (y+1) Y_1''-(2 y+1) Y_1'+V_1(y;m,\delta\hat{\omega},\lambda_1) Y_1=0, \label{NH:radialODEs1} \\
& \mathrm{ODE}_{Y_2}(Y_2;m,\delta\hat{\omega},\lambda_2)=0 \quad\ \Leftrightarrow \quad  y (y+1) Y_2''+V_2(y;m,\delta\hat{\omega},\lambda_2) Y_2=0,\label{NH:radialODEs2}
\end{align}
\end{subequations}
where the potentials are
\begin{subequations}\label{NH:radialODEsV}
\begin{align}
& V_1(y;m,\delta\hat{\omega},\lambda_1)=\frac{2 \left(1-4 \hat{a} ^2\right) \left(2-\hat{a} ^2\right)}{1+2\hat{a}^2}
+\frac{4 i \hat{a}  m \left[2+3 y-\hat{a} ^2 (5+3 y)\right]}{\left(1+\hat{a}^2\right) \left(1+2\hat{a}^2\right) (y+1)} \nonumber\\
& \hspace{9.5cm} +\frac{m^2 \left(1+y+8 \hat{a} ^4 y-4 \hat{a} ^2\right)}{\left(1+\hat{a}^2\right)^2 \left(1+2\hat{a}^2\right) (y+1)} \nonumber\\
& \hspace{2.8cm}+\frac{1}{y+1} \left(\frac{\left(1+\hat{a}^2\right)^2 \delta  \tilde{\omega }^2}{y}
+4\delta  \tilde{\omega } \left[\hat{a}  m+i \left(1+\hat{a}^2\right) \left(\frac{1}{2 y}+1\right) \right]\right)
 +\lambda_1\,, \\
& V_2(y;m,\delta\hat{\omega},\lambda_2)=\frac{1}{4}-\frac{4 \hat{a} ^2 m^2}{\left(1+\hat{a}^2\right)^2 (y+1)}-\frac{2 i \hat{a}  m}{\left(1+\hat{a}^2\right) (y+1)}  \nonumber\\
&  \hspace{2.8cm}
+\frac{\left(1+\hat{a}^2\right)^2 \delta  \tilde{\omega }^2+\delta  \tilde{\omega } \left[i \left(1+\hat{a}^2\right)+4 \hat{a}  m y+2 i \left(1+\hat{a}^2\right) y\right]}{y (y+1)} -\frac{\lambda_2}{4 \left(1+\hat{a}^2\right)^2}\,.
\end{align}
\end{subequations}

The reader will notice that in \eqref{NH:towardsSep1} and \eqref{NH:towardsSep2}, the terms that are spoiling the separation of variables are those proportional to $\rho_{12}X_1^{-1}Y_1^{-1}$ and $\rho_{21}X_2^{-1}Y_2^{-1}$, respectively. We can however separate these equations if the factor multiplying $\rho_{12}X_1^{-1}Y_1^{-1}$ in \eqref{NH:towardsSep1} is proportional to $Y_1(y)$ and if the factor multiplying $\rho_{21}X_2^{-1}Y_2^{-1}$ in \eqref{NH:towardsSep2} is proportional to $Y_2(y)$, \ie if
\begin{subequations}\label{NH:Starobinsky}
\begin{align}
& Y_1=K_{12} \bigg\{ y (y+1)Y_2'- \left[
1+2 y \left(1-\frac{i \hat{a}  m}{1+\hat{a}^2}\right) -i \left(1+\hat{a}^2\right) \delta\hat{\omega}\right]Y_2 \bigg\},  \label{NH:Starobinsky1} \\
& Y_2=K_{21} \bigg(  
Y_1'-\frac{i \left(2 \hat{a}  m y+\left(1+\hat{a}^2\right)^2 \delta\hat{\omega}\right)}{\left(1+\hat{a}^2\right) y (y+1)}Y_1 \bigg),\label{NH:Starobinsky2}
\end{align}
\end{subequations}
for constant $K_{12}$ and $K_{21}$ to be determined. If this is the case and \eqref{NH:Starobinsky1} holds,  then the first term in \eqref{NH:towardsSep1} gives the radial equation for $Y_1(y)$, namely \eqref{NH:radialODEs1}, while the term inside curly brackets yields the angular equation for $X_1$. Similarly,  if \eqref{NH:Starobinsky2} holds,  in \eqref{NH:towardsSep2} we clearly identify the radial equation for $Y_2(y)$, namely \eqref{NH:radialODEs2}, and the angular equation for $X_2$ inside the curly brackets. 
However, in order for the separation procedure to be consistent, \eqref{NH:Starobinsky} must still be supplemented by another two relations. Firstly, when we substitute \eqref{NH:Starobinsky1} into \eqref{NH:radialODEs1} we must certainly get a trivial identity after using \eqref{NH:radialODEs2} and its derivative. Similarly, we must get a trivial identity when we substitute \eqref{NH:Starobinsky2} into \eqref{NH:radialODEs2} and use \eqref{NH:radialODEs1} and its derivative. This is the case if and only if the two separation constants of the system are related in a specific way, $\lambda_1=\lambda_1(\lambda_2)$. Secondly, if we substitute \eqref{NH:Starobinsky1} into \eqref{NH:Starobinsky2} we must again obtain a trivial identity after using the equation of motion \eqref{NH:radialODEs2} for $Y_2$. Equivalently, we must also get a trivial identity if we substitute \eqref{NH:Starobinsky2} into \eqref{NH:Starobinsky1} and use the equation of motion \eqref{NH:radialODEs1} for $Y_1$. This is the case if and only if a specific relation $K_{21}=K_{21}(K_{12})$ holds.
Altogether, the two consistency conditions that must be imposed, together with \eqref{NH:Starobinsky}, to get a separated system of equations are: 
\begin{subequations}\label{NH:StarobinskyConstants}
\begin{align}
&\lambda_1=-\frac{\lambda _2}{4 \left(1+\hat{a}^2\right)^2}-\frac{\left(1+\hat{a}^2\right)^2 \left(32 \hat{a} ^4-90 \hat{a} ^2+7\right)+4 \left(8 \hat{a} ^4+1\right) m^2+48 i \hat{a}  \left(1-\hat{a} ^4\right) m}{4 \left(1+\hat{a}^2\right)^2 \left(1+2\hat{a}^2\right)}\,, \label{NH:StarobinskyConstants1}\\
& K_{21}=\frac{1}{K_{12} }\,\frac{\left[4 \hat{a}  m+3 i \left(1+\hat{a}^2\right)\right]^2+\lambda_2}{4 \left(1+\hat{a}^2\right)^2}  \,,
\label{NH:StarobinskyConstants2}
\end{align}
\end{subequations}
where, without loss of generality since this is a linear system, we can set $K_{12}\equiv 1$.

What is the meaning of \eqref{NH:Starobinsky} and \eqref{NH:StarobinskyConstants}?
Recall that in the case of the Teukolsky equation describing perturbations in the Kerr black hole \cite{Teukolsky:1972my}, it is well known that the so-called Starobinsky-Teukolsky relations relate perturbations with spin $s$ to those with spin $-s$ \cite{Chandra:1983} (see also Appendix of \cite{Cardoso:2013pza}). Thus, one interprets relations \eqref{NH:Starobinsky}-\eqref{NH:StarobinskyConstants} as being effectively a kind of Starobinsky-Teukolsky relations for the KN perturbations. In this case they relate the wavefunction of spin $s=-2$ with that of spin $s=-1$ because the perturbations for these two spins are coupled. 

After this long tour, we should recap what we have learned so far. The gravito-electromagnetic perturbations of the KN black are described by a coupled system \eqref{ChandraEqs} of two PDEs for $\{\psi_{-2},\psi_{-1}\}$.
However, if we take its near-horizon limit near extremality, as described in \eqref{def:sigma}-\eqref{fieldRedef}, we get two near-horizon coupled PDEs for $\{\Sigma_{-2},\Sigma_{-1}\}$ that can be solved assuming the separation of variables \eqref{NH:separationAnsatz}. After using the Starobinsky-Teukolsky$-$like relations \eqref{NH:Starobinsky}-\eqref{NH:StarobinskyConstants}, we verify that the system indeed separates. We get two decoupled ODEs  \eqref{NH:radialODEs} for the radial wavefunctions $Y_1(y)$ and $Y_2(y)$. (This decoupling reflects the fact that in NHEKN the radial perturbations are exactly described by the AdS$_2$ harmonics as explained below \eqref{NHEKN}). Once we know the separation constant $\lambda_2$, and thus $\lambda_1$ via \eqref{NH:StarobinskyConstants}, these two ODEs \eqref{NH:radialODEs}  can be solved independently as a quadratic eigenvalue problem for $\delta\hat{\omega}$ (for a given $m$). On the other hand, the angular equations for $X_1$ and $X_2$ -- given by the curly brackets of  \eqref{NH:towardsSep} after using \eqref{NH:Starobinsky}--\eqref{NH:StarobinskyConstants} -- do not decouple. Thus we have to solve this coupled system of two ODEs (that are independent of $\delta\hat{\omega}$) to find the eigenvalue $\lambda_2$ (and thus $\lambda_1$ given in \eqref{NH:StarobinskyConstants1}). This can be done numerically as we discuss later. But we can also solve this coupled ODE system analytically in a large $m$ WKB expansion. This is what we do next.

Substituting \eqref{NH:definitions1}--\eqref{NH:definitions3} and  \eqref{NH:Starobinsky}--\eqref{NH:StarobinskyConstants} into the curly brackets expressions of  \eqref{NH:towardsSep}, we find that \eqref{NH:towardsSep1} is a second order ODE for $X_1$ (hereafter we denote this as the `{\it first}' angular equation) that also depends on $X_2'$ and $X_2$ but not on $X_2''$. Similarly, \eqref{NH:towardsSep2} is a second order ODE for $X_2$ (henceforth denoted as the `{\it second}' angular equation) that also depends on $X_1'$ and $X_1$ but not on $X_1''$.  
If we redefine
\be
X_1(x)=\chi_1(x)\,,\qquad X_2(x)=K_{12}\,\chi_2(x)\,,
\ee
where $K_{12}$ was first introduced in \eqref{NH:Starobinsky1}, we can solve the equation for $\chi_1$ to express $\chi_2'=\chi_2'(\chi_2,\chi_1'',\chi_1',\chi_1)$. We substitute this relation and its derivative into the second angular equation to get a differential equation that can be solved to express $\chi_2=\chi_2(\chi_1''',\chi_1'',\chi_1',\chi_1)$. Substituting this back in the first angular equation we end up with a {\it fourth} order differential equation for $\chi_1$ that no longer depends on $\chi_2$. This is a non-polynomial eigenvalue problem for $\chi_1$ and $\lambda_2$; recall \eqref{NH:StarobinskyConstants1}. Perhaps remarkably, this fourth order differential equation can be solved analytically in a $|m| \gg 1$ WKB expansion to find  $\chi_1$ and $\lambda_2$.

We substitute the WKB {\it ansatz}
\begin{subequations}\label{NH:WKBansatz}
\begin{align}
&\chi_1(x)=e^{m\varphi(x)}\left[ \chi_{1,0}(x)+\frac{\chi_{1,1}(x)}{m}+\frac{\chi_{1,2}(x)}{m^2}+\mathcal{O}\left(1/m^3\right) \right], \label{NH:WKBansatzA} \\
& \lambda_2=\lambda_{2,0}\,m^2+\lambda_{2,1}\,m+\lambda_{2,2}+\frac{\lambda_{2,3}}{m}+\mathcal{O}\left(1/m^2\right) \,, \label{NH:WKBansatzB}
\end{align}
\end{subequations}
into the fourth order ODE and solve it order by order in a standard large $m$ expansion, requiring that the solution is regular at $x=\pm 1$.  
The leading order WKB wavefunction is  
\be
\varphi=\frac{\sqrt{2 \hat{a} ^2+\hat{a} ^4 x^2+1}}{\hat{a} ^2+1}-\tanh ^{-1}\left(\frac{\sqrt{2 \hat{a} ^2+\hat{a} ^4 x^2+1}}{\hat{a} ^2+1}\right)-\frac{\sqrt{2 \hat{a} ^2+1}}{\hat{a} ^2+1}+\tanh ^{-1}\left(\frac{\sqrt{2 \hat{a} ^2+1}}{\hat{a} ^2+1}\right)
\ee
and the separation constant $\lambda_2$ is given by \eqref{NH:WKBansatz} with WKB coeficients
\begin{subequations}\label{NH:WKBansatzCoef}
\begin{align}
&\lambda_{2,0}=4 \left(1-4 \hat{a} ^2\right),\qquad \qquad
\lambda_{2,1}= -4 \left(1+\hat{a}^2\right) \left(2 \sqrt{1-\hat{a} ^2}-\sqrt{1+2 \hat{a} ^2}\right),
   \\
&\lambda_{2,2}= \frac{3 \sqrt{1-\hat{a} ^2} \left(1+\hat{a}^2\right)^2 \left(3-726 \hat{a} ^{10}-253 \hat{a} ^8+128 \hat{a} ^6-74 \hat{a} ^4-50 \hat{a} ^2\right)}{\left(1+2\hat{a}^2\right) \left[\left(66 \hat{a} ^6-5 \hat{a} ^4-12 \hat{a} ^2+5\right) \sqrt{1-\hat{a} ^2}+4 \left(1-\hat{a} ^4\right) \sqrt{2 \hat{a} ^2+1}\right]},  
 \tag{\stepcounter{equation}\theequation}   \\
&\lambda_{2,3}=
\bigg[ 
4 \left(1+2\hat{a}^2\right)^{7/2}\bigg(
578577650112 \hat{a} ^{40}-338129795520 \hat{a} ^{38}-1042453021104 \hat{a} ^{36}
\nonumber \\
&\hspace{0.4cm} +1170932108544 \hat{a} ^{34}+243872180244 \hat{a} ^{32}-1092788709804 \hat{a} ^{30}+457571937931 \hat{a} ^{28}
\nonumber \\
&\hspace{0.4cm} 
+286639850738 \hat{a} ^{26}-371225227587 \hat{a} ^{24} +75821376048 \hat{a} ^{22}+83823143199 \hat{a} ^{20}
\nonumber \\
&\hspace{0.4cm} 
-64522516578 \hat{a} ^{18}+5397537793 \hat{a} ^{16}+11870759300 \hat{a} ^{14}-5939331087 \hat{a} ^{12}
\nonumber \\
&\hspace{0.4cm} 
+15670254 \hat{a} ^{10}+798959271 \hat{a} ^8-269248008 \hat{a} ^6-8868395 \hat{a} ^4+20327618 \hat{a} ^2-4782969
\bigg)
\nonumber \\
&\hspace{0.4cm} 
+4 \sqrt{1-\hat{a} ^2} \left(1+2\hat{a}^2\right)^3\bigg(
661231600128 \hat{a} ^{40}-788969522880 \hat{a} ^{38}-475886378880 \hat{a} ^{36}
\nonumber \\
&\hspace{0.4cm} 
+1029138506352 \hat{a} ^{34}-630648141552 \hat{a} ^{32}-452699156052 \hat{a} ^{30}+658166339168 \hat{a} ^{28}
\nonumber \\
&\hspace{0.4cm} 
-186975958943 \hat{a} ^{26}-249892000005 \hat{a} ^{24}+178743692406 \hat{a} ^{22}-3249242106 \hat{a} ^{20}
\nonumber \\
&\hspace{0.4cm} 
-56479482309 \hat{a} ^{18}+20902690721 \hat{a} ^{16}+3663601312 \hat{a} ^{14}-5845481340 \hat{a} ^{12} 
\nonumber \\
&\hspace{0.4cm} 
+1100552199 \hat{a} ^{10}+410656173 \hat{a} ^8-279409506 \hat{a} ^6+19829366 \hat{a} ^4
\nonumber \\
&\hspace{0.4cm}
+13153165 \hat{a} ^2-4782969 
\bigg)\bigg]^{-1}
\nonumber \\
&\hspace{0.4cm} 
\bigg[ 
3 \hat{a} ^2 \sqrt{1-\hat{a} ^2} \left(1+\hat{a}^2\right)^3 \sqrt{2 \hat{a} ^2+1} \bigg(
90588729217536 \hat{a} ^{46}+93586813404480 \hat{a} ^{44}
\nonumber \\
&\hspace{0.4cm} 
-64234642488192 \hat{a} ^{42} -54181551934224 \hat{a} ^{40}+14733709326864 \hat{a} ^{38}
\nonumber \\
&\hspace{0.4cm}
-34708141099764 \hat{a} ^{36}-8979094220672 \hat{a} ^{34}+34432474064505 \hat{a} ^{32} -10922161747605 \hat{a} ^{30}
\nonumber \\
&\hspace{0.4cm} 
-23041644949212 \hat{a} ^{28}+5136927583340 \hat{a} ^{26}+4733507876355 \hat{a} ^{24}-3578226571619 \hat{a} ^{22}
\nonumber \\
&\hspace{0.4cm} 
-898929274206 \hat{a} ^{20}+753565243446 \hat{a} ^{18}-135077374365 \hat{a} ^{16}-174223122235 \hat{a} ^{14}
\nonumber \\
&\hspace{0.4cm} 
+33089919120 \hat{a} ^{12}+8380363168 \hat{a} ^{10}-9890782275 \hat{a} ^8-803782461 \hat{a} ^6
\nonumber \\
&\hspace{0.4cm} 
+541670718 \hat{a} ^4-148272034 \hat{a} ^2-57395628  \bigg)
\nonumber \\
&\hspace{0.4cm} 
+3 \hat{a} ^2 \left(1+\hat{a}^2\right)^3 \bigg( 
158530276130688 \hat{a} ^{48}+192260601732672 \hat{a} ^{46}-226279077675552 \hat{a} ^{44}
\nonumber \\
&\hspace{0.4cm} 
-257580189150768 \hat{a} ^{42}+238634465705064 \hat{a} ^{40}+187478664334236 \hat{a} ^{38}
\nonumber \\
&\hspace{0.4cm} 
-167948153974214 \hat{a} ^{36}-79050787933609 \hat{a} ^{34}+69165996968940 \hat{a} ^{32}
\nonumber \\
&\hspace{0.4cm} 
+1562277529575 \hat{a} ^{30}-26149776558142 \hat{a} ^{28}+6310859786413 \hat{a} ^{26} +3820171951948 \hat{a} ^{24}
\nonumber \\
&\hspace{0.4cm} 
-4424582883901 \hat{a} ^{22}-417658252182 \hat{a} ^{20}+868831525263 \hat{a} ^{18}-249677209480 \hat{a} ^{16}
\nonumber \\
&\hspace{0.4cm} 
-170706582299 \hat{a} ^{14}+47404470046 \hat{a} ^{12}+4708012127 \hat{a} ^{10}-10932078636 \hat{a} ^8
\nonumber \\
&\hspace{0.4cm} 
-398469675 \hat{a} ^6+532105820 \hat{a} ^4-176969858 \hat{a} ^2-57395628
\bigg)\bigg].
\end{align}
\end{subequations}
Of course, now that we have the eigenpair $\lambda_2$ and $\chi_1(x)$ (in the WKB approximation) we can straightforwardly obtain $\lambda_1$ and $\chi_2(x)$ using \eqref{NH:StarobinskyConstants1} and the aforementioned relation $\chi_2=\chi_2(\chi_1''',\chi_1'',\chi_1',\chi_1)$. This terminates our WKB analysis of the angular equations.

\begin{figure}[th]
\centering
\includegraphics[width=.47\textwidth]{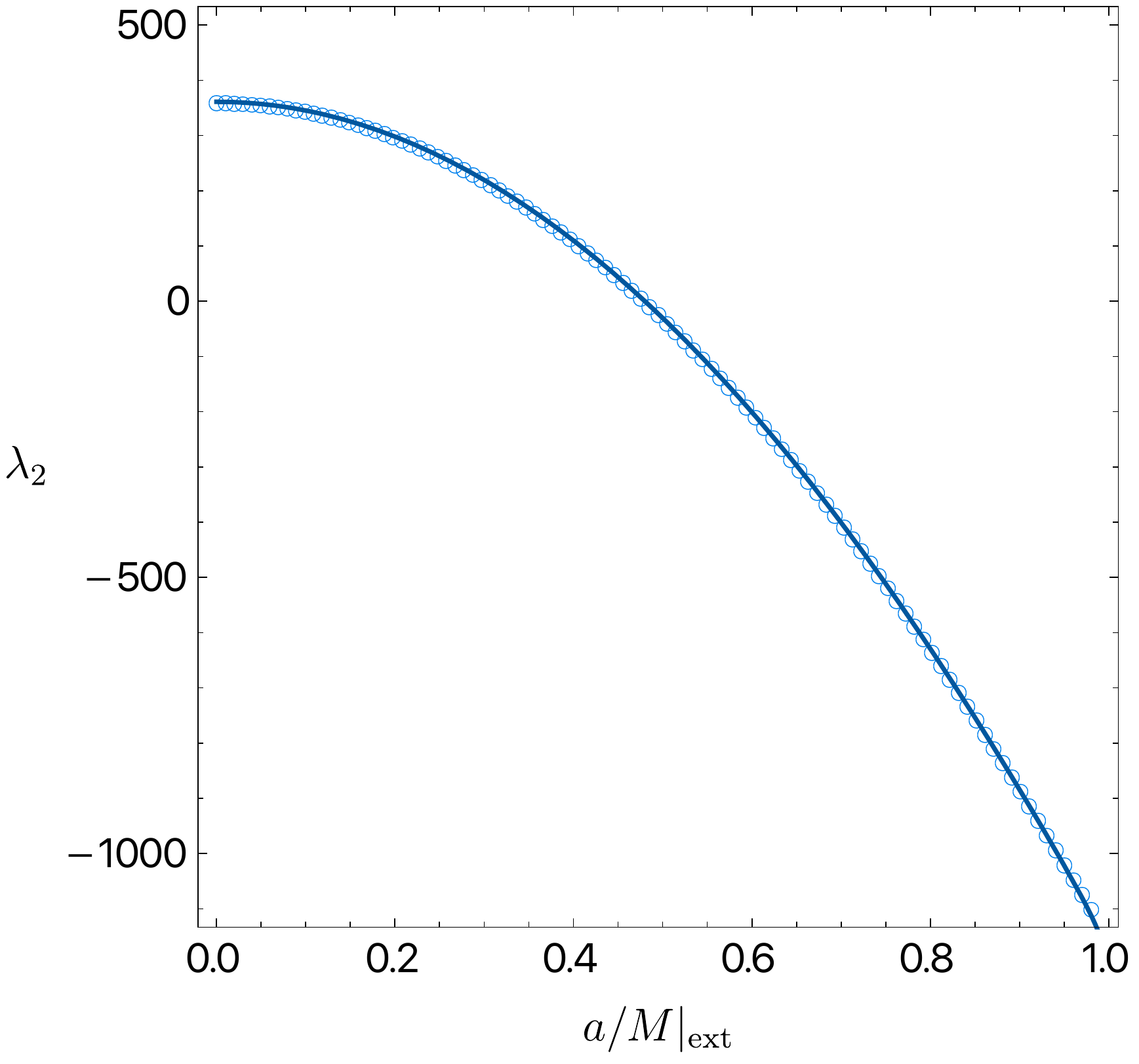}
\hspace{0.5cm}
\includegraphics[width=.45\textwidth]{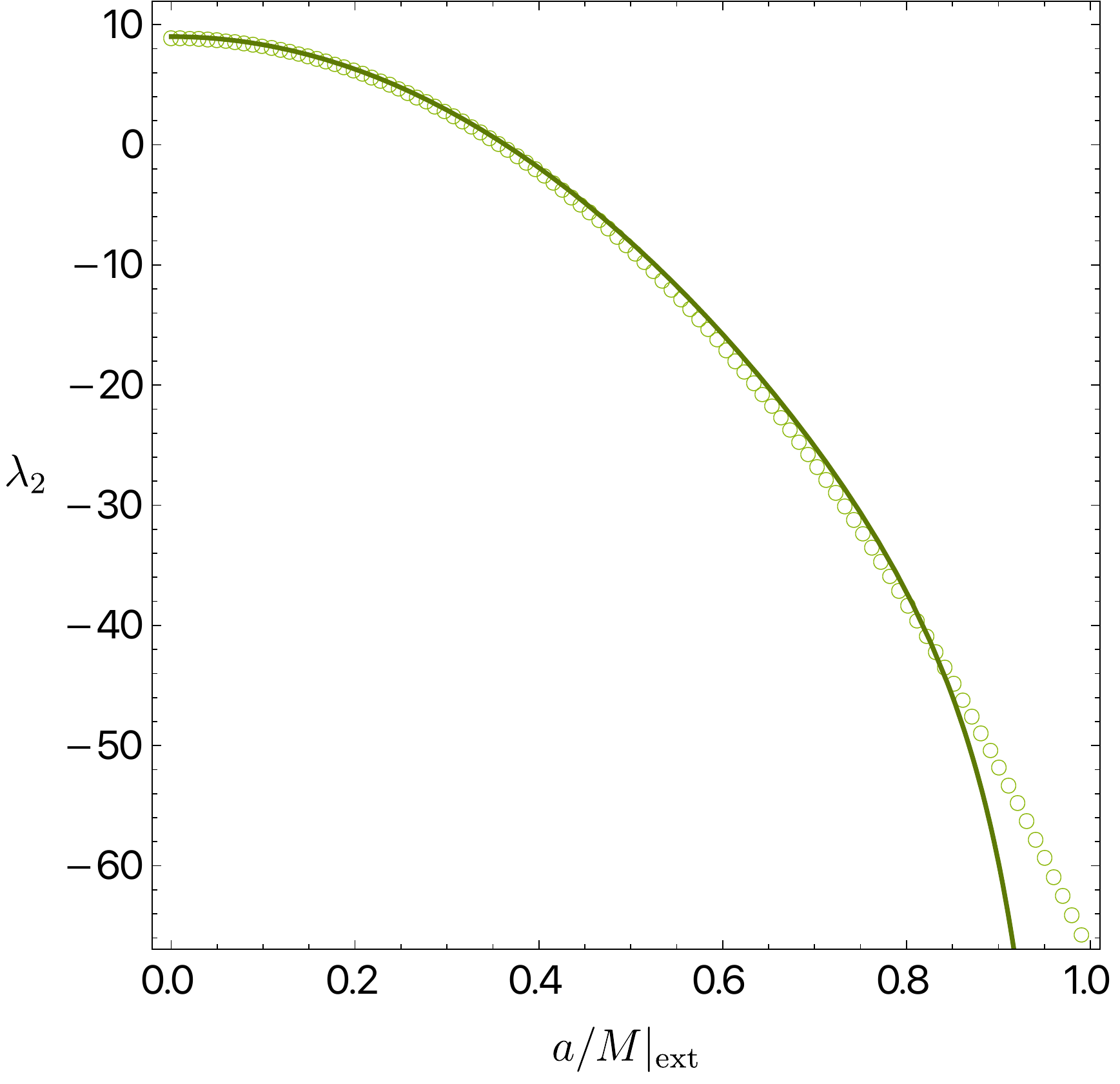}
\caption{Comparing the WKB result (continuous line) for $\lambda_2(m,\hat{a})$ with the exact result (circles).
{\bf Left panel:} $m=10$ case.  {\bf Right panel:} $m=2$ case.}
\label{Fig:WKBlambda2}
\end{figure}  

We can also solve numerically the coupled pair of angular ODEs for $X_1$ and $X_2$ to check that the WKB result is indeed a good approximation, even for $m=2$.
For $m\geq 2$, regularity at $x=\pm 1$ requires that we keep the $X_1, X_2$ solution that behaves as $(1-x)^{\frac{1}{2}(s+m)}$ at $x=1$ and as $(1+x)^{\frac{1}{2}(-s+m)}$  at $x=-1$ where $s=-2,-1$ for $X_1, X_2$, respectively. We can impose these boundary conditions straightforwardly  if we introduce the field redefinition
\be
X_1=(1-x)^{-1+\frac{m}{2}}(1+x)^{1+\frac{m}{2}}Q_1(x)\,, \qquad 
X_2=(1-x)^{-\frac{1}{2}+\frac{m}{2}}(1+x)^{\frac{1}{2}+\frac{m}{2}}Q_2(x)
\ee
and solve the two coupled second order ODEs for smooth $Q_1$ and $Q_2$ and the eigenvalue $\lambda_2$, after using \eqref{NH:StarobinskyConstants1}. As explained above, we use a Newton-Raphson algorithm with pseudospectral discretization \cite{Dias:2015nua}.  
In Fig.~\ref{Fig:WKBlambda2} we compare the WKB result \eqref{NH:WKBansatz}-\eqref{NH:WKBansatzCoef} with the numerical $\lambda_2$. We see that, as expected, for large $m$, $m=10$ (left panel), there is perfect agreement between the WKB result (continuous dark-blue curve) and the numerical result (blue circles). However, as the right panel demonstrates, the WKB approximation (continuous dark-green line) proves to be a good approximation to the exact result (green circles) already for $m=2$. Also note that as $\hat{a}$ increases from $\hat{a}=0$, $\lambda_2$ changes sign from positive to negative. This fact will be important later.

Having solved the angular equations we can now focus our attention on the radial ODEs \eqref{NH:radialODEs}-\eqref{NH:radialODEsV}. Recall that we can solve one of these, e.g.\  \eqref{NH:radialODEs2} for $Y_2$, and the solution for $Y_1$ is then straightforwardly given by the Starobinsky-Teukolsky  differential map \eqref{NH:Starobinsky1}. Further recall that \eqref{NH:radialODEs2} is a quadratic  eigenvalue problem in $\delta\hat{\omega}$. This ODE turns out to be a standard hypergeometric equation with most general solution given by
\begin{eqnarray}
&&\hspace{-0.4cm} Y_2=(y+1)^{\frac{2 i \hat{a}  m}{1+\hat{a}^2}-i \left(1+\hat{a}^2\right) \delta  \tilde{\omega }} \notag \\
&&\bigg[
c_1 y^{i \left(1+\hat{a}^2\right) \delta  \tilde{\omega }} \, _2F_1\left(\frac{4 i m \hat{a} -\sqrt{\lambda _2}}{2 \left(1+\hat{a}^2\right)}-\frac{1}{2},\frac{4 i m \hat{a} +\sqrt{\lambda _2}}{2 \left(1+\hat{a}^2\right)}-\frac{1}{2};2 i \left(1+\hat{a}^2\right) \delta  \tilde{\omega };-y\right)
\nonumber \\
&& \hspace{-0.4cm} +c_2 y^{1-i \left(1+\hat{a}^2\right) \delta  \tilde{\omega }} \, _2F_1\left(\frac{4 i m \hat{a} -\sqrt{\lambda _2}}{2 \left(1+\hat{a}^2\right)}+\frac{1}{2}-2 i \left(1+\hat{a}^2\right) \delta  \tilde{\omega }, \right. \notag \\
&& \hspace{40mm} \left.\frac{4 i m \hat{a} +\sqrt{\lambda _2}}{2 \left(1+\hat{a}^2\right)}+\frac{1}{2}-2 i \left(1+\hat{a}^2\right) \delta  \tilde{\omega }; 2-2 i \left(1+\hat{a}^2\right) \delta  \tilde{\omega };-y\right)
\bigg]\nonumber \\
\end{eqnarray}
where $c_1, c_2$ are arbitrary integration constants.
At the event horizon, $y=0$, this solution behaves as $Y_2|_{y=0}\sim c_1 y^{i \left(1+\hat{a}^2\right) \delta\hat{\omega}}+c_2 y^{1-i \left(1+\hat{a}^2\right) \delta\hat{\omega}}$. Regularity in ingoing Eddington-Finkelstein coordinates at the future event horizon  requires that we set $c_1=0$ to eliminate the outgoing modes. On the other hand, far away from the horizon, \ie at large $y$, the regular solution at the horizon behaves as 
\begin{eqnarray}\label{NH:farSol}
Y_2\big|_{y\gg 1}&\sim& c_2 \Gamma \Big(2-2 i \left(1+\hat{a}^2\right) \delta  \tilde{\omega }\Big) \notag \\
&& \hspace{-10mm}
\bigg[
\frac{\Gamma \left(-\frac{\sqrt{\lambda _2}}{1+\hat{a}^2}\right)}{\Gamma \left(\frac{3}{2}-\frac{4 i m \hat{a} +\sqrt{\lambda _2}}{2 \left(1+\hat{a}^2\right)}\right) \Gamma \left(\frac{1}{2}+\frac{4 i \left(m \hat{a} -\left(1+\hat{a}^2\right)^2 \delta  \tilde{\omega }\right)-\sqrt{\lambda _2}}{2 \left(1+\hat{a}^2\right)}\right)}\,y^{\frac{1}{2}\left(1-\frac{\sqrt{\lambda _2}}{1+\hat{a}^2}\right)}
\nonumber \\
&&
+ \frac{\Gamma \left(\frac{\sqrt{\lambda _2}}{1+\hat{a}^2}\right)}{\Gamma \left(\frac{3}{2}-\frac{4 i m \hat{a} -\sqrt{\lambda _2}}{2 \left(1+\hat{a}^2\right)}\right) \Gamma \left(\frac{1}{2}+\frac{4 i \left(m \hat{a} -\left(1+\hat{a}^2\right)^2 \delta  \tilde{\omega }\right)+\sqrt{\lambda _2}}{2 \left(1+\hat{a}^2\right)}\right)}\,y^{\frac{1}{2}\left(1+\frac{\sqrt{\lambda _2}}{1+\hat{a}^2}\right)}
\bigg].
\end{eqnarray}
Assume for now that $\lambda_2>0$. From Fig.~\ref{Fig:WKBlambda2}, this happens when $\hat{a}_{\hbox{\footnotesize ext}}=\sqrt{1-\hat{Q}^2}$ is small, which occurs for large $\hat{Q}$.
For $\lambda_2>0$, at large $y$, the solution $y^{\frac{1}{2}\left(1-\frac{\sqrt{\lambda _2}}{1+\hat{a}^2}\right)}$ in \eqref{NH:farSol} decays while $y^{\frac{1}{2}\left(1+\frac{\sqrt{\lambda _2}}{1+\hat{a}^2}\right)}$ diverges.\footnote{Note that the metric components that must be a regular 2-tensor behave as $y^{\pm \frac{1}{2}\frac{\sqrt{\lambda _2}}{1+\hat{a}^2}}$.}

In the context of a matched asymptotic expansion, the large behaviour of the near-region (near-horizon) solution \eqref{NH:farSol} must now be matched with the far-region solution  (near extremality). As explained at the beginning of this section, we expect the near-horizon modes we are looking into to have wavefunctions that die-off very quickly away from the black hole horizon (near extremality). This will be confirmed by our numerical analysis. Therefore, as a first approximation $-$ that we henceforth call a {\it `poor man's matched asymptotic expansion (MAE)} $-$ we take the far region to be described by a vanishing wavefunction. That is to say, in the overlapping region, we match the near-region solution \eqref{NH:farSol} with $Y_2|_{\hbox{\footnotesize far}}\simeq 0$.\footnote{Ideally, we would also solve the far-region equations to obtain the sub-leading far-region solution but in the KN background we cannot do this analytically.} It is important to emphasize that this is an {\it ansatz} or educated guess that we cannot argue for in a formal mathematical away that goes deeper than the above heuristics. It is ultimately only validated {\it a posteriori} by the fact that the final quantization agrees with the numerical results for the frequency spectra (indeed, $Y_2|_{\hbox{\footnotesize far}}$ is never  exactly zero and thus a small component of the divergent term in \eqref{NH:farSol} should be used in a proper matching). This {\it ansatz} requires that we kill the solution $y^{\frac{1}{2}\left(1+\frac{\sqrt{\lambda _2}}{1+\hat{a}^2}\right)}$ in  \eqref{NH:farSol} that  diverges for large $y$.
Since $\Gamma(-n)\to\infty $ for $n\in \mathbb{N}_0$, this is the case if we quantize the frequency correction to be such that the argument of the gamma function in the denominator of the divergent term is a non-positive integer $n$:
\be \label{NH:freqCorrection}
\delta\hat{\omega}\simeq \frac{m\hat{a}}{\left(1+\hat{a}^2\right)^2}-\frac{i}{4 \left(1+\hat{a}^2\right)}\left(1+2n+\frac{\sqrt{\lambda_2(m,\tilde{a})}}{1+\hat{a}^2}\right)\,,\quad n=0,1,2,3,\cdots 
\ee 
Inserting this frequency correction into the frequency expansion \eqref{NH:freqRedefA} one gets the final expression for the frequency in units of $r_+$:  $\hat{\omega} = m \hat{\Omega}_H + \sigma \,\delta\hat{\omega}$. We can now convert this into units of $M$ by multiplying this expression by $M/r_+$ (since $\hat{\omega}M/r_+=\tilde{\omega}=\omega M$) and expanding it in terms of $\sigma$ while keeping terms only up to $\mathcal{O}(\sigma)$ (since all our analysis is valid only up to this order). This yields the frequency quantization for the near-horizon (NH) QNMs which can be written as:
\begin{equation}\label{NH:freq}
 \tilde{\omega}_{\hbox{\tiny MAE}} \simeq  
 \frac{m\tilde{a}}{1+\tilde{a}^2}
+\sigma \bigg[\frac{m \tilde{a}(1-\tilde{a}^2)}{2(1+ \tilde{a}^2)^2}-\frac{i}{4} \frac{1+2n}{1+ \tilde{a}^2}-\,\frac{\sqrt{-\lambda_2(m,\tilde{a})}}{4(1+\tilde{a}^2)^2}\bigg]  +\mathcal{O}\left(\sigma^2\right) \,,\quad n=0,1,2,3,\cdots 
\end{equation}
where $\tilde{a}$ in this expression must be evaluated at extremality, \ie 
$\tilde{a}=\tilde{a}_{\hbox{\footnotesize ext}}$, the off-extremal parameter $\sigma$ is defined in \eqref{def:sigma}, and we have defined $\sqrt{z}$ to be such that $\mathrm{Re}(\sqrt{z})>0$ ($\mathrm{Im}(\sqrt{z})>0$) for positive (negative) values of $z$.

\begin{figure}[th]
\centering
\includegraphics[width=.472\textwidth]{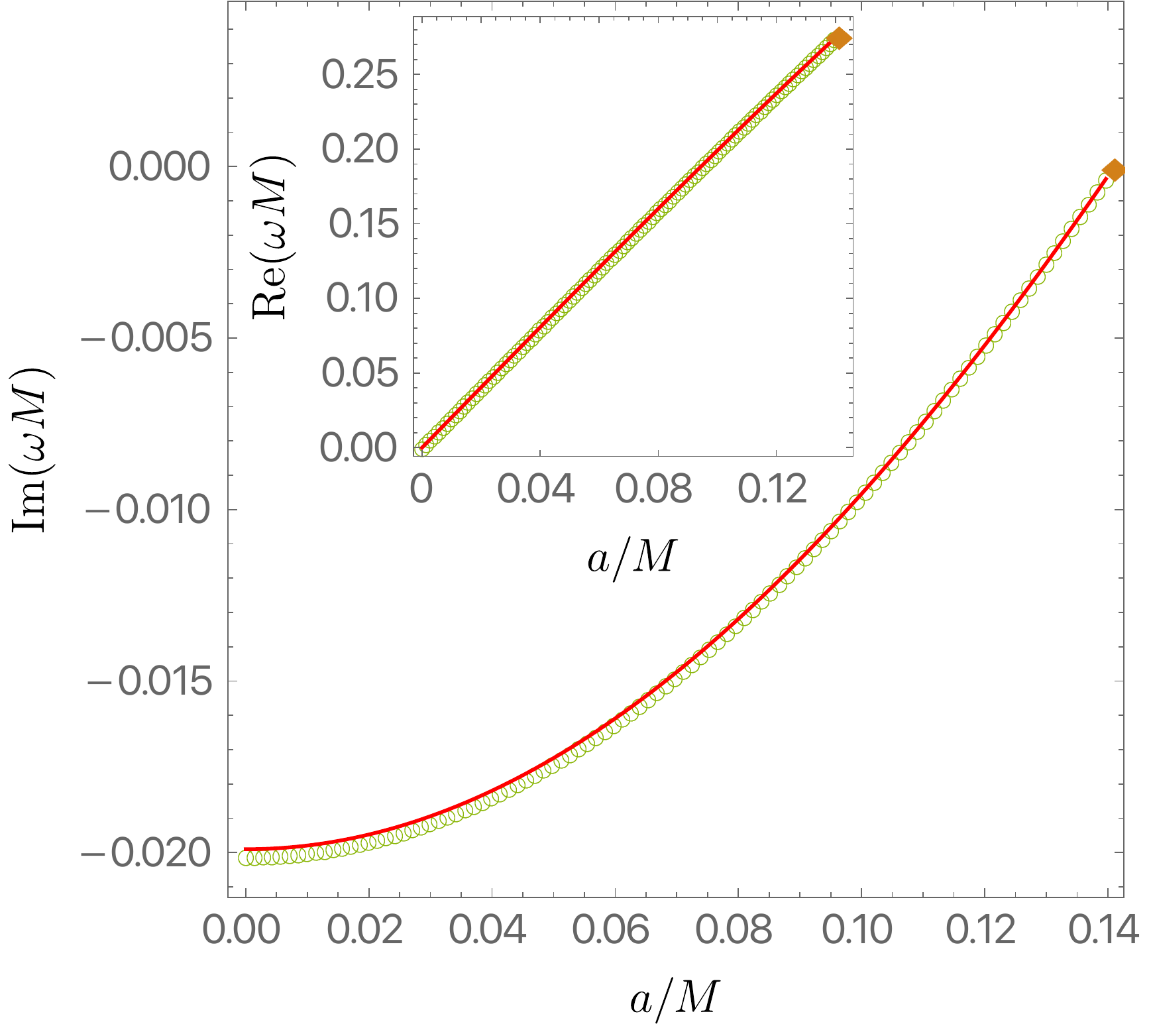}
\hspace{0.5cm}
\includegraphics[width=.46\textwidth]{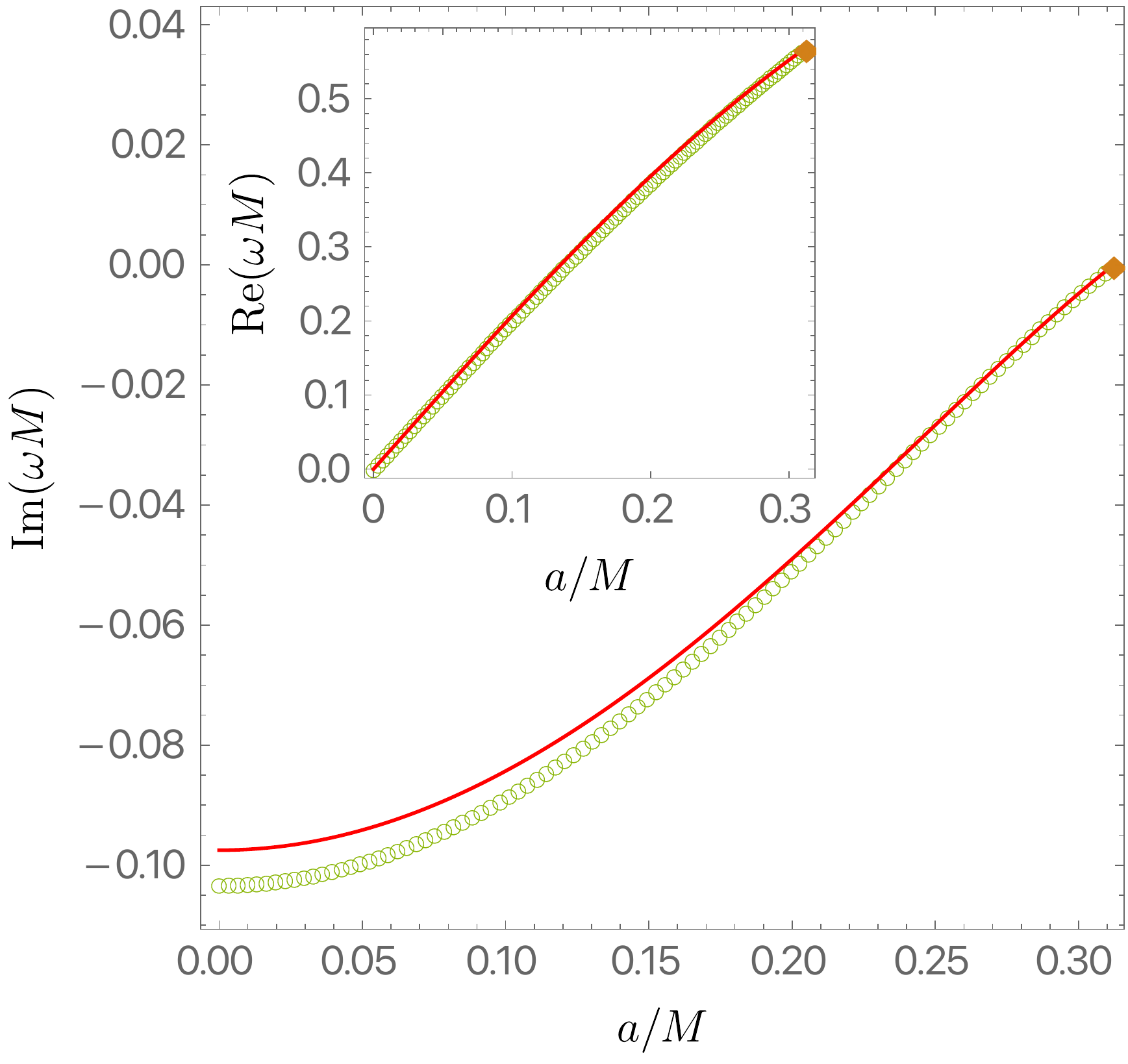}
\caption{Imaginary part and real part (inset plot) of the frequency as a function of the rotation for the NH QNM family with $Q/r_+=0.99$ ({\bf left panel}) and $Q/r_+=0.95$ ({\bf right panel}). The numerical results are given by the green circles while the  red line is the analytical result \eqref{NH:freq}. The brown diamond is the value $\tilde{\omega}=m\tilde{\Omega}_H^{\hbox{\footnotesize ext}}$ at extremality. For large $\tilde{Q}$ (left panel),  $\tilde{\omega}_{\hbox{\tiny MAE}}$ is an excellent approximation even away from extremality but it becomes less good away from extremality for smaller $\tilde{Q}$ (right panel).}
\label{Fig:NHq099-q095}
\end{figure}

How good an approximation is \eqref{NH:freq}?
It is in excellent agreement with the numerical NH frequencies near extremality, as will be discussed in the analysis of Figs.~\ref{Fig:spectraFix-a}-\ref{Fig:spectraFix-a2}. This is further illustrated in the left panel of Fig.~\ref{Fig:NHq099-q095} where we take a KN BH family with $Q/r_+=0.99$ and compare the numerical results (green circles) with the red curve $ \tilde{\omega}_{\hbox{\tiny MAE}} $ given by \eqref{NH:freq}. It turns out that for very large $\hat{Q}$ the agreement is excellent not only near-extremality but also far away from it down to small $\hat{a}$. So much so that we can basically use \eqref{NH:freq} for any astrophysical application that requires the knowledge of the dominant NH frequencies for $0.99<Q/r_+<1$, say. Accordingly, the reader will later find that we have not felt the need to collect numerical data in the window $0.99<Q/r_+<1$ in our plots: see e.g.\ the gap between the green surface and extremal brown curve in Fig.~\ref{Fig:Z2l2m2n0+} and the similar gaps in Figs.~\ref{Fig:Z2l3m3n0+}$-$\ref{Fig:Z2l5m5n0+}.
 Naturally, as we decrease $\hat{Q}$ the approximation \eqref{NH:freq} becomes increasingly less accurate when we move away from extremality. This is demonstrated in the right panel of Fig.~\ref{Fig:NHq099-q095} where we do the comparison between  \eqref{NH:freq} (red line) and the numerical data for  $Q/r_+=0.95$. 

  \begin{figure}[t]
\centering
\includegraphics[width=.472\textwidth]{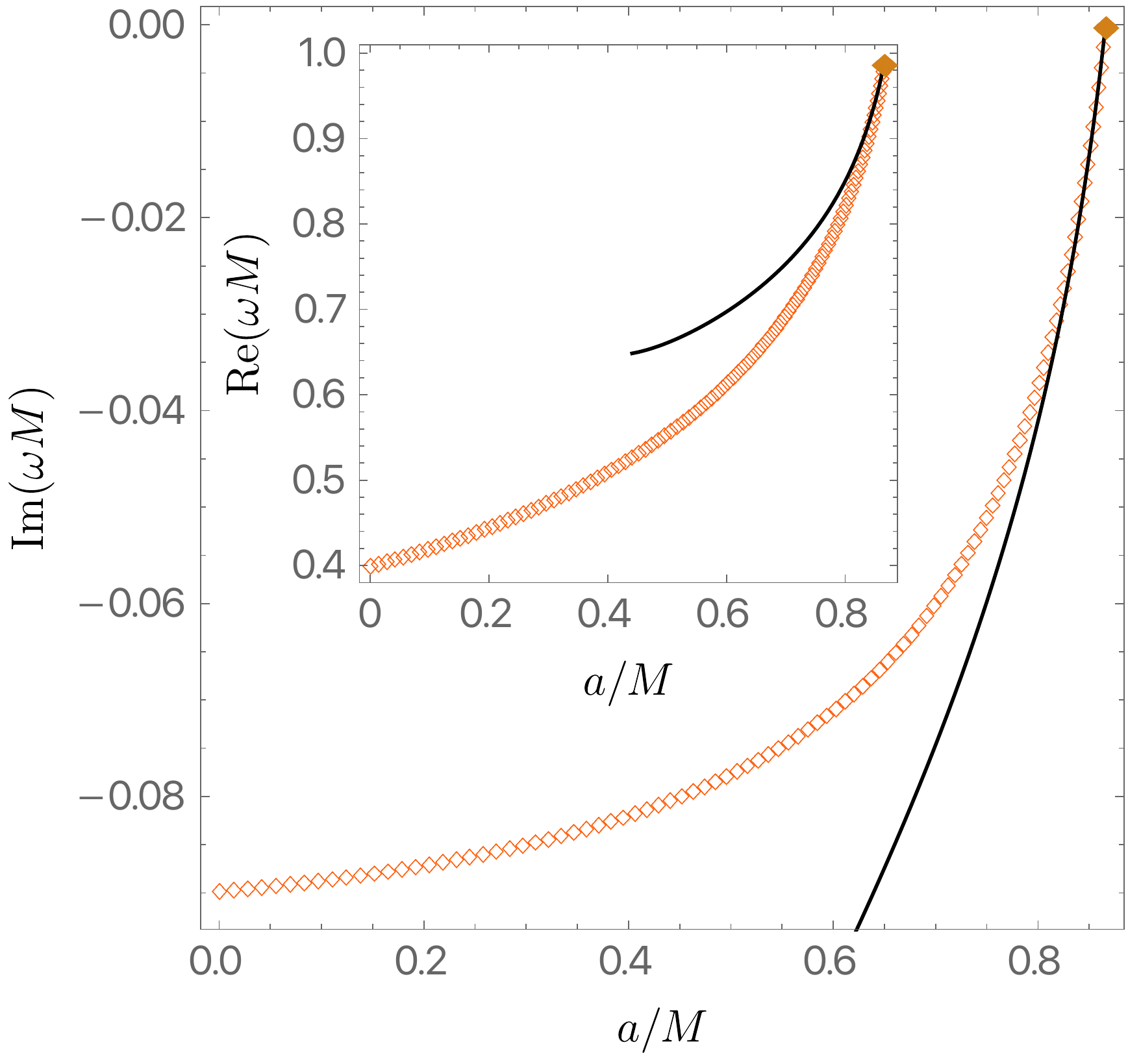}
\hspace{0.5cm}
\includegraphics[width=.47\textwidth]{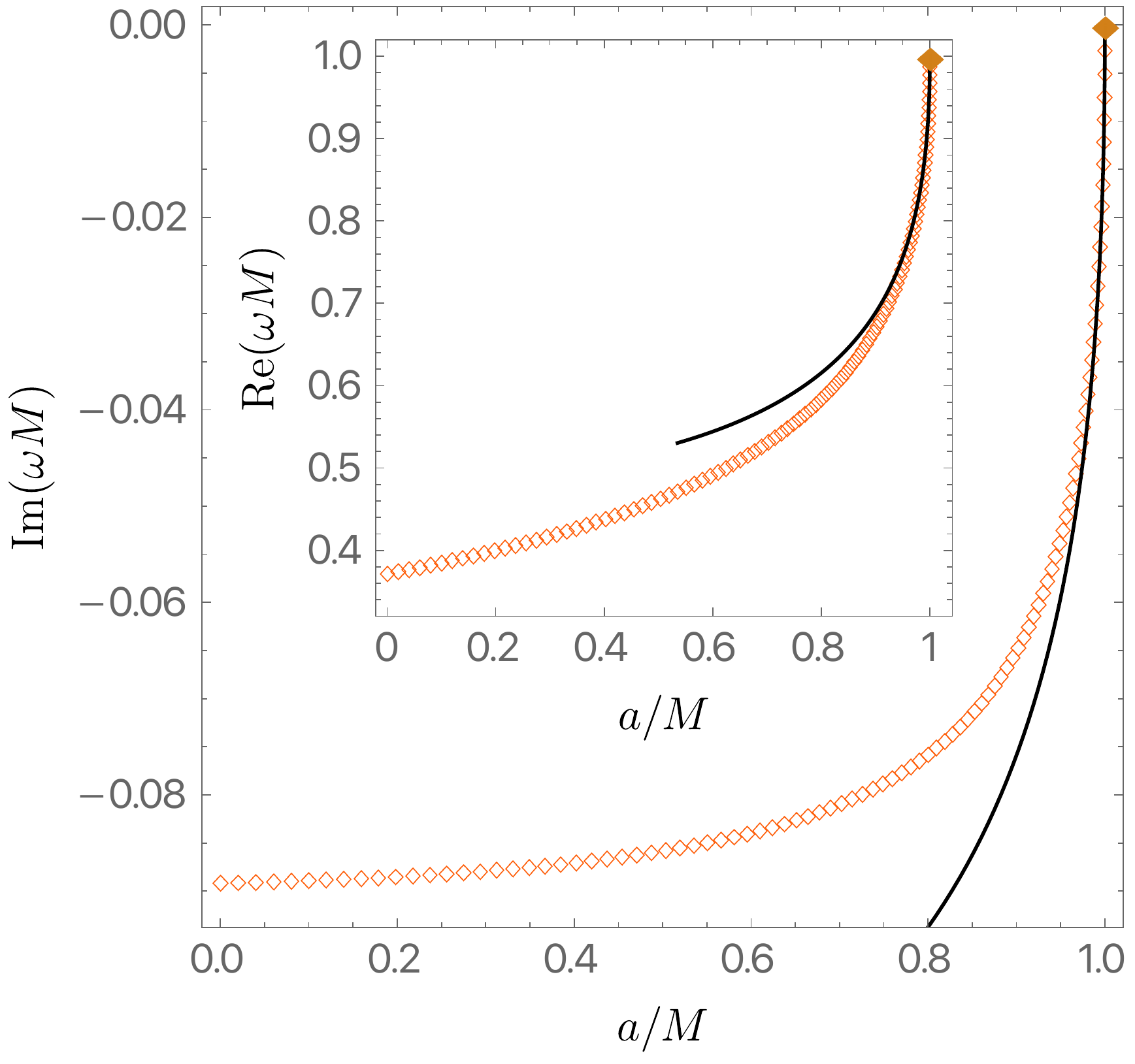}
\caption{Imaginary and real (inset plot) part of the frequency as a function of the rotation for the PS QNM family with $Q/r_+=0.5$ ({\bf left panel}) and $Q=0$ ({\bf right panel}). The numerical results are given by the orange diamonds while the black line is the analytical result \eqref{NH:freq}. The brown diamond is the value $\tilde{\omega}=m\tilde{\Omega}_H^{\hbox{\footnotesize ext}}$ at extremality.
$\tilde{\omega}_{\hbox{\tiny MAE}}$ in \eqref{NH:freq}  is a good  analytical approximation for those PS modes that approach $\mathrm{Im}\,\tilde{\omega}=0$ at extremality. The approximation \eqref{NH:freq} to the PS modes improves as $\tilde{Q}$ decreases. 
}
\label{Fig:PSanalyticNH}
\end{figure}

In the final steps leading to \eqref{NH:freq}, we assumed that $\lambda_2>0$. From Fig.~\ref{Fig:WKBlambda2}, this happens when $\tilde{a}_{\hbox{\footnotesize ext}}=\sqrt{1-\tilde{Q}^2}$ is small, which occurs for large $\tilde{Q}$, as is the case in Fig.~\ref{Fig:NHq099-q095}. This also includes the extremal RN limit, $(\tilde{Q}, \tilde{a})=(1,0)$ in which case \eqref{NH:freq} reduces to the expression first found in \cite{Zimmerman:2015trm}.  
However, nothing impedes us from extending the application of \eqref{NH:freq} also to the case where $\lambda_2<0$. From Fig.~\ref{Fig:WKBlambda2}, this happens for large $\tilde{a}_{\hbox{\footnotesize ext}}=\sqrt{1-\tilde{Q}^2}$, and thus for small $\tilde{Q}$. In particular, this includes the extremal Kerr limit, $(\tilde{Q}, \tilde{a})=(1,0)$. 
Interestingly, when $\lambda_2<0$ (unlike for $\lambda_2>0$), one is effectively in a region of the parameter space where the PS family terminates at  $\mathrm{Im}\,\tilde{\omega}=0$ and $\mathrm{Re}\,\tilde{\omega}=m\tilde{\Omega}_H^{\hbox{\footnotesize ext}}$ and, quite importantly, it dominates over the NH family. Therefore, by construction \eqref{NH:freq} should  be able to capture (also, or in this case) the frequency of the dominant PS modes
  near extremality. And indeed it does so, as illustrated in Fig.~\ref{Fig:PSanalyticNH} where we compare \eqref{NH:freq} (black curve) against the numerical PS frequency (orange diamonds) for the KN families with $Q/r_+=0.5$ (left panel) and $Q=0$ (right panel). The latter case is the Kerr solution, where \eqref{NH:freq} reduces to the expression first found in  \cite{Yang:2012pj,Yang:2013uba}. Thus,  $\tilde{\omega}_{\hbox{\tiny MAE}}$ in \eqref{NH:freq}  (also) provides an analytical approximation for PS modes when they approach $\mathrm{Im}\,\tilde{\omega}=0$ at extremality that complements, and is independent of, the eikonal analytical approximation $\omega^{\hbox{\tiny eikn}}_{\hbox{\tiny PS}}$ given in \eqref{PS:eikonal}. It has the added value of being very accurate near extremality already for $m=2$ (\ie well outside the $|m| \gg 1$ eikonal regime of validity). Interestingly, the approximation \eqref{NH:freq} for the PS modes improves as $\tilde{Q}$ decreases, as can  be inferred from the two cases presented in  Fig.~\ref{Fig:PSanalyticNH}. 
  
 Altogether, and to summarize, we find that \eqref{NH:freq} is an excellent approximation for the dominant modes (which always approach $\mathrm{Im}\,\tilde{\omega}=0$ and $\mathrm{Re}\,\tilde{\omega}=m\tilde{\Omega}_H^{\hbox{\footnotesize ext}}$ at extremality) when we are close to extremality, \ie when $a/\tilde{a}_{\hbox{\footnotesize ext}} \lesssim 1$, {\it independently} of the QNM family that dominates, as best illustrated in Figs.~\ref{Fig:NHq099-q095}-\ref{Fig:PSanalyticNH}. For large $\tilde{Q}$ the dominant modes are the NH modes and \eqref{NH:freq} describes them. For small  $\tilde{Q}$ the dominant modes are instead the PS modes and \eqref{NH:freq} also describes them. This might sound a bit puzzling: how can it be that the near-horizon MAE analysis captures sometimes the PS modes? This is because, away from the RN limit, the distinction between the PS and NH families becomes less clean and actually the dominant QNM family is better described by a combination of the PS and NH modes (that we will denote as a PS$-$NH family) due to the phenomenon of eigenvalue repulsion. This statement will be clarified and made accurate when discussing the results of Figs.~\ref{Fig:spectraFix-a}$-$\ref{Fig:spectraFix-a2} so we postpone further discussion till then.

In the analysis of the eikonal expression \eqref{PS:eikonal} and associated Fig.~\ref{Fig:PSwkb-m2-extremality}, we have already pointed out that when we are at extremality, \eg if we place ourselves on the extremal brown curve of Fig.~\ref{Fig:PSwkb-m2-extremality} (or of  Fig.~\ref{Fig:Z2l2m2n0+}) and  move along it from $\tilde{a}_{\hbox{\footnotesize ext}}=1$ down to $\tilde{a}_{\hbox{\footnotesize ext}}=0$ (or, equivalently, from $\tilde{Q}_{\hbox{\footnotesize ext}}=0$ to $\tilde{Q}_{\hbox{\footnotesize ext}}=1$), there is a critical rotation $\tilde{a}_{\hbox{\footnotesize ext}}=\tilde{a}_\star$ (or, equivalently, a critical charge $\tilde{Q}_\star=\sqrt{1-\tilde{a}_\star^2}$).  
For $\tilde{a}_\star<\tilde{a}_{\hbox{\footnotesize ext}}\leq 1$ (\ie $0\leq \tilde{Q}_{\hbox{\footnotesize ext}}< \tilde{Q}_\star$) the PS family terminates at  $\mathrm{Im} \,\tilde{\omega}=0$ and $\mathrm{Re}\, \tilde{\omega}=m \tilde{\Omega}_H^{\hbox{\footnotesize ext}}$ at extremality (\eg in the Kerr limit where $\tilde{a}_{\hbox{\footnotesize ext}}=1$), but it fails to do so otherwise (\eg in the RN limit where $\tilde{a}_{\hbox{\footnotesize ext}}=0$ and $\tilde{Q}_{\hbox{\footnotesize ext}}=1$). 

\begin{figure}[th]
\centering
\includegraphics[width=.45\textwidth]{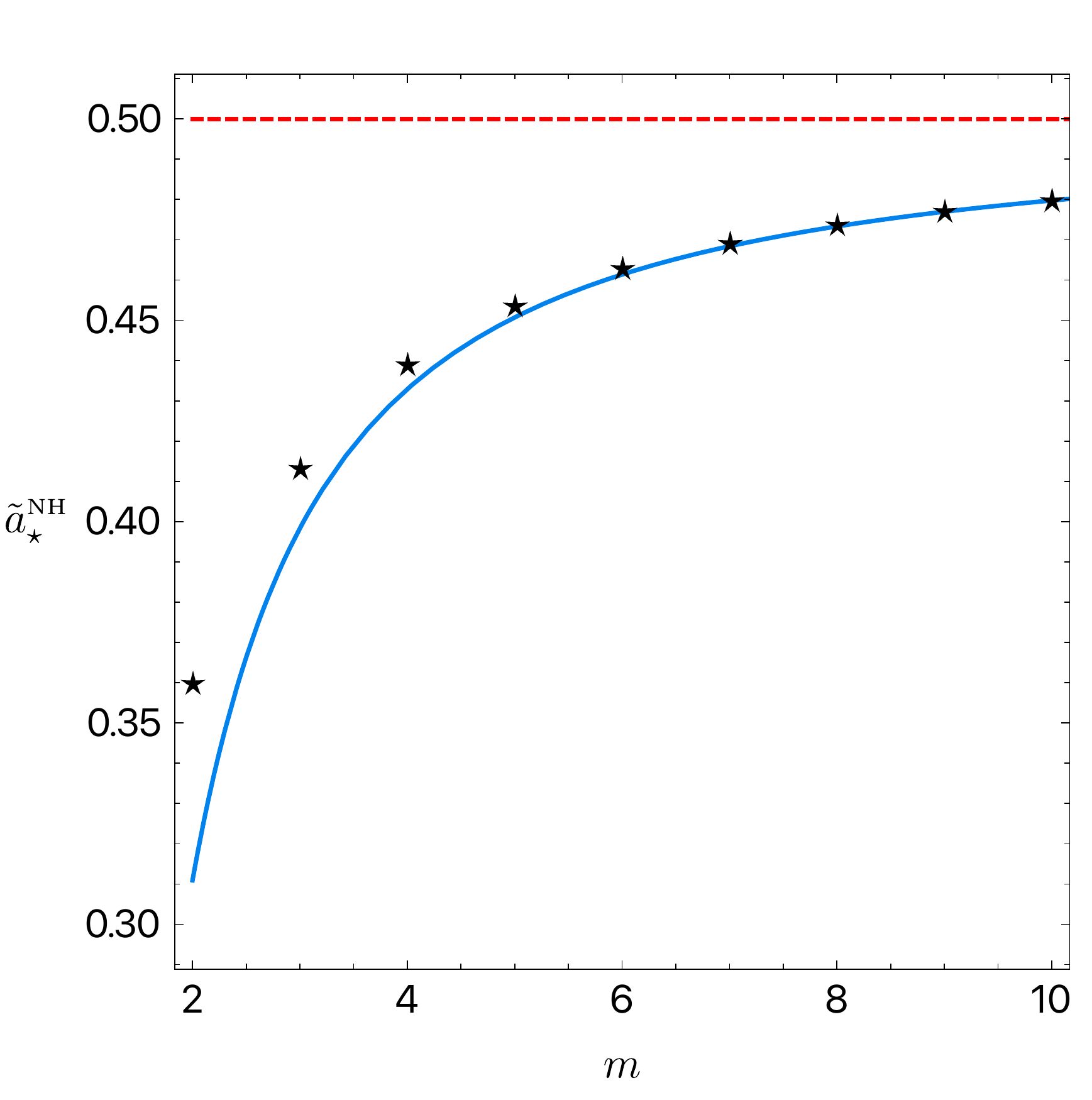}
\hspace{0.5cm}
\includegraphics[width=.46\textwidth]{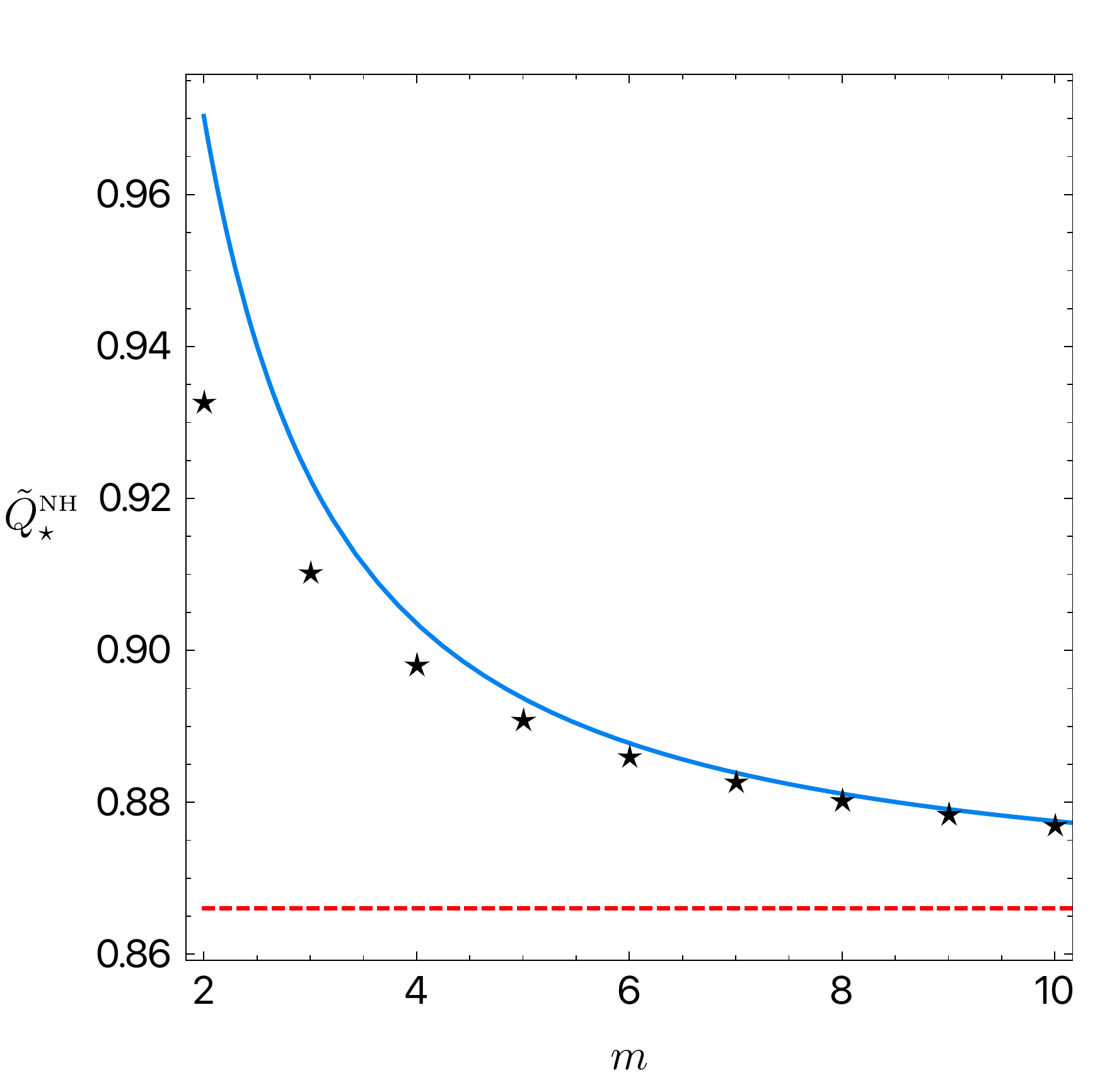}
\caption{Critical values of $\tilde{a}_\star^{\hbox{\tiny NH}}$ (left panel) and $\tilde{Q}_\star^{\hbox{\tiny NH}}$  (right panel) for which $\lambda_2$ vanishes as a function of $m$. The WKB approximation \eqref{NH:starWKB} (blue line) already gives an excellent agreement with the exact results (black $\star$'s) for values of $m$ as low as 5 or 6. Further note that as $m$ increases, $\tilde{a}_\star^{\hbox{\tiny NH}}\to \tilde{a}_\star^{\hbox{\tiny eikn}}=\frac{1}{2}$ (left panel)  and $\tilde{Q}_\star^{\hbox{\tiny NH}}\to \tilde{Q}_\star^{\hbox{\tiny eikn}}=\frac{\sqrt{}3}{2}$ (right panel) $-$ see the red dashed lines $-$ where the latter eikonal values were  discussed in Fig.~\ref{Fig:PSwkb-m2-extremality}  (see its red $\star$ point).
}
\label{Fig:WKBlambda2-m}
\end{figure}  

Interestingly, we find that this $\star$ transition point turns out to be very well approximated (if not exactly given) by the point where the separation constant $\lambda_2(m,\tilde{a}_{\hbox{\footnotesize ext}})$ in \eqref{NH:freq} vanishes:  $\lambda_2(m,\tilde{a}_\star^{\hbox{\tiny NH}})=0$. For $\tilde{a}_{\hbox{\footnotesize ext}}<\tilde{a}_\star^{\hbox{\tiny NH}}$ (or equivalently,  $\tilde{Q}_{\hbox{\footnotesize ext}}>\tilde{Q}_\star^{\hbox{\tiny NH}}$) one has $\lambda_2>0$ and for $\tilde{a}_{\hbox{\footnotesize ext}}>\tilde{a}_\star^{\hbox{\tiny NH}}$ we have $\lambda_2<0$. To get the accurate values for $\tilde{a}_\star^{\hbox{\tiny NH}}$ $-$ which are displayed as black $\star$'s in Fig.~\ref{Fig:WKBlambda2-m} $-$ we use the numerical solution for $\lambda_2(m,\tilde{a}_{\hbox{\footnotesize ext}})$ as displayed in Fig.~\ref{Fig:WKBlambda2}.  Alternatively,  since  $\lambda_2$ has the WKB expansion  \eqref{NH:WKBansatzB} and \eqref{NH:WKBansatzCoef}, we can use it to find $\tilde{a}_\star^{\hbox{\tiny NH}}|_{\hbox{\tiny WKB}}$ or $\tilde{Q}_\star^{\hbox{\tiny NH}}|_{\hbox{\tiny WKB}}$,  yielding
\begin{subequations}\label{NH:starWKB}
\begin{align}
&\tilde{a}_\star^{\hbox{\tiny NH}}|_{\hbox{\tiny WKB}}\simeq \frac{1}{2}-\frac{5 \sqrt{3} \left(2-\sqrt{2}\right)}{32 \,m}+\frac{5 \left(69-176 \sqrt{2}\right)}{2048 \,m^2}+\mathcal{O}\left(1/m^3\right), \label{NH:starWKBA} \\
& \tilde{Q}_\star^{\hbox{\tiny NH}}|_{\hbox{\tiny WKB}} \simeq \frac{\sqrt{3}}{2}+\frac{5 \left(2-\sqrt{2}\right)}{32 m}
+ \frac{5 \sqrt{3} \left(112 \sqrt{2}-103\right)}{2048 m^2} +\mathcal{O}\left(1/m^3\right). \label{NH:starWKBB}
\end{align}
\end{subequations}
Using our numerical data for $\lambda_2$ (Fig.~\ref{Fig:WKBlambda2}), when $m=2$ we get $\{\tilde{a}_\star,\tilde{Q}_\star\}^{\hbox{\tiny NH}}\simeq \{0.360, 0.932\}$ while the WKB approximation \eqref{NH:starWKB} yields $\{\tilde{a}_\star,\tilde{Q}_\star\}^{\hbox{\tiny NH}}_{\hbox{\tiny WKB}}\sim \{0.311, 0.970\}$. Being a WKB approximation, \eqref{NH:starWKB} is expected to be accurate only as $m\to \infty$. To confirm this, we compute these critical rotations for $m=2$ to $m=10$ and Fig.~\ref{Fig:WKBlambda2-m}  shows that $\tilde{a}_\star^{\hbox{\tiny NH}}|_{\hbox{\tiny WKB}}$ as given by  \eqref{NH:starWKB} (the solid blue line) indeed approaches increasingly the value of $\tilde{a}_\star^{\hbox{\tiny NH}}$ (the black $\star$'s) as $m$ grows, with excellent agreement already for $m=10$ (or even $m=6$). Further note that as $m$ increases, $\tilde{a}_\star^{\hbox{\tiny NH}}$ and $\tilde{a}_\star^{\hbox{\tiny NH}}|_{\hbox{\tiny WKB}}$ approach from below the eikonal value $\tilde{a}_\star^{\hbox{\tiny eikn}}=\frac{1}{2}$ (or  from above the eikonal $\tilde{Q}_\star^{\hbox{\tiny eikn}}=\sqrt{1-(\tilde{a}_\star^{\hbox{\tiny eikn}})^2}=\sqrt{3}/2\simeq 0.866025$) discussed in Fig.~\ref{Fig:PSwkb-m2-extremality} (see its red $\star$).

\begin{figure}[ht]
\centering
\includegraphics[width=.465\textwidth]{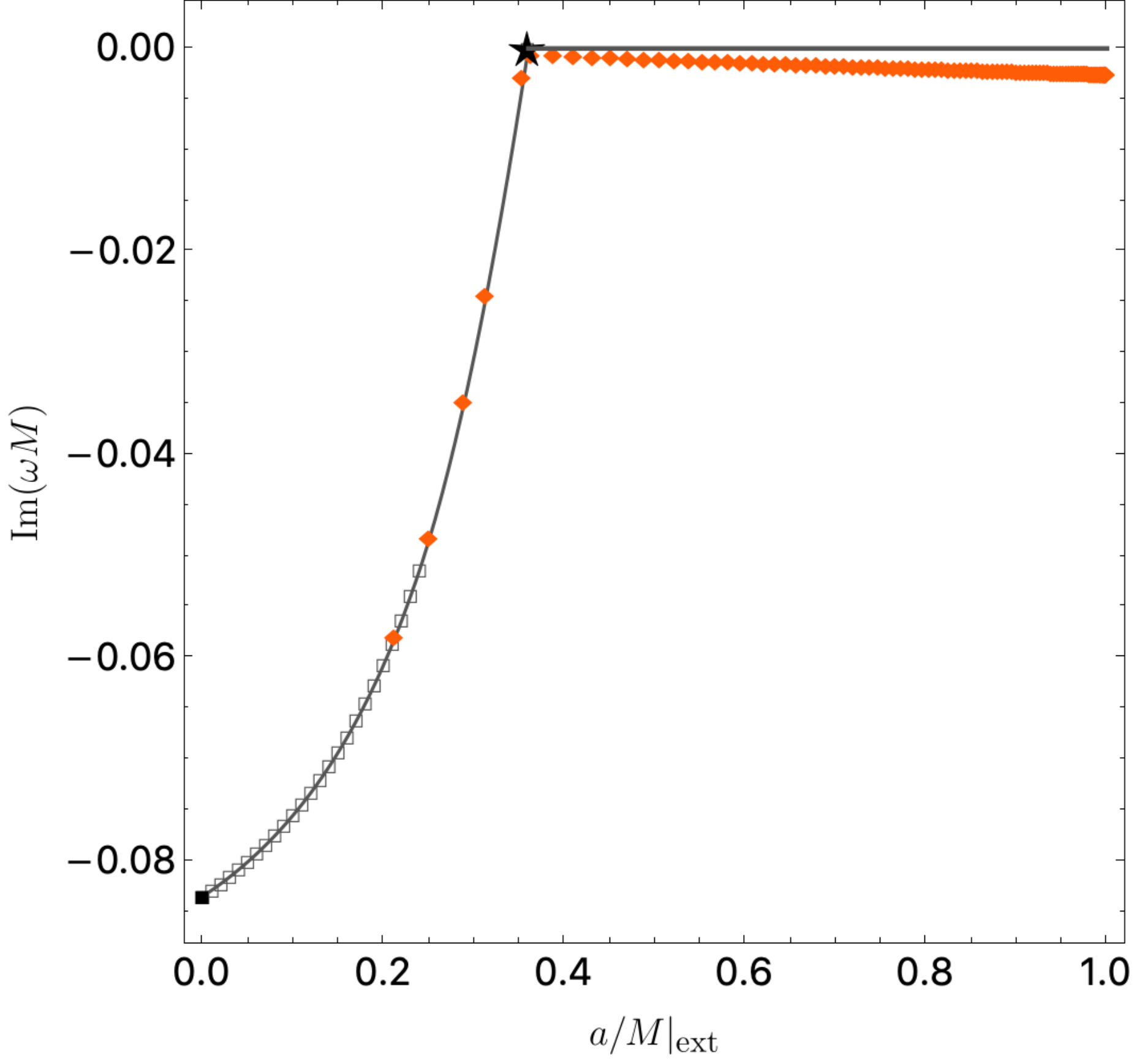}
\hspace{0.5cm}
\includegraphics[width=.445\textwidth]{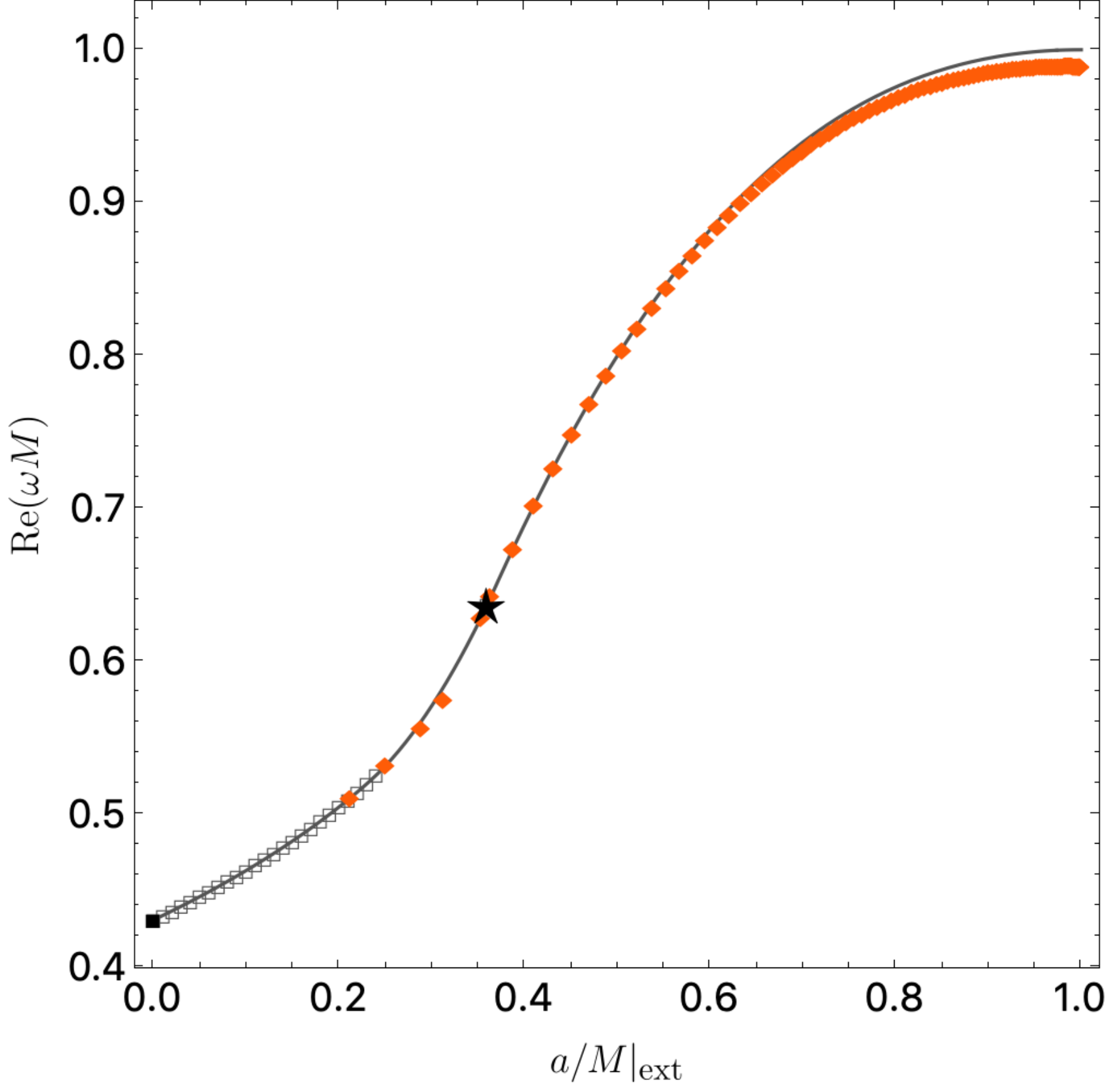}
\caption{The PS modes at extremality. The black $\star$  at $\tilde{a}_{\hbox{\footnotesize ext}}=\tilde{a}_\star\simeq 0.360$ is the one shown in Fig.~\ref{Fig:WKBlambda2-m}. The grey squares in the range $\tilde{a}_{\hbox{\footnotesize ext}}\in [0,0.24]$ describe data  obtained  solving the gravito-electromagnetic PDEs {\it directly at extremality}. The grey line in the range  $\tilde{a}_{\hbox{\footnotesize ext}}\in [0,\tilde{a}_\star]$ is an interpolation of the grey square and $\star$ points. On the other hand, for $\tilde{a}_{\hbox{\footnotesize ext}}>\tilde{a}_\star$ it is simply described by $\mathrm{Im}\,\tilde{\omega}= 0$ and $\mathrm{Re}\,\tilde{\omega}= m\tilde{\Omega}_H^{\hbox{\footnotesize ext}}$. The orange diamonds describe the closest point to extremality we obtained using the non-extremal code.     
{\bf Left panel:} Imaginary part of the PS frequency.  {\bf Right panel:} Real part of the PS frequency.}
\label{Fig:PS-extremality}
\end{figure}  

So  our numerical results indicate that the critical rotation/charge $\{\tilde{a}_\star, \tilde{Q}_\star\}$ seem to be given to very good accuracy by the values $\{\tilde{a}_\star^{\hbox{\tiny NH}}, \tilde{Q}_\star^{\hbox{\tiny NH}}\}$ discussed above and displayed in Fig.~\ref{Fig:WKBlambda2-m}. This is further demonstrated in Fig.~\ref{Fig:PS-extremality}. 
In these plots we show the imaginary and real part of the PS frequency as a function of the rotation at extremality, $\tilde{a}_{\hbox{\footnotesize ext}}$,  for $Z_2$ $\ell=m=2,n=0$ modes. The black $\star$ at $\tilde{a}_{\hbox{\footnotesize ext}}=\tilde{a}_\star^{\hbox{\tiny NH}}\simeq 0.360$  (\ie $\tilde{Q}_{\hbox{\footnotesize ext}}=\tilde{Q}_\star^{\hbox{\tiny NH}}\simeq 0.932$) is the point already displayed in Fig.~\ref{Fig:WKBlambda2-m}. The set of black squares displayed only for $\tilde{a}_{\hbox{\footnotesize ext}}\in [0,0.24]$ describe data we obtained by solving the gravito-electromagnetic PDEs {\it directly at extremality} (numerically, it is very hard to extend the computation for higher $\tilde{a}_{\hbox{\footnotesize ext}}$; recall that at extremality we have a degenerate horizon and thus the boundary conditions differ from the non-extremal case). On the other hand, the auxiliary grey line that joins these black squares and connects to the black $\star$ point at $\tilde{a}_{\hbox{\footnotesize ext}}\sim 0.360$, is an interpolation curve built from the black square and $\star$ points. Finally, the PS modes closest to extremality that we found using our non-extremal code are identified with orange diamonds (with $\tilde{a}_{\hbox{\footnotesize ext}}\gtrsim 0.2$ since it is hard to obtain data when $\tilde{a}_{\hbox{\footnotesize ext}}\to 0$). For $0<\tilde{a}_{\hbox{\footnotesize ext}}<\tilde{a}_\star \simeq 0.360$ they are just below the interpolation grey  line. Altogether this indicates that PS modes indeed terminate at the grey interpolation line for $0 \leq \tilde{a}_{\hbox{\footnotesize ext}}\leq \tilde{a}_\star$. On the other hand, for $ \tilde{a}_\star < \tilde{a}_{\hbox{\footnotesize ext}}\leq 1$, the grey horizontal line displayed in Fig.~\ref{Fig:PS-extremality} has $\mathrm{Im}\,\tilde{\omega}= 0$ and $\mathrm{Re}\,\tilde{\omega}= m\tilde{\Omega}_H^{\hbox{\footnotesize ext}}$. The orange diamonds in this region are the closest PS modes we obtained using the non-extremal PS numerical code; these points are at 99\% of extremality. Again we see that they indeed approach the grey horizontal line. To find even further approach we need to extend our collection of data closer to extremality, say up to 99.9\% of extremality. We did this for a few constant charge families (not shown) to confirm it is indeed the case (and these are very accurately described by $\tilde{\omega}_{\hbox{\tiny MAE}}$ in \eqref{NH:freq} as discussed previously).

\section{Eigenvalue repulsions (also known as level repulsions or avoided crossing)}\label{sec:repulsion}

The analytical analyses of Section~\ref{sec:AnalyticalPSNH} allowed us to find corners of the 2-parameter space of KN where we can obtain good analytical approximations for the QNMs of KN.
Importantly, they gave us evidence for the existence of not one but two main families of QNMs: the photon sphere and near-horizon families. These are distinct families because the analytical analyses reveal different origins: the PS family is associated with properties of null orbits in the eikonal limit, while the NH family is related to modes whose wavefunction is very localized about the horizon near extremality. 
Our numerical search of QNMs, whose findings will be presented in this section and in Sections~\ref{sec:DominantSpectra}$-$\ref{sec:FullSpectra}, confirm that KN indeed has two families of QNMs and not more, and all the numerical QNM frequencies are well approximated by \eqref{PS:eikonal} and/or \eqref{NH:freq} in the regimes where the latter are valid. 

However, the distinction between the two QNM families of KN becomes very fuzzy as we move along the 2-parameter space of KN. In the RN limit ($a=0,Q\neq 0$) this distinction is very sharp: one of the families is the PS family well approximated by \eqref{PS:eikonal} and the second one is  the NH family  
well described by \eqref{NH:freq}; recall Fig.~\ref{Fig:RN}. But when we switch on the rotation and allow it to increase we find that the PS and NH families lose their individual identities. Instead branches of  these two families combine with each other to produce a combined family that we can appropriately call PS$-$NH modes (and their radial overtone families). This occurs because the KN spectra has
a novel phenomenon that is special to the KN QNM system (i.e.\ present neither in Kerr nor RN), namely {\it eigenvalue repulsion} between QNM families. In subsection~\ref{sec:repulsionSub} we will describe in detail this phenomenon in the KN QNM spectra. Although, in the context of black hole QNMs eigenvalue repulsions are particular to KN (see also footnote \ref{ft:ds}), such a feature is common in some eigenvalue problems, notably: 1) in solid state physics where e.g.\ it is responsible for energy bands/gaps in the spectra of electrons moving in certain Schr\"odinger potentials, and in 2) in quantum mechanical eigenvalue systems with the so-called avoided crossing phenomenon. Therefore, before discussing eigenvalue repulsions in KN, in subsection~\ref{sec:repulsionTextbook} we will present a simple textbook example of eigenvalue repulsions that will allow us to understand from first principles what occurs in the KN eigenfrequency spectra.  

\subsection{Complexified eigenvalue repulsion\label{sec:repulsionTextbook}}
Eigenvalue repulsion is a phenomenon that occurs in simple quantum mechanical models (albeit it can also occur in classical physics, most notably when two levels of a classical harmonic oscillator are coupled). In quantum mechanics this phenomenon is also known as the Wigner-Teller effect, \emph{avoided crossing} or \emph{level repulsion}  \cite{Landau1981Quantum,Cohen-Tannoudji:1977}).

To explain the similarities and differences between what is observed in standard quantum mechanics textbooks and the phenomenon that we observe numerically in the QNM spectra of KN, we will start by reviewing the simplest textbook example exhibiting avoided crossing (see for instance $\S 79$ of \cite{Landau1981Quantum} and/or $\S \rm{IV.C}$ of \cite{Cohen-Tannoudji:1977})).

For concreteness consider a two-level system with Hamiltonian $H_0$, orthonormal eigenstates $\ket{\psi_i}$ and energy levels $E_i$, so that
\begin{equation}
H_0 \ket{\psi_i}=E_i\,\ket{\psi_i}\, \quad \text{ and } \quad \braket{\psi_i|\psi_j}=\delta_{ij}\,, \qquad \text{ with }\:i,j=1,2\,.
\end{equation}

Let us imagine perturbing $H_0$ with an interaction $W$, such that the full Hamiltonian is given by $H=H_0+W$. Here $W$ can be thought as coupling the two eigenstates $\psi_i$. In the $\{\ket{\psi_1},\ket{\psi_2}\}$ basis, the coupling is given as a $2\times2$ matrix $\mathcal{W}$ with entries $\mathcal{W}_{ij}\equiv \braket{\psi_i|W|\psi_j}$.

In the $\{\ket{\psi_1},\ket{\psi_2}\}$ basis the perturbed Hamiltonian matrix $\mathcal{H}_{ij}= \braket{\psi_i|H|\psi_j}$ can be written as
\begin{equation}
\mathcal{H}=\left[\begin{array}{cc}
E_1+\mathcal{W}_{11} & \mathcal{W}_{12}
\\
\mathcal{W}_{21} & E_2+\mathcal{W}_{22}
\end{array}
\right]\,.
\label{eq:ham}
\end{equation}
Self-adjointness of the perturbed Hamiltonian then demands $\mathcal{H}$ to be Hermitian, and thus $\mathcal{W}_{21}=\overline{\mathcal{W}_{12}}$, where the bar denotes complex conjugation.

It is a rather standard exercise to diagonalise $\mathcal{H}$ given in \eqref{eq:ham} and find that the eigenvalues of the perturbed Hamiltonian are:
\begin{equation}
E_{\pm}=\frac{\tilde{E}_1+\tilde{E}_2}{2}\pm\sqrt{\frac{(\tilde{E}_1-\tilde{E}_2)^2}{4}+|\mathcal{W}_{12}|^2}\,,
\label{eq:splitting}
\end{equation}
where $\tilde{E}_i=E_i-\mathcal{W}_{ii}$ (with no Einstein summation convention on the last term). \emph{Eigenvalue crossing} (i.e.\ $E_-=E_+$) will only occur if the argument of the square root vanishes. Since the argument of the square root is given by a sum of \emph{two} positive definite terms, we must demand each to be zero \emph{separately}:
\begin{equation}
\mathcal{W}_{12}=0\,\quad\text{and}\quad \tilde{E}_1=\tilde{E}_2\,.
\end{equation}

Let us now imagine that $W$ is a function of a number of real parameters, say $N$. Since we have two conditions to be satisfied in order for crossing to occur, we expect that crossing can only happen over a subspace of the $N$ real variables parametrised by $N-2$ real variables.\footnote{This is indeed the case, so long as $\mathcal{W}_{12}$ does not vanish for some symmetry reasons \cite{Landau1981Quantum}.} Except at this special subspace, \eqref{eq:splitting} predicts that eigenvalues do not cross under the effect of perturbations $W$ (since $E_- <  E_+$ for $\mathcal{W}_{12}\neq0$). This is known as \emph{avoided crossing}.

However, the case at hand (QNMs of KN), is more complicated than this standard textbook example because the perturbation operator is not self-adjoint. However, we shall see that progress can nevertheless be made to understand the properties of its intricate QNM spectra in terms of avoided crossing. Let us denote by $L_0$ the operator whose eigenspectrum yields the QNM spectrum of a RN black hole, which is non-degenerate: see Fig.~\ref{Fig:RN}\footnote{Note that in Fig.~\ref{Fig:RN}, the imaginary part of the PS mode crosses the imaginary part of the NH mode. Nevertheless, the real parts of the PS and NH frequencies are distinct. Thus, RN has \emph{no} crossing in the complex eigenfrequency plane.}. Let us label such QNMs as $\{\psi_i,\omega_i\}$ with $i=1,2$ (in the simplest case, we should regard these as the two slowest decaying QNMs for a given value of $Q/M$ as shown in Fig.~\ref{Fig:RN}). We would like to investigate what will happen to these two QNMs as we turn on $J/M^2$. The operator governing the eigenspectrum will change to $L=L_0+K$, so that $K=0$ at $J=0$.

For quasinormal modes, $L$ is \emph{not} self-adjoint, but one can nevertheless introduce a non-degenerate bilinear form $\braket{\braket{\cdot | \cdot}}$ with respect to which the $\psi_i$ are orthogonal \cite{Leaver:1986gd}. However, in general, $\braket{\braket{\cdot | \cdot}}$ will be complex. We will choose the normalisation of the $\psi_i$ to be such that $\braket{\braket{\overline{\psi}_i | \psi_j}}=\delta_{ij}$\footnote{Here we use the fact that combining the non-degeneracy of the spectrum of $L_0$ with the non-degeneracy of $\braket{\braket{\cdot | \cdot}}$ requires $\braket{\braket{\overline{\psi}_i | \psi_j}}\neq 0$.}. Note that we cannot choose $\braket{\braket{\psi_i | \psi_j}}=\delta_{ij}$ since it could well be that $\braket{\braket{\psi_i | \psi_i}}=0$.

As with the Hermitian case, we define $\mathcal{L}_{ij}=\braket{\braket{\overline{\psi}_i || \psi_j}}$, which leads to the perturbed matrix
\begin{equation}
\mathcal{L}=\left[\begin{array}{cc}
\omega_1+\mathcal{K}_{11} & \mathcal{K}_{12}
\\
\mathcal{K}_{21} & \omega_2+\mathcal{K}_{22}
\end{array}
\right]\,,
\end{equation}
with $\mathcal{K}_{ij}=\braket{\braket{\overline{\psi}_i |K| \psi_j}}$. $\mathcal{L}$ can also be straightforwardly diagonalised as
\begin{equation}
\omega_{\pm}=\frac{\tilde{\omega}_1+\tilde{\omega}_2}{2}\pm\sqrt{\frac{(\tilde{\omega}_1-\tilde{\omega}_2)^2}{4}+\mathcal{K}_{12}\mathcal{K}_{21}}\,,
\label{eq:splittingcomplex}
\end{equation}
where $\tilde{\omega}_i=\omega_i-\mathcal{K}_{ii}$ (with no Einstein summation convention on the last term).  Eigenvalue crossing will only occur if the argument of the square root vanishes. Unlike the Hermitian case, this time this gives only one condition
\begin{equation}
\frac{(\tilde{\omega}_1-\tilde{\omega}_2)^2}{4}+\mathcal{K}_{12}\mathcal{K}_{21}=0\,.
\label{eq:condcomp}
\end{equation}
Let us now imagine that $\mathcal{K}$ depends on $N$ \emph{real} parameters. Since the  condition \eqref{eq:condcomp} is in general complex, it provides a restriction on two of the $N$ parameters. This means eigenvalue crossing can only occur on a $N-2$ subspace, just as in the Hermitian case. This is the reason why we need also at least \emph{two} real parameters to see avoided crossing in the non-Hermitian case. In the black hole context, this justifies why we can see this phenomenon in Kerr-Newman \cite{Dias:2021yju}, RN-dS \cite{Dias:2020ncd}, Myers-Perry-dS \cite{Davey:2022vyx}, but not in RN or Kerr black holes.

The analysis above also shows that level crossing (in the complex frequency plane) will only occur at most at a \emph{point} in the full Kerr-Newman space of parameters (which has $N=2$ adimensional parameters, namely $Q/M$ and $J/M^2$). Our numerical analysis of the KN QNM spectra (mainly of of Sections~\ref{sec:NHanalytics} and \ref{sec:repulsionSub}) provide us with strong evidence to conjecture that this level crossing point lies precisely at extremality when the PS modes reach $\mathrm{Im}(\omega)=0$. This is the $\star$ point in Fig.~\ref{Fig:PS-extremality} (in the case of $n=0$ PS and NH modes). This conjecture is backed up not only by our numerical studies, but also by our approximate analytic form of the near-horizon matching asymptotic expansion frequency \eqref{NH:freq}, which has the same elements as \eqref{eq:splittingcomplex}, with $-\lambda_2(m,\tilde{a})$ playing the role of $\frac{(\tilde{\omega}_1-\tilde{\omega}_2)^2}{4}+\mathcal{K}_{12}\mathcal{K}_{21}$.

The study of level crossing for non-Hermitian systems remains an active topic of research particularly when more than two-levels are considered (see \cite{Rotter_2009} for an excellent topical review on the subject). For instance, in \eqref{eq:condcomp} we could have demanded that the real (imaginary) part vanishes, but let the imaginary (real) part be arbitrary. This would lead to avoided crossing in the imaginary (real) part, but would allow for crossing in the real (imaginary) part. This example shows that avoided crossing for non-Hermitian matrices can indeed be a richer phenomenon than its Hermitian cousin.

\subsection{Eigenvalue repulsions in the frequency spectra of KN \label{sec:repulsionSub}}

\begin{figure*}
\includegraphics[width=.47\textwidth]{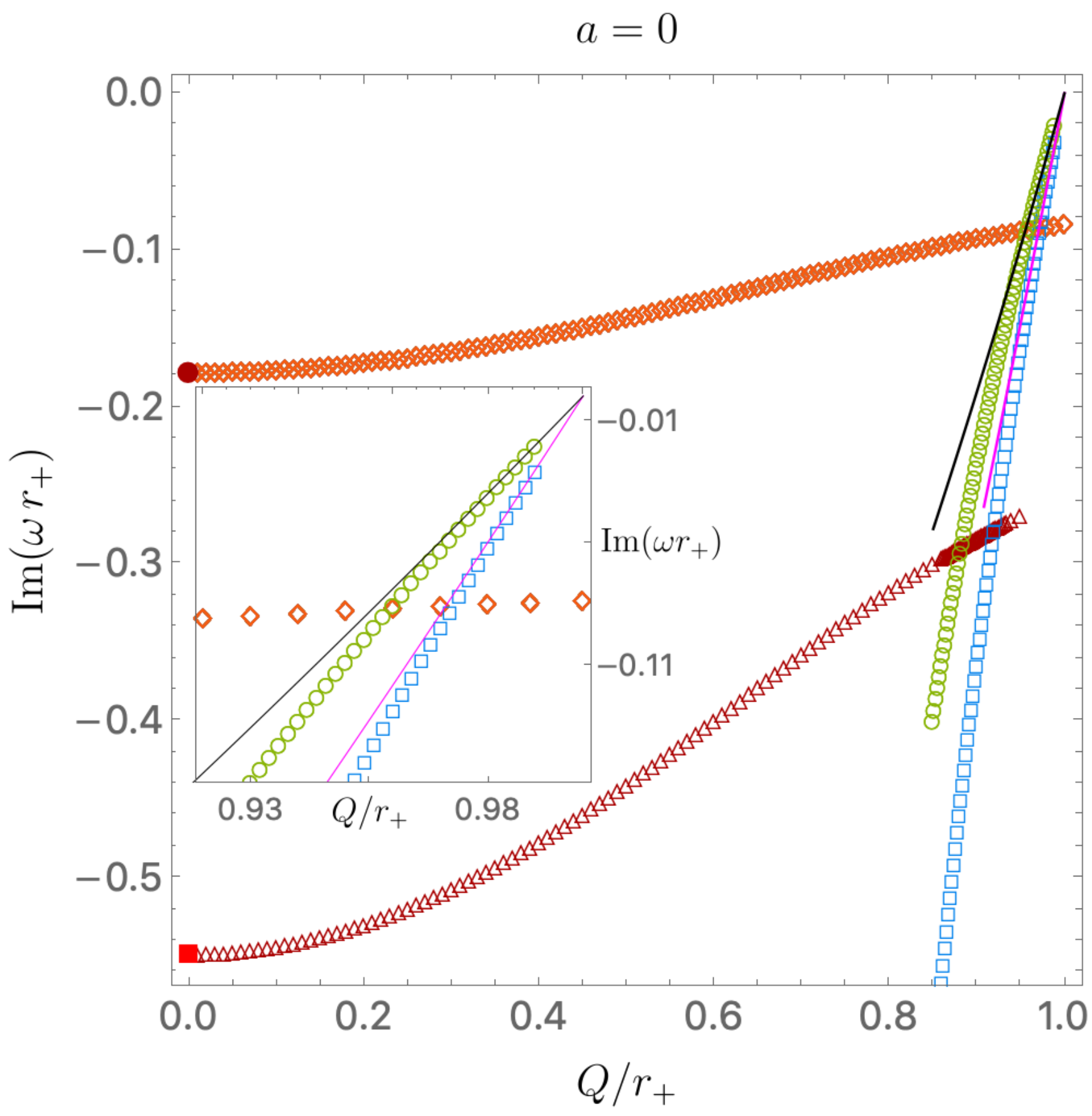}
\hspace{0.5cm}
\includegraphics[width=.48\textwidth]{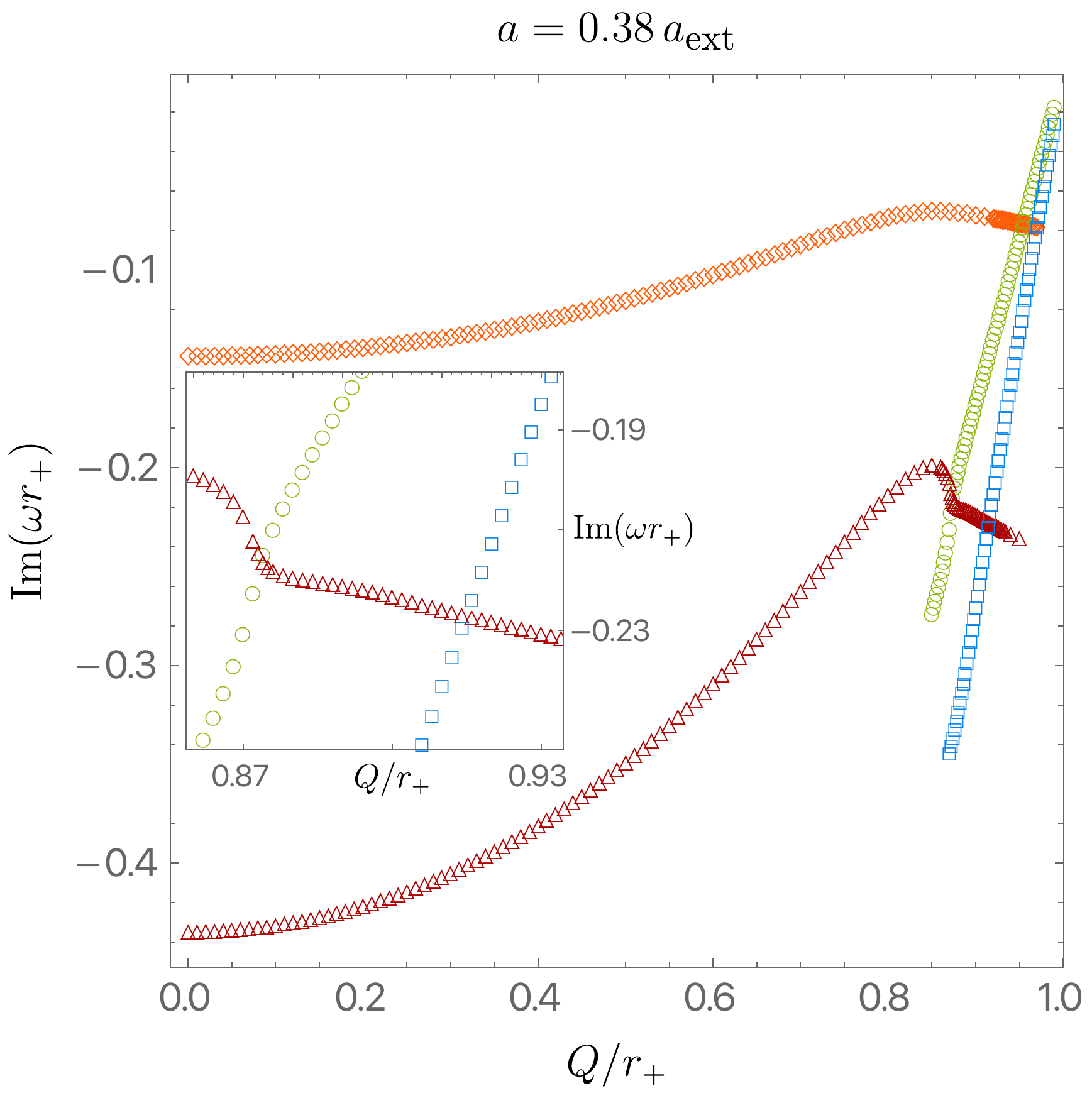}
\vskip 0.2cm
\includegraphics[width=.47\textwidth]{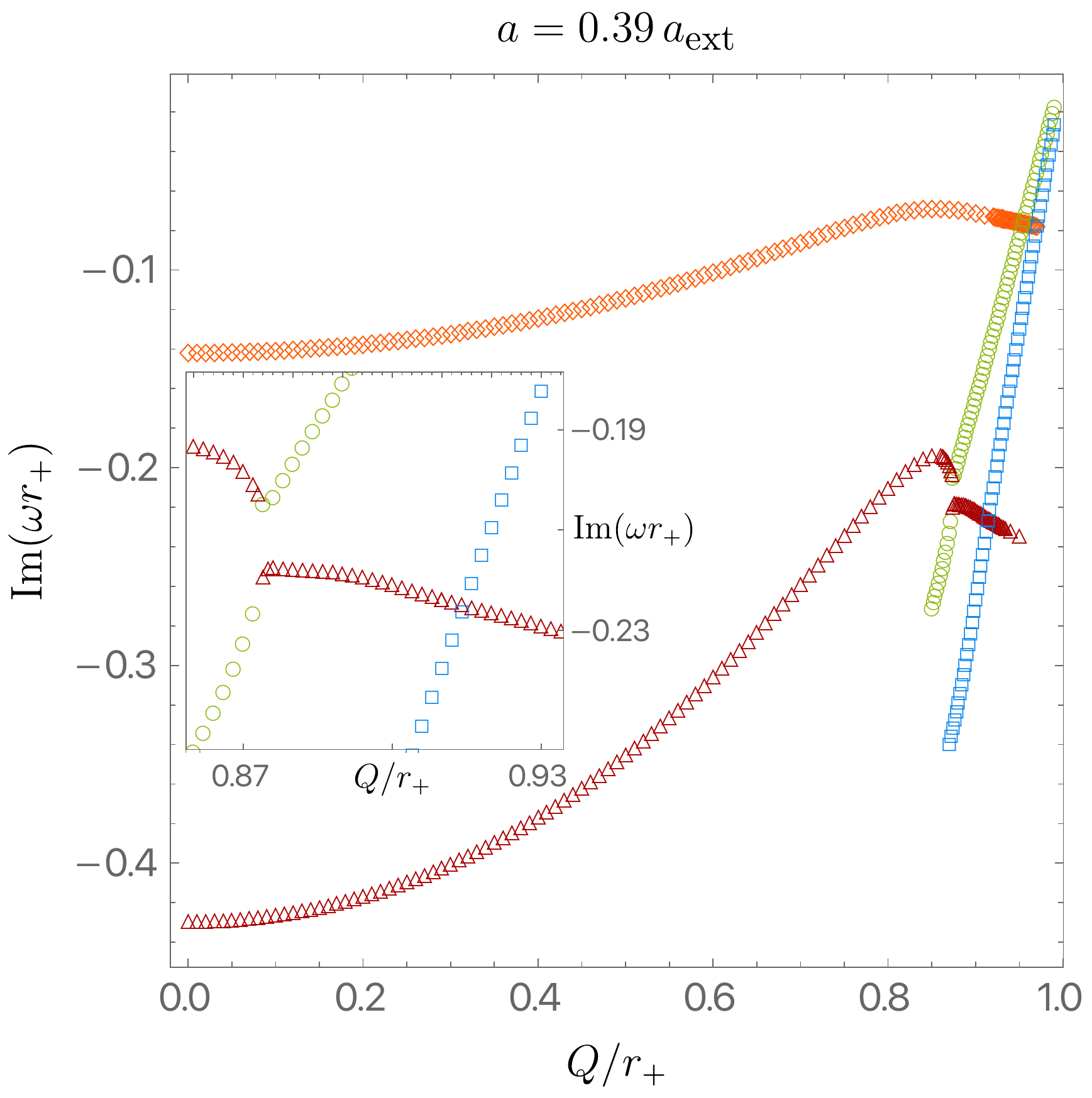}
\hspace{0.5cm}
\includegraphics[width=.48\textwidth]{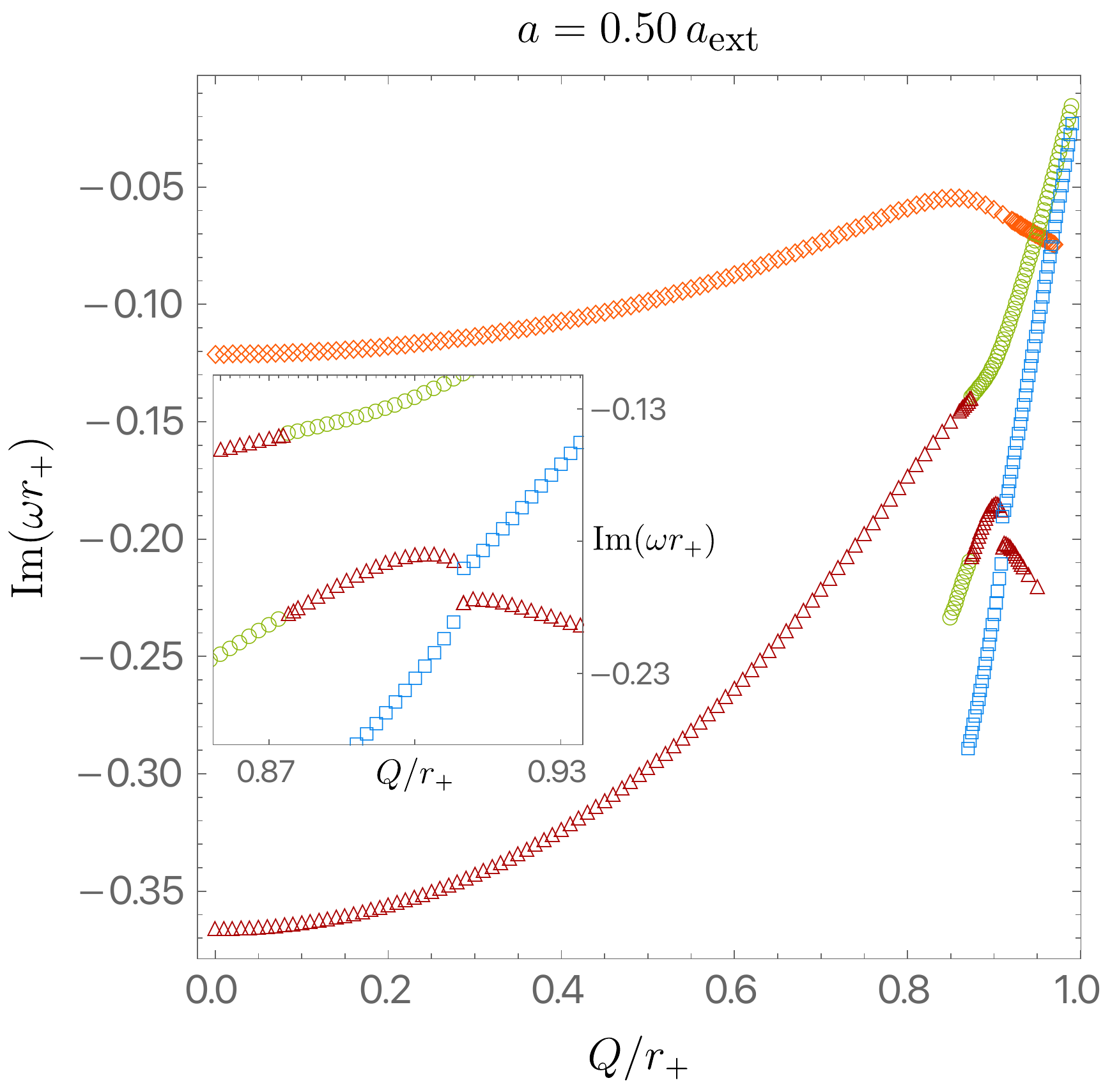}
\caption{QNM spectra for KN BHs with $a/a_{\hbox{\footnotesize ext}}=0$ (top left), $0.38$ (top right), $0.39$ (bottom left) and $0.50$ (bottom right). In the RN case, there is an unambiguous QNM family classification: the orange diamond (dark-red triangle) curve is the $n=0$ ($n=1$) PS family which reduces to the dark-red disk $\omega\, r_+=0.74734337 - 0.17792463\, i $ (red square $\omega \, r_+=0.69342199 - 0.54782975\, i$) in the Schwarzschild limit  \cite{Chandra:1983,Leaver:1985ax}. The green circle (blue square) curve is the $n=0$ ($n=1$) NH family (not shown: for $\hat{Q}<0.85$ these curves extend to lower $\mathrm{Im}\,\hat{\omega}$). In the middle panels one observes eigenvalue repulsions unique to the KN QNM spectra. In the RN case, we also show the frequency  $\tilde{\omega}_{\hbox{\tiny MAE}}$ given by \eqref{NH:freq} for $n=0$ (black curve) and for $n=1$ (magenta curve).
}
\label{Fig:spectraFix-a}
\end{figure*}  

\begin{figure*}
\includegraphics[width=.48\textwidth]{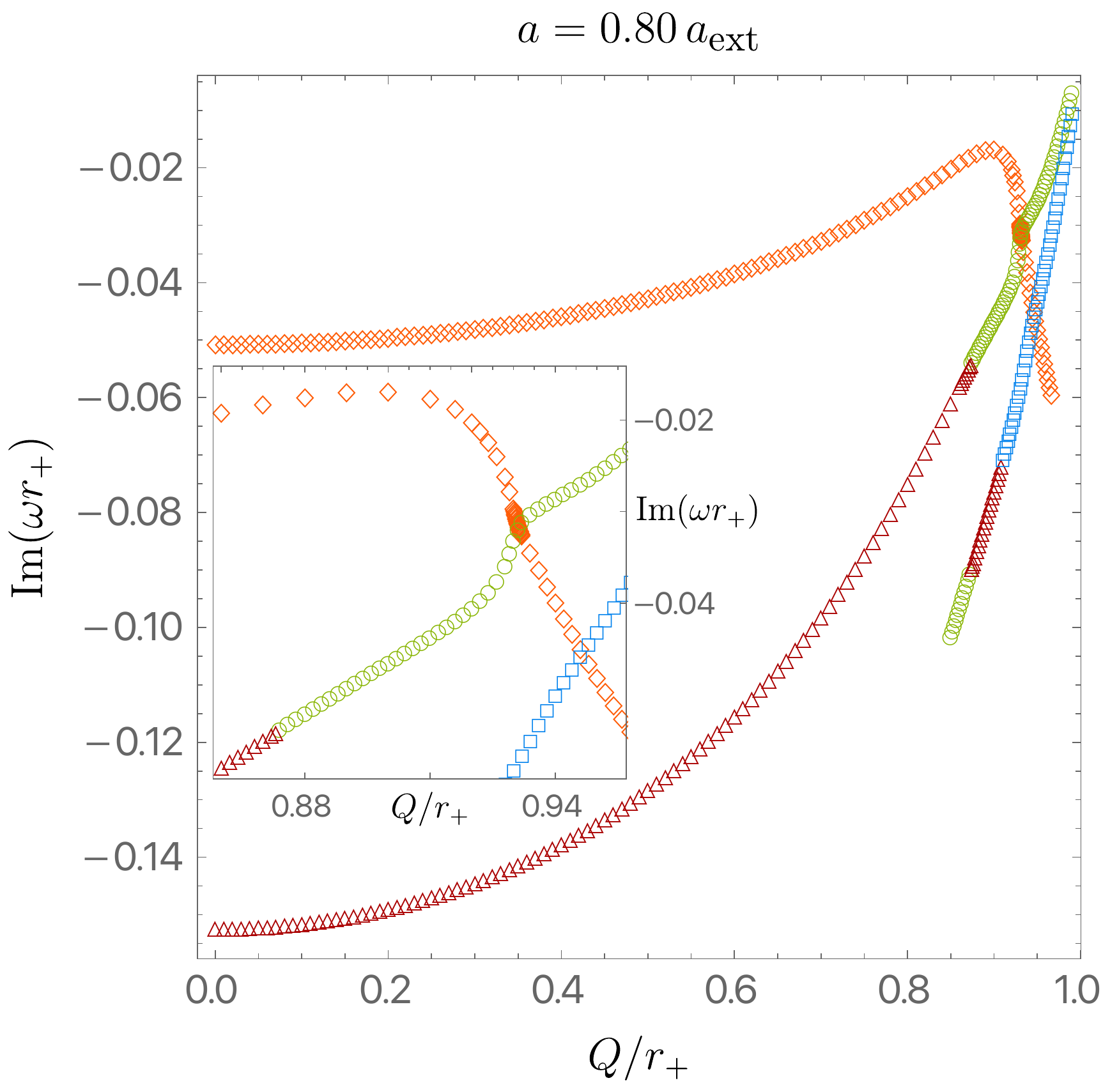}
\hspace{0.4cm}
\includegraphics[width=.48\textwidth]{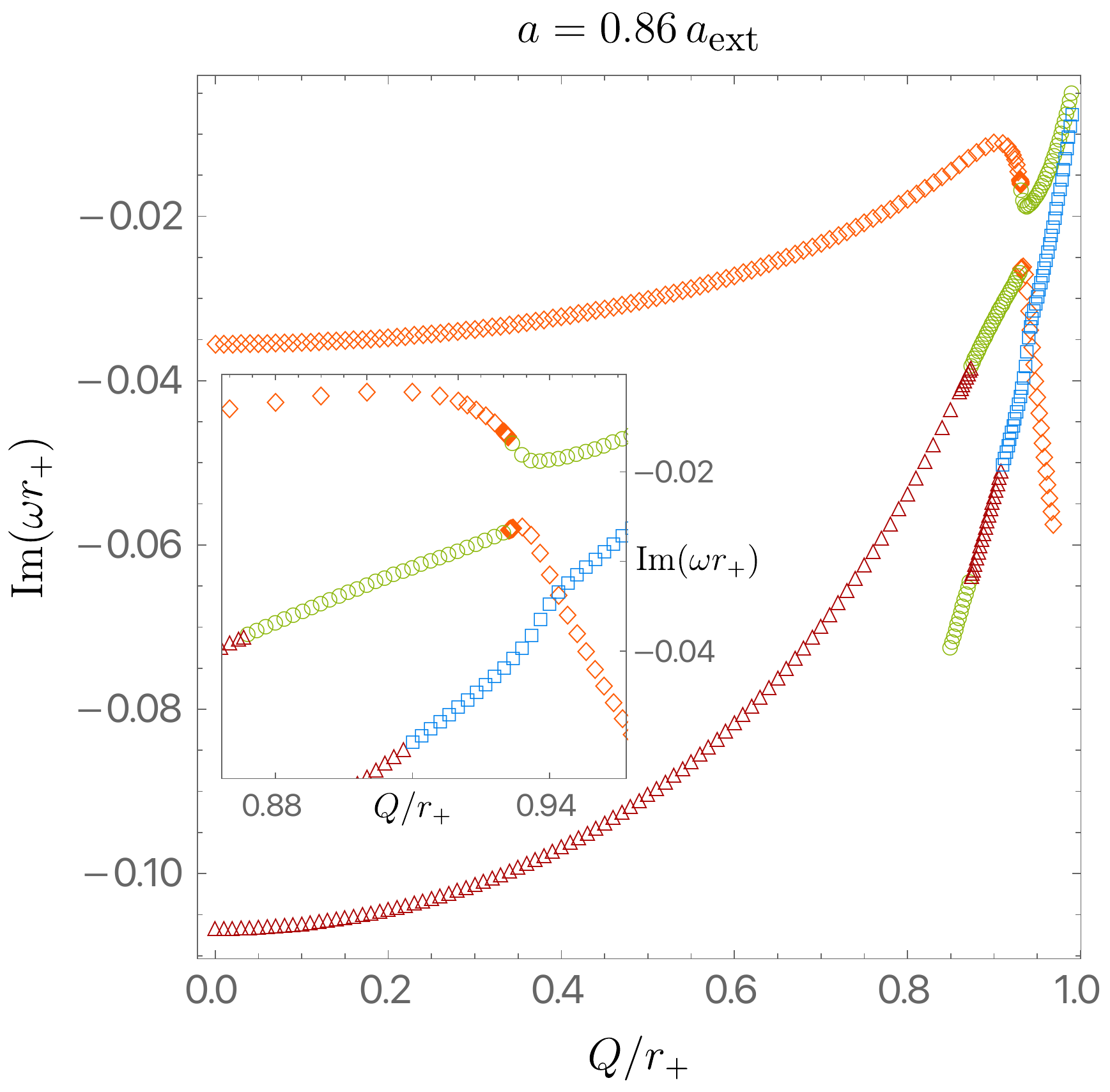}
\vskip 0.2cm
\includegraphics[width=.48\textwidth]{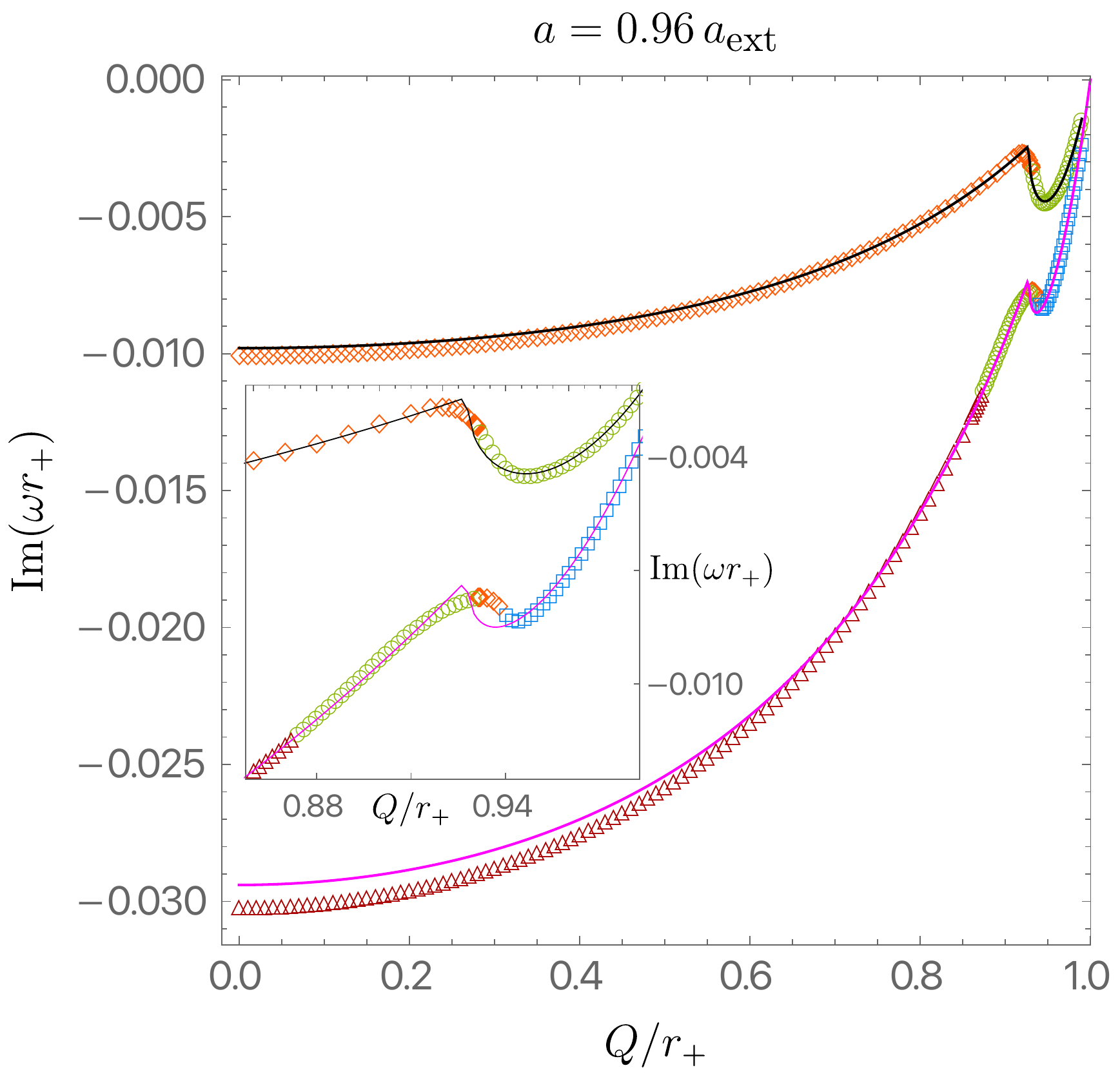}
\caption{QNM spectra for KN BHs with $a/a_{\hbox{\footnotesize ext}}=0.80$ (top left), $0.39$ (top right) and $0.96$ (bottom). One observes further eigenvalue repulsions unique to the KN QNM spectra. On the bottom panel we also show the frequency  $\tilde{\omega}_{\hbox{\tiny MAE}}$ given by \eqref{NH:freq} for $n=0$ (black curve) and for $n=1$ (magenta curve).
}
\label{Fig:spectraFix-a2}
\end{figure*}  

Perhaps surprisingly at first sight, but certainly not after the discussions in subsection~\ref{sec:repulsionTextbook}, in the numerical search of the KN QNM spectra we find eigenvalue repulsions between the two distinct families of QNMs of the system. These eigenvalue repulsions are unique to the KN QNM system since they are not observed in the spectra of Schwarzschild, RN nor Kerr black holes (for reasons that were understood in subsection~\ref{sec:repulsionTextbook}).
Our strategy to describe and discuss further these eigenvalue repulsions is the following. In Figs.~\ref{Fig:spectraFix-a}$-$\ref{Fig:spectraFix-a2} we display a series of panels. Each one of them plots the imaginary part of the dimensionless frequency, $\mathrm{Im}(\omega r_+)$, as a function of the dimensionless charge, $\hat{Q}=Q/r_+$, at fixed $a/a_{\hbox{\footnotesize ext}}$.  We choose to use units of $r_+$ since some curves change too much in a small range of charge if we use units of $M$. Different plots of this series are for different values of fixed $a/a_{\hbox{\footnotesize ext}}$. Namely, moving from top-left into bottom-right panels of Figs.~\ref{Fig:spectraFix-a}$-$\ref{Fig:spectraFix-a2} we have fixed $a/a_{\hbox{\footnotesize ext}}=0, 0.38, 0.39, 0.5, 0.8, 0.86$ and $0.96$ (see legend on top of each panel). So we start at the RN family with $a=0$ and progressively increase $a/a_{\hbox{\footnotesize ext}}$ till we reach a KN black hole family parametrized by $0\leq \hat{Q}\leq 1$ where the whole KN family is at 96\% of extremality. We have chosen these particular $a/a_{\hbox{\footnotesize ext}}$ cases because they are representative of what happens to the system in a window of $a/a_{\hbox{\footnotesize ext}}$ centred at the given $a/a_{\hbox{\footnotesize ext}}$. When we move to the next panel a new major feature appears that justifies introducing a new plot to illustrate it. We only display the imaginary part of the frequency. This is because the plots for the real part of the frequency are not very illuminating since the curves for the different modes quickly become  very close to each other as we approach extremality i.e, as  $a/a_{\hbox{\footnotesize ext}}$ increases. We will add  comments about the real part of the frequency whenever appropriate and at the end of this section. 

We can now describe in detail the content of each plot in Figs.~\ref{Fig:spectraFix-a}$-$\ref{Fig:spectraFix-a2}.
In the top-left panel of Fig.~\ref{Fig:spectraFix-a} we start with the RN black hole ($a=0$). This describes what happens to the system with $a=0$ but it is also representative of small rotation cases with $a/a_{\hbox{\footnotesize ext}}$  below $0.38$. We plot the first two overtones ($n=0,1$) of the PS QNM family (that we denote by PS$_0$ and PS$_1$ or, more generically, as PS$_n$ modes) and the first two overtones ($n=0,1$) of the NH QNM family (denoted as NH$_0$ and NH$_1$ or simply as NH$_n$ modes). The PS$_0$ and PS$_1$ curves are described by orange diamonds and dark-red triangles, respectively. In the Schwarzschild limit ($\hat{Q}= 0$), the PS$_0$  family reduces to the dark-red disk $\omega\, r_+=0.74734337 - 0.17792463\, i $ while the PS$_1$ curve reduces to red square $\omega \, r_+=0.69342199 - 0.54782975\, i$, first computed by \cite{Chandra:1983,Leaver:1985ax}. 
On the other hand, the NH$_0$ and NH$_1$ families  are the green circle and blue square curves, respectively (not shown: for $\hat{Q}<0.85$ these curves plunge quickly to lower $\mathrm{Im}\,\hat{\omega}$). Note that this plot contains the same PS$_0$ and NH$_0$ information as the one of Fig.~\ref{Fig:RN} (although here we use units of $r_+$ instead of $M$). Moreover, w.r.t.\ Fig.~\ref{Fig:RN}, in the top-left panel of  Fig.~\ref{Fig:spectraFix-a} we also display the near-horizon matched asymptotic expansion frequency $\hat{\omega}_{\hbox{\tiny MAE}}$ as given by \eqref{NH:freq} for $n=0$ (solid black curve) and for $n=1$ (solid magenta curve); these are better seen in the inset plot where one finds that \eqref{NH:freq} gives the correct slopes near-extremality at $\hat{Q}\lesssim 1$. As emphasized already in subsection~\ref{sec:NHanalytics}, these analytical $\hat{\omega}_{\hbox{\tiny MAE}}$ are in excellent agreement with the numerical NH QNM frequencies, as long as we are near extremality (which for RN occurs at $\hat{Q}=1$). Actually this time we demonstrate that \eqref{NH:freq}  is an excellent approximation (near extremality) not only for the first overtone NH$_0$ but also for NH$_1$ (and higher overtones $n$ although not shown). 
A major feature of this $a=0$ plot is that the PS$_n$ and NH$_n$ curves are very well defined and clearly distinct from each other, with the PS$_n$ frequencies well approximated by $\hat{\omega}^{\hbox{\tiny eikn}}_{\hbox{\tiny PS}}$ in \eqref{PS:eikonal}, and the NH$_n$ frequencies in excellent agreement with $ \hat{\omega}_{\hbox{\tiny MAE}} $ as given by \eqref{NH:freq}. It is also important to emphasize that the imaginary part of the PS$_0$ and NH$_0$ curves (in particular) cross each other but, as best displayed in the right panel of Fig.~\ref{Fig:RN}, this is not the case for the real part of the frequency. This will be a common feature in all the cases displayed in  Figs.~\ref{Fig:spectraFix-a}$-$\ref{Fig:spectraFix-a2}: whenever we see crossing between two curves describing the imaginary part of the frequency there is no crossing between the curves that represent the real part of the frequency. 

As far as it is possible, we will keep the same colour/shape code for the PS$_n$ and NH$_n$ QNM families displayed in the top-left panel as we move to the other plots with increasing $a/a_{\hbox{\footnotesize ext}}$. However, at a certain point we will no longer be able to assign the PS or NH nomenclatures to the QNM curves of the system. 

As we switch on $a$ and increase $a/a_{\hbox{\footnotesize ext}}$, the QNM spectra remains similar to the one on the top-left panel but the PS$_1$ (dark-red triangles) and NH$_0$ (green circles) curves start getting deformed in the region where they intersect as a simple crossover in the imaginary part. It is as if each of these curves starts feeling the presence of the other and they start interacting. This is particularly seen in the top-right panel of Fig.~\ref{Fig:spectraFix-a} for $a/a_{\hbox{\footnotesize ext}}=0.38$. Then, increasing a little bit the rotation, at $a/a_{\hbox{\footnotesize ext}}=0.39$ (bottom-left panel of Fig.~\ref{Fig:spectraFix-a}) a dramatic new feature occurs. The `old' PS$_1$ (by `old' we mean w.r.t.\ the previous plot or, ultimately, w.r.t. the $a=0$ plot) dark-red triangle curve breaks into two pieces, and the same occurs for the `old' NH$_0$ curve. This occurs for $\hat{Q}\sim 0.875$ as best seen in the inset plot. Not less remarkably, the left-branch ($\hat{Q}\lesssim 0.875$) of the `old' PS$_1$ curve merges with the right-branch ($\hat{Q}\gtrsim 0.875$) of the `old' NH$_0$ curve. That is to say, the PS$_1$ and NH$_0$ families lose their individual identity and they combine into what we now can call the PS$_1$$-$NH$_0$ family of QNMs. Similarly, the left-branch of the `old'  NH$_0$ ($\hat{Q}\lesssim 0.875$) curve joins with the right-branch of the `old'  PS$_1$ ($\hat{Q}\gtrsim 0.875$) curve to form together a new QNM family that we denote as the NH$_0$$-$PS$_1$ family of QNMs. These breakups and subsequent mergers are even more surprising because they glue two sub-families that were, for lower rotations, assigned different radial overtones $n$. At this rotation parameter we can say that we have 4 families of QNMs (from top-left to bottom-right): the PS$_0$, the PS$_1$$-$NH$_0$,  the NH$_0$$-$PS$_1$ and the NH$_1$.

Altogether, these features and frequency gaps are characteristic of the phenomenon of {\it eigenvalue repulsion} that we discussed in subsection~\ref{sec:repulsionTextbook}. In particular, in the breakup region, there is a a `{\it frequency gap}' between the new PS$_1$$-$NH$_0$ and NH$_0$$-$PS$_1$ curves. This `frequency gap' is zero exactly at the breakup rotation (somewhere in the window $a/a_{\hbox{\footnotesize ext}} \in \, [0.38,0.39]$), and then it grows as $a/a_{\hbox{\footnotesize ext}}$ increases. This is what is seen e.g.\ when we move to $a/a_{\hbox{\footnotesize ext}}=0.5$ case shown in the bottom-right panel of Fig.~\ref{Fig:spectraFix-a}. In this plot we see that a further eigenvalue repulsion episode happened  in the window $a/a_{\hbox{\footnotesize ext}} \in \, [0.39,0.5]$. Indeed, the NH$_0$$-$PS$_1$ curve (green circles plus dark-red triangles) broke up around $\hat{Q}\sim 0.91$ and the same happened to the NH$_1$ curve (blue squares). The left-branch of the `old' NH$_0$$-$PS$_1$ curve is now merged with the right-branch of the `old' NH$_1$ curve to form what we can call a NH$_0$$-$PS$_1$$-$NH$_1$ family of QNMs. Simultaneously, the left-branch of the `old' NH$_1$ curve (blue squares) is now merged with the right-branch of the `old' NH$_0$$-$PS$_1$ curve (or with a portion of the even `older' PS$_1$ curve since it only contains dark-red triangles) to form what we can call a NH$_1$$-$PS$_1$ curve.

So far the original PS$_0$ family escaped eigenvalue repulsion phenomena, but this changes when we keep increasing $a/a_{\hbox{\footnotesize ext}}$ even further as seen in Fig.~\ref{Fig:spectraFix-a2} (in this figure we drop the subdominant NH$_1$$-$PS$_1$ curve). Indeed, at $a/a_{\hbox{\footnotesize ext}}=0.8$ we already notice that  the PS$_0$ curve (orange diamonds) and NH$_0$ portion (green circles) of the PS$_1$$-$NH$_0$ curve are getting deformed by each other in the region where they  intersect as a simple crossover. Again, it is as if each of these curves feels the presence of the other and reacts to the interaction (see the inset plot). This is similar to the eigenvalue repulsion observed before between the PS$_1$ and NH$_0$ modes and, inevitably, the PS$_0$ and the PS$_1$$-$NH$_0$ curves break up in the window $a/a_{\hbox{\footnotesize ext}} \in \, [0.8,0.86]$. Indeed, in the top-right panel of Fig.~\ref{Fig:spectraFix-a2}, we see that at $a/a_{\hbox{\footnotesize ext}}=0.86$ these two curves break at $\hat{Q}\sim 0.93$. The left-branch (orange diamonds) of the `old' PS$_0$ curve merges with the right-hand branch (green circles) of the NH$_0$ portion (green circles) of the PS$_1$$-$NH$_0$ curve to produce what we denote as the PS$_0$$-$NH$_0$ family of QNMs. At the same $\hat{Q}\sim 0.93$, the left-branch of the `old' PS$_1$$-$NH$_0$ is now merged with the right branch of the `old' PS$_0$ curve to give birth to what we call a  PS$_1$$-$NH$_0$$-$PS$_0$ family.

Similar eigenvalue repulsions keep occurring when we increase $a/a_{\hbox{\footnotesize ext}}$ towards extremality. For example, already very close to extremality, namely at $a/a_{\hbox{\footnotesize ext}}=0.96$, the two most dominant QNM families are shown in the bottom panel of Fig.~\ref{Fig:spectraFix-a2} (we do not show data for even higher overtones). Here, we identify  the PS$_0$$-$NH$_0$ curve already observed in the previous plot. This is the family that has the lowest $|\mathrm{Im}\,\hat{\omega}|$ for all $\hat{Q}$. Additionally, we see that the `old' PS$_1$$-$NH$_0$$-$PS$_0$ curve of the $a/a_{\hbox{\footnotesize ext}}=0.86$ broke again (around $\hat{Q}\sim 0.94$) and merged with the right branch of the `old' NH$_1$ (blue squares) to form a four colour PS$_1$$-$NH$_0$$-$PS$_0$-NH$_1$ curve (see inset plot). 

To conclude by summarizing the key aspects of our findings, the first plot of Fig.~\ref{Fig:spectraFix-a} together with the last plot of Fig.~\ref{Fig:spectraFix-a2}, are those that probably best illustrate the main conclusion of our study. There is no doubt that $a=0$, the RN black hole, has two clearly distinct families of QNMs: the PS and NH families, together with overtones for each of them (first plot of Fig.~\ref{Fig:spectraFix-a}). Here, the PS$_n$ frequencies are well approximated by $\hat{\omega}^{\hbox{\tiny eikn}}_{\hbox{\tiny PS}}$ in \eqref{PS:eikonal}, and the NH$_n$ frequencies are in excellent agreement with $ \hat{\omega}_{\hbox{\tiny MAE}} $ as given by \eqref{NH:freq}. However, as the rotation increases, several eigenvalue repulsions progressively appear that increasingly break and combine pieces of the `old' PS$_n$ and NH$_n$ curves. Very close to extremality, we end up with a QNM landscape that is definitely very different from the RN one. Indeed,  as best illustrated in the last plot of Fig.~\ref{Fig:spectraFix-a2}, instead of having the PS$_n$ and NH$_n$ curves, one now has what we can simply call the `PS$-$NH' family and its radial overtones (with higher  $|\mathrm{Im}\, \hat{\omega}|$). Interestingly, the near-horizon matched asymptotic expansion frequency $\hat{\omega}_{\hbox{\tiny MAE}} $  given by \eqref{NH:freq} describes accurately this PS$-$NH family (and its overtones) for the whole range of $\hat{Q}$ at a fixed $a/a_{\hbox{\footnotesize ext}} $ that is close to extremality. Indeed, in the bottom panel of Fig.~\ref{Fig:spectraFix-a2}, the solid black curve describes \eqref{NH:freq} with $n=0$ and the solid magenta line represents \eqref{NH:freq} with $n=1$. And these match very well the numerical frequencies for the $n=0$ and $n=1$ PS$-$NH modes, respectively. This is a conclusion that we had already reached when discussing \eqref{NH:freq} and Figs.~\ref{Fig:NHq099-q095}-\ref{Fig:PSanalyticNH} of subsection~\ref{sec:NHanalytics}. Notice, that this matching between $\hat{\omega}_{\hbox{\tiny MAE}}$ and the numerical data includes the region of the QNM curve that we can trace back as descending from the RN PS modes (i.e.\ the orange diamond section of the $n=0$ PS$-$NH curve), in agreement with the discussion of the extension of \eqref{NH:freq} to negative values of $\lambda_2$ and associated  Fig.~\ref{Fig:PSanalyticNH} that we had in subsection~\ref{sec:NHanalytics}.  In particular, this means that the PS$-$NH overtone curves (including the two shown in the in bottom panel of Fig.~\ref{Fig:spectraFix-a2}) approach $\mathrm{Im}\,\omega=0$ and $\mathrm{Re}\,\omega=m \Omega_H^{\hbox{\footnotesize ext}}$ as $a/a_{\hbox{\footnotesize ext}}\to 1$ for any value of $\hat{Q}$.

In the analysis of this section we have not discussed much the behaviour of the real part of the frequency. This is because nothing of relevance happens to this quantity as we evolve though the 2-parameter space of KN black holes away from the level crossing point that occurs in the imaginary part. Take for example the $n=0$ PS and NH modes. As we move away from the level repulsion point at $\{\hat{a},\hat{Q}\}|_{\hbox{\footnotesize ext}}=\{\hat{a}_\star,\hat{Q}_\star\} \simeq \{0.360,0.932\}$, during a good neighbourhood the real part of these two modes is very similar (parallel to each other) but they do not cross. Then, sufficiently far away from the level crossing point the two $\mathrm{Re}\,\hat{\omega}$ surfaces become clearly distinct. These properties will be observed in the right panel of Fig.~\ref{Fig:Z2l2m2n0n1-rp}. 
It turns out that the eigenvalue repulsions induce strong effects at the level of the imaginary part of the frequencies but leave no (notably visible) imprint on the real part of the frequencies. 
In more detail, whenever there is crossing between two curves describing the imaginary part of the frequency there is no crossing between the curves that represent the real part of the frequency; that is, the crossing in the imaginary part of the frequency never extends to the full complex frequency plane, with one exception. 
For each pair of modes,  this exception occurs when we approach the particular extremal KN black hole with $\{\hat{a},\hat{Q}\}|_{\hbox{\footnotesize ext}}=\{\hat{a}_\star,\hat{Q}_\star\}$. 

To summarise, for definiteness  consider again the PS$_0$ and NH$_0$ pair of modes. In this case, the star point at extremality has  $\{\hat{a}_\star,\hat{Q}_\star\} \simeq \{0.360,0.932\}$ and is represented with a $\star$ in Fig.~\ref{Fig:PS-extremality}. At this $\star$ point,  the PS and NH modes both have $\mathrm{Im}\,\omega=0$ and $\mathrm{Re}\,\omega=m \Omega_H^{\hbox{\footnotesize ext}}$. That is, they have the same complex frequency and, as discussed in subsection~\ref{sec:repulsionTextbook}, this is the only level crossing point of the system. As we move away from this $\star$ point, avoided crossing effects emerge and Figs.~\ref{Fig:spectraFix-a}$-$\ref{Fig:spectraFix-a2} illustrate that these repulsion effects can be strong and induce intricate features in the behaviour of the $\mathrm{Im}\,\hat\omega$ curves (but not in the $\mathrm{Re}\,\hat\omega$ curves) in a neighbourhood of the level crossing point but they become unnoticed far away from this point.

\section{Full frequency spectra of the QNMs with slowest decay rate\label{sec:DominantSpectra}}

\begin{figure}[h]
\centering
\includegraphics[width=.50\textwidth]{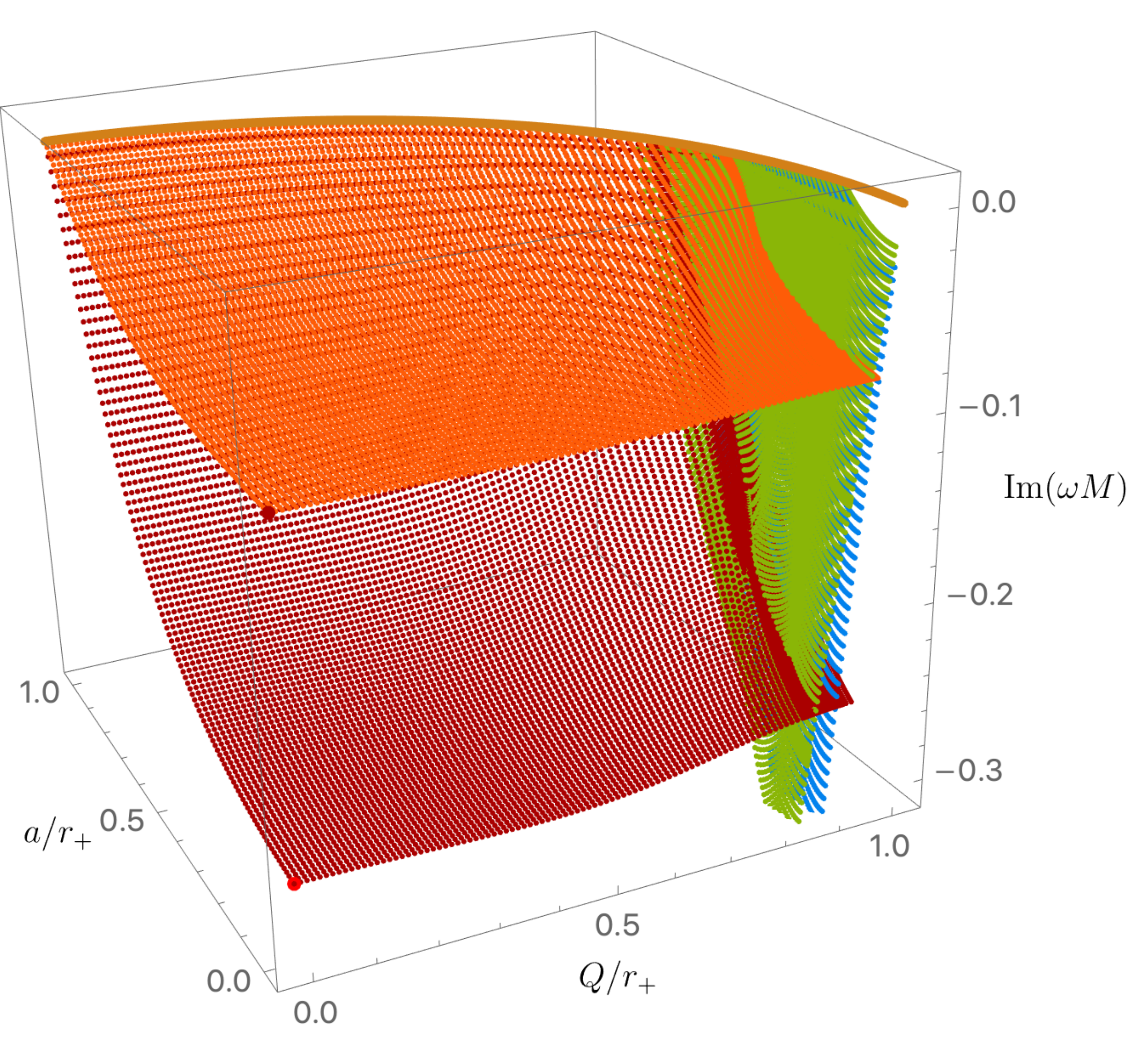}
\includegraphics[width=.48\textwidth]{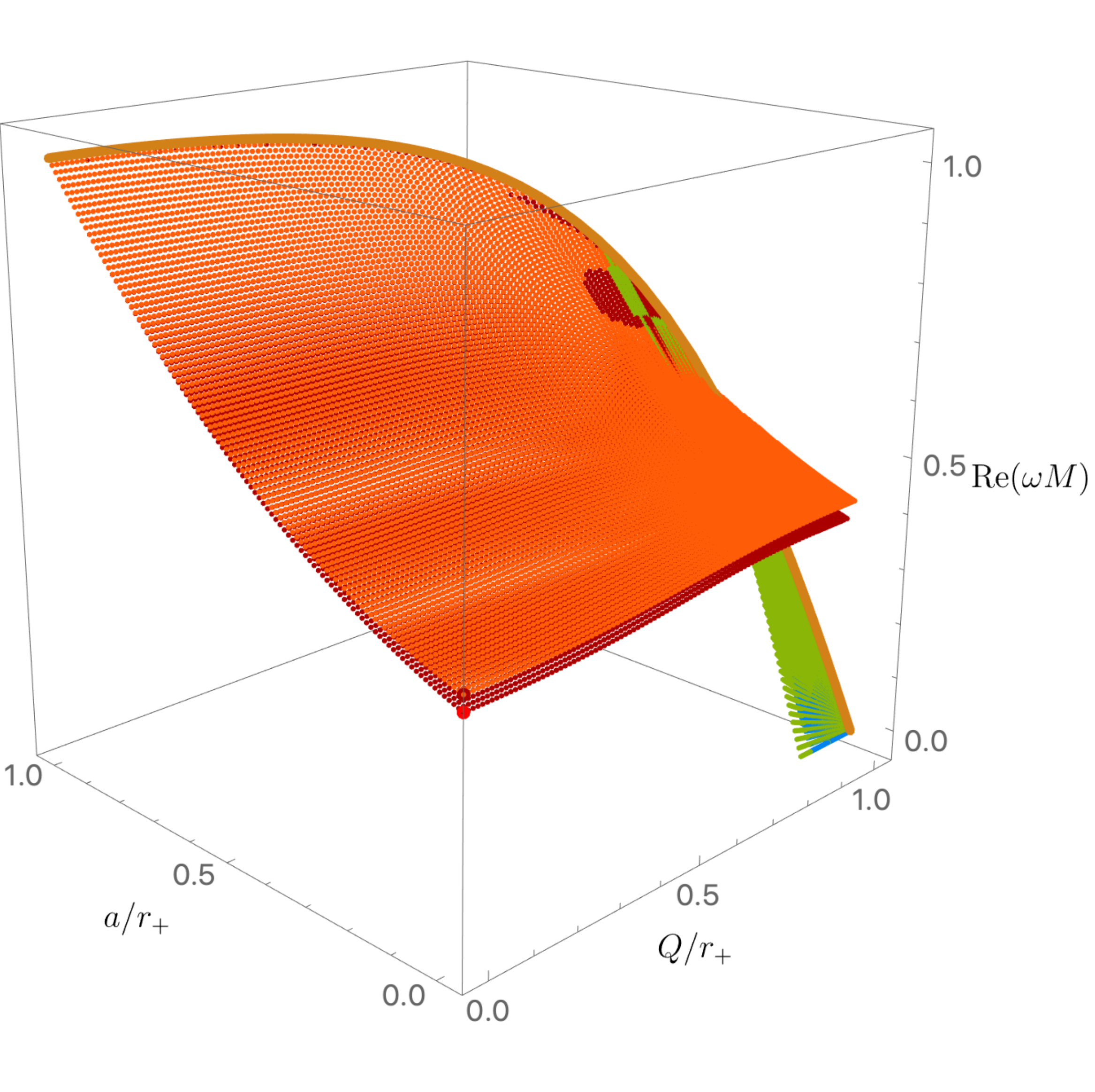}
\caption{Imaginary (left panel) and real (right panel) part of the frequency for the $Z_2$ $\ell=m=2$ KN QNMs. The orange and the green surfaces are the PS$_0$ and NH$_0$ families (respectively), while the dark-red and blue  surfaces are the PS$_1$ and NH$_1$ families (respectively). When $\{Q,a\}=\{0,0\}$, The PS$_0$ surface reduces to $\omega\, M=0.37367168 - 0.08896232\, i $, while the PS$_1$ surface reduces to $\omega \, M=0.34671099 - 0.27391487\, i$ \cite{Chandra:1983,Leaver:1985ax}.
The extremal KN frequencies are described by the solid brown line which has $\mathrm{Im}\,\tilde{\omega}=0$ and $\mathrm{Re}\,\tilde{\omega}=m \tilde{\Omega}_H^{\hbox{\footnotesize ext}}$.
}\label{Fig:Z2l2m2n0n1-rp}
\end{figure} 

We have done a fairly good survey (having in mind the associated computational cost; more in Section~\ref{sec:FullSpectra}) of several gravito-electromagnetic QNMs of KN and we conclude that, as expected, the modes that have the slowest decay rate are those that are the $\{Q,a\}\neq \{0,0\}$ extension of the Schwarzschild mode that Chandraseckar classifies as the $Z_2$, $\ell=m=2$, $n=0$ mode; see Table V, page 262  of \cite{Chandra:1983} and associated discussion. These are also the modes we discussed in Section~\ref{sec:repulsionTextbook}, together with the $n=1$ overtone of the same family (which was first studied by Leaver \cite{Leaver:1985ax}).

Therefore, before doing a general survey of other modes of interest in Section~\ref{sec:FullSpectra}, in this section we display the QNM spectra of the $Z_2$, $\ell=m=2$, $n=0$ modes and the $n=1$ overtones (the latter will allow us to complement or even complete the analysis of Section~\ref{sec:repulsionSub}). Unlike in Section~\ref{sec:repulsionSub} where we made a judicious choice of 2-dimensional plots at fixed $a/a_{\hbox{\footnotesize ext}}$ to exhibit and explain eigenvalue repulsions, in this section we plot the QNM frequencies as a function of the full 2-dimensional parameter space of KN. As discussed previously, we can take these 2 dimensionless parameters to be  $\{\tilde{a},\tilde{Q}\}\equiv \{a/M,Q/M\}$ or $\{\hat{a},\hat{Q}\}\equiv \{a/r_+,Q/r_+\}$.
From an astrophysical perspective, it is appropriate to work in units of $M$ and this is how we present  many of our physical results, in particular the frequency $\omega M$. However, in practice we have scanned the 2-dimensional parameter space in units of $r_+$: typically (except when we needed a finer grid to study a particular feature), we divided our numerical grid for $\{\hat{Q}, \hat{a} \}\equiv \{Q/r_+, a/r_+\}$ with $100\times 100$ points with $0\leq \hat{Q}\leq 1$ and   $0\leq \hat{a}\leq \hat{a}_{\hbox{\footnotesize ext}} $ with  $\hat{a}_{\hbox{\footnotesize ext}}= \sqrt{1-\hat{Q}^2}$. This is because 
some features of the QNM spectra (e.g.\  the crossovers or eigenvalue repulsions between modes) occur in small windows of $(Q/M,a/M)$ which translate into wider windows of $(Q/r_+,a/r_+)$.
For this reason, some of the fine details of the frequency spectra that we discuss in this section are better displayed if we present our results in 3-dimensional plots $\{Q/r_+, a/r_+,\omega M\}$. 
In the figures of this section, the left panel always displays the imaginary part of the frequency, $\mathrm{Im}(\omega M)$, while the right panel plots the  real part of the frequency, $\mathrm{Re}(\omega M)$.

In Fig.~\ref{Fig:Z2l2m2n0n1-rp} we present the raw data that we collected for the two QNM families that have the slowest rate, namely the families that we identify with the PS$_0$ and NH$_0$ modes in the RN limit and their $n=1$ overtone cousins PS$_1$ and NH$_1$ (for the $Z_2$ $\ell=m=2$ modes). Namely,  the  orange and the green surfaces are the $n=0$ PS ($\mathrm{PS}_0$) and $n=0$ NH ($\mathrm{NH}_0$)  families, respectively. 
On the other hand, the dark-red surface and the blue surface describe the  $n=1$ PS ($\mathrm{PS}_1$) and $n=1$ NH ($\mathrm{NH}_1$)  families, respectively. 
Thus, we are using the same colour code that was employed in Figs.~\ref{Fig:spectraFix-a}$-$\ref{Fig:spectraFix-a2} of section~\ref{sec:repulsionSub}. 
The solid brown curves are at extremality. They are parametrized by $ \hat{a}=a_{\hbox{\footnotesize ext}} = \sqrt{1-\hat{Q}^2}$ and have $\mathrm{Im}\,\tilde{\omega}=0$ and $\mathrm{Re}\,\tilde{\omega}=m \tilde{\Omega}_H^{\hbox{\footnotesize ext}}$ (we will use the same colour code for this extremal curve in all other 3-dimensional plots where this curve plays a relevant role).
Note that the NH$_{0,1}$ surfaces have a very large slope and plunge into very large negative $\mathrm{Im}(\omega M)$ as we move away from the $\hat{a}= \hat{a}_{\hbox{\footnotesize ext}}(\hat{Q})$ extremal curve or, in the RN case, away from the $\hat{Q}=1$ extremal solution. Therefore we only plot these families at large $\hat{Q}$ (say, for $\hat{Q}\gtrsim 0.8$) where they can have  $|\mathrm{Im}(\omega M)|$ of the order as or smaller than those for the PS$_{0,1}$ surfaces. 

From the analysis of  Fig.~\ref{Fig:Z2l2m2n0n1-rp}, several properties emerge. First, as expected, the $n=0$ overtones always have the slowest decay rate of their families. Namely, in the left panel, the orange PS$_0$ surface is above the dark-red PS$_1$  surface and the green NH$_0$ surface is above the blue NH$_1$  surface. The PS$_0$ and PS$_1$ surfaces reduce to the Schwarzschild QNMs (red points) at $\{Q,a\}=\{0,0\}$, whose frequencies where first computed in \cite{Chandra:1983,Leaver:1985ax}. The plane with $a=0$ in Fig.~\ref{Fig:Z2l2m2n0n1-rp} coincides with the RN plots of Fig.~\ref{Fig:RN} or, equivalently, with the RN plot in the top-left panel of  Fig.~\ref{Fig:spectraFix-a}, after we do the required conversion between $r_+$ and $M$ units.
Similarly, ``snapshots" at constant $a/a_{\hbox{\footnotesize ext}}=0, 0.38, 0.39, 0.5, 0.8, 0.86,0.96$ of   Fig.~\ref{Fig:Z2l2m2n0n1-rp} yields the series of 2-dimensional plots displayed in Figs.~\ref{Fig:spectraFix-a}$-$\ref{Fig:spectraFix-a2}, after we do the units conversion  $\omega M \to \omega r_+$. 

{\it Naively},  a ``bird's-eye" view of the  left panel of  Fig.~\ref{Fig:Z2l2m2n0n1-rp} {\it seems} to suggest that the four surfaces intersect each other with simple crossovers. For example, the orange $\mathrm{PS}_0$  and the green $\mathrm{NH}_0$ surfaces seem to intersect along a curve $\hat{Q}=\hat{Q}_c(\hat{a})$.  
In the RN limit $\hat{a}\to 0$, this intersection curve gives the RN intersection point, \ie $\hat{Q}_c(\hat{a}=0)=\hat{Q}_c^{\hbox{\tiny RN}}\simeq 0.959227$ (which corresponds, in units of $M$, to $\tilde{Q}_c^{\hbox{\tiny RN}}\simeq 0.9991342$) already displayed in the left panel of Fig.~\ref{Fig:RN}. On the opposite end, at extremality (on top of the solid brown curve), we should have $\hat{Q}_c(\hat{a}=\hat{a}_{\hbox{\footnotesize ext}}) \simeq \hat{Q}_\star$ where the $\star$ point was defined in the discussion that leads to \eqref{NH:starWKB}. This intersection curve $\hat{Q}=\hat{Q}_c(\hat{a})$ between the $\mathrm{PS}_0$  and $\mathrm{NH}_0$ surfaces indeed is well defined for $0\leq a/a_{\hbox{\footnotesize ext}}\lesssim 0.82$ but, a fine-tuning or zoom-in analysis of the left panel proves that this is definitely no longer the case for $0.82<a/a_{\hbox{\footnotesize ext}}\leq 1$. Indeed, this fine-tuned analysis was already performed in Figs.~\ref{Fig:spectraFix-a}$-$\ref{Fig:spectraFix-a2}: in all plots of Fig.~\ref{Fig:spectraFix-a} and in the top-left panel of Fig.~\ref{Fig:spectraFix-a2}  the PS$_0$ and NH$_0$ families indeed intersect with a simple crossover (but only the imaginary part of the frequency cross). However, between the top-left (for $a/a_{\hbox{\footnotesize ext}}=0.8$) and top-right (for $a/a_{\hbox{\footnotesize ext}}=0.86$) panels of Fig.~\ref{Fig:spectraFix-a2} we concluded that the PS$_0$ and NH$_0$ families, instead of intersecting, suffer eigenvalue repulsions that effectively destroy their individual identities and leads to the formation of PS-NH families of modes. Coming back to the  left panel of  Fig.~\ref{Fig:Z2l2m2n0n1-rp}, these eigenvalue repulsions occur roughly for 
 $0.82<a/a_{\hbox{\footnotesize ext}}\leq 1$ and in the charge window $0.928 \lesssim \hat{Q}\lesssim 0.960$. Again, the eigenvalue repulsions in this region are not visible in the ``bird's-eye" view of the left panel of  Fig.~\ref{Fig:Z2l2m2n0n1-rp}; we need to zoom-in to make this noticeable very much like we did in Figs.~\ref{Fig:spectraFix-a}$-$\ref{Fig:spectraFix-a2}. However, in Fig.~\ref{Fig:Z2l2m2n0n1-rp} there is a particular point that stands-out. The is the level crossing point located at the extremal brown curve with  $\{\hat{a}_\star,\hat{Q}_\star\} \simeq \{0.360,0.932\}$ and $\mathrm{Im}\,\omega=0$ and $\mathrm{Re}\,\omega=m \Omega_H^{\hbox{\footnotesize ext}}$ where the orange and green surfaces meet (see subsection~\ref{sec:repulsionTextbook}).
 
 Similarly, a zoom-in of the  left panel of Fig.~\ref{Fig:Z2l2m2n0n1-rp} (illustrated again in Figs.~\ref{Fig:spectraFix-a}$-$\ref{Fig:spectraFix-a2}) shows that the dark-red PS$_1$ and green NH$_0$ surfaces intersect with simple crossovers in the window $0\leq a/a_{\hbox{\footnotesize ext}}\lesssim 0.38$  (but only the imaginary part of the frequency cross), but this is replaced by eigenvalue repulsions between the two families roughly in the window  $0.38< a/a_{\hbox{\footnotesize ext}}\leq 1$  and $0.870 \lesssim\hat{Q}\lesssim 0.885$. Finally, other eigenvalue repulsions, e.g.\  between the PS$_1$ and NH$_1$ surfaces, also occur in the left panel of Fig.~\ref{Fig:Z2l2m2n0n1-rp} as identified in Figs.~\ref{Fig:spectraFix-a}$-$\ref{Fig:spectraFix-a2}.

The evolution and intersections of the four QNM surfaces is much simpler and much less dramatic at the level of the real part of the frequency, which is plotted in the right panel of  Fig.~\ref{Fig:Z2l2m2n0n1-rp}.
We see that the $\mathrm{Re}(\omega M)$ of the orange PS$_0$ and dark-red PS$_1$ families is very similar and the same happens for  $\mathrm{Re}(\omega M)$ of the green NH$_0$ and blue NH$_1$ families. So much so that one barely distinguishes the PS$_0$ and PS$_1$ surfaces and, even less, the NH$_0$ and NH$_1$ surfaces. Moreover, nothing special happens to the real part of the frequency in the regions where the eigenvalue repulsions happen in the imaginary part of the frequency (see further discussions about this in the end of subsection~\ref{sec:repulsionSub}).  
 
 \begin{figure}
\includegraphics[width=.50\textwidth]{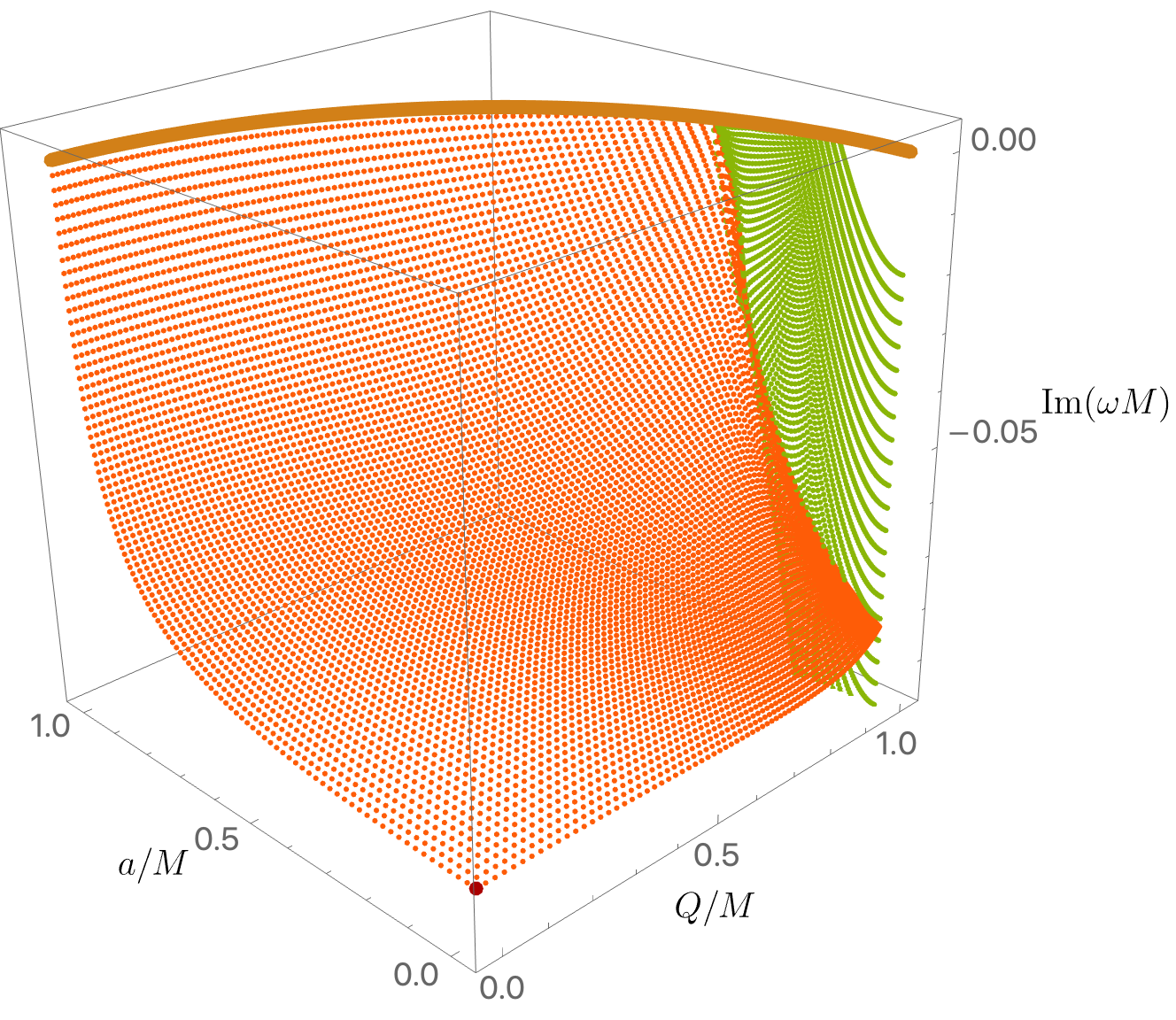}
\includegraphics[width=.49\textwidth]{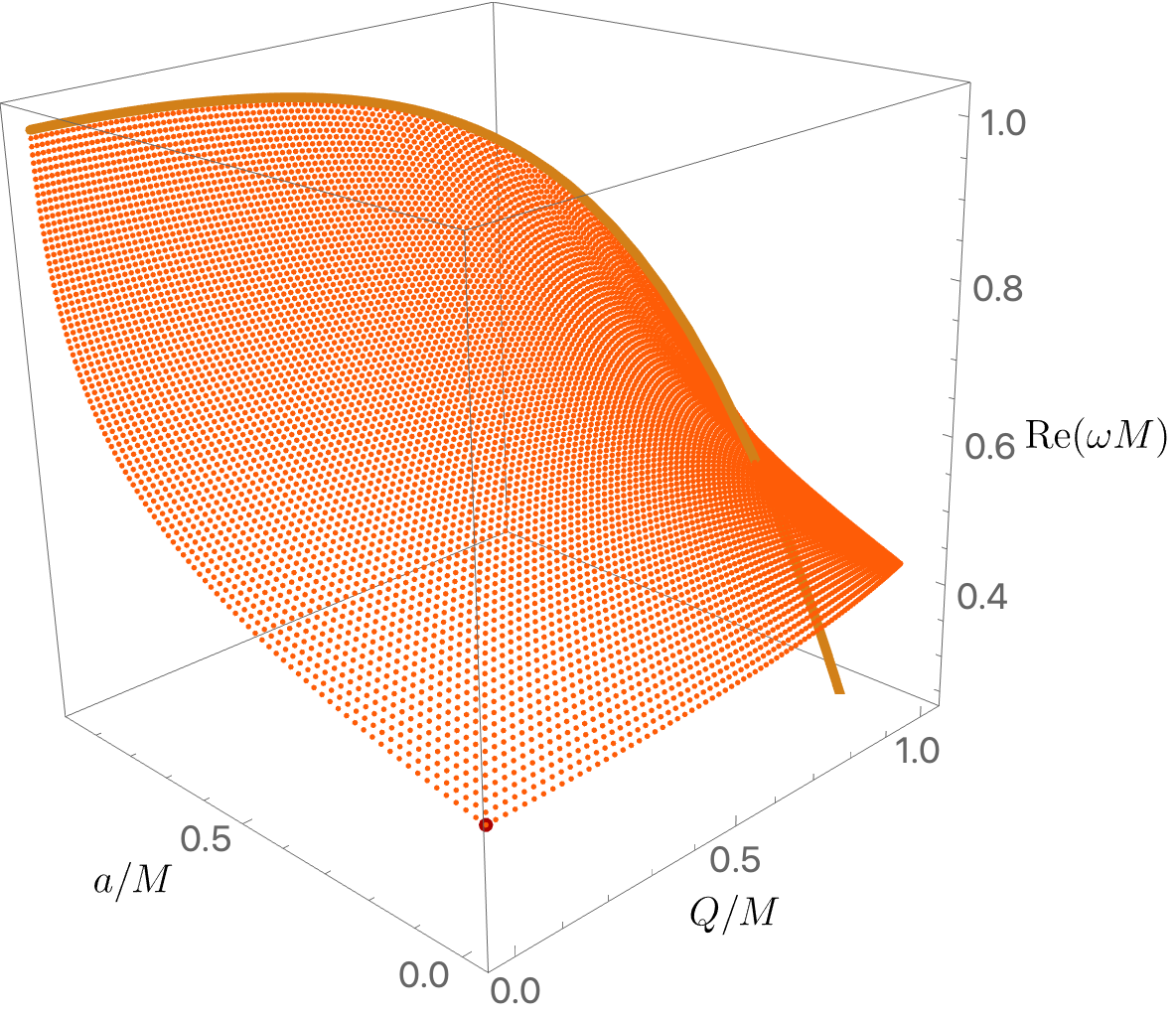}
\caption{Imaginary (left panel) and real (right panel) parts of the frequency for the $Z_2$, $(\ell,m,n)=(2,2,0)$ KN PS QNM. The dark-red point ($a=0=Q$), $\tilde{\omega}\simeq 0.37367168 - 0.08896232\, i $, is the gravitational QNM of Schwarzschild  \cite{Chandra:1983,Leaver:1985ax}.
The orange surface describes $\mathrm{PS}_0$ modes while the green surface corresponds to the $\mathrm{NH}_0$ modes.
}
\label{Fig:Z2l2m2n0+}
\end{figure}  

From the analysis of both plots in Fig.~\ref{Fig:Z2l2m2n0n1-rp} we see that the NH$_{0,1}$ surfaces always approach the solid brown curve with  $\mathrm{Im}\,\omega=0$ and $\mathrm{Re}\,\tilde{\omega}=m \tilde{\Omega}_H^{\hbox{\footnotesize ext}}$ at extremality. On the other hand, the PS$_{0,1}$ curves  approach this solid brown curve  if and only if $\hat{a}_{\hbox{\footnotesize ext}}(\hat{Q})>\hat{a}_{\star}$  where the $\star$ point was introduced in the discussion that leads to \eqref{NH:starWKB}, pinpointed  in Fig.~\ref{Fig:PS-extremality}, and identified as the \emph{level crossing point} of the system in subsection~\ref{sec:repulsionTextbook}. For  $\hat{a}_{\hbox{\footnotesize ext}}(\hat{Q})< \hat{a}_{\star}$ which happens for $\hat{Q}_{\star} < \hat{Q}\leq 1$ this is no longer the case, in agreement with the discussions of \eqref{NH:starWKB} and of Figs.~\ref{Fig:WKBlambda2-m}$-$\ref{Fig:PS-extremality}.
 
To explicitly demonstrate/justify why we have chosen to display many of our plots in units of $r_+$ in the plots of Figs.~\ref{Fig:spectraFix-a}$-$\ref{Fig:Z2l2m2n0n1-rp}, in Fig.~\ref{Fig:Z2l2m2n0+} we plot the $n=0$ PS and NH families, so the same as in Fig.~\ref{Fig:Z2l2m2n0n1-rp}, but this time without including the $n=1$ families and all quantities in units of $M$, i.e.\ the plot $\{Q/M,a/M,\omega M\}$.
In the left panel, the slope of the green NH$_0$ surface is now even more vertical than in Fig.~\ref{Fig:Z2l2m2n0n1-rp}, which indicates that the eigenvalue repulsions occur in windows of $Q/M$ that are much narrower than in $Q/r_+$. Furthermore,  in the right panel the NH$_{0,1}$ surfaces exist in such a narrow region that they are not visible: they are too close to the extremal solid brown line with a width extremely small and invisible to the naked eye.

\begin{figure*}
\includegraphics[width=.50\textwidth]{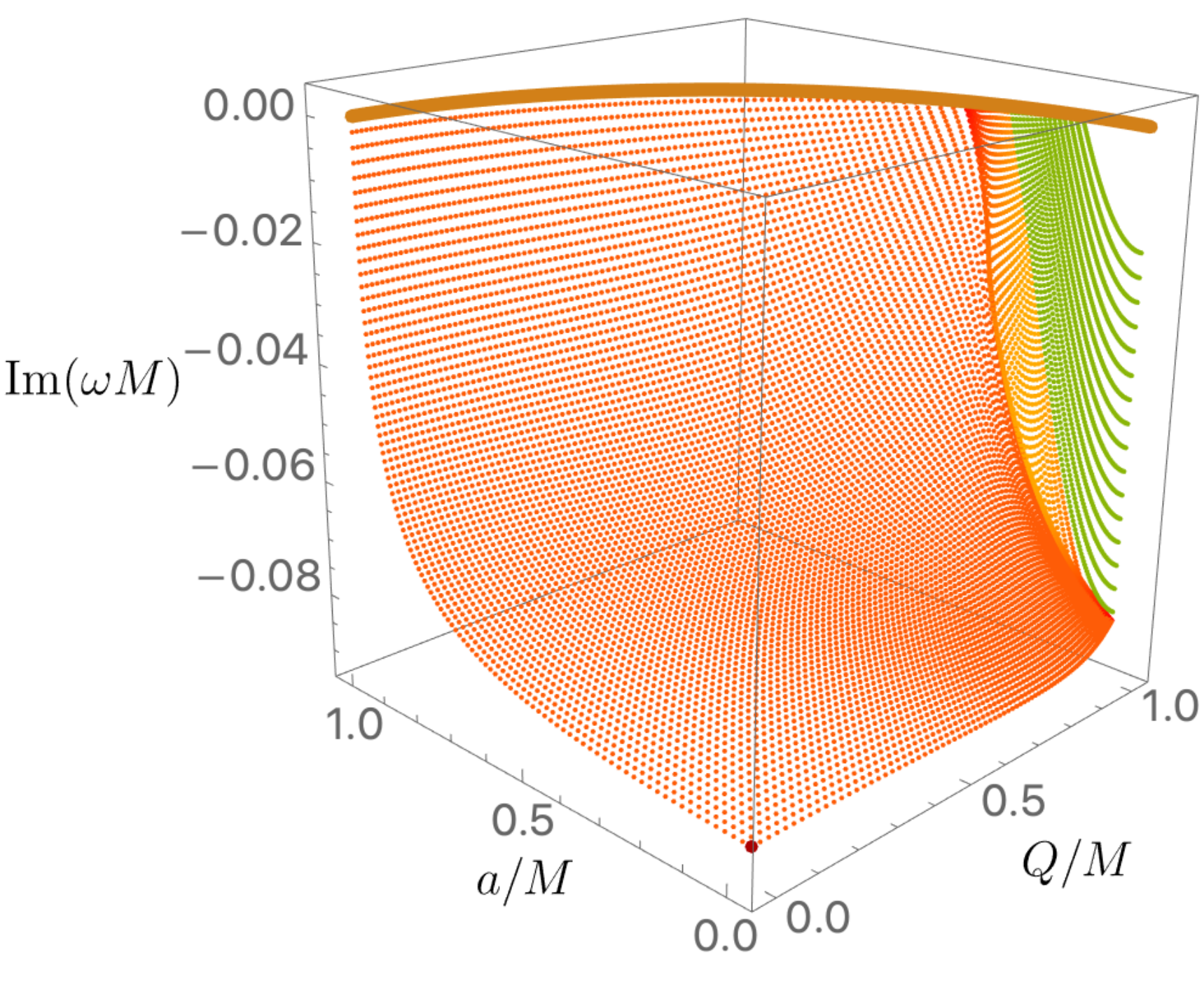}
\includegraphics[width=.49\textwidth]{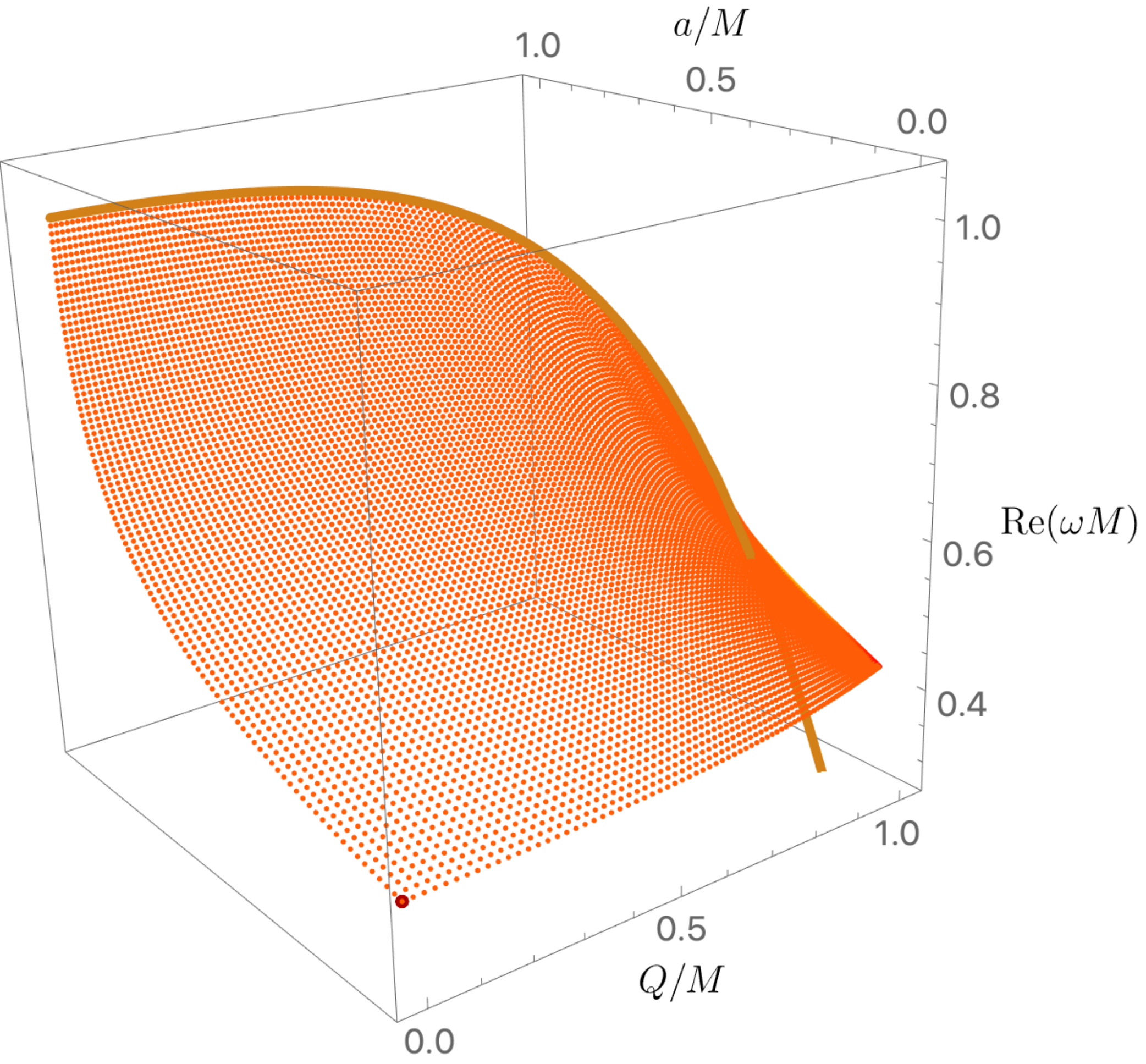}
\caption{Imaginary (left panel) and real (right panel) parts of the frequency for the $Z_2$, $\ell=m=2, n=0$ KN QNM with lowest $\mathrm{Im}\,|\tilde{\omega}|$. At extremality, the dominant mode always starts at $\mathrm{Im}\,\tilde{\omega}=0$ and $\mathrm{Re}\,\tilde{\omega}=m\tilde{\Omega}_H^{\hbox{\footnotesize ext}}$ (brown curve). The dark-red point ($a=0=Q$), $\tilde{\omega}\simeq 0.37367168 - 0.08896232\, i $, is the gravitational QNM of Schwarzschild ~\cite{Chandra:1983,Leaver:1985ax}. In the right panel, the orange and green regions are so close to the extremal brown curve that they are not visible.}
\label{Fig:Z2l2m2n0-SlowestDecay}
\end{figure*}

Finally note that if we are interested on the numerical value of the frequency of the slowest decaying mode of KN, we simply need to take the mode with minimum $|\mathrm{Im}(\omega M)|$ for a given $\{Q/M,a/M\}$ in Fig.~\ref{Fig:Z2l2m2n0+}. For completeness, we display the result of this operation in Fig.~\ref{Fig:Z2l2m2n0-SlowestDecay}, which was first presented in \cite{Dias:2021yju}.
The $Z_2$ $\ell=m=2$, $n=0$ KN modes with slowest decay rate always terminate at extremality along the extremal solid brown curve, with the frequencies off extremality well approximated by \eqref{NH:freq} as best illustrated in Figs.~\ref{Fig:NHq099-q095}$-$\ref{Fig:PSanalyticNH} and in the bottom panel of Fig.~\ref{Fig:spectraFix-a2}.
The red surface family, continuously connected to the Schwarzschild mode (dark-red point ~\cite{Chandra:1983,Leaver:1985ax}), is the $\mathrm{PS}_0$ QNM family as we unambiguously identify it in the RN limit. It dominates the spectra for most of the parameter space. However, for large $\tilde{Q}$ it is instead the green surface $\mathrm{NH}_0$ QNM family (as clearly identified in the RN limit) that has the  lowest $|\mathrm{Im}\,\tilde{\omega}|$. In between these orange/green regions there is a yellowish zone. This is where either simple crossovers  (that trade mode dominance) or eigenvalue repulsions between the $\mathrm{PS}_0$ and $\mathrm{NH}_0$ modes occurs. These were analysed in the discussion of Figs.~\ref{Fig:spectraFix-a}$-$\ref{Fig:spectraFix-a2} where we also found that as we approach extremality it is appropriate to drop the PS and NH classifcation and adopt the nomenclature PS$-$NH families and their overtones.

In the three Figs.~\ref{Fig:Z2l2m2n0n1-rp}$-$\ref{Fig:Z2l2m2n0-SlowestDecay}, at very large charge, namely for $\hat{Q}>0.99$ there is a gap between the last green NH line (with $\hat{Q}=0.99$) and the extremal solid brown curve. We have not collected data in this region because we already know (see Fig.~\ref{Fig:NHq099-q095} and the bottom  panel of Fig.~\ref{Fig:spectraFix-a2}) that in this region so close to extremality, the analytical near-horizon MAE frequency 
$\tilde{\omega}_{\hbox{\tiny MAE}}$ $-$ as given by \eqref{NH:freq} $-$ provides an excellent approximation that can be used for any physical application where such high charge values might be needed.

\section{QNM spectra: a survey of key modes \label{sec:FullSpectra}}

So far we have been assuming that the least damped gravito-electromagnetic QNMs of the KN black holes are the $Z_2$ $\ell=m=2$ modes with $n=0$. But we have not yet provided evidence that this is the case. It is certainly the case for the RN black hole subfamily ($a=0$) and for the Kerr black hole ($Q=0$) since several $\{\ell, m\}$ modes have already been computed in the literature for these cases. But strictly speaking the $Z_2$ $\ell=m=2$ does not necessarily need to be the mode with slowest decay in the {\it whole} parameter space $\{Q/M, a/M\}$ of the KN black hole away from the RN and Kerr sufamilies. Therefore, in this section we do a survey of what should be the QNMs of KN that could eventually challenge the dominance of the $Z_2$ $\ell=m=2$ mode. Mainly these are all families with $\ell=1,2$ and $|m|\leq \ell$ that are not pure gauge modes and the $Z_2$ families with $\ell=m=2,3,4,5$. For each mode, we only display the data for the first radial overtone ($n=0$) because higher overtones always have larger $|\mathrm{Im}(\omega M)|$ that the $n=0$ one.

To classify and identify more precisely the QNMs families that we will study,  note that for $Q,a\to 0$ we must recover the Schwarzschild QNMs. In this limit, it is well known that there are two families of QNMs, namely the Regge-Wheeler (aka odd or axial) modes \cite{Regge:1957td}  and the Zerilli (aka even or polar) modes \cite{Zerilli:1974ai,Moncrief:1974am} . These families are isospectral, i.e.\ they have exactly the same spectrum \cite{Chandra:1983}. Ultimately, we only need  to distinguish the gravitational modes of Schwarzschild (described in Table V of page 262 \cite{Chandra:1983}---hereafter Table of \cite{Chandra:1983}---by the eigenfunction $Z_2$) from the electromagnetic modes of Schwarzschild (described in Table of \cite{Chandra:1983} by the eigenfunction $Z_1$). In recent decades, these QNMs were computed more accurately as detailed in the review  \cite{Berti:2009kk}. Each of these $Z_2$ and $Z_1$ modes  in Schwarzschild  can be found by solving a single pair of ODEs that constitute an eigenvalue problem for the angular separation constant and frequency  \cite{Regge:1957td,Zerilli:1974ai,Moncrief:1974am}. The Schwarzschild modes are specified by the harmonic number $\ell=0,1,2,3,\cdots$ that essentially fixes the separation constant of the problem after requiring regularity of its spherical harmonic eigenfunctions  ($Z_2$ perturbations with $\ell=0$ and $\ell=1$ are modes that change the mass and the angular momentum of the black hole, respectively; thus we do not discuss these further). When the black hole has charge and rotation, we have to scan a two parameter space in $\{Q/M,a/M\}$. The above two families become coupled gravito-electromagnetic QNMs and the Schwarzschild eigenvalue $\ell$ does not appear explicitly in the KN PDEs \eqref{ChandraEqs}. However, we can still count the number of nodes along the polar direction of the eigenfunctions of \eqref{ChandraEqs} and this gives $\ell$. So, when $Q\neq 0$ and $a\neq 0$, we can still assign to a given mode the value of $\ell$ that the mode has when we trace it back continuously  to the Schwarzschild limit. This is what we will do to catalogue the modes we study.
In Table~\ref{Table:SchwGravZ2Z1} we give the list of all modes we present. The first table is for $Z_2$ modes while the second is for $Z_1$ modes. In both tables the first column specifies $\{\ell,m\}$, the second column gives the value of the frequency for the Schwarzschild case. It matches the frequencies first computed and listed in  Table V of page 262  or in Table IV of page 202 of \cite{Chandra:1983}. Finally, in the third column we identify the figure of our manuscript where the QNMs of KN with the given $\{\ell,m\}$ of the first column are displayed. In the plots of all these figures, the QNM surfaces reduce to the values of the second column of Table~\ref{Table:SchwGravZ2Z1} in the Schwarzschild limit (see red points at $Q=a=0$ in our figures).

\begin{table}[ht]
\begin{eqnarray}
\nonumber
\begin{array}{||c||c|c||}\hline\hline
\bm{Z_2} & \hbox{Schwarzschild ($Q=a=0$)} & \hbox{Kerr-Newman}\\
       & \hbox{{\bf Gravitational} QNMs}           & \\
\hline \hline
\ell=3, m=3 & \omega M \simeq 0.59944329 - 0.09270305\, i  &  \hbox{Fig.~\ref{Fig:Z2l3m3n0+}} \\
\hline
\ell=4, m=4 & \omega M \simeq 0.80917838 - 0.09416396\, i  &  \hbox{Fig.~\ref{Fig:Z2l4m4n0+}} \\
\hline
\ell=5, m=5 & \omega M \simeq 1.01229531 - 0.09487052\, i &  \hbox{Fig.~\ref{Fig:Z2l5m5n0+}} \\
\hline
\ell=6, m=6 & \omega M \simeq 1.21200982 - 0.09526585\, i&  \hbox{Fig.~\ref{Fig:PSwkb-m6}} \\
 \hline
\ell=2, m=2 & \omega M \simeq 0.37367168 - 0.08896232\, i  &  \hbox{Fig.~\ref{Fig:Z2l2m2n0+}} \\
\hline
\ell=2, m=1 & \omega M \simeq 0.37367168 - 0.08896232\, i  &  \hbox{Fig.~\ref{Fig:Z2l2m1n0+}} \\
\hline
\ell=2, m=0 & \omega M \simeq 0.37367168 - 0.08896232\, i  &  \hbox{Fig.~\ref{Fig:Z2l2m0n0}} \\
\hline
\ell=2, m=-1 & \omega M \simeq 0.37367168 - 0.08896232\, i &  \hbox{Fig.~\ref{Fig:Z2l2m1n0-}} \\
 \hline
\ell=2, m=-2 & \omega M \simeq 0.37367168 - 0.08896232\, i  &  \hbox{Fig.~\ref{Fig:PSwkb-m2neg}} \\
\hline
\ell=6, m=-6 & \omega M \simeq 1.21200982 - 0.09526585\, i&  \hbox{Fig.~\ref{Fig:PSwkb-m6neg}} \\
\hline\hline
\end{array}
\end{eqnarray}
\begin{eqnarray}
\nonumber
\begin{array}{||c||c|c||}\hline\hline
\bm{Z_1} & \hbox{Schwarzschild ($Q=a=0$)} & \hbox{Kerr-Newman}\\
       & \hbox{{\bf Electromagnetic} QNMs}           & \\
\hline \hline
\ell=2, m=2 & \omega M \simeq 0.45759551 - 0.09500443\, i  &  \hbox{Fig.~\ref{Fig:Z1l2m2n0+}} \\
\hline
\ell=2, m=1 & \omega M \simeq 0.45759551 - 0.09500443\, i   &  \hbox{Fig.~\ref{Fig:Z1l2m1n0+}} \\
\hline
\ell=2, m=0 & \omega M \simeq 0.45759551 - 0.09500443\, i   &  \hbox{Fig.~\ref{Fig:Z1l2m0n0}} \\
\hline
\ell=2, m=-1 & \omega M \simeq 0.45759551 - 0.09500443\, i &  \hbox{Fig.~\ref{Fig:Z1l2m1n0-}} \\
\hline
\ell=2, m=-2 & \omega M \simeq 0.45759551 - 0.09500443\, i  &  \hbox{Fig.~\ref{Fig:Z1l2m2n0-}} \\
\hline
\ell=1, m=1 & \omega M \simeq 0.24826326 - 0.09248772\, i  &  \hbox{Fig.~\ref{Fig:Z1l1m1n0+}} \\
\hline
\ell=1, m=0 & \omega M \simeq 0.24826326 - 0.09248772\, i  &  \hbox{Fig.~\ref{Fig:Z1l1m0n0}} \\
\hline
\ell=1, m=-1 & \omega M \simeq 0.24826326 - 0.09248772\, i  &  \hbox{Fig.~\ref{Fig:Z1l1m1n0-}} \\
\hline\hline
\end{array}
\end{eqnarray}
\caption{List of most relevant gravitational ($Z_2$)  and electromagnetic ($Z_1$) QNMs of Schwarzschild (all with $n=0$). Note that $Z_2$ $\ell=1$ modes are pure gauge. The Schwarzschild frequencies displayed in this table agree with the values listed in  Table V of page 262  or in table IV of page 202 of \cite{Chandra:1983}. In the last column of each Table we indicate the figure that extends the Schwarzschild result to the $\tilde{Q}\neq 0$, $\tilde{a}\neq 0$ case. Note that for a given $\ell$, modes with $|m|\leq \ell$ are degenerate in the Schwarzschild limit but this degeneracy is broken once we switch on $\tilde{Q}$ and $\tilde{a}$.} \label{Table:SchwGravZ2Z1}
\end{table}

\begin{figure}
\includegraphics[width=.4\textwidth]{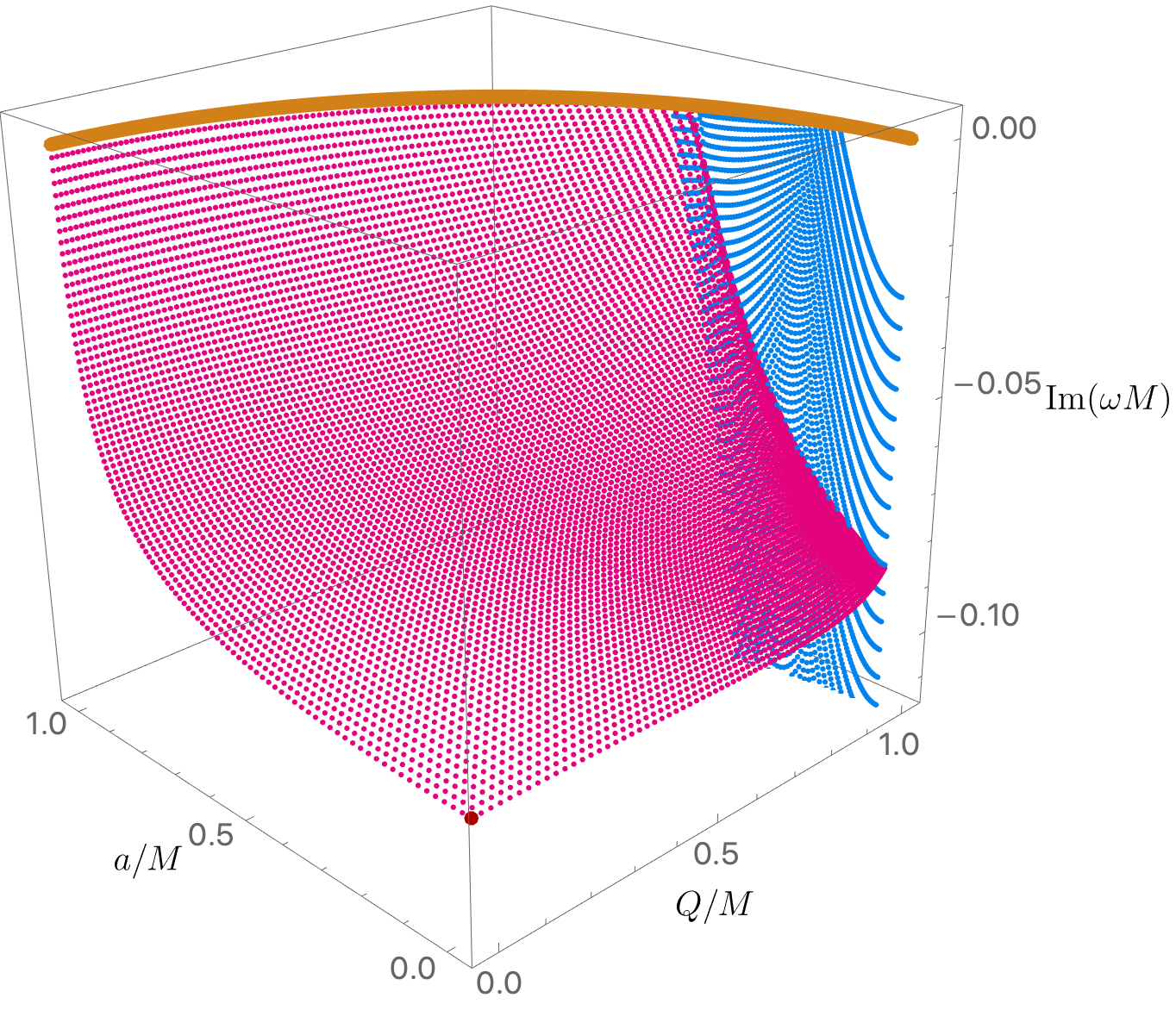}
\hspace{1.5cm}
\includegraphics[width=.38\textwidth]{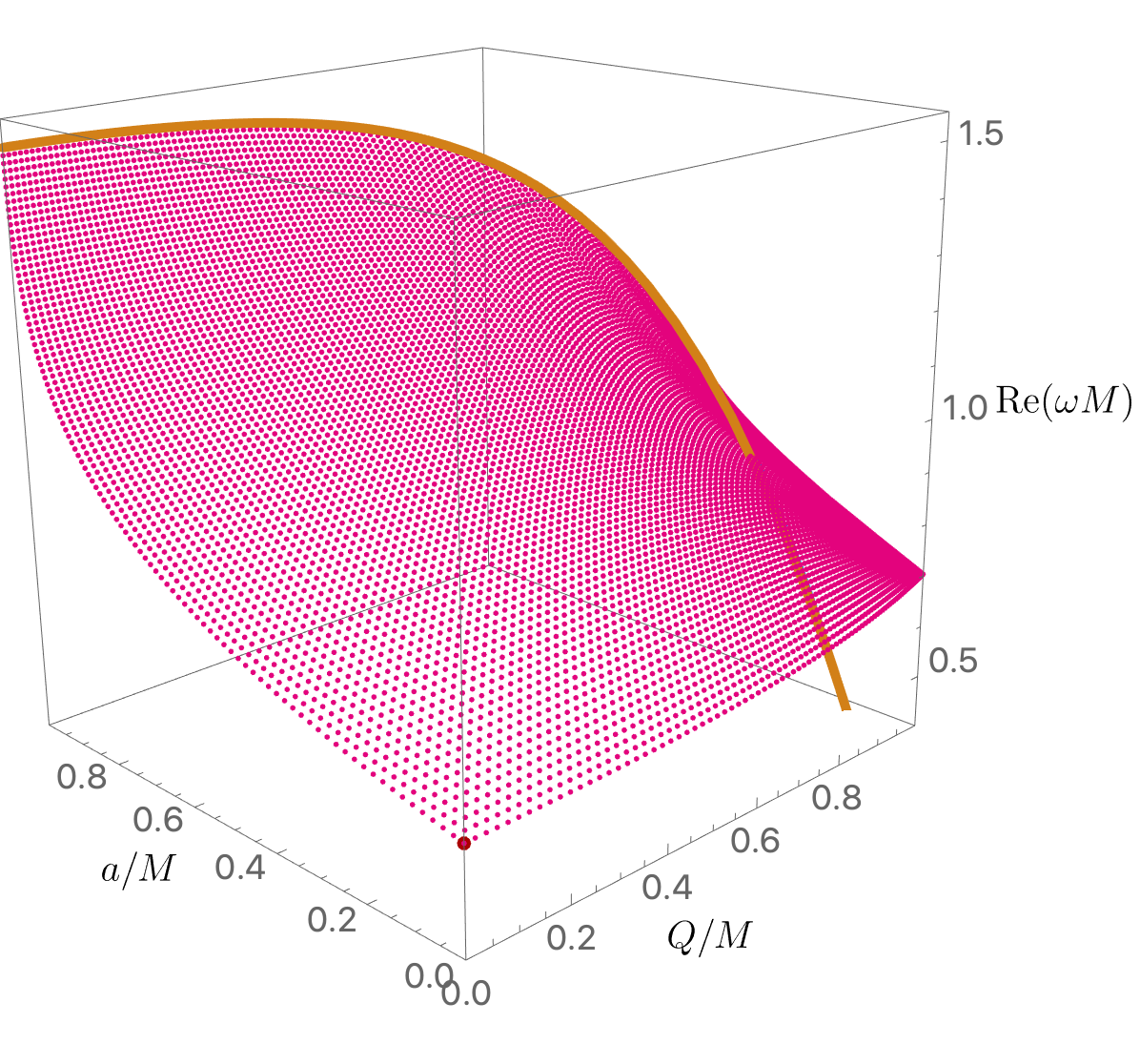}
\caption{Imaginary (left panel) and real (right panel) parts of the frequency for the $Z_2$, $(\ell,m,n)=(3,3,0)$ KN PS QNM. The dark-red point ($a=0=Q$), $\tilde{\omega}\simeq 0.59944329 - 0.09270305\, i $, is the gravitational QNM of Schwarzschild  \cite{Chandra:1983,Leaver:1985ax}.
The magenta surface describes $\mathrm{PS}_0$ modes while the blue surface corresponds to the $\mathrm{NH}_0$ modes.
}
\label{Fig:Z2l3m3n0+}
\end{figure}  

\begin{figure}
\includegraphics[width=.4\textwidth]{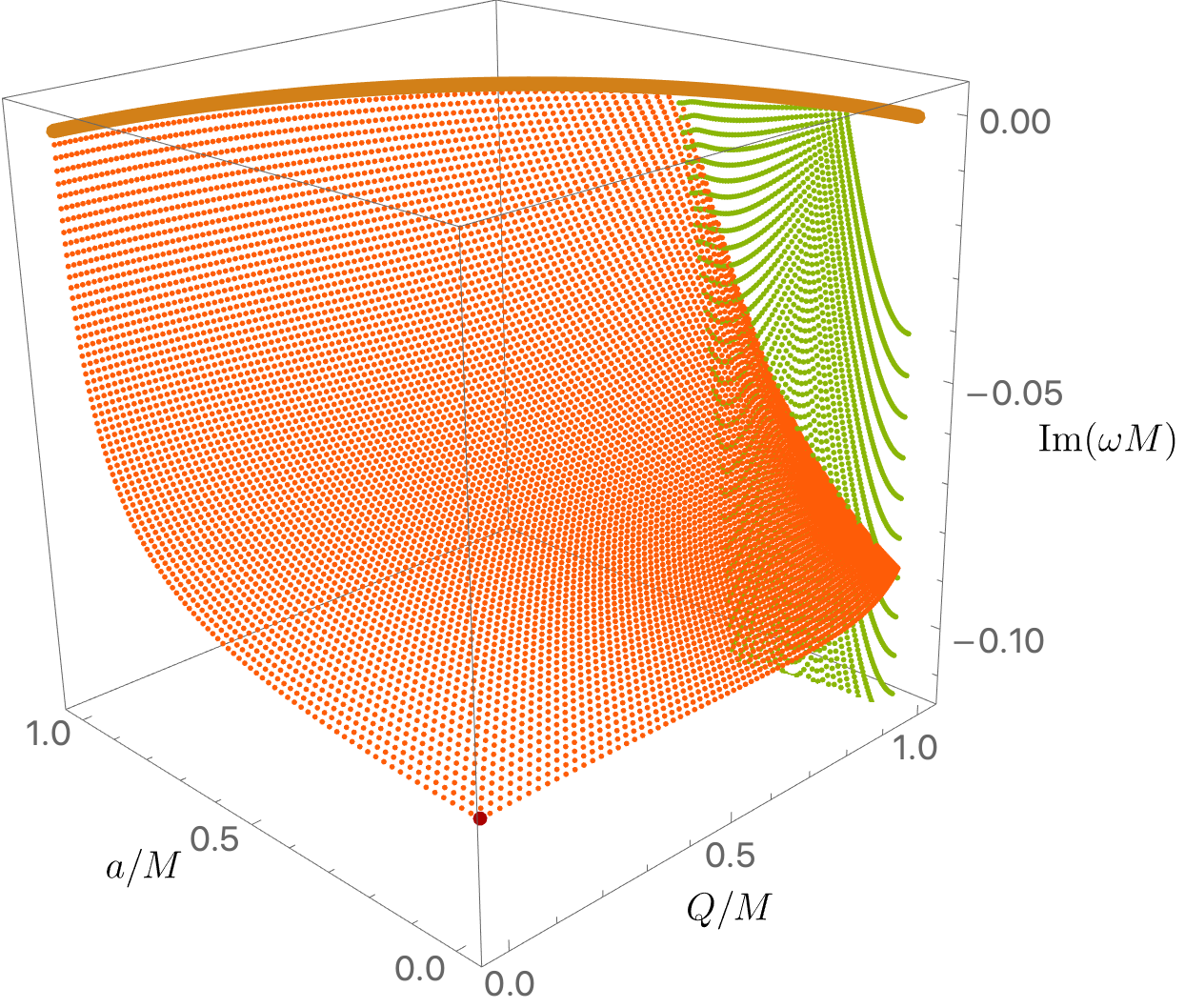}
\hspace{1.5cm}
\includegraphics[width=.38\textwidth]{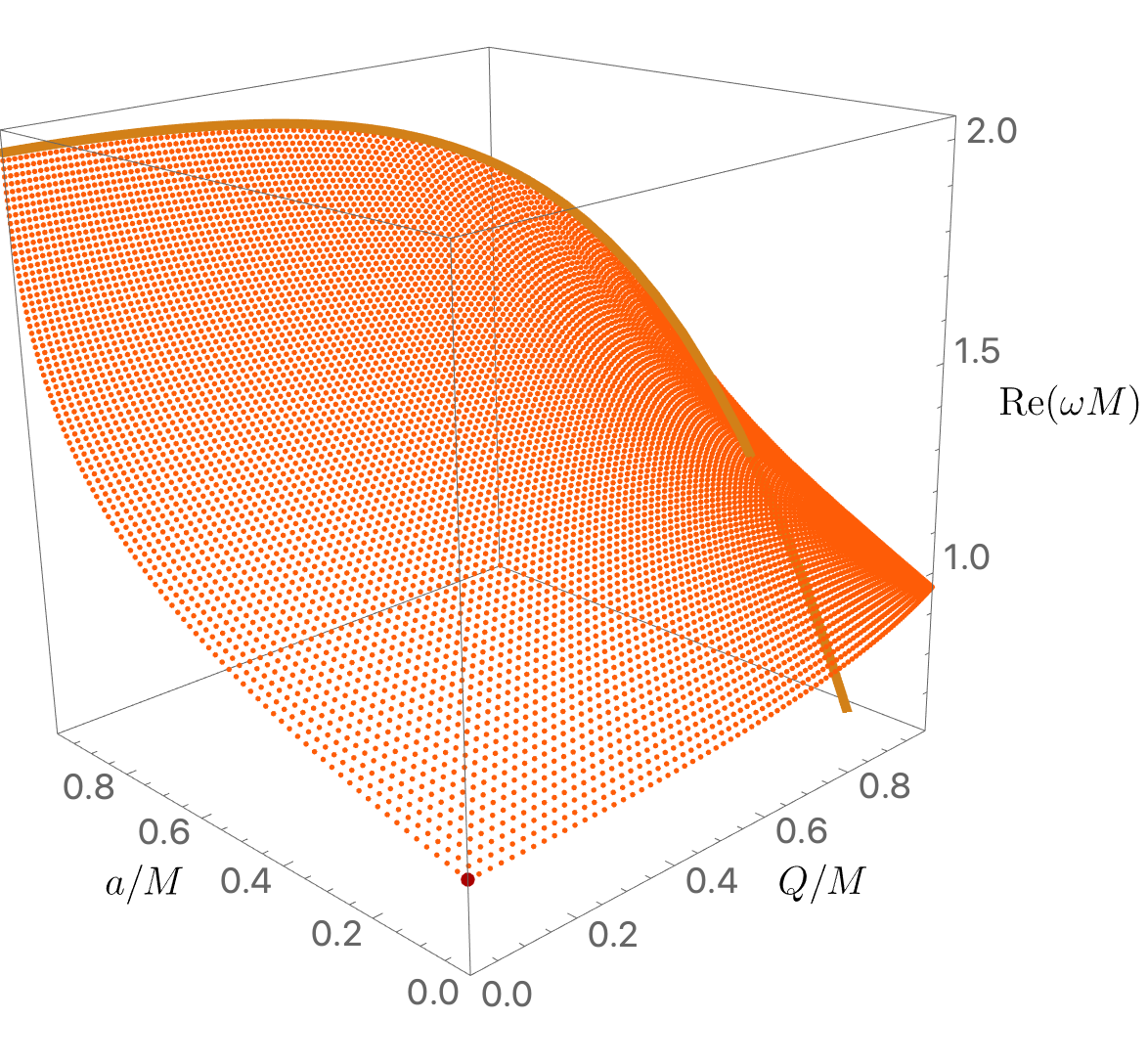}
\caption{Imaginary (left panel) and real (right panel) parts of the frequency for the $Z_2$, $(\ell,m,n)=(4,4,0)$ KN PS QNM. The dark-red point ($a=0=Q$), $\tilde{\omega}\simeq 0.80917838 - 0.09416396\, i $, is the gravitational QNM of Schwarzschild  \cite{Chandra:1983,Leaver:1985ax}.
The orange surface describes $\mathrm{PS}_0$ modes while the green surface corresponds to the $\mathrm{NH}_0$ modes.
}
\label{Fig:Z2l4m4n0+}
\end{figure}  

\begin{figure}
\includegraphics[width=.4\textwidth]{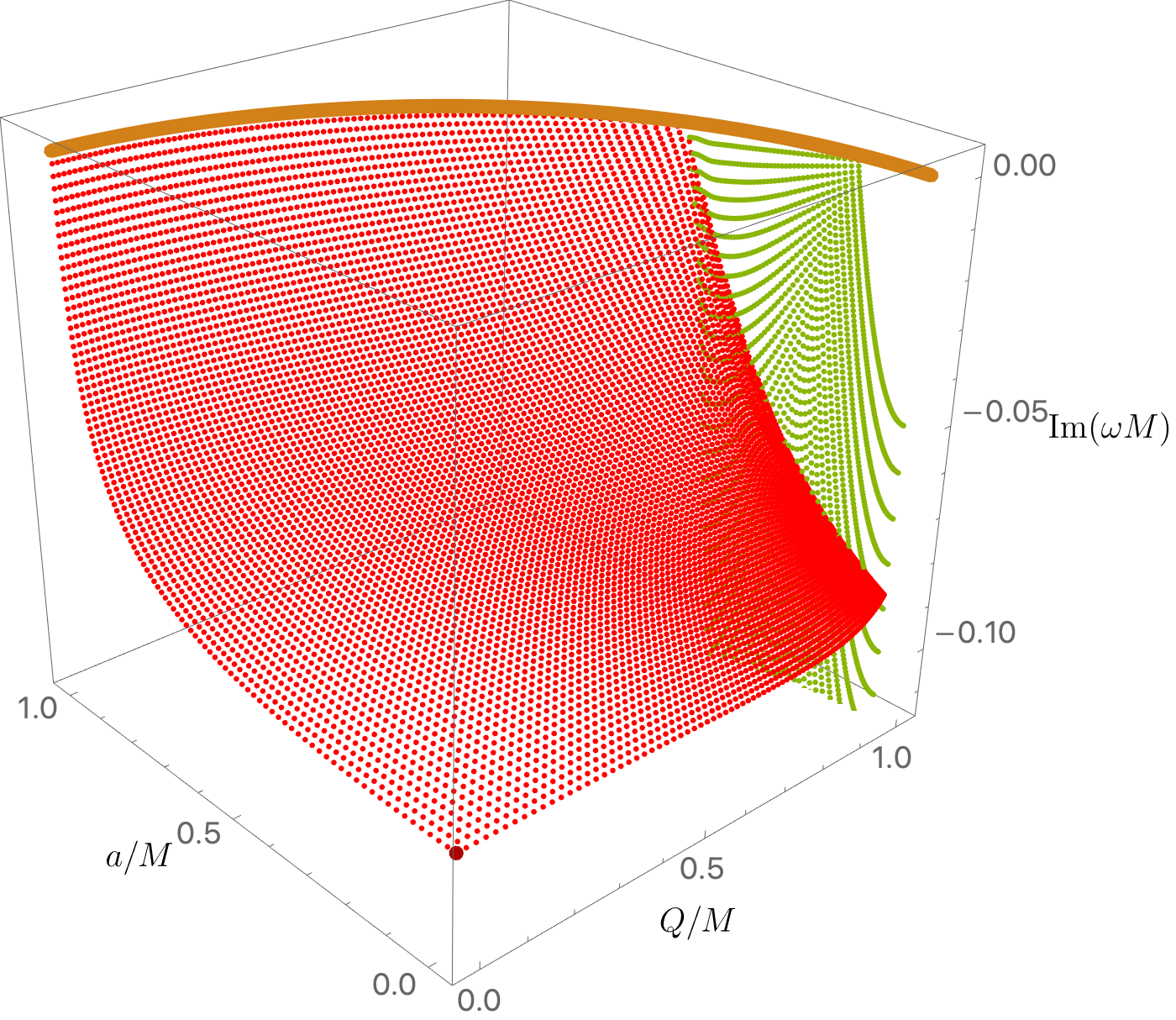}
\hspace{1.5cm}
\includegraphics[width=.38\textwidth]{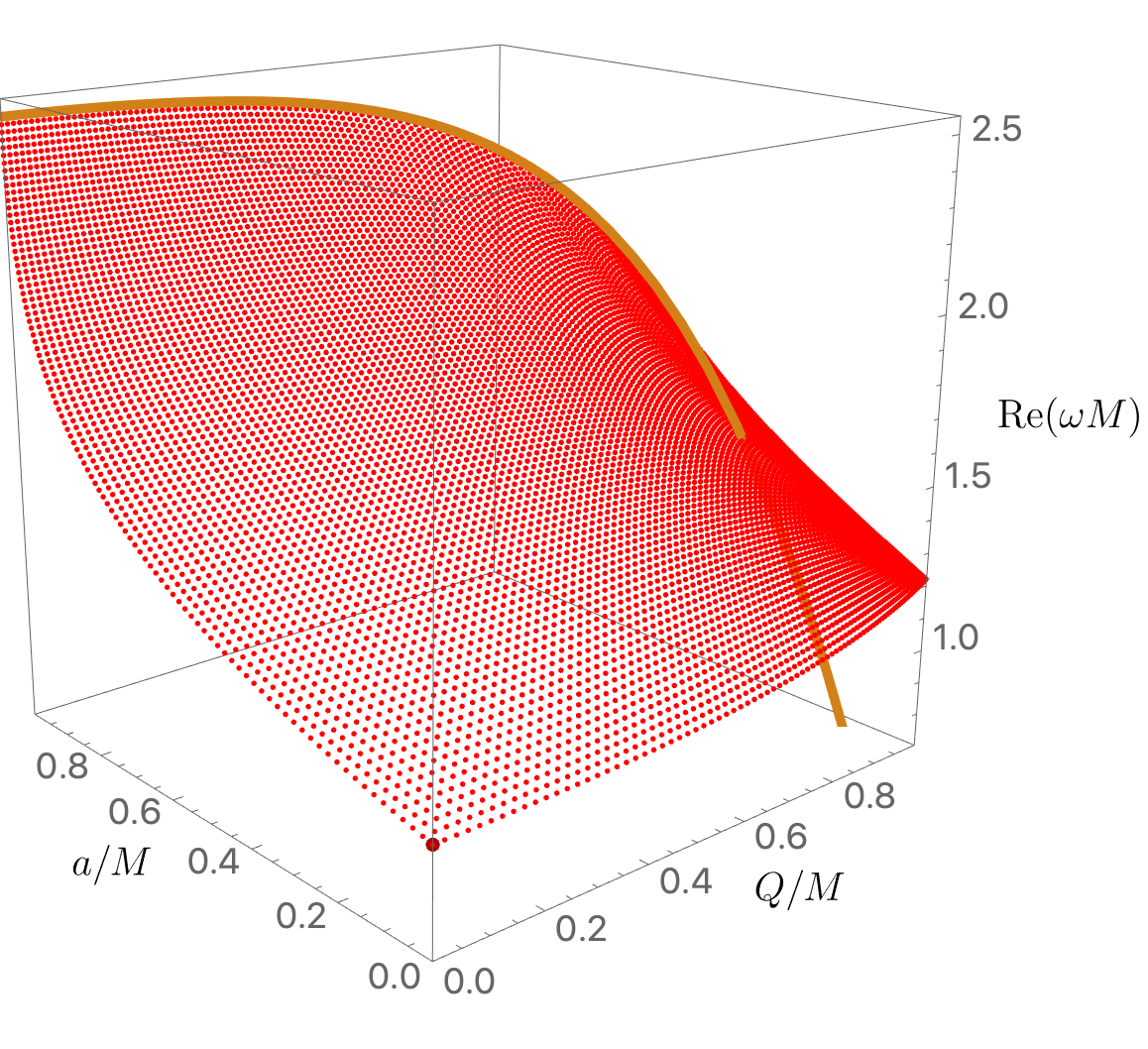}
\caption{Imaginary (left panel) and real (right panel) parts of the frequency for the $Z_2$, $(\ell,m,n)=(5,5,0)$ KN PS QNM. The dark-red point ($a=0=Q$), $\tilde{\omega}\simeq 1.01229531 - 0.09487052\, i $, is the gravitational QNM of Schwarzschild  \cite{Chandra:1983,Leaver:1985ax}.
The red surface describes $\mathrm{PS}_0$ modes while the green surface corresponds to the $\mathrm{NH}_0$ modes.
}
\label{Fig:Z2l5m5n0+}
\end{figure}  

We start by analysing what happens to the QNM $Z_2$ spectra with $\ell=m$ when  $\ell=m$ progressively grows from $\ell=m=2$ (Fig.~\ref{Fig:Z2l2m2n0+}), to $\ell=m=3$ (Fig.~\ref{Fig:Z2l3m3n0+}), to $\ell=m=4$ (Fig.~\ref{Fig:Z2l4m4n0+}) and, finally, to $\ell=m=5$ (Fig.~\ref{Fig:Z2l5m5n0+}). As for the  $\ell=m=2$ case of Fig.~\ref{Fig:Z2l2m2n0+}, the solid brown curves  at extremality are parametrized by $ \hat{a}=a_{\hbox{\footnotesize ext}} = \sqrt{1-\hat{Q}^2}$ and have $\mathrm{Im}\,\tilde{\omega}=0$ and $\mathrm{Re}\,\tilde{\omega}=m \tilde{\Omega}_H^{\hbox{\footnotesize ext}}$.  We see that the main features of Figs.~\ref{Fig:Z2l3m3n0+}$-$\ref{Fig:Z2l5m5n0+} for $\ell=m=3,4,5$ are very similar to those of Fig.~\ref{Fig:Z2l2m2n0+} for the $\ell=m=2$ mode that was already analysed in much detail in Sections~\ref{sec:repulsionSub} and \ref{sec:DominantSpectra}.\footnote{As with the  $\ell=m=2$ case of Fig.~\ref{Fig:Z2l2m2n0+}, note that in the right panels of Figs.~\ref{Fig:Z2l3m3n0+}$-$\ref{Fig:Z2l5m5n0+}  the NH$_0$ surface exist in such a narrow width around the solid brown extremal line that they are not visible to the naked eye.} In particular, we identify the PS$_0$ and NH$_0$ surfaces (as unambiguously identified in the RN limit) and a zoom in of Figs.~\ref{Fig:Z2l3m3n0+}$-$\ref{Fig:Z2l5m5n0+} (not shown in our figures) shows that these surfaces intersect with simple crossovers or with eigenvalue repulsions very much similar to those detailed in Figs.~\ref{Fig:spectraFix-a}$-$\ref{Fig:spectraFix-a2} for the $\ell=m=2$ case. Therefore, the key features of  Figs.~\ref{Fig:Z2l3m3n0+}$-$\ref{Fig:Z2l5m5n0+} are as discussed before. However, we highlight three features. First, the $\ell=m=3,4,5$ $\mathrm{Im}(\omega M)$ surfaces are always below and thus more damped than the $\ell=m=2$ one, and the damping increases as $\ell=m$ increases. Second, the NH$_0$ surfaces only dominate the spectra for very large $Q/M$ and close to extremality, again very much like in the $\ell=m=2$ case. In fact, since black holes with very large charge are not expected to have any astrophysical interest, in the plots for the other modes listed in Table~\ref{Table:SchwGravZ2Z1} we will no longer display the NH families (when they exist; this is certainly the case for the $Z_1$ $\ell=m=1,2$ modes). Finally, note that  for $\ell=m\geq 3$ it is still true that the PS$_0$ frequencies are well approximated by \eqref{PS:eikonal} (in fact the approximation gets  better as $m$ increases and we approach the eikonal limit; see also the $\ell=m=6$ case in Fig.~\ref{Fig:PSwkb-m6}) and the NH$_0$ frequencies are in excellent agreement with \eqref{NH:freq} near extremality. Complementing the analysis reported here, note that in \cite{Dias:2015wqa} we have reported our findings for $Z_2$ $\ell=3$ with $m=-3,-2,-1,0,1,2,3$ and we have concluded that their $|\mathrm{Im}\,\tilde{\omega}|$ is aways higher than the $Z_2$ $\ell=m=2$ modes.  

\begin{figure}
\includegraphics[width=.4\textwidth]{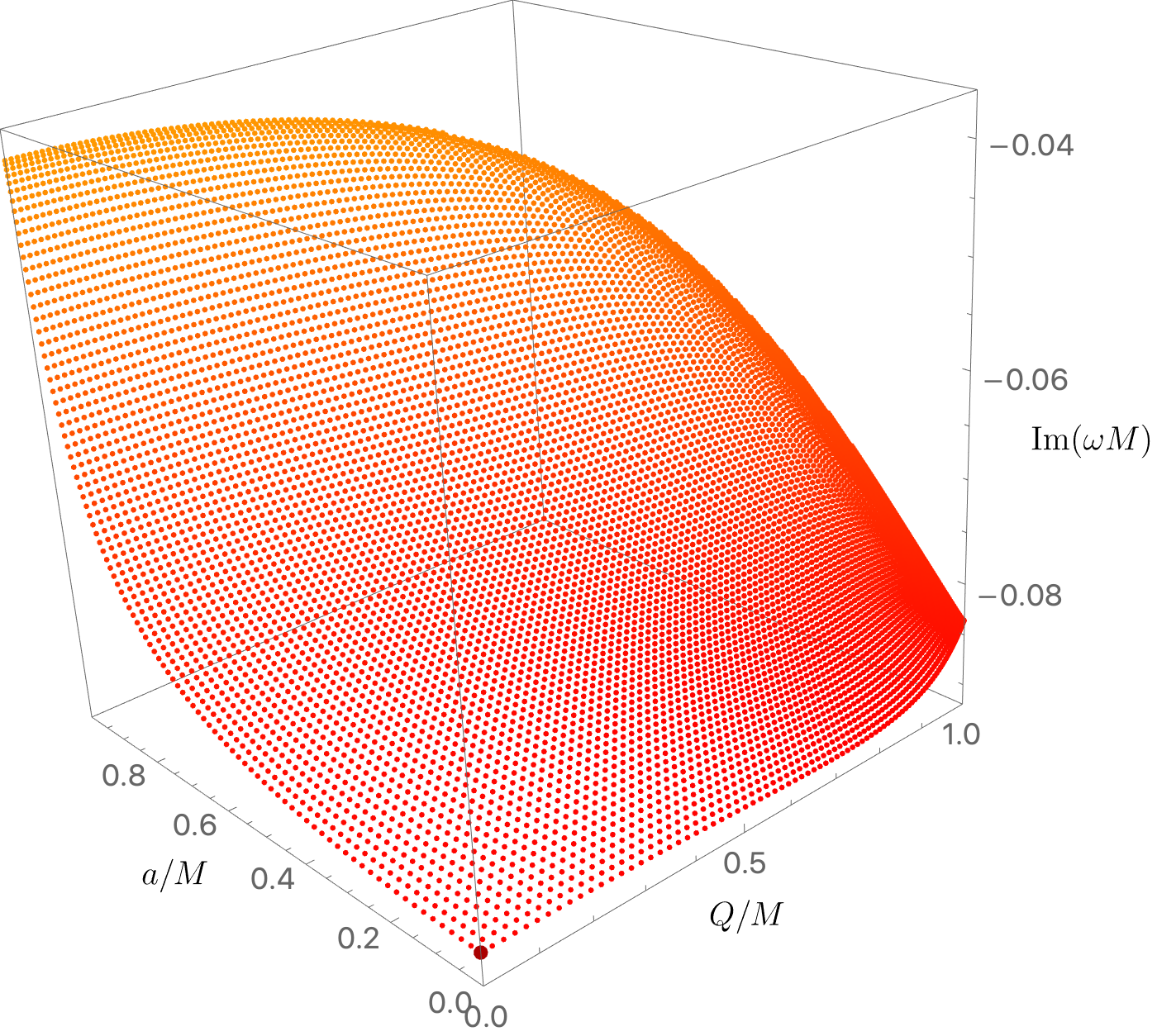}
\hspace{1.5cm}
\includegraphics[width=.38\textwidth]{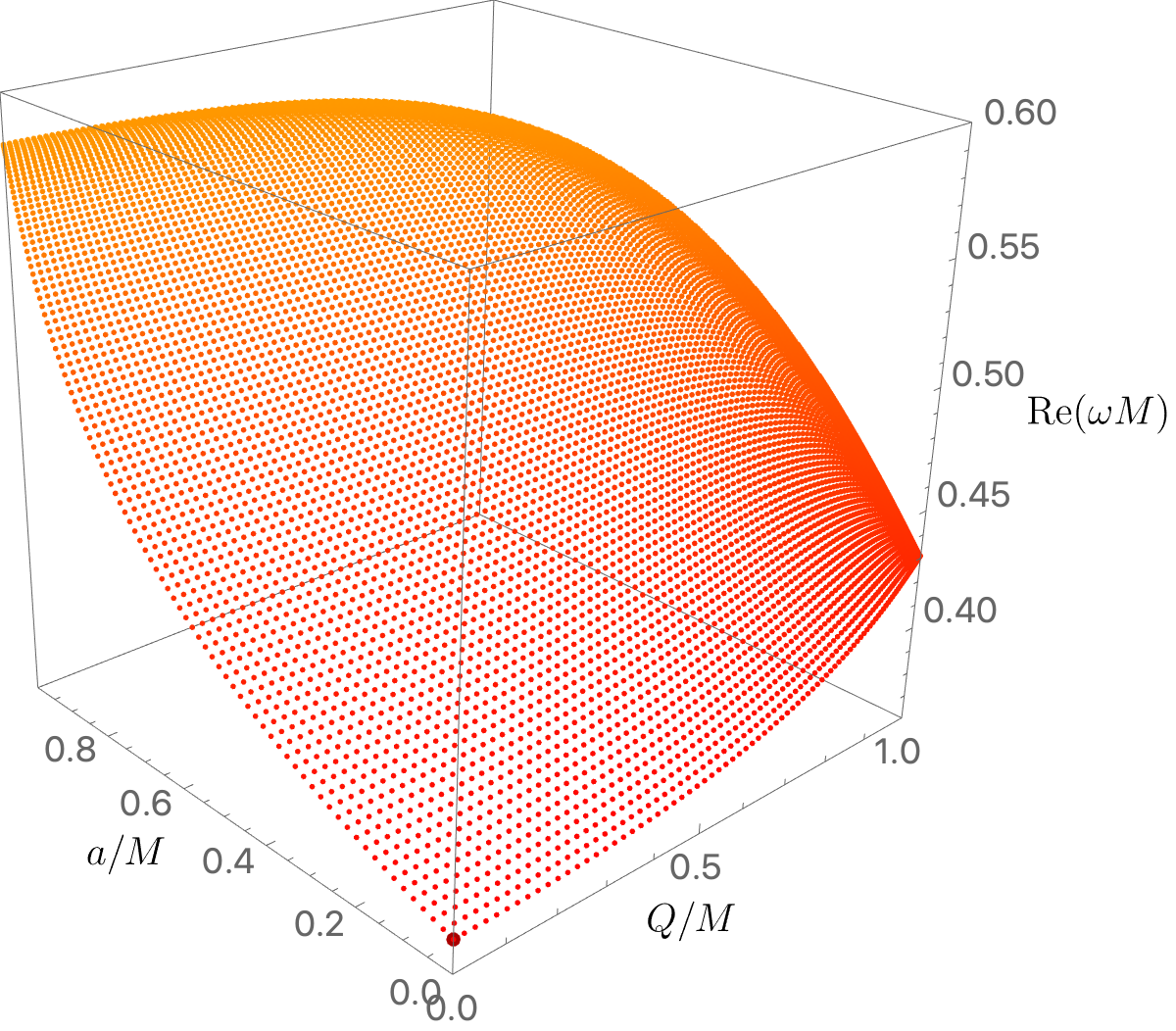}
\caption{Imaginary (left panel) and real (right panel) parts of the frequency for the $Z_2$, $(\ell,m,n)=(2,1,0)$ KN PS QNM. The dark-red point ($a=0=Q$), $\tilde{\omega}\simeq 0.37367168 - 0.08896232\, i $, is the gravitational QNM of Schwarzschild  \cite{Chandra:1983,Leaver:1985ax}.
}
\label{Fig:Z2l2m1n0+}
\end{figure}  

\begin{figure}
\includegraphics[width=.4\textwidth]{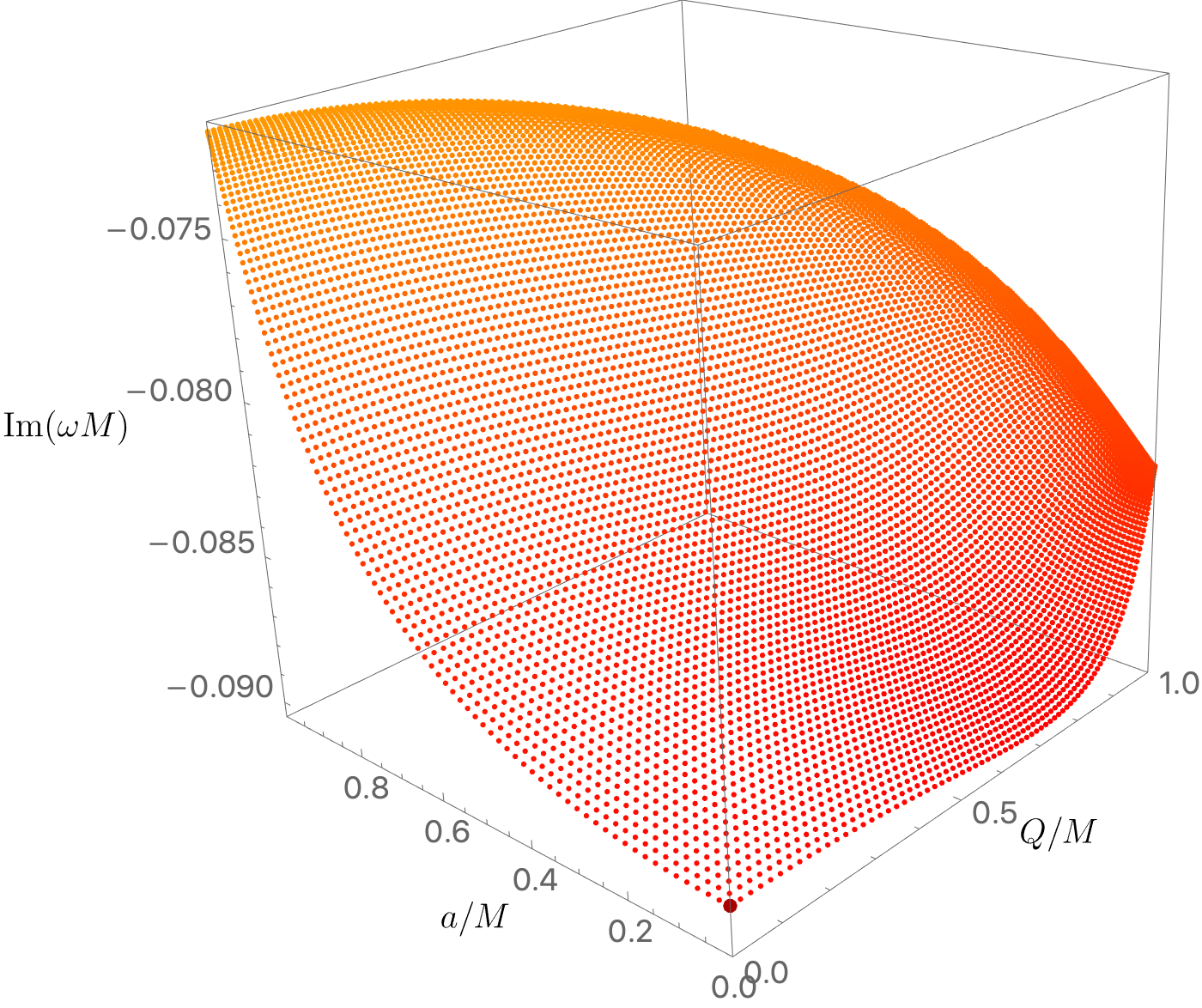}
\hspace{1.5cm}
\includegraphics[width=.38\textwidth]{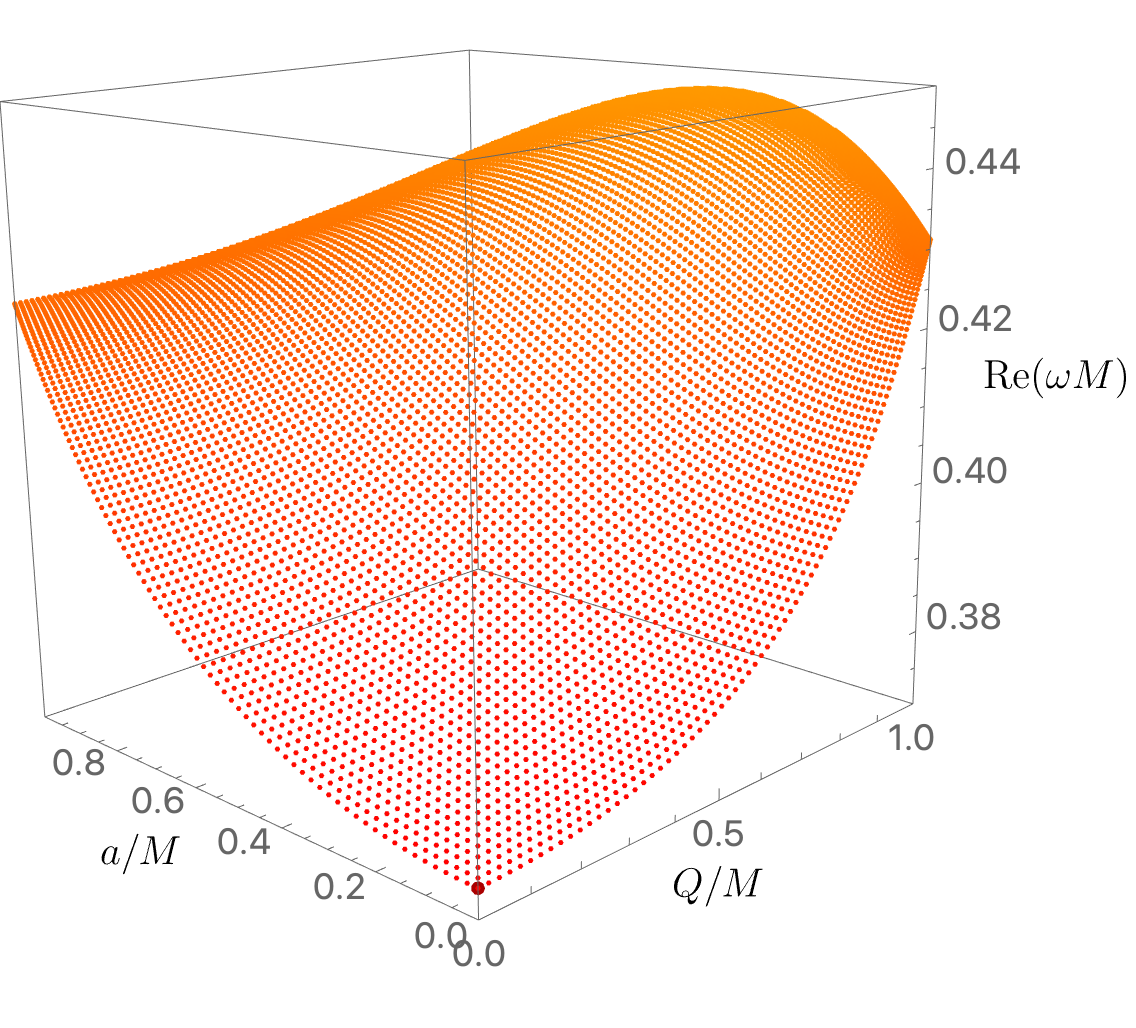}
\caption{Imaginary (left panel) and real (right panel) parts of the frequency for the $Z_2$, $(\ell,m,n)=(2,0,0)$ KN PS QNM. The dark-red point ($a=0=Q$), $\tilde{\omega}\simeq 0.37367168 - 0.08896232\, i $, is the gravitational QNM of Schwarzschild  \cite{Chandra:1983,Leaver:1985ax}.}
\label{Fig:Z2l2m0n0}
\end{figure}  

\begin{figure}
\includegraphics[width=.4\textwidth]{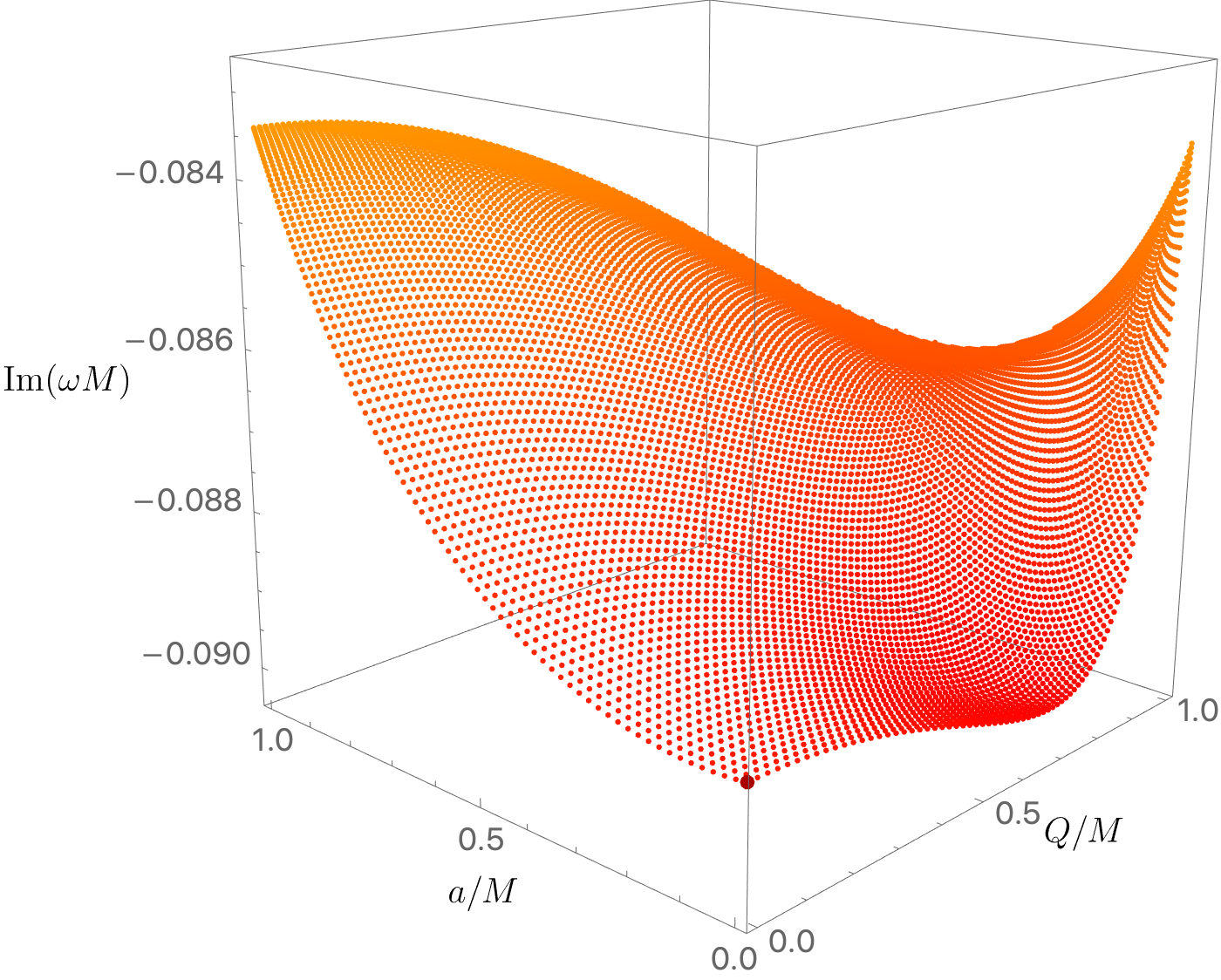}
\hspace{1.5cm}
\includegraphics[width=.38\textwidth]{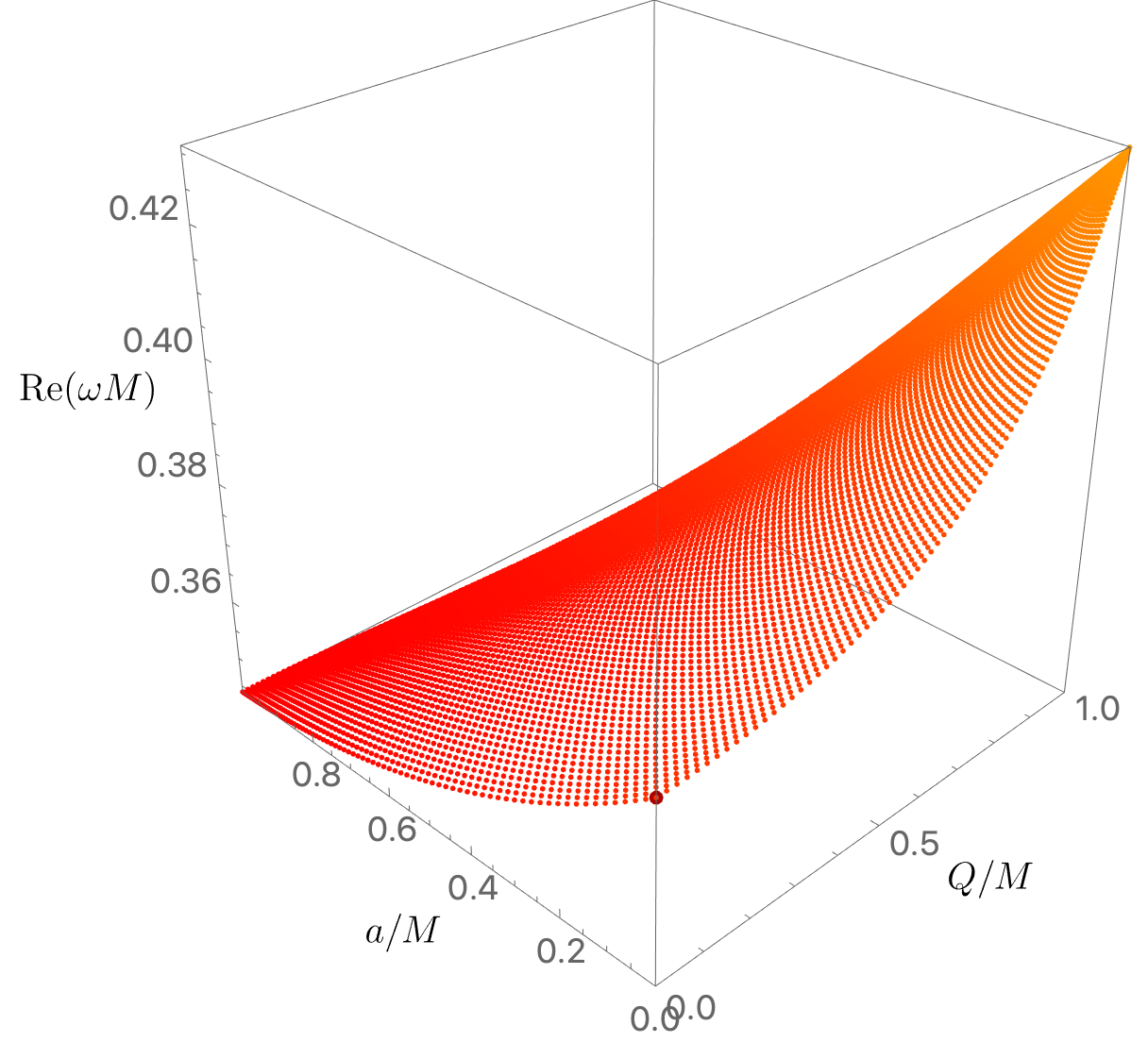}
\caption{Imaginary (left panel) and real (right panel) parts of the frequency for the $Z_2$, $(\ell,m,n)=(2,-1,0)$ KN PS QNM. The dark-red point ($a=0=Q$), $\tilde{\omega}\simeq 0.37367168 - 0.08896232\, i $, is the gravitational QNM of Schwarzschild  \cite{Chandra:1983,Leaver:1985ax}.}
\label{Fig:Z2l2m1n0-}
\end{figure}

Next, we consider the several cases of $Z_2$ modes with $|m|\leq \ell=2$ in Fig.~\ref{Fig:Z2l2m2n0+} ($m=2$), Fig.~\ref{Fig:Z2l2m1n0+} ($m=1$), Fig.~\ref{Fig:Z2l2m0n0} ($m=0$), Fig.~\ref{Fig:Z2l2m1n0-} ($m=-1$), and Fig.~\ref{Fig:PSwkb-m2neg} ($m=-2$). In the Schwarzschild limit all these $\ell=2$ modes are degenerate with $\tilde{\omega}\simeq 0.45759551 - 0.09500443\, i $, but this degeneracy is broken once we switch on $\tilde{Q}$ and $\tilde{a}$. The figures speak for themselves and we refrain from describing them further. We note simply that the surfaces for positive $m$ have a qualitative shape that is significantly distinct from the ones for negative $m$ (notably, $m\geq 0$ cases have a monotonic behaviour that is not observed in the $m<0$ cases) and, as expected, further note that for $m\neq \ell$ the PS$_0$ surfaces no longer approach   $\mathrm{Im}\,\tilde{\omega}=0$ and $\mathrm{Re}\,\tilde{\omega}=m \tilde{\Omega}_H^{\hbox{\footnotesize ext}}$ at extremality (hence we do not display these solid brown curves in the associated plots). This sequence of figures demonstrates, as previously claimed, that $Z_2$ modes with $\ell=m=2$ are the dominant ones among the $|m|\leq \ell=2$ families (and all others).

\begin{figure}
\includegraphics[width=.4\textwidth]{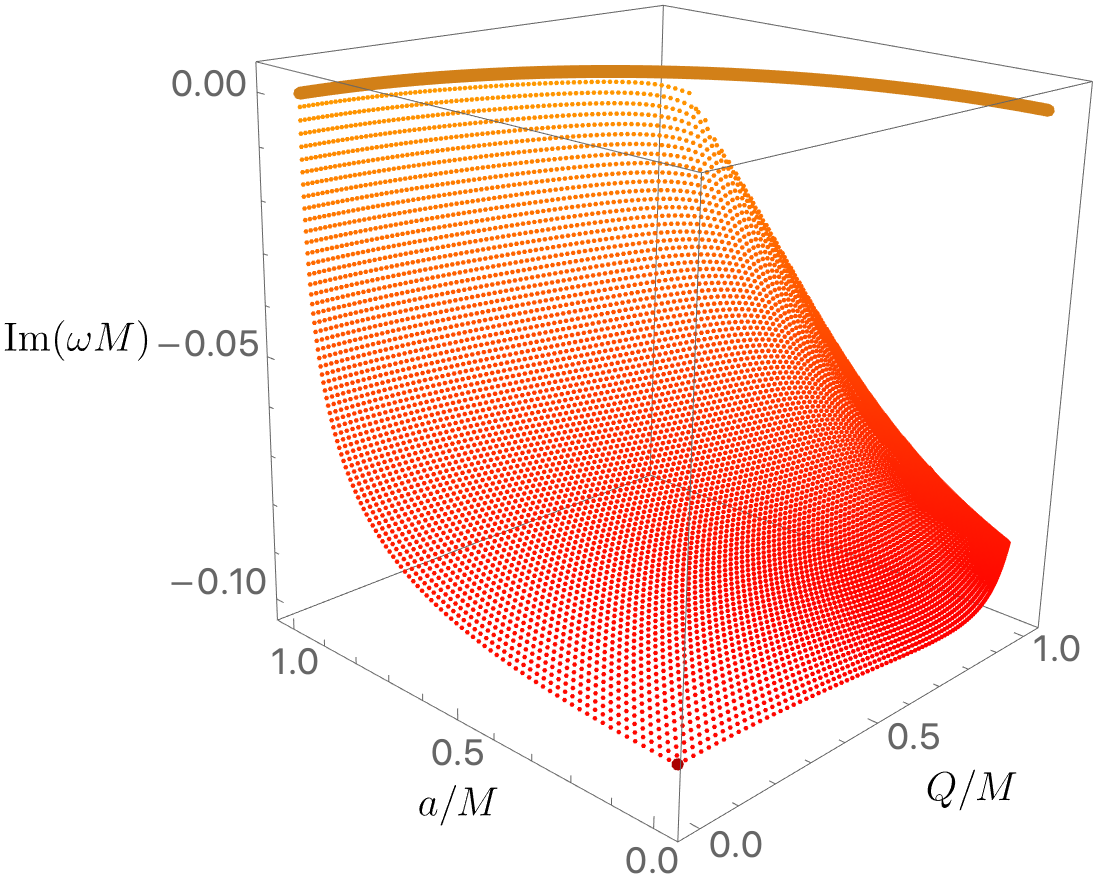}
\hspace{1.5cm}
\includegraphics[width=.38\textwidth]{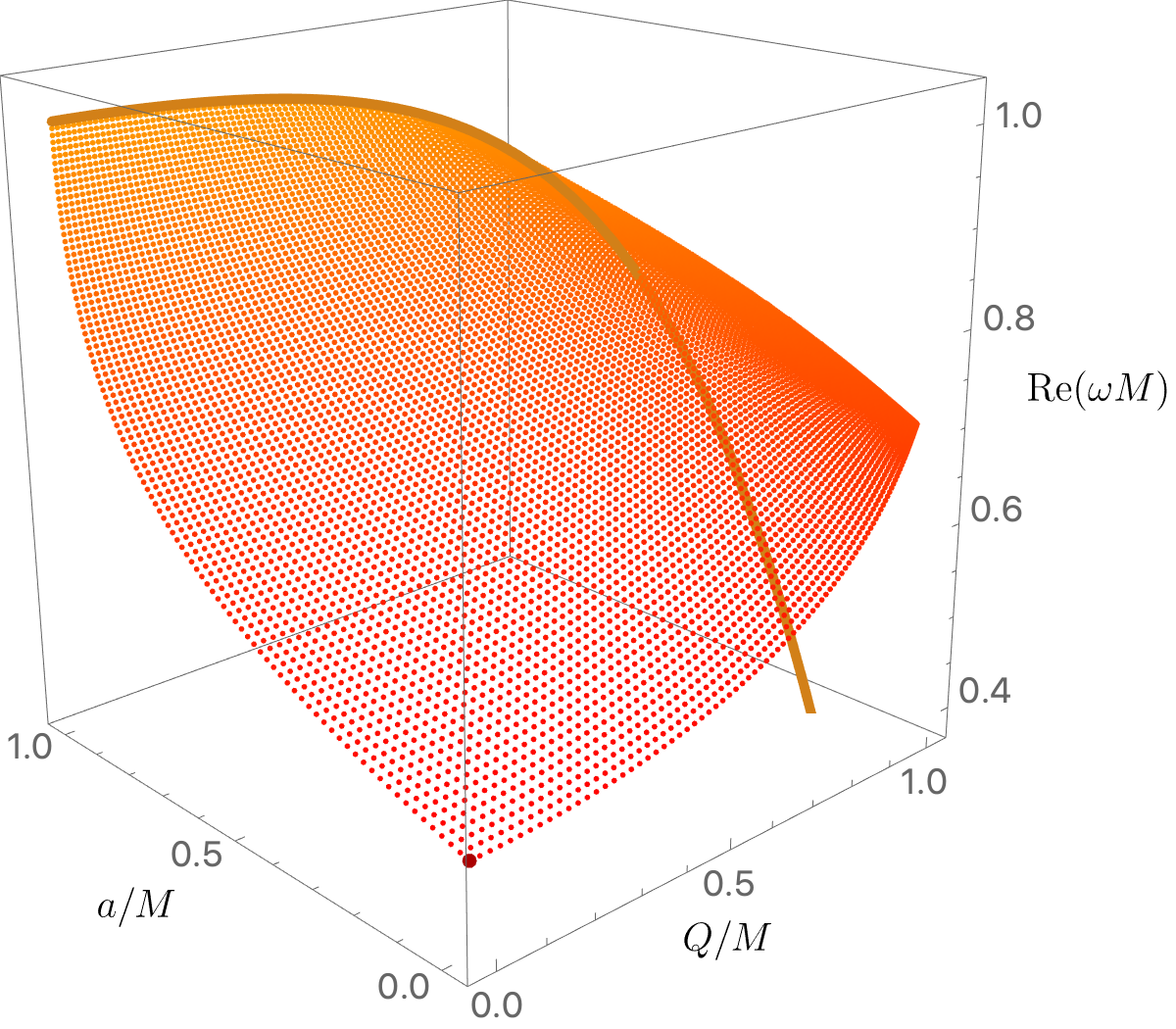}
\caption{Imaginary (left panel) and real (right panel) parts of the frequency for the $Z_1$, $(\ell,m,n)=(2,2,0)$ KN PS QNM. The dark-red point ($a=0=Q$), $\tilde{\omega}\simeq 0.45759551 - 0.09500443\, i $, is the gravitational QNM of Schwarzschild  \cite{Chandra:1983,Leaver:1985ax}.
}
\label{Fig:Z1l2m2n0+}
\end{figure}

\begin{figure}
\includegraphics[width=.4\textwidth]{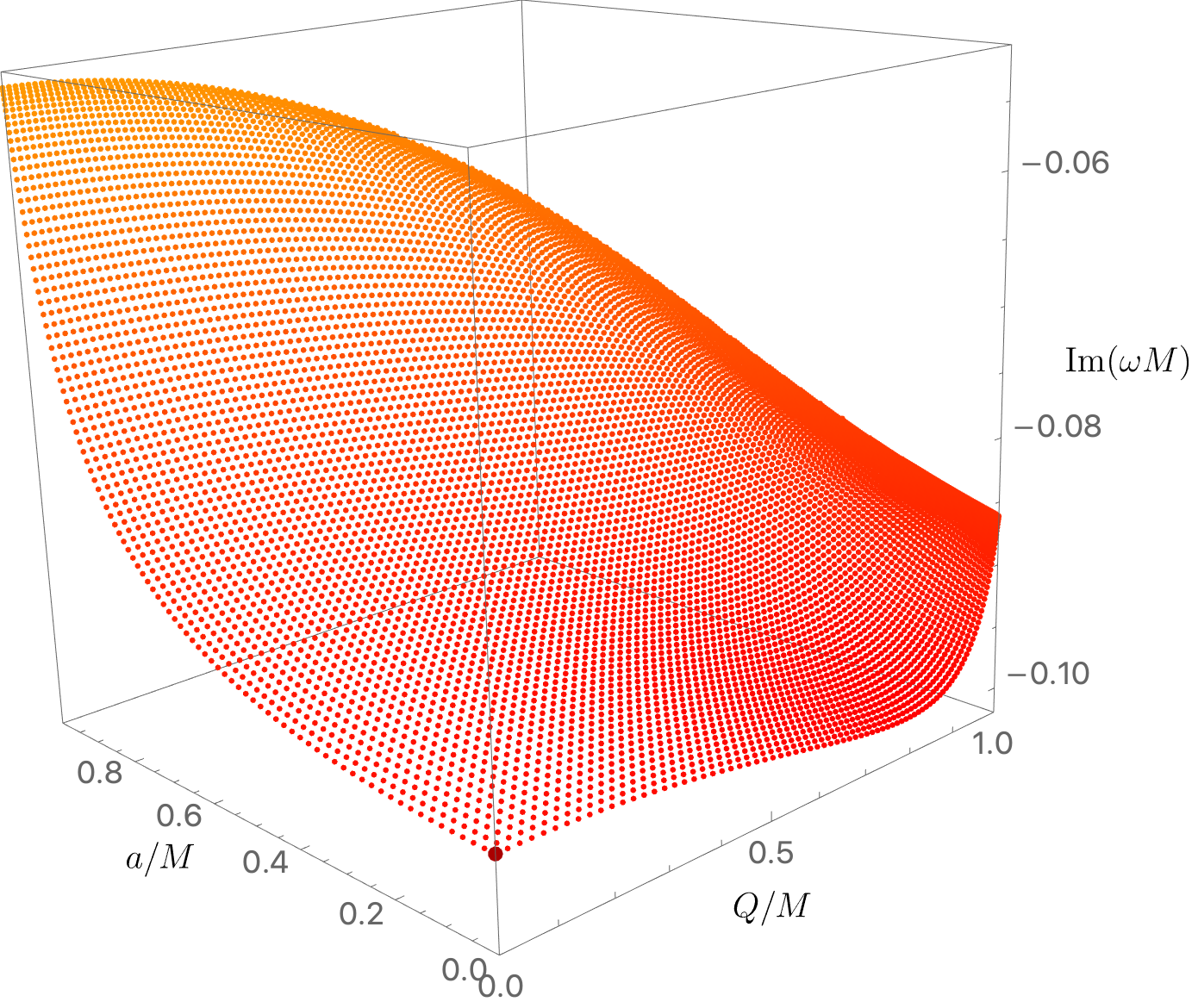}
\hspace{1.5cm}
\includegraphics[width=.38\textwidth]{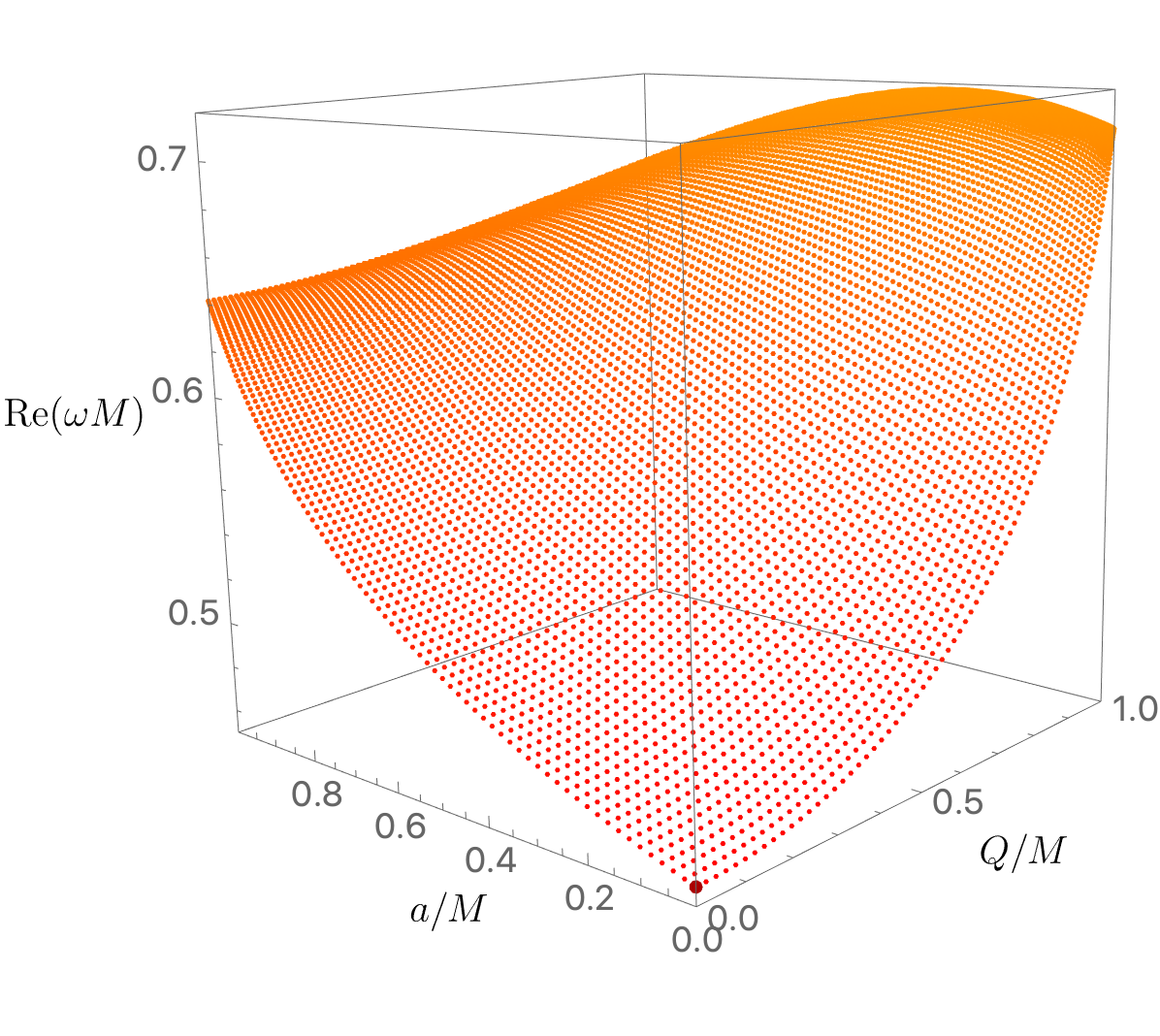}
\caption{Imaginary (left panel) and real (right panel) parts of the frequency for the $Z_1$, $(\ell,m,n)=(2,1,0)$ KN PS QNM. The dark-red point ($a=0=Q$), $\tilde{\omega}\simeq 0.45759551 - 0.09500443\, i $, is the gravitational QNM of Schwarzschild  \cite{Chandra:1983,Leaver:1985ax}.
}
\label{Fig:Z1l2m1n0+}
\end{figure}

\begin{figure}
\includegraphics[width=.4\textwidth]{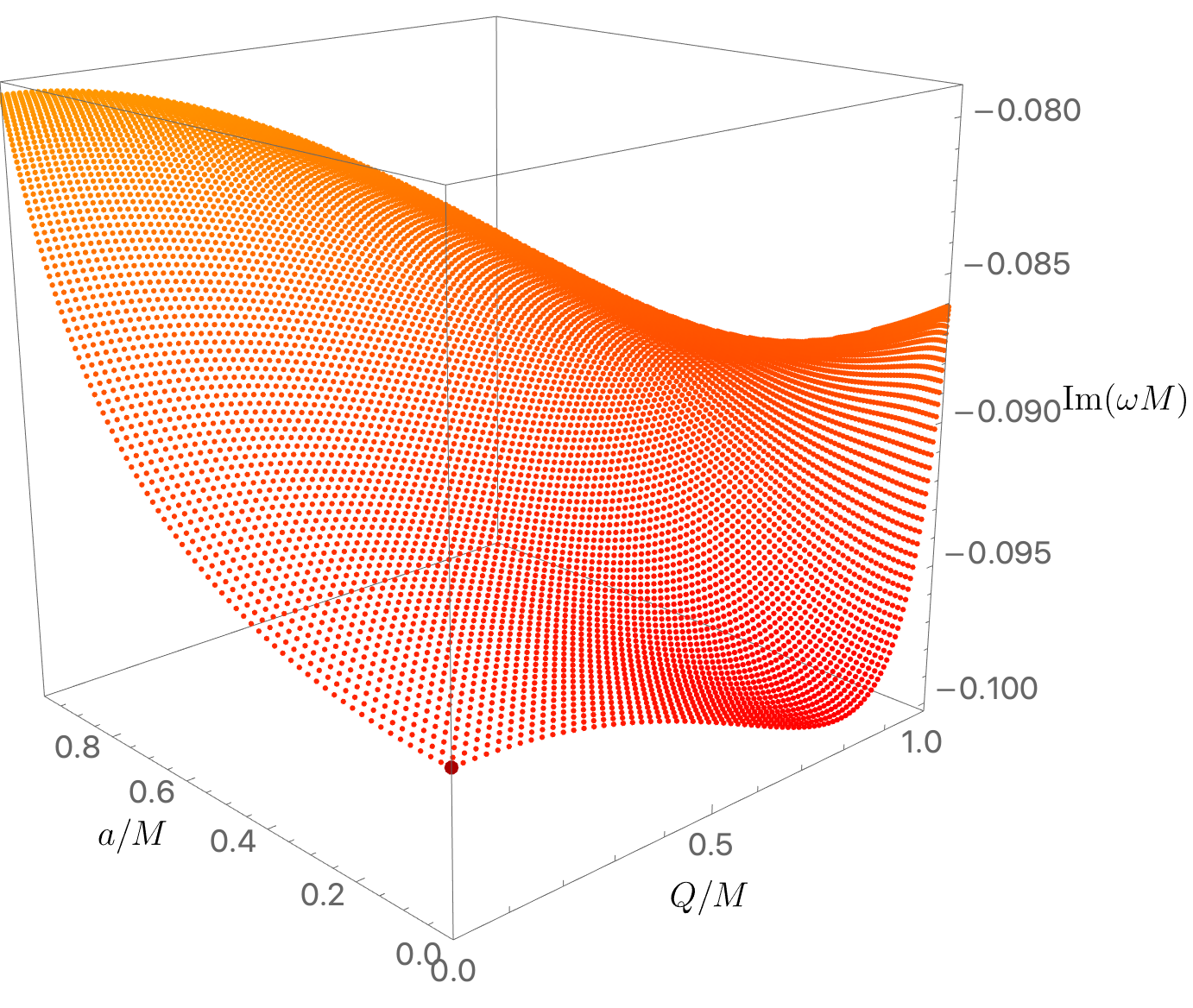}
\hspace{1.5cm}
\includegraphics[width=.38\textwidth]{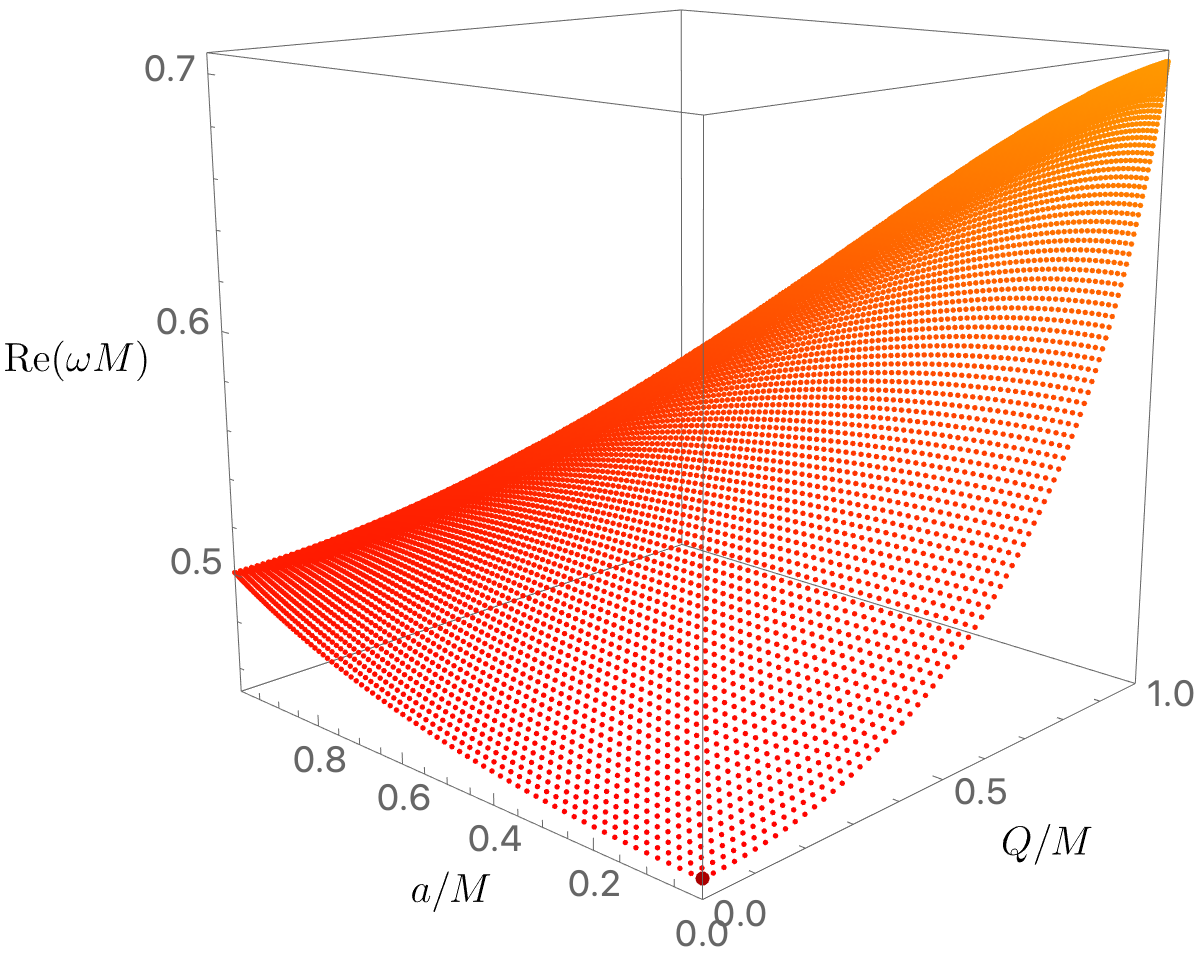}
\caption{Imaginary (left panel) and real (right panel) parts of the frequency for the $Z_1$, $(\ell,m,n)=(2,0,0)$ KN PS QNM. The dark-red point ($a=0=Q$), $\tilde{\omega}\simeq 0.45759551 - 0.09500443\, i $, is the gravitational QNM of Schwarzschild  \cite{Chandra:1983,Leaver:1985ax}.
}
\label{Fig:Z1l2m0n0}
\end{figure}

\begin{figure}
\includegraphics[width=.4\textwidth]{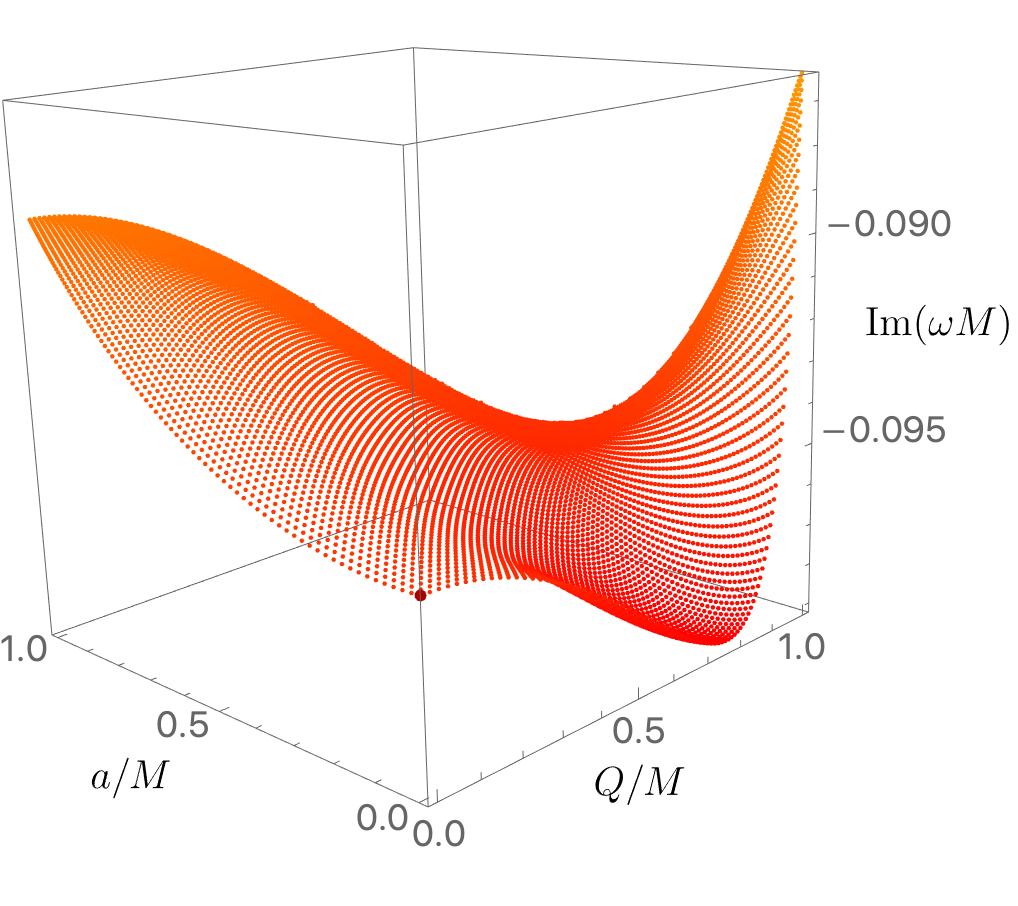}
\hspace{1.5cm}
\includegraphics[width=.38\textwidth]{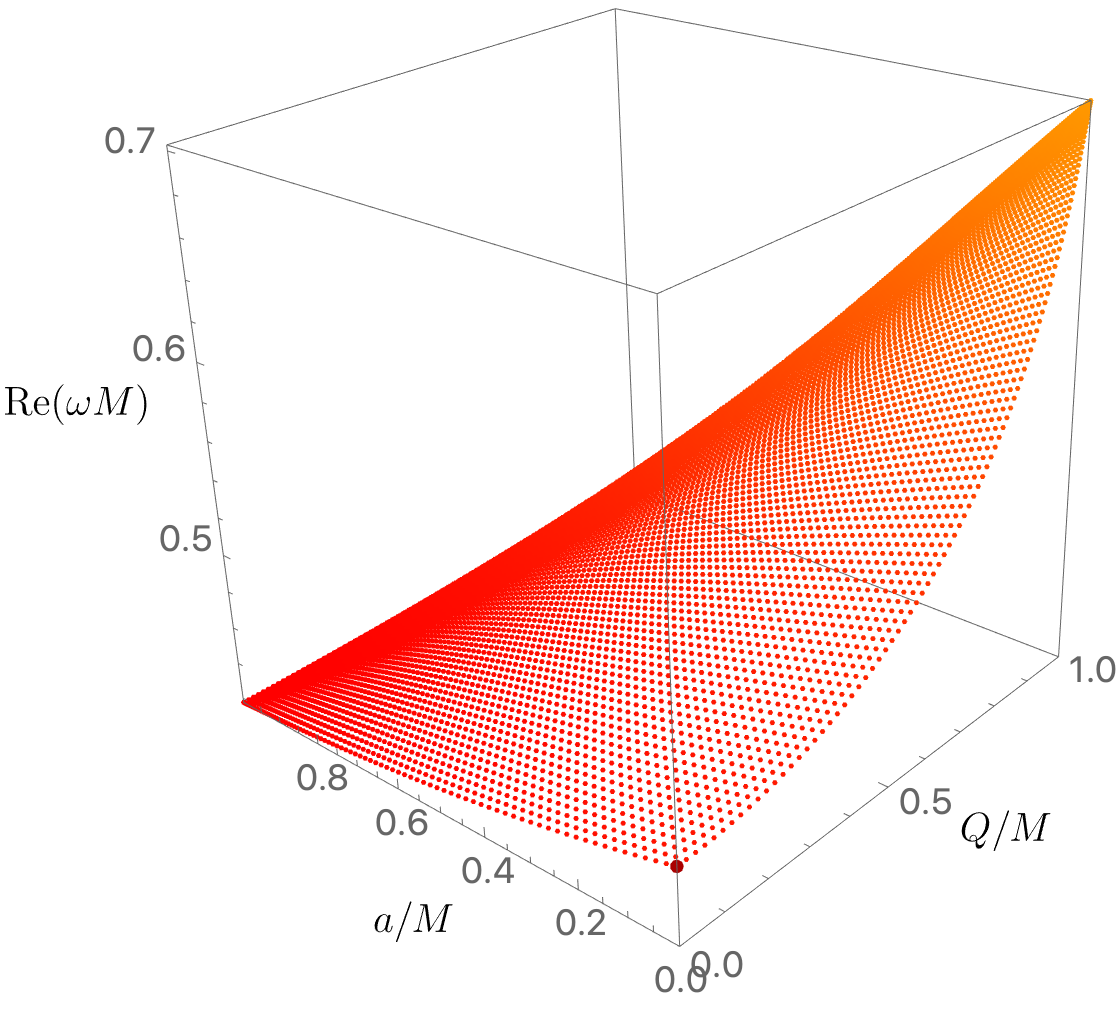}
\caption{Imaginary (left panel) and real (right panel) parts of the frequency for the $Z_1$, $(\ell,m,n)=(2,-1,0)$ KN PS QNM. The dark-red point ($a=0=Q$), $\tilde{\omega}\simeq 0.45759551 - 0.09500443\, i $, is the gravitational QNM of Schwarzschild  \cite{Chandra:1983,Leaver:1985ax}.
}
\label{Fig:Z1l2m1n0-}
\end{figure}

\begin{figure}
\includegraphics[width=.4\textwidth]{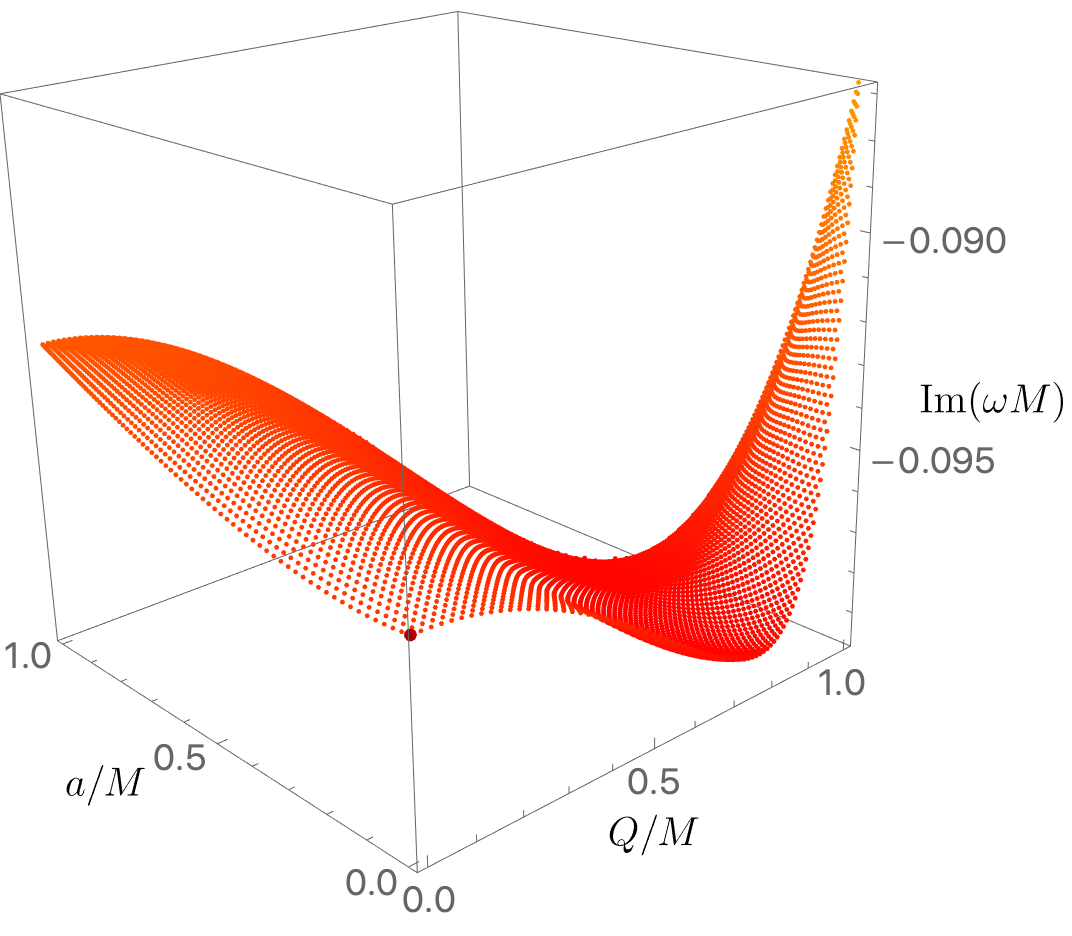}
\hspace{1.5cm}
\includegraphics[width=.38\textwidth]{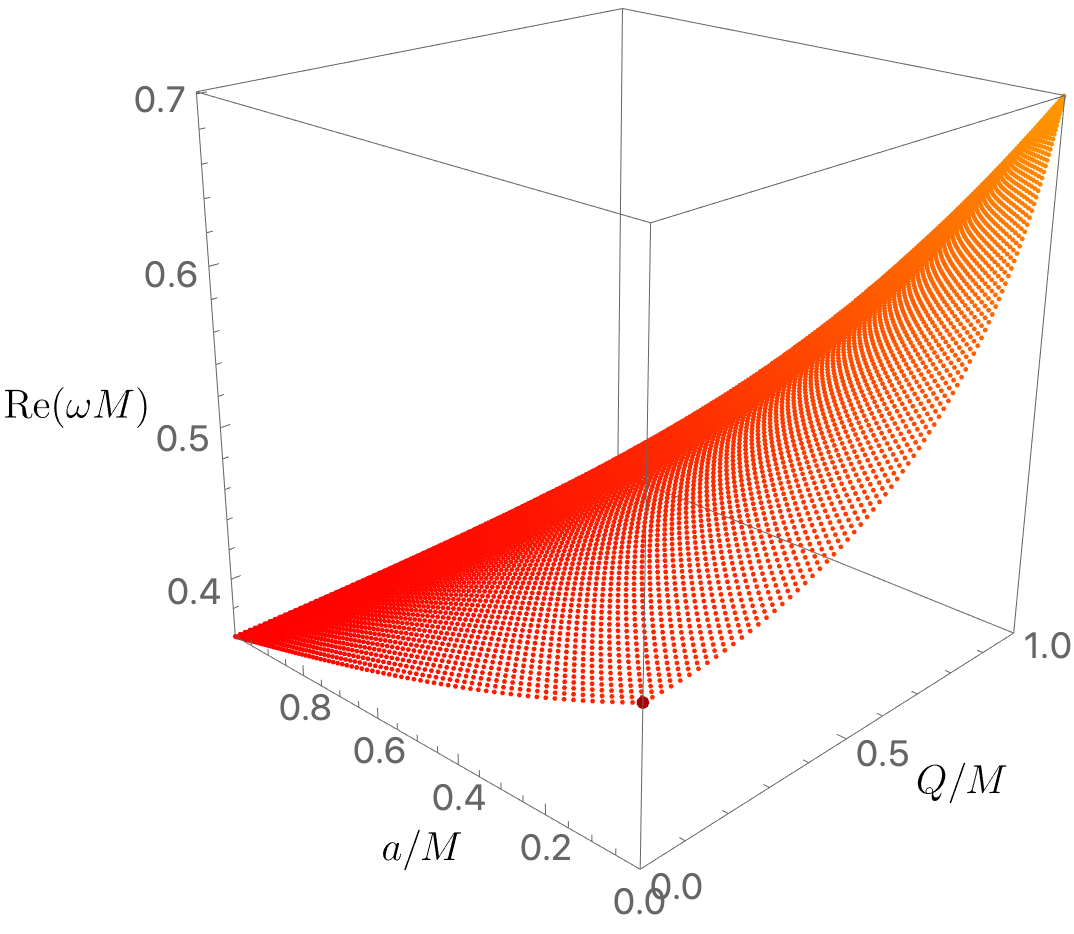}
\caption{Imaginary (left panel) and real (right panel) parts of the frequency for the $Z_1$, $(\ell,m,n)=(2,-2,0)$ KN PS QNM. The dark-red point ($a=0=Q$), $\tilde{\omega}\simeq 0.45759551 - 0.09500443\, i $, is the gravitational QNM of Schwarzschild  \cite{Chandra:1983,Leaver:1985ax}.
}
\label{Fig:Z1l2m2n0-}
\end{figure}

\begin{figure}
\includegraphics[width=.4\textwidth]{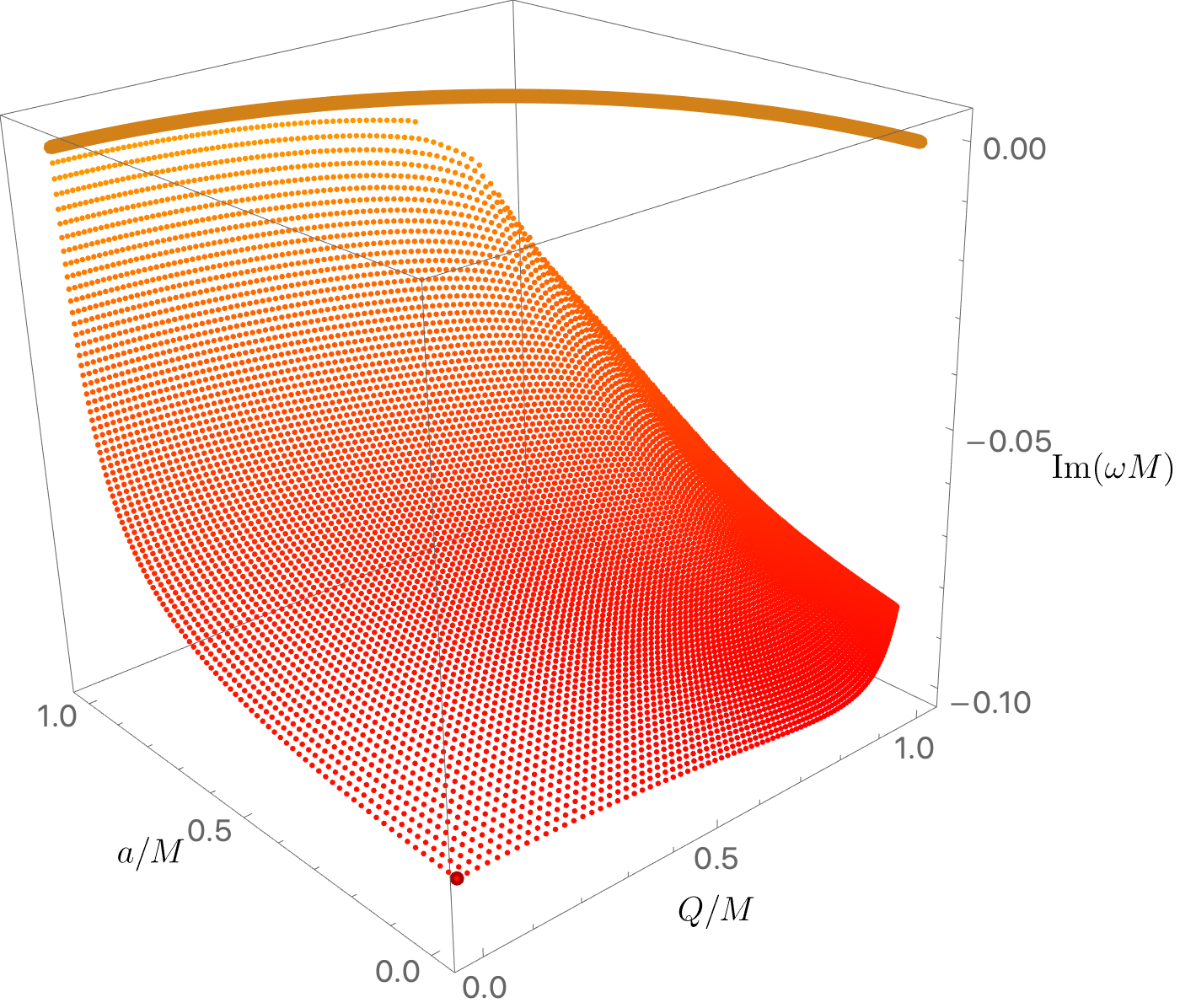}
\hspace{1.5cm}
\includegraphics[width=.38\textwidth]{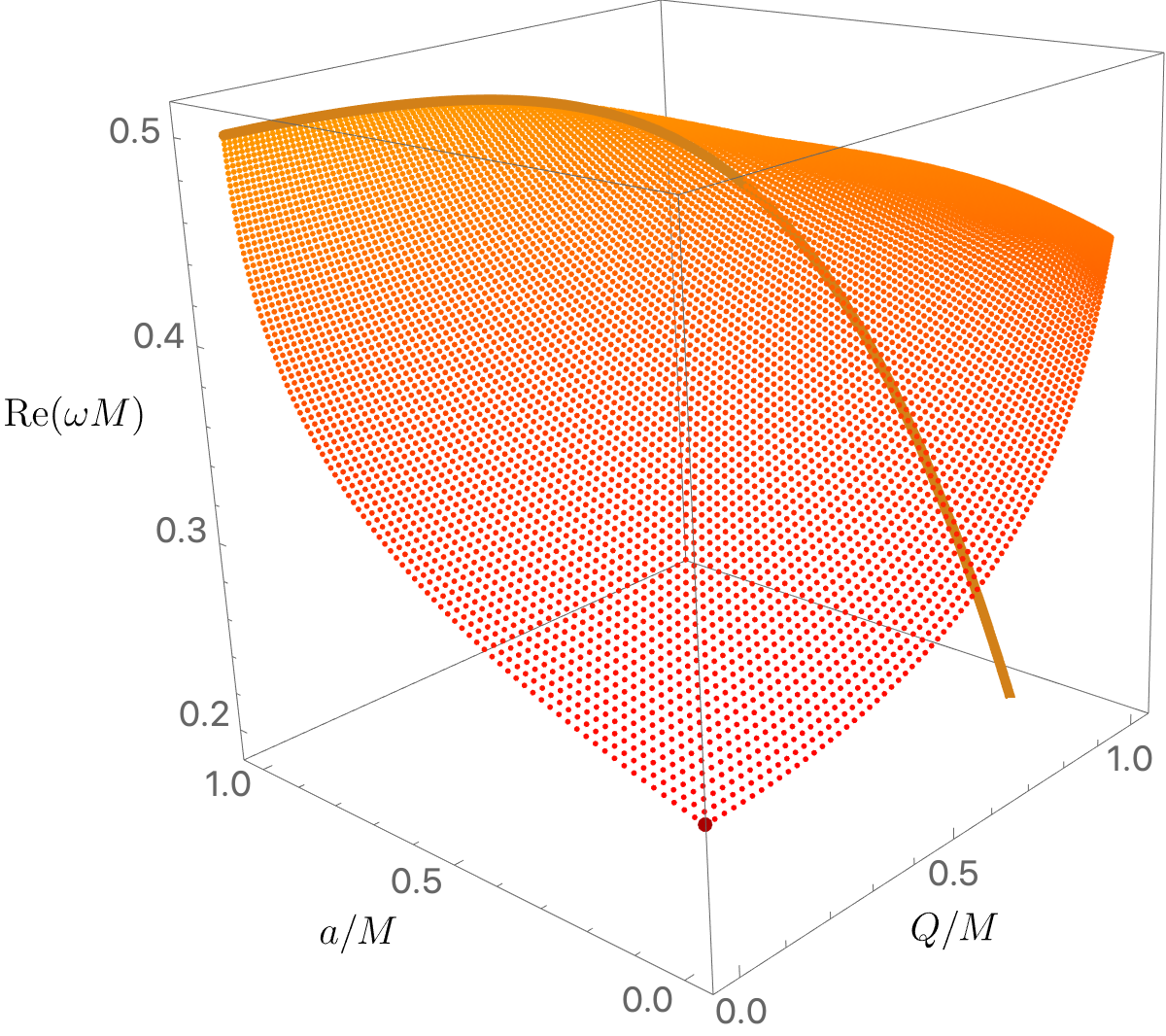}
\caption{Imaginary (left panel) and real (right panel) parts of the frequency for the $Z_1$, $(\ell,m,n)=(1,1,0)$ KN PS QNM. The dark-red point ($a=0=Q$), $\tilde{\omega}\simeq 0.24826326 - 0.09248772\, i $, is the gravitational QNM of Schwarzschild  \cite{Chandra:1983,Leaver:1985ax}.
}
\label{Fig:Z1l1m1n0+}
\end{figure}

\begin{figure}
\includegraphics[width=.4\textwidth]{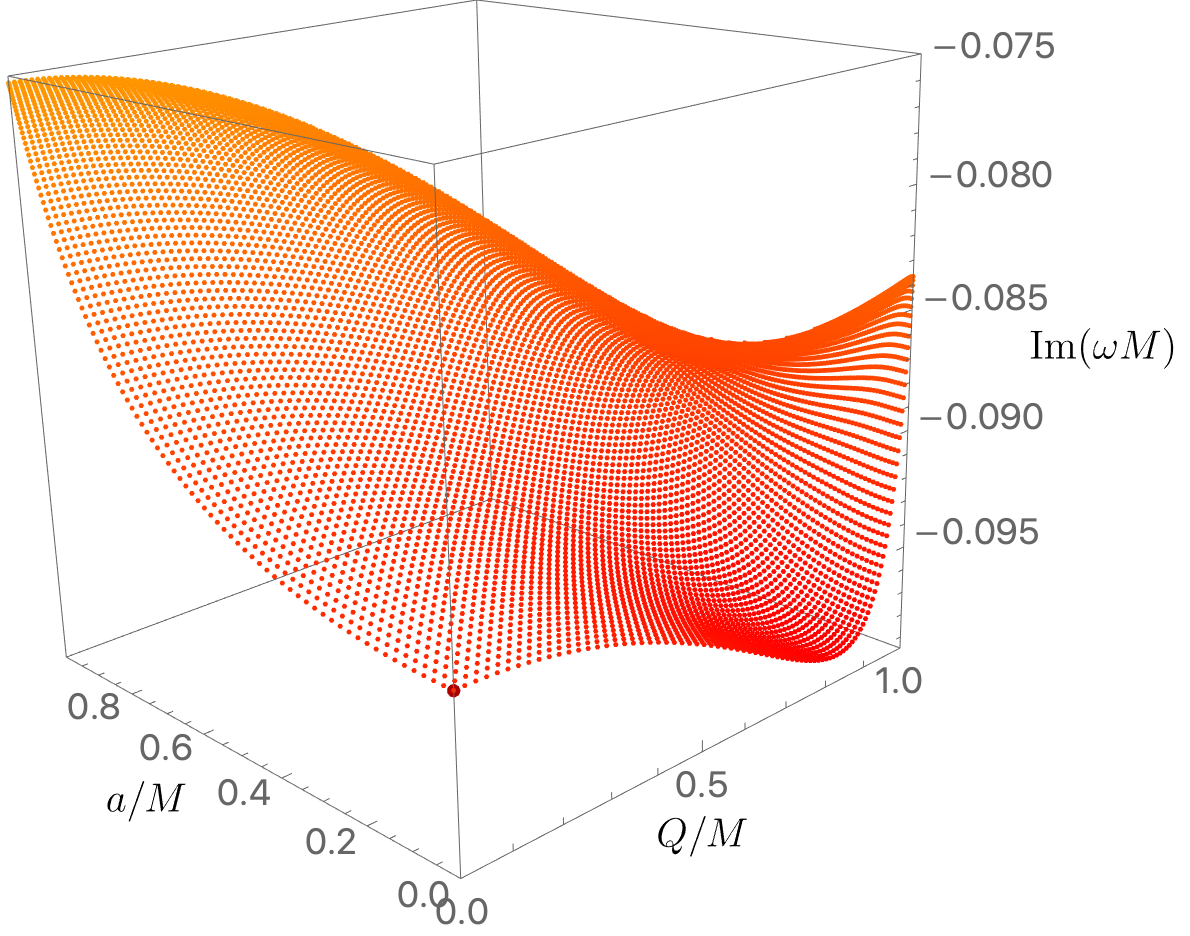}
\hspace{1.5cm}
\includegraphics[width=.38\textwidth]{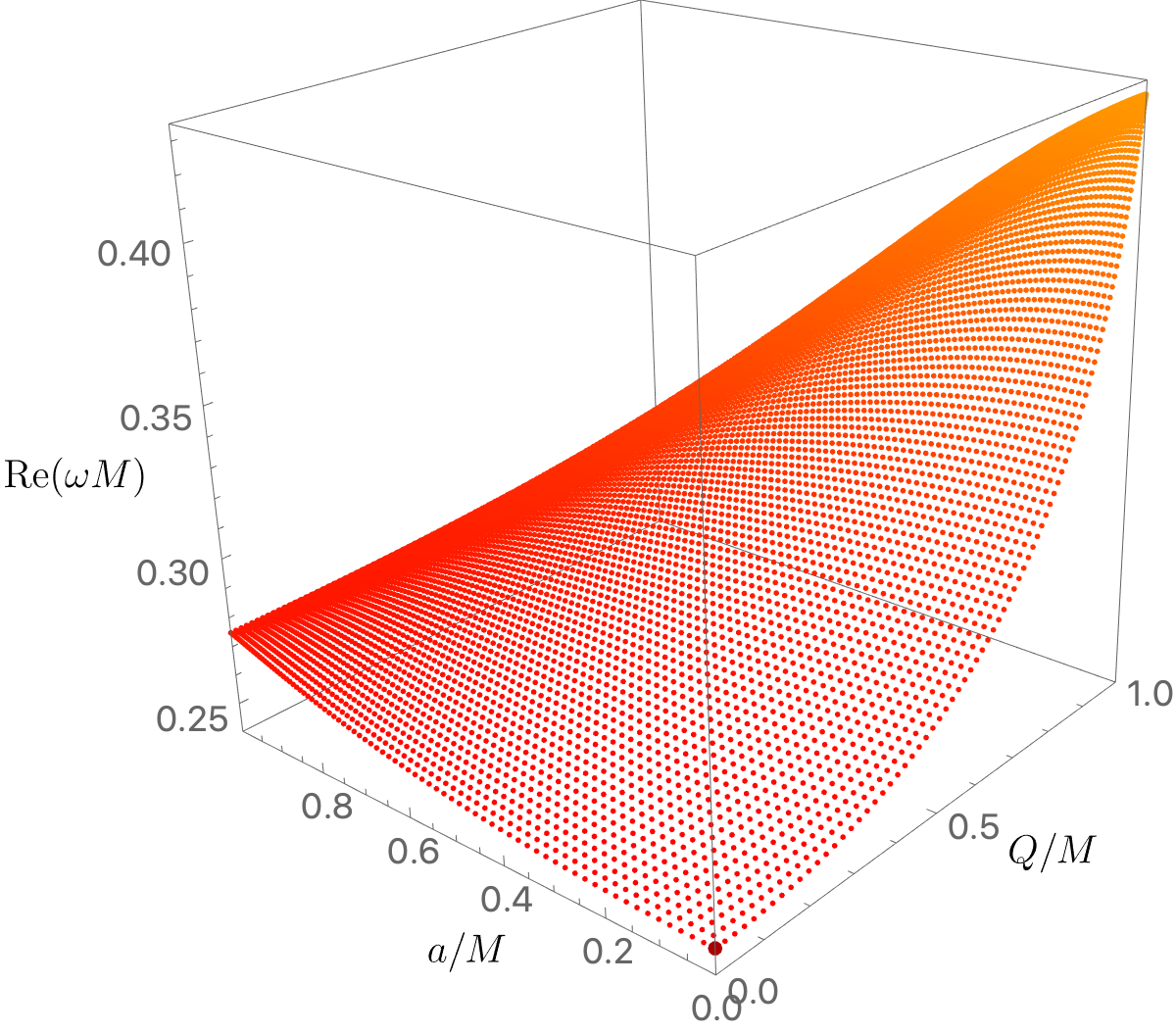}
\caption{Imaginary (left panel) and real (right panel) parts of the frequency for the $Z_1$, $(\ell,m,n)=(1,0,0)$ KN PS QNM. The dark-red point ($a=0=Q$), $\tilde{\omega}\simeq 0.24826326 - 0.09248772\, i $, is the gravitational QNM of Schwarzschild  \cite{Chandra:1983,Leaver:1985ax}.
}
\label{Fig:Z1l1m0n0}
\end{figure}

We can now consider the $Z_1$ modes which  are purely electromagnetic modes in the Schwarzschild (and Kerr) limit.\footnote{For $Z_1$ modes we do not attempt to extend our numerical data collection to large values of $\tilde{Q}$ and $\tilde{a}$ because it is computationally very costly and it does not add much to our physical discussions.} In Figs.~\ref{Fig:Z1l2m2n0+}, \ref{Fig:Z1l2m1n0+}, \ref{Fig:Z1l2m0n0}, \ref{Fig:Z1l2m1n0-} and \ref{Fig:Z1l2m2n0-}, we display the $\ell=2$ PS surfaces of this family for $m=2,1,0,-1,-2$, respectively. Moreover, in Figs.~\ref{Fig:Z1l1m1n0+}, \ref{Fig:Z1l1m0n0} and \ref{Fig:Z1l2m1n0-}, we display the $Z_1$ $\ell=1$ PS surfaces for $m=1,0,-1$, respectively. Comparing  $Z_1$ modes with the same $\{\ell,m\}$ as $Z_2$ modes, we see that the qualitative shape of the surfaces is similar but $Z_1$ modes are typically more damped than the $Z_2$ modes. Moreover, $Z_1$ modes with $\ell=m$ also approach $\mathrm{Im}\,\tilde{\omega}=0$ and $\mathrm{Re}\,\tilde{\omega}=m \tilde{\Omega}_H^{\hbox{\footnotesize ext}}$ at extremality if and only if $\hat{a}_{\hbox{\footnotesize ext}}(\hat{Q})>\hat{a}_{\star}$  (see Fig.~\ref{Fig:Z1l2m2n0+} for $\ell=m=2$ and Fig.~\ref{Fig:Z1l1m1n0+} for $\ell=m=1$) where the $\star$ point was defined in the discussion leading up to \eqref{NH:starWKB}. For  $\hat{a}_{\hbox{\footnotesize ext}}(\hat{Q})< \hat{a}_{\star}$ which occurs for $\hat{Q}_{\star} < \hat{Q}\leq 1$ this is no longer the case, very much like in the $Z_2$  $\ell=m$ discussions of \eqref{NH:starWKB} and of Figs.~\ref{Fig:WKBlambda2-m}$-$\ref{Fig:PS-extremality}. For a given $\ell=m$, the value of  $\hat{a}_{\star}$ for $Z_1$ modes tends to be higher than the one for $Z_2$ modes.

\begin{figure}
\includegraphics[width=.4\textwidth]{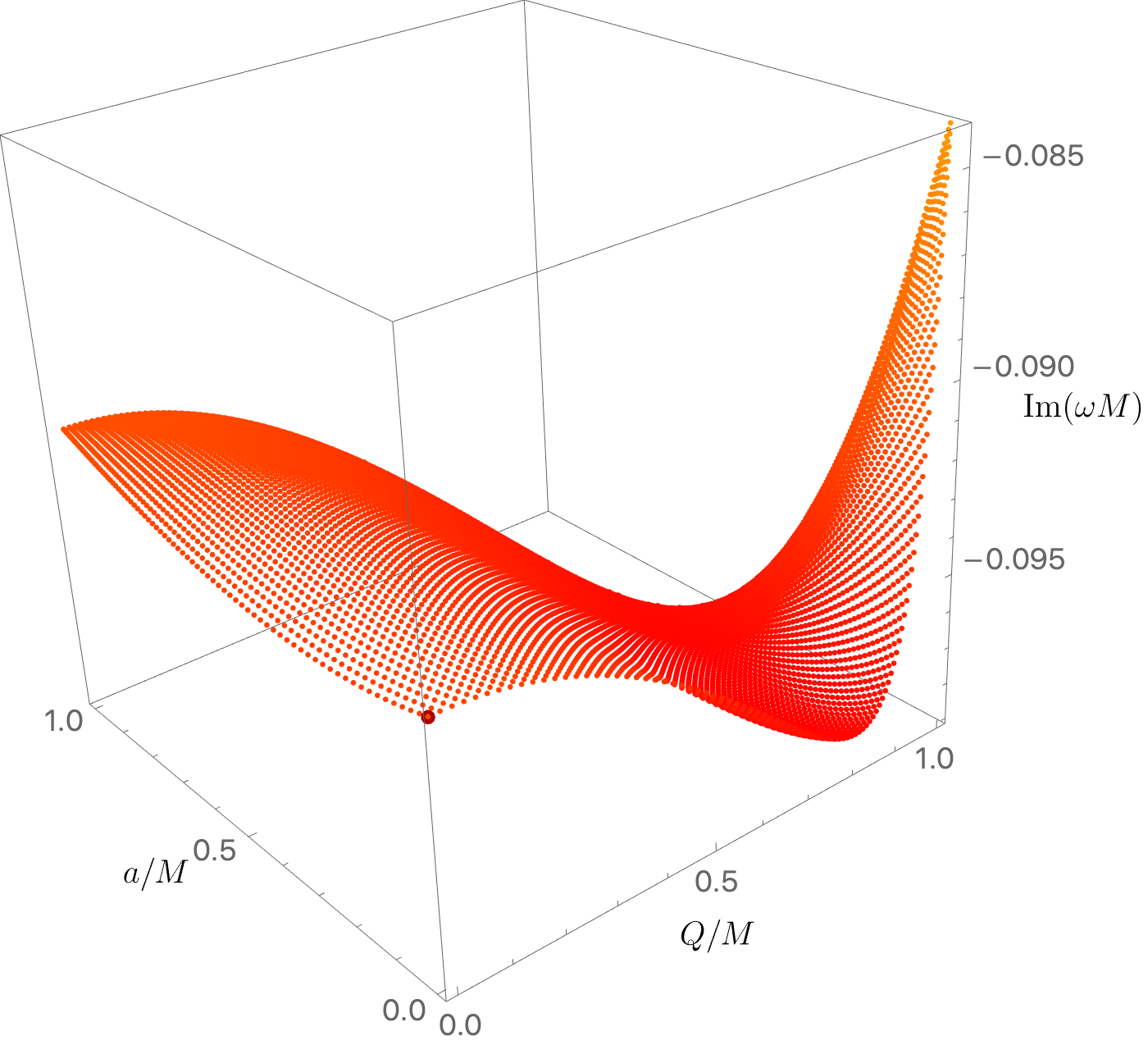}
\hspace{1.5cm}
\includegraphics[width=.38\textwidth]{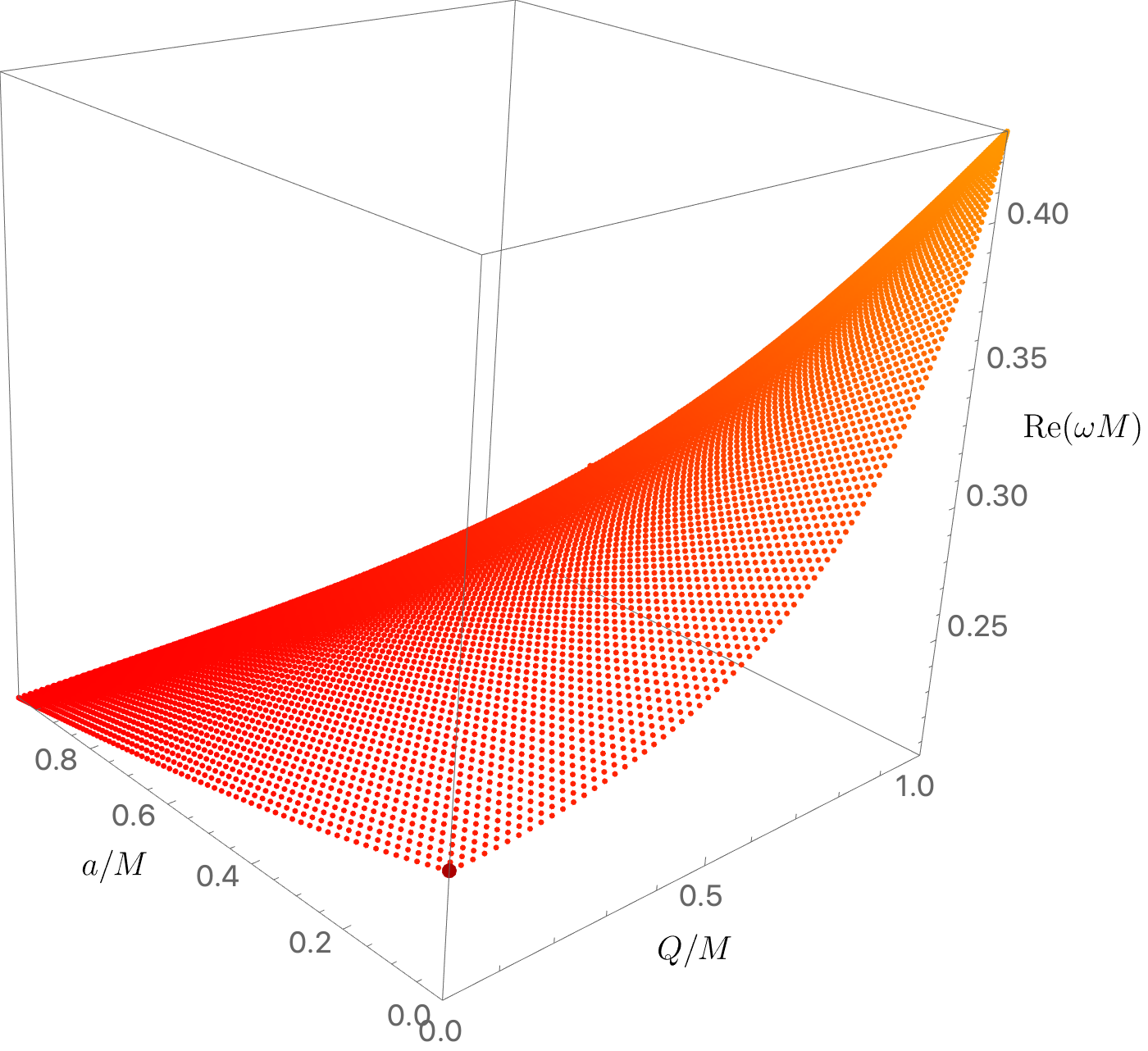}
\caption{Imaginary (left panel) and real (right panel) parts of the frequency for the $Z_1$, $(\ell,m,n)=(1,-1,0)$ KN PS QNM. The dark-red point ($a=0=Q$), $\tilde{\omega}\simeq 0.24826326 - 0.09248772\, i $, is the gravitational QNM of Schwarzschild  \cite{Chandra:1983,Leaver:1985ax}.
}
\label{Fig:Z1l1m1n0-}
\end{figure}

\smallskip

\noindent{\bf Acknowledgments.}

We warmly acknowledge Gregorio Carullo, Danny Laghi, Nathan~K.~Johnson-McDaniel and Walter Del Pozzo for our collaborations in references \cite{Dias:2021yju,AstroConstraints} that partially motivated the present study and for useful comments. The authors acknowledge the use of the cluster `Baltasar-Sete-S\'ois', and associated support services at CENTRA/IST, in the completion of this work. The authors further acknowledge the use of the IRIDIS High Performance Computing Facility, and associated support services at the University of Southampton, in the completion of this work.
 O.~C.~D.\ acknowledges financial support from the STFC ``Particle Physics Grants Panel (PPGP) 2018" Grant No.~ST/T000775/1. M.~G.\ is supported by a Royal Society University Research Fellowship. J.~E.~S.\ has been partially supported by STFC consolidated grants ST/P000681/1, ST/T000694/1. The research leading to these results has received funding from the European Research Council under the European Community's Seventh Framework Programme (FP7/2007-2013) / ERC grant agreement no. [247252]. 




\bibliographystyle{JHEP}
\bibliography{refsKN}

\end{document}